\documentclass[10pt,twoside]{amundi_article}

\usepackage[table]{xcolor}
\usepackage[latin9]{inputenc}
\usepackage{amssymb}
\usepackage{amsfonts}
\usepackage{amsmath}
\usepackage{fancybox}
\usepackage{fancyhdr}
\usepackage{graphicx}
\usepackage{subcaption}
\usepackage[hyphens]{url}
\usepackage{arydshln}
\usepackage{multirow}
\usepackage{pdflscape}
\usepackage{tikz}
\usepackage{comment}
\usepackage{nicefrac}
\usepackage{subcaption}
\usepackage{dsfont}
\usepackage{natbib}
\usepackage[colorlinks = true, linkcolor = blue, urlcolor = blue,
            citecolor = blue, anchorcolor = blue]{hyperref}
\usepackage{adjustbox}
\usepackage{array}
\usepackage{lipsum}
\usepackage{eurosym}
\usepackage{morefloats}
\usepackage{empheq}
\usepackage[most]{tcolorbox}

\newcolumntype{R}[2]{%
    >{\adjustbox{angle=#1,lap=\width-(#2)}\bgroup}%
    l%
    <{\egroup}%
}

\newlength{\figurewidth}
\newlength{\figureheight}
\def\tableskip{\vskip 10pt plus 2pt minus 2pt\relax}
\def\figureskip{\vskip 10pt plus 2pt minus 2pt\relax}

\newtheorem{remark}{Remark}

\def\limfunc#1{\mathop{\rm #1}}

\newcommand{\mr}[1]{\multirow{2}{*}{#1}}
\newcommand{\mrm}[2]{\multirow{#1}{*}{#2}}

\newcommand{\TsIII}{\hspace{3pt}}
\newcommand{\TsV}{\hspace{5pt}}

\newcommand{\TsVIII}{\hspace{8pt}}
\newcommand{\TsX}{\hspace{10pt}}
\newcommand{\TsXIII}{\hspace{13pt}}
\newcommand{\TsXV}{\hspace{15pt}}

\DeclareMathAlphabet{\mathpzc}{OT1}{pzc}{m}{it}
\newcommand{\redemption}{\ensuremath{\boldsymbol{\mathpzc{R}}}}

\newcommand{\spread}{\ensuremath{\boldsymbol{\mathpzc{s}}}}
\newcommand{\turnover}{\boldsymbol{\tau}}
\newcommand{\nshares}{\ensuremath{\boldsymbol{\mathpzc{n}}}}
\newcommand{\mcap}{\ensuremath{\boldsymbol{\mathpzc{M}}}}
\newcommand{\relative}{\ensuremath{\boldsymbol{\mathpzc{r}}}}

\newcommand{\cost}{\pmb{c}}
\newcommand{\impact}{\pmb{\pi}}
\newcommand{\Cost}{\pmb{C}}

\newcommand{\TC}{\mathcal{TC}}
\newcommand{\BAS}{\mathcal{BAS}}
\newcommand{\PI}{\mathcal{PI}}

\newcommand{\dts}{\limfunc{DTS}\nolimits}
\newcommand{\cspread}{\mathfrak{s}}

\definecolor{myblue}{rgb}{0.61,0.87,1.00}
\tcbset{highlight math style={boxsep=0mm, colframe=red, colback=myblue!50!white}}

\usetikzlibrary{positioning}
\usetikzlibrary{patterns,arrows,decorations.pathreplacing}
\usetikzlibrary{backgrounds}
\usetikzlibrary{shapes.symbols}

\newcommand{\Ns}{Normal}
\newcommand{\Ss}{Stress}

\newif\ifResearchVersion
\ResearchVersiontrue

\begin{document}

\ifResearchVersion

\title{\textbf{\color{amundi_blue}Liquidity Stress Testing in Asset Management\\Part 2. Modeling the Asset Liquidity Risk}%
\footnote{We are grateful to Ugo Girard, Charles Kalisz and Nermine Moussi for their helpful comments. This research has
also benefited from the support of Amundi Asset Management, which has
provided the data. However, the opinions expressed in this article are those
of the authors and are not meant to represent the opinions or official
positions of Amundi Asset Management.}}
\author{
{\color{amundi_dark_blue} Thierry Roncalli} \\
Quantitative Research \\
Amundi Asset Management, Paris \\
\texttt{thierry.roncalli@amundi.com} \and
{\color{amundi_dark_blue} Amina Cherief} \\
Quantitative Research \\
Amundi Asset Management, Paris \\
\texttt{amina.cherief@amundi.com} \and
{\color{amundi_dark_blue} Fatma Karray-Meziou} \\
Risk Management \\
Amundi Asset Management, Paris \\
\texttt{fatma.karraymeziou@amundi.com} \and
{\color{amundi_dark_blue} Margaux Regnault} \\
Statistics \& Economics \\
ENSAE, Paris \\
\texttt{margaux.regnault@ensae.fr}}

\date{\color{amundi_dark_blue}May 2021}

\maketitle

\begin{abstract}
This article is part of a comprehensive research project on liquidity risk
in asset management, which can be divided into three dimensions. The first
dimension covers liability liquidity risk (or funding
liquidity) modeling, the second dimension focuses on asset
liquidity risk (or market liquidity) modeling, and the third dimension considers
the asset-liability management of the liquidity gap risk (or asset-liability
matching). The purpose of this research is to propose a methodological
and practical framework in order to perform liquidity stress testing
programs, which comply with regulatory guidelines \citep{ESMA-2019,
ESMA-2020} and are useful for fund managers. The review of the academic
literature and professional research studies shows that there is a lack of
standardized and analytical models. The aim of this research project is
then to fill the gap with the goal of developing mathematical and statistical
approaches, and providing appropriate answers.\smallskip

In this second article focused on asset
liquidity risk modeling, we propose a market impact model to estimate
transaction costs. After presenting a toy model that helps to
understand the main concepts of asset liquidity, we consider a two-regime
model, which is based on the power-law property of price impact.
Then, we define several asset liquidity
measures such as liquidity cost, liquidation ratio and shortfall or
time to liquidation in order to assess the different dimensions of
asset liquidity. Finally, we apply this asset liquidity framework to
stocks and bonds and discuss the issues of calibrating the
transaction cost model.
\end{abstract}

\medskip


\clearpage

\else

\setcounter{page}{7}

\fi

\section{Introduction}

Since September 2020, the  European Securities and Markets Authority
(ESMA) has required asset managers to adopt a
liquidity stress testing (LST) policy for their investment funds
\citep{ESMA-2020}. More precisely, each asset manager must assess
the liquidity risk factors across their funds in order to ensure
that stress testing is tailored to the liquidity risk profile of
each fund. The issue of liquidity stress testing is that the
analysis should include both sides of the equation: liability (or
funding) liquidity and asset (or market) liquidity. This issue is not specific
to the asset management industry, because it is a general problem
faced by financial firms including the banking industry:
\begin{quote}
\textquotedblleft \textsl{A liquidity stress test is the process of
assessing the impact of an adverse scenario on institution's cash
flows as well as on the availability of funding sources, and on
market prices of liquid assets}\textquotedblright\ \citep[page
60]{BCBS-2017}.
\end{quote}
However, the main difference between the asset management and
banking sectors is that banks have a longer experience than asset
managers, both in the field of stress testing and liquidity
management \citep{BCBS-2013b}. Another difference is that the
methodology for computing the liquidity coverage ratio and the
monitoring tools are precise, comprehensive and very detailed by the
regulator \citep{BCBS-2013a}. This is not the case for the redemption
coverage ratio, since the regulatory text only contains guidelines
and no methodological aspects. Certainly, these differences can be
explained by the lack of maturity of this topic in the asset
management industry.\smallskip

The aim of this research is to provide a methodological
support for managing liquidity risk of investment funds. Since it is
a huge project, we have divided it into three dimensions: (1)
liability liquidity risk modeling, (2) asset
liquidity risk measurement and (3) asset-liability liquidity risk management.
This article only covers the second dimension and proposes a
framework for assessing the liquidity of a portfolio given a redemption
scenario\footnote{The liability liquidity risk is studied in
\citet{Roncalli-lst1-2020}, whereas the asset-liability management
tools are presented in \citet{Roncalli-lst3-2021}}.\smallskip

Assessing the asset liquidity risk is equivalent to measuring the
transaction cost of liquidating a portfolio. This means estimating
the bid-ask spread component, the price impact of the transaction,
the time to liquidation, the implementation shortfall, etc. This
also implies defining a liquidation policy. Contrary to the
liability liquidity risk where the academic literature is poor and
not helpful, there are many quantitative works on the aspects of
asset liquidity risk. This is particularly true for the modeling of
transaction costs, much less for liquidation policies. The challenge
is then to use the most interesting studies that are relevant from a
professional point of view, and to cast them into a practical stress
testing framework. This means simplifying and defining a few
appropriate parameters that are useful to assess the asset liquidity
risk.\smallskip

This paper is organized as follows. Section Two deals with
transaction cost modeling. A toy model will be useful to define
the concepts of price impact and liquidation policies. Then, we
consider a two-regime transaction cost model based on the power-law
property of the price impact. In Section Three, we present the asset
liquidity measures such as the liquidation ratio, the time to
liquidation or the implementation shortfall. The implementation of a
stress testing framework is developed in Section Four. In particular,
we consider an approach that distinguishes invariant parameters and
risk parameters that are impacted by a stress regime. We also
discuss the portfolio distortion that may be induced by a
liquidation policy, which does not correspond to the proportional
rule. Finally, Section Five applies the analytical framework to
stocks and bonds, and Section Six offers some concluding remarks.

\clearpage

\section{Transaction cost modeling}

In this section, we develop a transaction cost model that incorporates both
the bid-ask spread and the market impact. For that, we first define these two
concepts and explain the difference between real and nominal variables. Then,
we present a toy model that allows to understand the main characteristics of
a transaction cost function. Using the power-law property of price impact, we
derive the square-root-linear model and show how this model can be
calibrated.

\subsection{Definition}

\subsubsection{Unit transaction cost}

In what follows, we break down the unit transaction cost into two parts:
\begin{equation}
\cost\left( x\right) =\spread + \impact\left( x\right)
\label{eq:unit-cost-1}
\end{equation}%
where $\spread$ does not depend on the trade size and represents
half of the bid-ask spread of the security, and $\impact\left( x\right) $
depends on the trade size $x$ and represents the price impact (or
PI) of the trade. The trade size $x$ is an invariant variable and is
the ratio between the number of traded shares $q$ (sold or purchased) and the
daily trading volume $v$:
\begin{equation}
x=\frac{q}{v}  \label{eq:trade-size}
\end{equation}
It is also called the participation rate.

\begin{remark}
If we express the quantities in nominal terms, we have:
\begin{equation*}
x=\frac{Q}{V}=\frac{q\cdot P}{v\cdot P}=\frac{q}{v}
\end{equation*}%
where $P$ is the security price that is observed for the current date, and
$Q=q\cdot P$ and $V=v\cdot P$ are the nominal values of $q$ and $v$
(expressed in USD or EUR). In the sequel, lowercase symbols generally
represent quantities or numbers of shares whereas uppercase symbols are
reserved for nominal values. For example, the unit transaction cost
$\Cost\left( Q,V\right)$ is defined by:
\begin{equation}
\Cost\left( Q,V\right) =\cost\left( \frac{Q}{V}\right) =\cost\left( \frac{q}{v}\right)
\label{eq:unit-cost-2}
\end{equation}%
\end{remark}

\subsubsection{Total transaction cost}

The total transaction cost of the trade is the product of the unit
transaction cost and the order size expressed in dollars:%
\begin{equation}
\mathcal{TC}\left( q\right) =q\cdot P\cdot \cost\left( x\right) =
Q\cdot \cost\left( x\right)   \label{eq:total-cost-1}
\end{equation}%
where $P$ is the price of the security. Again, we can break down $\TC\left(
q\right) $ into two components:
\begin{equation}
\TC\left( q\right) = \BAS\left( q\right) + \PI\left( q\right)
\label{eq:total-cost-2}
\end{equation}%
where $\BAS\left( q\right) =Q \cdot \spread$ is the trading cost due
to the bid-ask spread and $\PI\left( q\right) = Q\cdot
\impact\left( x\right) $ is the trading cost due to the market
impact.

\begin{remark}
By construction, we have:
\begin{equation}
\mathcal{TC}\left( Q, V\right) = Q\cdot \Cost\left( Q,V\right)
\label{eq:total-cost-3}
\end{equation}%
\end{remark}

\subsubsection{Trading limit}
\label{section:trading-limit}

The previous framework only assumes that $x\geq 0$. However, this is not
realistic since we cannot trade any values of $x$ in practice. From a
theoretical point of view, we have $q\leq v$, meaning that $x\leq 1$
and $x$ is a participation rate. From a
practical point of view, $q$ is an ex-post quantity whereas $v$ is an ex-ante
quantity, implying that $x$ is a relative trading size and can be larger than one.
Nevertheless, it is highly
unlikely that the fund manager will trade a quantity larger than the ex-ante
daily trading volume. It is more likely that the asset
manager's trading policy imposes a trading limit $x^{+}$ beyond which the fund
manager cannot trade:%
\begin{equation}
0\leq x\leq x^{+}<1 \label{eq:trading-limit-1}
\end{equation}%
This is equivalent to say that the unit transaction cost becomes infinite when
the trade size is larger than the trading limit. It follows that the
unit transaction cost may be designed in the following way:%
\begin{equation}
\cost\left( x\right) =\left\{
\begin{array}{ll}
\spread+\impact\left( x\right)  & \text{if }x\in \left[ 0,x^{+}%
\right]  \\
+\infty  & \text{if }x>x^{+}%
\end{array}%
\right.   \label{eq:unit-cost-3}
\end{equation}
In this case, the concept of total transaction cost (or trading cost)
only makes sense if the trade size $x$ is lower than the trading limit
$x^{+}$. Therefore, we will see later that the trading (or liquidation) cost
must be completed by liquidation measures such as liquidation ratio or
liquidation time.

\begin{remark}
The trading limit $x^{+}$ is expressed in \%. For instance, it is generally
set at $10\%$ for equity trading desks. This means that the trader can sell any
volume up to $10\%$ of the average daily volume without any permissions.
Above the $10\%$ trading limit, the trader must inform the risk manager and
obtain authorization to execute its sell order. This trading limit
$x^{+}$ can be expressed as a maximum number of shares $q^{+}$. The advantage of this trading
policy is that it does not depend on the daily volume, which
is time-varying. Another option is to express the trading limit in nominal
terms. Let $Q^{+}$ be the nominal trading limit. We have the following
relationship:
\begin{equation}
x^{+}=\frac{q^{+}}{v}=\frac{Q^{+}}{V} \label{eq:trading-limit-2}
\end{equation}
\end{remark}

\subsection{A toy model of transaction cost}

\begin{figure}[tbph]
\centering
\caption{Simple modeling of unitary transaction costs}
\label{fig:toy1}
\figureskip
\includegraphics[width = \figurewidth, height = \figureheight]{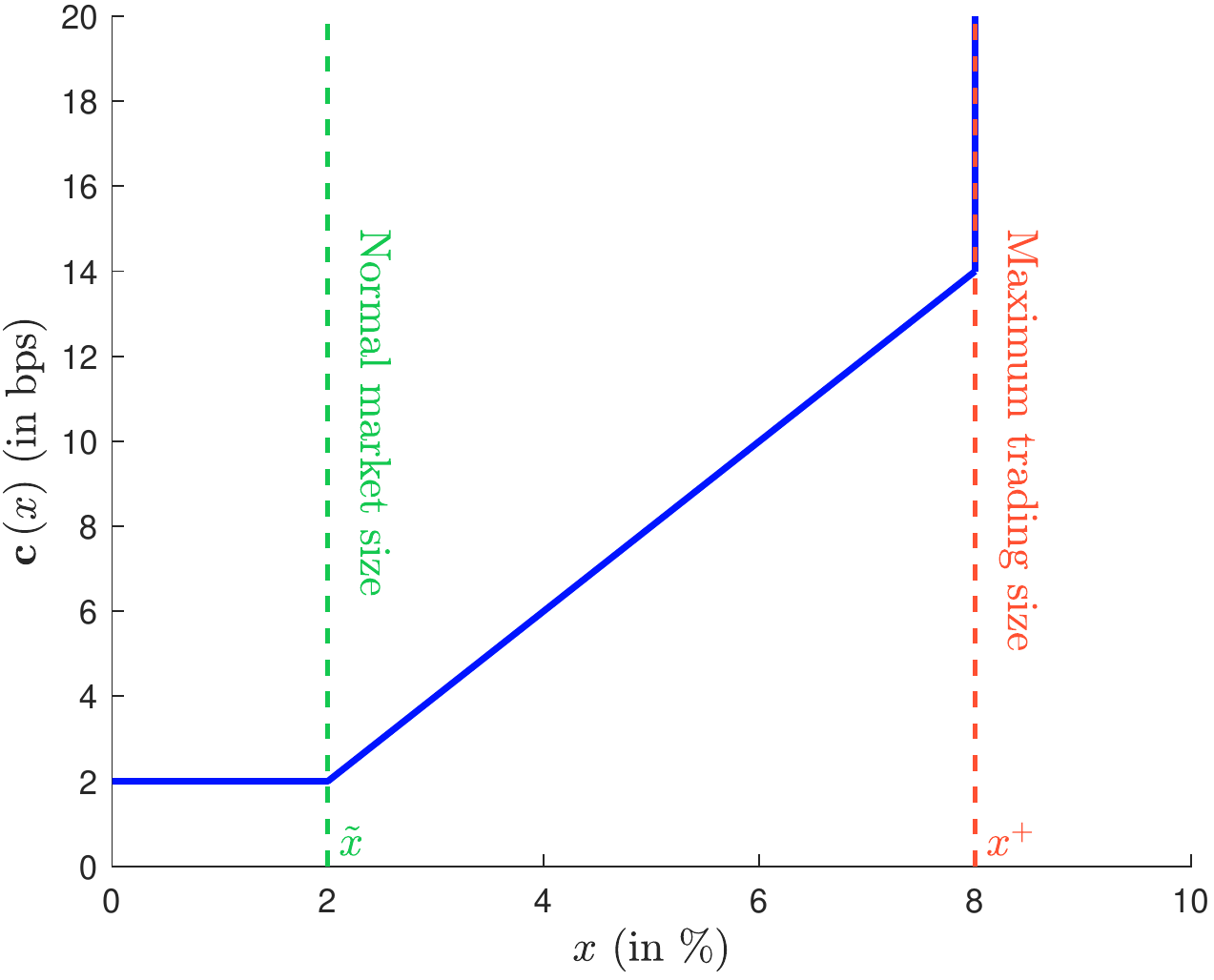}
\end{figure}

Let us consider a simple model where the unit transaction cost has the
functional form given in Figure \ref{fig:toy1}. In this toy model, we assume
that the unit transaction cost corresponds to the bid-ask spread if the selling
amount $x$ is lower than a threshold $\tilde{x}$. Beyond this normal market
size, the transaction cost includes a market impact. This market impact is
linear and is an increasing function of $x$. Moreover, we generally assume that
market impact becomes infinite if the selling amount is larger than $x^{+}$,
which is known as the maximum trading size or the trading limit. It follows
that the unit transaction cost may be parameterized by this function:
\begin{equation}
\cost^{\prime }\left( x\right) =\left\{
\begin{array}{ll}
\spread & \text{if }x\leq \tilde{x} \\
\spread+\alpha \left( x-\tilde{x}\right)  & \text{if }\tilde{x}\leq x \leq x^{+} \\
+\infty  & \text{if }x > x^{+}%
\end{array}%
\right.   \label{eq:toy1}
\end{equation}%
It depends on four parameters: the bid-ask spread $\spread$, the slope
$\alpha $ of the market impact and two thresholds: the normal size
$\tilde{x}$ and the maximum trading size $x^{+}$. For example, we obtain
Figure \ref{fig:toy1} with the following set of parameters: $\spread=2$ bps,
$\alpha =2\%$, $\tilde{x}=2\%$ and $x^{+}=8\%$. The unit transaction cost is
equal to 2 bps for small orders and reaches 14 bps when the trade size
equals to the trading limit that is equal to $8\%$.\smallskip

For each security $i$, the unit transaction cost is then
defined by the 4-tuple $\left( \spread_{i},\alpha _{i},\tilde{x}%
_{i},x_{i}^{+}\right) $ where $\spread_{i}$ is a security-specific parameter and
$\alpha _{i}$ is a model parameter. This means that $\alpha _{i}$ is
the same for all securities that belong to the same
liquidity bucket $\mathcal{LB}_{j}$. For instance, $\mathcal{LB}_{j}$
may group all large cap US stocks. $\tilde{x}_{i}$ and
$x_{i}^{+}$ may be security-specific parameters, but they are generally
considered as model parameters in order to simplify the calibration of
the unit transaction cost.\smallskip

The previous approach may be simplified by considering that the market impact
begins at $x=\tilde{x}=0$. In this case, the unit transaction cost becomes:
\begin{equation}
\cost^{\prime \prime }\left( x\right) =\left\{
\begin{array}{ll}
\spread+\alpha x & \text{if }x \leq x^{+} \\
+\infty  & \text{if }x > x^{+}%
\end{array}%
\right.   \label{eq:toy2}
\end{equation}%
The interest of this parametrization is to reduce the number of parameters
since this unit transaction cost function is then defined by the triplet $\left(
\spread_{i},\alpha _{i},x_{i}^{+}\right) $ for each security
$i$. An example is provided in Figure \ref{fig:toy2}
on page \pageref{fig:toy2}.

\begin{remark}
The parameterization $\cost^{\prime \prime }\left( x\right) $ allows us to use
the traditional mean-variance framework based on QP optimization
\citep{Chen-2019}. This explains the practitioners' great interest in the
function $\cost^{\prime \prime }\left( x\right) $ because it is highly
tractable and is compatible with the Markowitz approach with low
computational complexity\footnote{Nevertheless, this parameterization is less
frequent than the simple approach that only considers the bid-ask spread
\citep{Scherer-2007}: $\cost^{\prime \prime \prime }\left( x\right) =\spread
$.}.
\end{remark}

\begin{remark}
In Appendix \ref{appendix:relationship-toy-model} on page
\pageref{appendix:relationship-toy-model}, we show how to transform the
function $\cost^{\prime }\left( x\right) $ into the function $\cost^{\prime
\prime }\left( x\right) $, and vice versa. However, the right issue is to
estimate $\hat{\cost}^{\prime \prime }\left( x\right) $ or more precisely the
slope $\hat{\alpha}$ of the market impact. In this case, we use Equations
(\ref{eq:app-alpha1}) and (\ref{eq:app-alpha2}) on page \pageref{eq:app-alpha1}
to transform $\hat{\alpha}$ into $\alpha $ for the functions
$\cost^{\prime}\left( x\right) $ and $\cost^{\prime \prime }\left( x\right) $.
\end{remark}

\subsection{The power-law model of price impact}

\subsubsection{General formula for the market impact}

The previous trading cost model is useful for portfolio optimization, but price
impact is certainly too simple from a trading or risk management perspective.
Nevertheless, price impact has been extensively studied by
academics\footnote{See for instance the survey articles of
\citet{Bouchaud-2010} and \citet{Kyle-2018}.}, and it is now well-accepted that
market impact is power-law:
\begin{equation}
\impact\left( x\right) := \impact\left( x; \gamma\right) = \varphi _{\gamma }\sigma
x^{\gamma }  \label{eq:market-impact1}
\end{equation}
where $\gamma >0$ is a scalar, $\sigma $ is the daily volatility of the
security\footnote{The daily volatility is equal to the annualized volatility
divided by the factor $\sqrt{260}$. In the sequel, we use the symbol $\sigma$
to name both the daily and annualized volatilities. When the volatility is
used in a transaction cost formula, it corresponds to a daily volatility. In
the text, the volatility is always expressed on an annual basis.} and
$\varphi _{\gamma }$ is a scaling factor\footnote{The value of $\varphi
_{\gamma }$ depends on the value taken by the exponent $\gamma $.}. In
particular, Equation (\ref{eq:market-impact1}) is valid under a no-arbitrage
condition \citep{Jusselin-2020}. Empirical studies showed that $\gamma \in
\left[0.3,0.7\right]$. For example, the seminal paper of \citet{Loeb-1983}
has been extensively used by \citet{Torre-1997} to develop the MSCI Barra
market impact model, which considers that $\gamma = 0.5$.
\citet{Almgren-2005} concluded that $\gamma = \nicefrac{3}{5}$ is a better
figure than $\gamma = \nicefrac{1}{2}$. On the contrary, \citet{Engle-2012}
found that $\gamma \approx 0.43$ for NYSE stocks and $\gamma \approx 0.37$
for NASDAQ stocks, while \citet{Frazzini-2018} estimated that the average
exponent is equal to $0.35$ for developed equity markets. \citet{Bacry-2015}
confirmed a square root temporary impact in the daily participation and
observed a power-law pattern with an exponent between $0.5$ and $0.8$.
However, the results obtained by academics are generally valid for small
values of $x$. For instance, the median value of $x$ is equal to $0.6\%$ in
\citet{Almgren-2005}, \citet{Toth-2011} have used trades\footnote{See Figure
1 in \citet{Toth-2011}.}, which are smaller than $0.01\%$,
\citet{Zarinelli-2015} have considered a database of seven million
metaorders, implying that data with small values of $x$ dominate data with
large values of $x$, etc.\smallskip

Even though there is an academic consensus\footnote{For instance,
the square-root model is used by \citet{Garleanu-2013}, \citet{Frazzini-2018}
and \citet{Briere-2020}.} that $\gamma \approx 0.5$, this
assumption is not satisfactory from a practical point of view when we have to
sell or buy a large order ($x \gg 0.5\%$). Some academics have also exhibited
that $\gamma$ is an increasing function of $x$. For instance,
\citet{Moro-2009} found that $\gamma$ is equal to $0.64$ for LSE stocks when
there is a low fraction of market orders, but $\gamma$ is equal to $0.72$
when there is a high fraction of market orders. Similarly, \citet{Cont-2014}
estimated that $\gamma$ is equal to $1$ when we aggregate trades and consider
order flow imbalance instead of single trade sizes. \citet{Breen-2002} used a linear regression model
for estimating the price impact. We also recall that the
seminal paper of \citet{Kyle-1985} assumes that $\gamma = 1$. In fact, these
two concepts of transaction cost are not necessarily exclusive:
\begin{quotation}
\textquotedblleft \textsl{Empirically, both a linear model and a square root
model explain transaction costs well. A square-root model explains transaction
costs for orders in the 90th to 99th percentiles better than a linear model; a
linear model explains transaction costs for the largest $1\%$ of orders
slightly better than the square-root model}\textquotedblright\ \citep[page
1347]{Kyle-2016}.
\end{quotation}
This finding is shared by \citet{Boussema-2002} and \citet{Dhondt-2008}, who
observed that market impact increases significantly when trade size is greater
than $1\%$ or turnover is lower than $0.03\%$.

\begin{remark}
According to \citet{Bucci-2019}, the relationship between trade size and
market impact is close to a square-root function for intermediate trading
volumes (i.e. when $0.1\% \leq x \leq 10\%$), but shows an approximate linear
behavior for smaller trading volumes (i.e. when $0.001\% \leq x \leq 0.1\%$).
These different results demonstrate that there is no consensus on a unique
functional form for computing the price impact.
\end{remark}

\begin{figure}[tbph]
\centering
\caption{Convexity measure of the power-law model}
\label{fig:power_law1}
\figureskip
\includegraphics[width = \figurewidth, height = \figureheight]{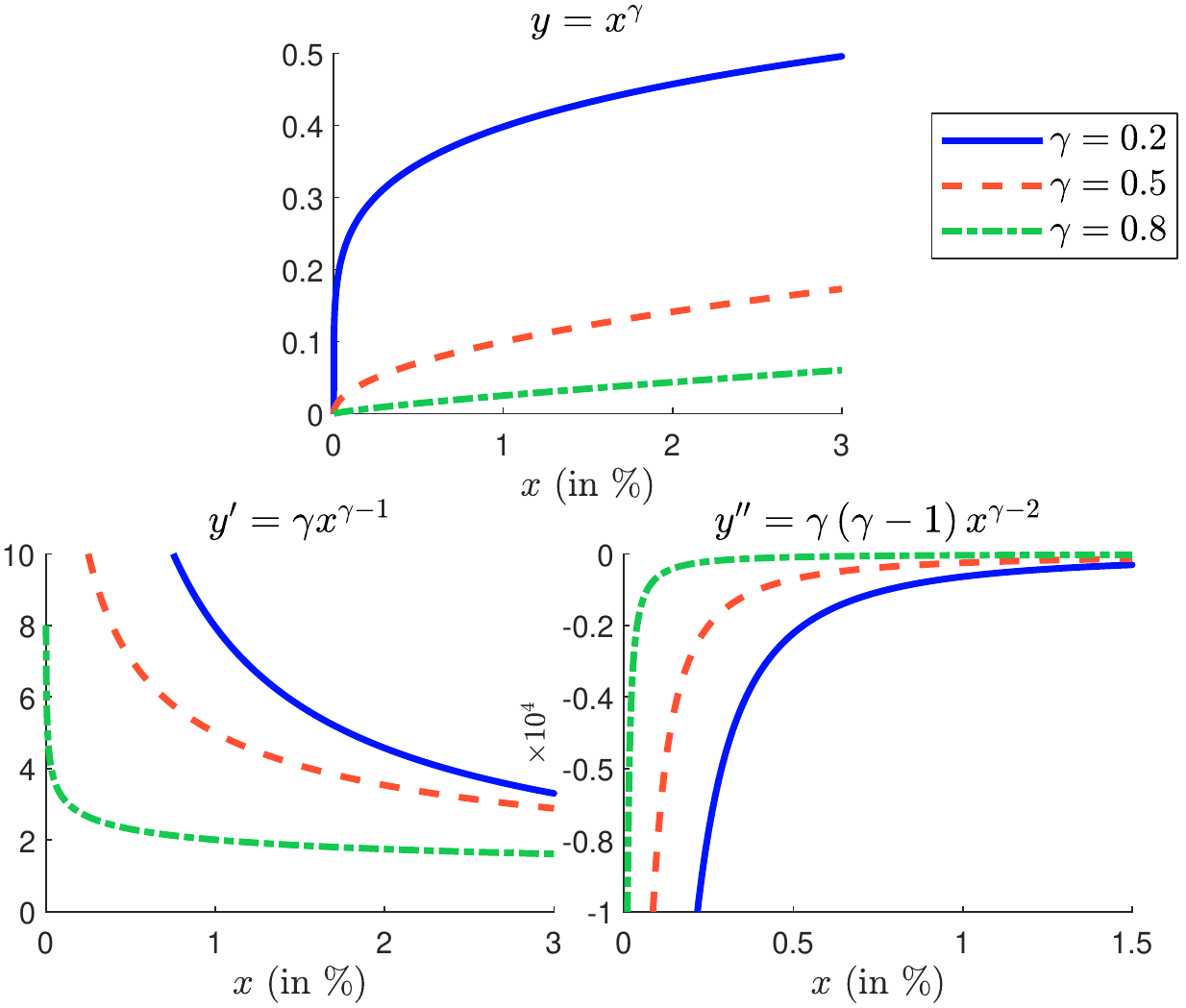}
\end{figure}

In Figure \ref{fig:power_law1}, we report the power function $y = x^{\gamma}$
and its first and second derivatives for three exponents $\gamma$.
We deduce that the concavity is larger for low values of $\gamma$ and $x$.
When $x$ is equal to $1$, the power function converges to the same value $y = 1$
whatever the value of $\gamma$. It follows that the choice of $\gamma$ primarily
impacts small trading sizes.

\subsubsection{Special cases}

From Equation (\ref{eq:market-impact1}), we deduce the two previous competing
approaches of \citet{Loeb-1983} and \citet{Kyle-1985}, and also the constant
(or bid-ask spread) model:
\begin{itemize}
\item The square-root model ($\gamma = \nicefrac{1}{2}$):
\begin{equation}
\impact\left( x;\nicefrac{1}{2}\right) \approx \varphi _{\nicefrac{1}{2}}
\sigma \sqrt{x} \label{eq:market-impact2}
\end{equation}%
Generally, we assume that the scaling factor $\varphi _{\nicefrac{1}{2}}$ is
close to one, implying that the multiplicative factor is equal to the daily
volatility.

\item The linear model ($\gamma = 1$):
\begin{equation}
\impact\left( x;1\right) \approx \varphi _{1}\sigma x
\label{eq:market-impact3}
\end{equation}%
In this case, the scaling factor $\varphi _{1}$ may be calibrated with
respect to $\varphi _{\nicefrac{1}{2}}$ by considering that the two price
impact functions coincide at a threshold $\tilde{x}$. We deduce
that\footnote{We have:
\begin{eqnarray*}
\impact\left( \tilde{x};1\right) =\impact\left( \tilde{x};\nicefrac{1}{2}%
\right)  &\Leftrightarrow &\varphi _{1}\sigma \tilde{x}%
=\varphi _{\nicefrac{1}{2}}\sigma \sqrt{\tilde{x}} \\
&\Leftrightarrow &\varphi _{1}=\frac{\varphi _{\nicefrac{1}{2}}}{\sqrt{%
\tilde{x}}}
\end{eqnarray*}%
}:%
\begin{equation}
\impact\left( x;1\right) \approx \varphi _{\nicefrac{1}{2}}\sigma
\frac{x}{\sqrt{\tilde{x}}} \label{eq:market-impact4}
\end{equation}

\item The constant model ($\gamma = 0$):
\begin{equation}
\impact\left( x;0\right) \approx \varphi _{0}\sigma
\label{eq:market-impact5}
\end{equation}%
By assuming that $\varphi _{0}=0$, we obtain the bid-ask spread model:
\begin{equation*}
\cost\left( x\right) = \spread
\end{equation*}

\end{itemize}

In Tables \ref{tab:power_law2a} and \ref{tab:power_law2b}, we have reported
the values taken by the price impact function $\impact\left( x\right)$ for
different values of the annualized volatility $\sigma$ and trade size $x$. We
assume that $\varphi _{\nicefrac{1}{2}}=1$ and $\tilde{x}=1\%$. It follows
that $\varphi _{1}=10$. Results must be read as follows: a trade size of
$0.50\%$ has a price impact of $4.4$ bps when the asset volatility is $10\%$
in the case of the square-root model, whereas the price impact becomes $3.1$
bps if we consider the linear model.\smallskip

\begin{table}[tbph]
\centering
\caption{Price impact in bps when $\gamma = \nicefrac{1}{2}$ (square-root model)}
\label{tab:power_law2a}
\begin{tabular}{cc|ccccccccc}
\hline
\multicolumn{2}{c}{$x$} & $0.01\%$ & $0.05\%$ & $0.10\%$ & $0.50\%$ & $1\%$ & $2\%$ & $5\%$ & $10\%$ & $15\%$ \\
\hline
\multirow{8}{*}{$\sigma$}
 & ${\TsV}1\%$ & $0.1$ & $0.1$ & $0.2$ & ${\TsV}0.4$ & ${\TsV}0.6$ & ${\TsV}0.9$ & ${\TsV}1.4$ & ${\TsV}2.0$ &  ${\TsX}2.4$ \\
 & ${\TsV}5\%$ & $0.3$ & $0.7$ & $1.0$ & ${\TsV}2.2$ & ${\TsV}3.1$ & ${\TsV}4.4$ & ${\TsV}6.9$ & ${\TsV}9.8$ & ${\TsV}12.0$ \\
 &      $10\%$ & $0.6$ & $1.4$ & $2.0$ & ${\TsV}4.4$ & ${\TsV}6.2$ & ${\TsV}8.8$ &      $13.9$ &      $19.6$ & ${\TsV}24.0$ \\
 &      $15\%$ & $0.9$ & $2.1$ & $2.9$ & ${\TsV}6.6$ & ${\TsV}9.3$ &      $13.2$ &      $20.8$ &      $29.4$ & ${\TsV}36.0$ \\
 &      $20\%$ & $1.2$ & $2.8$ & $3.9$ & ${\TsV}8.8$ &      $12.4$ &      $17.5$ &      $27.7$ &      $39.2$ & ${\TsV}48.0$ \\
 &      $25\%$ & $1.6$ & $3.5$ & $4.9$ &      $11.0$ &      $15.5$ &      $21.9$ &      $34.7$ &      $49.0$ & ${\TsV}60.0$ \\
 &      $30\%$ & $1.9$ & $4.2$ & $5.9$ &      $13.2$ &      $18.6$ &      $26.3$ &      $41.6$ &      $58.8$ & ${\TsV}72.1$ \\
 &      $50\%$ & $3.1$ & $6.9$ & $9.8$ &      $21.9$ &      $31.0$ &      $43.9$ &      $69.3$ &      $98.1$ &      $120.1$ \\
\hline
\end{tabular}
\end{table}

\begin{table}[tbph]
\centering
\caption{Price impact in bps when $\gamma = 1$ (linear model)}
\label{tab:power_law2b}
\begin{tabular}{cc|ccccccccc}
\hline
\multicolumn{2}{c}{$x$} & $0.01\%$ & $0.05\%$ & $0.10\%$ & $0.50\%$ & $1\%$ & $2\%$ & $5\%$ & $10\%$ & $15\%$ \\
\hline
\multirow{8}{*}{$\sigma$}
 & ${\TsV}1\%$ & $0.0$ & $0.0$ & $0.1$ & ${\TsV}0.3$ & ${\TsV}0.6$ & ${\TsV}1.2$ &  ${\TsX}3.1$ &  ${\TsX}6.2$ &  ${\TsX}9.3$ \\
 & ${\TsV}5\%$ & $0.0$ & $0.2$ & $0.3$ & ${\TsV}1.6$ & ${\TsV}3.1$ & ${\TsV}6.2$ & ${\TsV}15.5$ & ${\TsV}31.0$ & ${\TsV}46.5$ \\
 &      $10\%$ & $0.1$ & $0.3$ & $0.6$ & ${\TsV}3.1$ & ${\TsV}6.2$ &      $12.4$ & ${\TsV}31.0$ & ${\TsV}62.0$ & ${\TsV}93.0$ \\
 &      $15\%$ & $0.1$ & $0.5$ & $0.9$ & ${\TsV}4.7$ & ${\TsV}9.3$ &      $18.6$ & ${\TsV}46.5$ & ${\TsV}93.0$ &      $139.5$ \\
 &      $20\%$ & $0.1$ & $0.6$ & $1.2$ & ${\TsV}6.2$ &      $12.4$ &      $24.8$ & ${\TsV}62.0$ &      $124.0$ &      $186.1$ \\
 &      $25\%$ & $0.2$ & $0.8$ & $1.6$ & ${\TsV}7.8$ &      $15.5$ &      $31.0$ & ${\TsV}77.5$ &      $155.0$ &      $232.6$ \\
 &      $30\%$ & $0.2$ & $0.9$ & $1.9$ & ${\TsV}9.3$ &      $18.6$ &      $37.2$ & ${\TsV}93.0$ &      $186.1$ &      $279.1$ \\
 &      $50\%$ & $0.3$ & $1.6$ & $3.1$ &      $15.5$ &      $31.0$ &      $62.0$ &      $155.0$ &      $310.1$ &      $465.1$ \\
\hline
\end{tabular}
\end{table}

\begin{figure}[tbph]
\centering
\caption{Square-root model versus linear model ($\sigma = 10\%$)}
\label{fig:power_law3}
\figureskip
\includegraphics[width = \figurewidth, height = \figureheight]{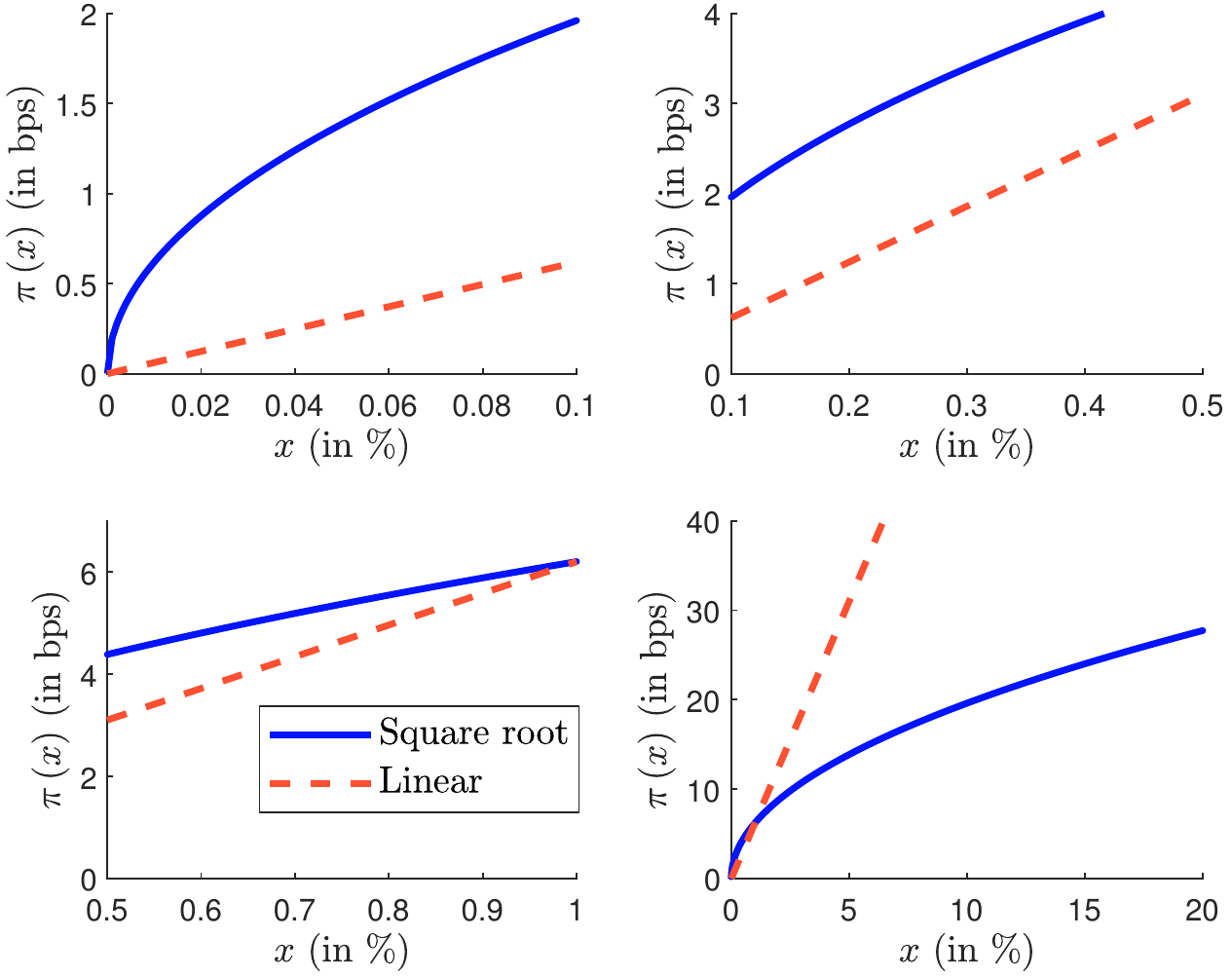}
\end{figure}

Figure \ref{fig:power_law3} shows the differences between the two models when
the annualized volatility is set to $10\%$. First, we notice that the
concavity of the square-root model is mainly located for small values of $x$,
since the trading cost function $\impact\left( x;\nicefrac{1}{2}\right)$
may be approximated by a piecewise linear function with only three or four
knots. Second, the square-root model implies higher trading costs than the
linear model when trade sizes are \textit{small}, and we verify that
$\impact\left( x;\nicefrac{1}{2}\right) \geq \impact\left( x;1\right) $
when $x\leq \tilde{x}=1\%$. For large trade sizes, it is the linear model
that produces higher trading costs compared to the square-root model%
\footnote{This large difference between square-root and linear models has been already
observed by \citet{Frazzini-2018}.}:
$\impact\left( x;1\right) \gg \impact\left( x;\nicefrac{1}{2}\right)$.

\subsection{A two-regime transaction cost model}
\label{section:two-regime}

\subsubsection{General formula}

In the toy model, we distinguish two market impact regimes. The first one
corresponds to small trading sizes --- $x\in \left[ 0,\tilde{x}\right] $,
which generate a low price impact. In the second regime, trading sizes are
larger --- $x\in \left[ \tilde{x},x^{+}\right] $, and the price impact has a
significant contribution to the transaction cost. The research studies on the
power-law model also show that there may be several regimes of market impact
depending on the value of $\gamma $. Therefore, we can generalize the
toy model where the two regimes correspond to two power functions:%
\begin{equation}
\impact\left( x\right) =\left\{
\begin{array}{ll}
\varphi _{1}\sigma x^{\gamma _{1}} & \text{if }x\leq
\tilde{x} \\
\varphi _{2} \sigma x^{\gamma _{2}} & \text{if }\tilde{x}%
\leq x\leq x^{+} \\
+\infty  & \text{if }x>x^{+}%
\end{array}%
\right.   \label{eq:two-regime1}
\end{equation}%
where $\gamma _{1}$ and $\gamma _{2}$ are the exponents of the two market
impact regimes. Moreover, the scalars $\varphi _{1}$ and $\varphi _{2}$ are
related since the cost function $\impact\left( x\right) $ is continuous.
This implies that $\varphi _{2}=\varphi _{1}\tilde{x}^{\gamma _{1}-\gamma
_{2}}$. In this case, the price impact model is defined by the 5-tuple $%
\left( \varphi _{1},\gamma _{1},\gamma _{2},\tilde{x},x^{+}\right) $ since $%
\varphi _{2}$ is computed from these parameters. An alternative approach is
to define the model by the parameter set $\left( \gamma _{1},\gamma _{2},%
\tilde{x},x^{+},\impact\left( \tilde{x}\right) \right) $. Here, we fix the
market impact at the inflection point, and we have $\varphi _{1}= \sigma
^{-1}\tilde{x}^{-\gamma _{1}}\cdot \impact\left( \tilde{x}\right) $ and
$\varphi _{2}=\varphi _{1}\tilde{x}^{\gamma _{1}-\gamma _{2}}$.

\begin{remark}
Another parameterization of the two-regime model may be:%
\begin{equation}
\impact\left( x\right) =\left\{
\begin{array}{ll}
\varphi _{1}\sigma x^{\gamma _{1}} & \text{if }x\leq
\tilde{x} \\
\impact\left( \tilde{x}\right) +\varphi _{2}\sigma
\left( x-\tilde{x}\right) ^{\gamma _{2}} & \text{if }\tilde{x}<x\leq x^{+}
\\
+\infty  & \text{if }x>x^{+}%
\end{array}%
\right.   \label{eq:two-regime2}
\end{equation}%
where $\impact\left( \tilde{x}\right) =\varphi _{1}\sigma
\tilde{x}^{\gamma _{1}}$. This model is defined by the parameter set $%
\left( \varphi _{1},\gamma _{1},\varphi _{2},\gamma _{2},\tilde{x}%
,x^{+}\right) $.
\end{remark}

\begin{remark}
The model of \citet{Bucci-2019} is obtained with the two parameterizations by
setting $\gamma _{1}=1$, $\tilde{x}=0.1\%$ and $\gamma _{2}=\nicefrac{1}{2}$.
\end{remark}

\begin{figure}[tbph]
\centering
\caption{Two-regime model (annualized volatility $\sigma = 10\%$)}
\label{fig:two_regime1}
\figureskip
\includegraphics[width = \figurewidth, height = \figureheight]{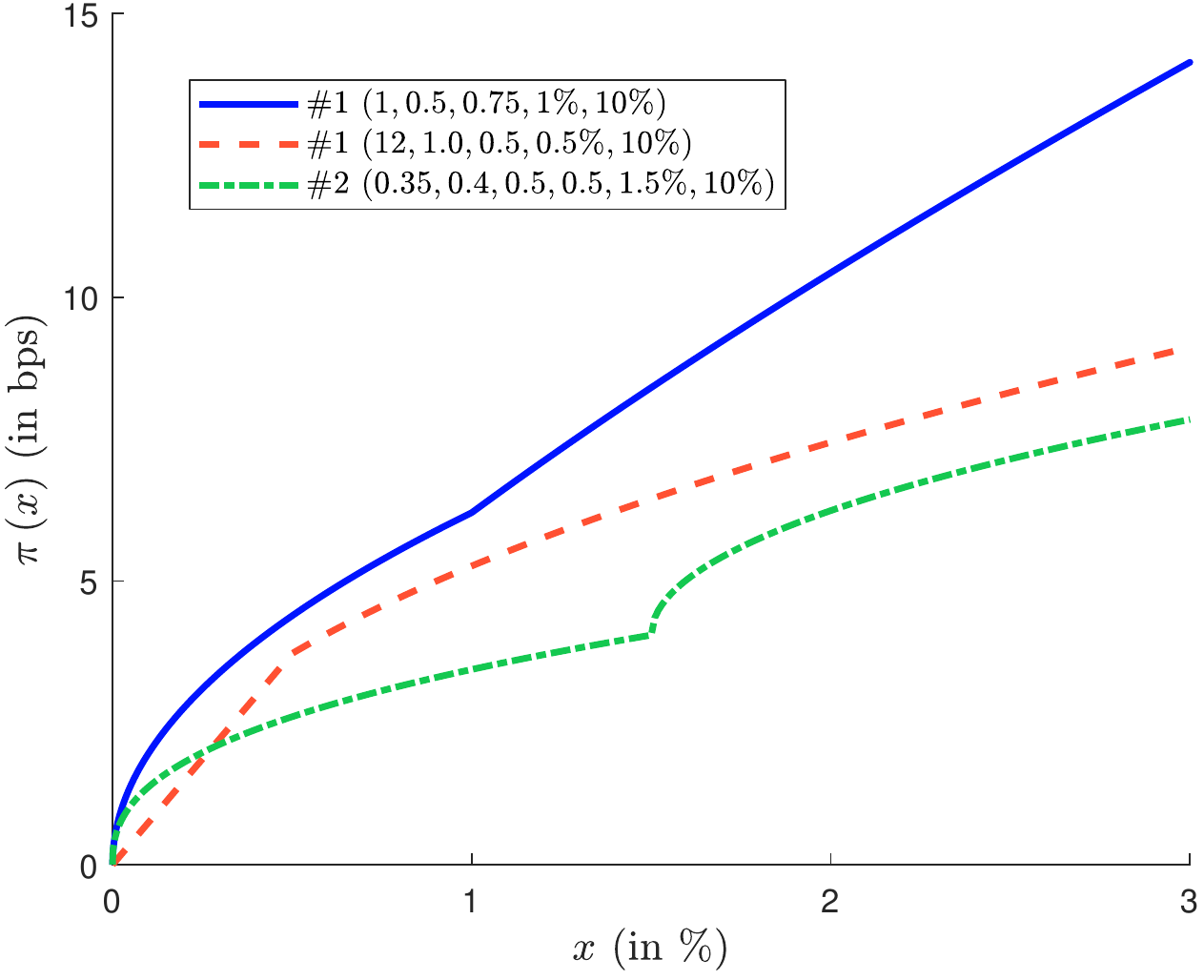}
\end{figure}

In Figure \ref{fig:two_regime1}, we report three examples of the two-regime
model. The first two examples correspond to the first parameterization,
whereas the last example uses the second parameterization. In this last case,
we observe a step effect due to the high concavity\footnote{This step effect
has been illustrated in Figure \ref{fig:power_law1} on page \pageref{fig:power_law1}.} applied
to the small values of $x-\tilde{x}$. Therefore, it is better to use
the first parameterization.

\begin{figure}[tbph]
\centering
\caption{Two-regime model ($\sigma = 10\%, \varphi_1 = 1$)}
\label{fig:two_regime2}
\figureskip
\includegraphics[width = \figurewidth, height = \figureheight]{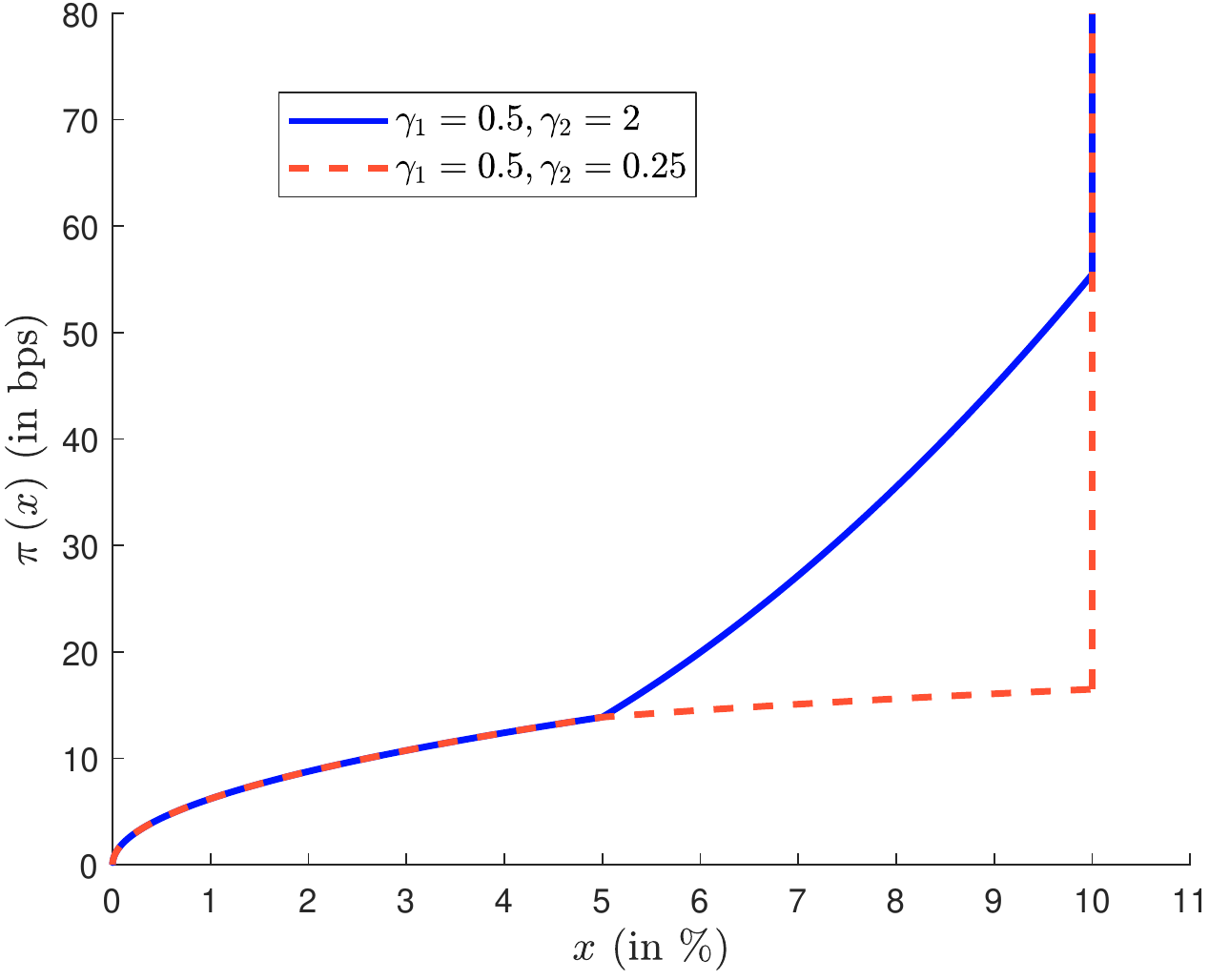}
\end{figure}

\begin{remark}
One of the questions which emerges with the calibration of the two-regime
model is the effective difference between the two regimes. In particular, we
have the choice between $\gamma_1 > \gamma_2$ and $\gamma_2 > \gamma_1$. In
other words, we have the choice to decrease or increase the convexity beyond
the inflection point $\tilde{x}$. The \textquotedblleft \textit{small size
effect}\textquotedblright\ described by \citet{Bucci-2019} is not really an
issue, because the impact is so small. Indeed, the order of magnitude of
the price impact for $x \leq 0.1\%$ is one or two basis points in the
power-law model\footnote{For instance, we have $\impact\left(0.01\%\right) =
0.62$ bps and $\impact\left(0.1\%\right) = 1.92$ bps when $\sigma = 10\%$ and
$\gamma = 0.5$.}. The significant issue is more to have a coherent approach
when the trading size is close to the trading limit $x^{+}$. An example is
provided in Figure \ref{fig:two_regime2} when the annualized volatility
$\sigma$  is $10\%$ and $\varphi_1 = 1$. We recall that
$\impact\left(x\right) = \infty$ when $x
> x^{+}$ because of the order execution policy imposed by the asset manager.
Therefore, it is obvious that the right choice is $\gamma_2 > \gamma_1$,
implying that the convexity must increase. Otherwise, it is not consistent to
impose a low convexity below $x^{+}$ and an infinite convexity beyond
$x^{+}$.
\end{remark}

\subsubsection{The square-root-linear model}

\begin{figure}[tbph]
\centering
\caption{Square-root-linear model ($\sigma = 10\%$)}
\label{fig:sqrl1}
\figureskip
\includegraphics[width = \figurewidth, height = \figureheight]{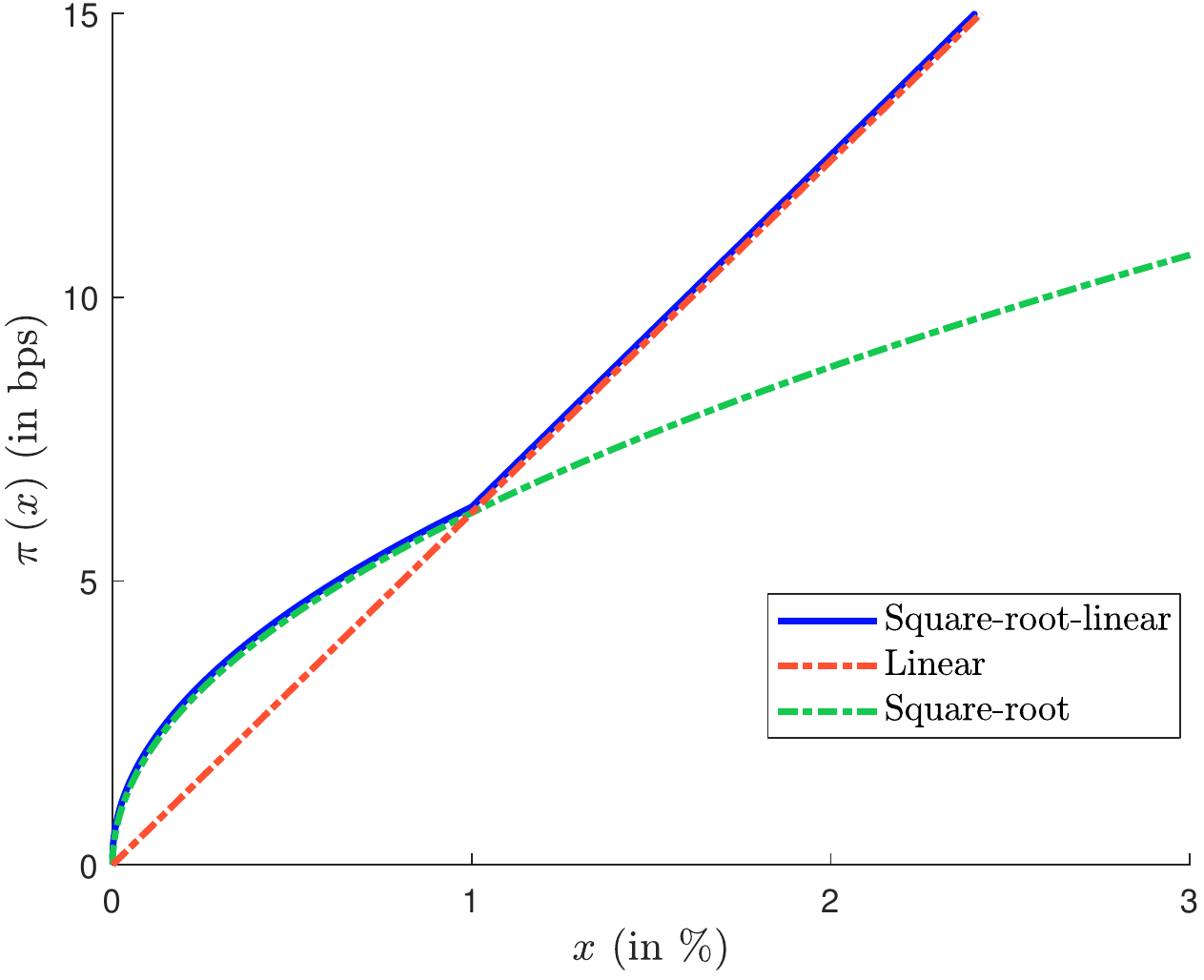}
\end{figure}

From the two-regime model, we can define the square-root-linear (SQRL)
model which has been suggested by \citet{Kyle-2016}:
\begin{equation}
\impact\left( x\right) =\left\{
\begin{array}{ll}
\varphi_{1}\sigma \sqrt{x} & \text{if }%
x\leq \tilde{x} \\
\varphi_{1} \sigma \dfrac{x}{\sqrt{\tilde{x}}%
} & \text{if }\tilde{x}\leq x\leq x^{+} \\
+\infty  & \text{if }x>x^{+}%
\end{array}%
\right.   \label{eq:sqrl1}
\end{equation}%
In this case, we assume that the square-root model is valid for
\textit{small} trade sizes ($x\leq \tilde{x}$), whereas the linear model is
better for \textit{large} trade sizes ($\tilde{x}\leq \tilde{x}\leq x^{+}$).
However, beyond the threshold value $x^{+}$, we consider that trading costs
are prohibitive and infinite. As for the toy model, the value $x^{+}$ may be
interpreted as a trading limit. We have represented the SQRL model in Figure
\ref{fig:sqrl1} for the previous parameters ($\sigma = 10\%$ and $\varphi_1=
1$) when the inflection point $\tilde{x}$ is equal to $1\%$.\smallskip

In Table \ref{tab:sqrl2}, we report the price impact of this model for
several values of the annualized volatility $\sigma$. We can compare these
figures with those given in Tables \ref{tab:power_law2a} and
\ref{tab:power_law2b} on page \pageref{tab:power_law2a}. Let us consider the
case when the volatility is equal to $20\%$, which corresponds to the typical
volatility observed for single stocks. We observe that there is an
acceleration of the price impact beyond the inflection point. For instance,
the price impact is equal to $24.8$ bps for $x=2\%$, $62.0$ bps for $x=5\%$,
etc.

\begin{table}[tbph]
\centering
\caption{Price impact in bps (square-root-linear model)}
\label{tab:sqrl2}
\begin{tabular}{cc|ccccccccc}
\hline
\multicolumn{2}{c}{$x$} & $0.01\%$ & $0.05\%$ & $0.10\%$ & $0.50\%$ & $1\%$ & $2\%$ & $5\%$ & $10\%$ & $15\%$ \\
\hline
\multirow{8}{*}{$\sigma$}
 & ${\TsV}1\%$ & $0.1$ & $0.1$ & $0.2$ & ${\TsV}0.4$ & ${\TsV}0.6$ & ${\TsV}1.2$ &  ${\TsX}3.1$ &  ${\TsX}6.2$ &  ${\TsX}9.3$ \\
 & ${\TsV}5\%$ & $0.3$ & $0.7$ & $1.0$ & ${\TsV}2.2$ & ${\TsV}3.1$ & ${\TsV}6.2$ & ${\TsV}15.5$ & ${\TsV}31.0$ & ${\TsV}46.5$ \\
 &      $10\%$ & $0.6$ & $1.4$ & $2.0$ & ${\TsV}4.4$ & ${\TsV}6.2$ &      $12.4$ & ${\TsV}31.0$ & ${\TsV}62.0$ & ${\TsV}93.0$ \\
 &      $15\%$ & $0.9$ & $2.1$ & $2.9$ & ${\TsV}6.6$ & ${\TsV}9.3$ &      $18.6$ & ${\TsV}46.5$ & ${\TsV}93.0$ &      $139.5$ \\
 &      $20\%$ & $1.2$ & $2.8$ & $3.9$ & ${\TsV}8.8$ &      $12.4$ &      $24.8$ & ${\TsV}62.0$ &      $124.0$ &      $186.1$ \\
 &      $25\%$ & $1.6$ & $3.5$ & $4.9$ &      $11.0$ &      $15.5$ &      $31.0$ & ${\TsV}77.5$ &      $155.0$ &      $232.6$ \\
 &      $30\%$ & $1.9$ & $4.2$ & $5.9$ &      $13.2$ &      $18.6$ &      $37.2$ & ${\TsV}93.0$ &      $186.1$ &      $279.1$ \\
 &      $50\%$ & $3.1$ & $6.9$ & $9.8$ &      $21.9$ &      $31.0$ &      $62.0$ &      $155.0$ &      $310.1$ &      $465.1$ \\
\hline
\end{tabular}
\end{table}

\begin{remark}
The SQRL model and more generally the two-regime model can
be used as an incentive trading model, since
trades are penalized when they are larger than $\tilde{x}$.
In this case, $x^{+}$ is a hard threshold limit
while $\tilde{x}$ can be considered as a soft threshold limit.
Indeed, the asset manager does not explicitly
prohibit the fund manager from trading between $\tilde{x}$ and $x^{+}$,
but he is clearly not encouraged to trade, because
the transaction costs are high\footnote{For instance, the price impact is
equal to $35.1$ bps for $x=2\%$ and $138.7$ bps for $x=5\%$ when we use
a two-regime model with the following parameters: $\sigma=20\%$, $\varphi_{1}= 1$,
$\gamma_1 = \nicefrac{1}{2}$, $\gamma_2 = \nicefrac{3}{2}$ and $\tilde{x} = 1\%$.}.
This is particularly true if the asset manager has a centralized trading desk
and ex-ante trading costs are charged to the fund manager.
\end{remark}

\section{Asset liquidity measures}

Since liquidity is a multi-faceted concept, we must use several measures in
order to encompass the different dimensions \citep[page 347]{Roncalli-2020}.
If we focus on asset liquidity, we generally distinguish two types of
measurement. The first category assesses the liquidity risk profile and
includes the liquidation ratio, the time to liquidation and the liquidation
shortfall. The second category concerns liquidity costs such as transaction
costs and effective costs. The main difference between the two categories is
that the first one focuses on the volume, while the second one mixes both
volume and price dimensions.

\subsection{Redemption scenario}

In \citet{Roncalli-lst1-2020}, we have developed several methods and
tools in order to define a redemption shock $\redemption$ for a given
investment fund. This redemption shock is expressed as a percentage
of the fund's total net asset $\limfunc{TNA}$. Therefore, we
can deduce the stress liability outflow:
\begin{equation*}
\mathcal{F}^{-}\left( t\right) := \mathbb{R} = \redemption\cdot \limfunc{TNA}\left(
t\right)
\end{equation*}%
The asset structure of the fund is given by the vector $\omega =\left(
\omega _{1},\ldots ,\omega _{n}\right) $ where $\omega _{i}$ is the
number of shares of security $i$ and $n$ is the number of assets that make up the asset
portfolio. By construction, we have:
\begin{equation*}
\limfunc{TNA}\left( t\right) =\sum_{i=1}^{n}\omega _{i}\cdot P_{i}\left(
t\right)
\end{equation*}%
where $P_{i}\left( t\right) $ is the current price of security $i$. The
redemption shock $\redemption$ must be translated into the redemption
scenario $q=\left( q_{1},\ldots ,q_{n}\right) $, where $q_{i}$ is
the number of shares of security $i$ that must be sold. After the sell
order, we must have the following equality\footnote{We notice that
the dollar value of the redemption is equal to
$\sum_{i=1}^{n}q_{i}\cdot P_{i}\left(t\right)$.}:
\begin{equation}
\limfunc{TNA}\left( t^{+}\right) := \limfunc{TNA}\left( t\right)
-\mathcal{F}^{-}\left( t\right) = \sum_{i=1}^{n}\left( \omega
_{i}-q_{i}\right) \cdot P_{i}\left( t\right)
\label{eq:redemption-scenario1}
\end{equation}
where $t^{+}$ means $t+\mathrm{d}t$ and $\mathrm{d}t$ is a small time step.
Generally, we assume that the portfolio composition remains the same,
meaning that:%
\begin{equation*}
\frac{q_{i}\cdot P_{i}\left( t\right) }{q_{j}\cdot P_{j}\left( t\right) }=%
\frac{\omega _{i}\cdot P_{i}\left( t\right) }{\omega _{j}\cdot P_{j}\left(
t\right) }
\end{equation*}%
It follows that the solution is simple and is equal to the proportional rule:%
\begin{equation}
q_{i}=\redemption\cdot \omega _{i}  \label{eq:redemption-scenario2}
\end{equation}%
It is called the vertical slicing approach (or pro-rata liquidation).
Nevertheless, since $\omega _{i}-q_{i}$ must be a natural number, $q_{i}$
must also be a natural number. Therefore, due to round-off errors, the final
redemption shock may not match the proportional rule.

\begin{remark}
In Section \ref{section:portfolio-distortion} on page
\pageref{section:portfolio-distortion}, we discuss
the construction of the redemption scenario in more detail, in particular how to
manage the distortion of the portfolio allocation weights.
\end{remark}

\subsection{Liquidity risk profile}

We first consider volume-related liquidity measures. One of the most popular
measures is the liquidation ratio $\mathcal{LR}\left( q;h\right) $, which
measures the proportion of a portfolio $q$ that can be liquidated after $h$
trading days. This statistic depends on the size of each exposure $q_{i}$
and the liquidation policy, which is defined by the trading limit $q_{i}^{+}$.
Another interesting statistic is the liquidation time (or time to
liquidation) $\mathcal{LT}\left( q;p\right) $, which is the inverse function
of the liquidity ratio. It indicates the number of required trading days in
order to liquidate a proportion $p$ of the portfolio.

\subsubsection{Liquidation ratio}

For each security that makes up the portfolio, we recall that $q_{i}^{+}$
denotes the maximum number of shares that can be sold during a trading day
for the asset $i$. The number of shares $q_{i}\left( h\right) $ liquidated
after $h$ trading days is defined as follows:
\begin{equation}
q_{i}\left( h\right) =\min \left( \left( q_{i}-\sum_{k=0}^{h-1}q_{i}\left(
k\right) \right) ^{+},q_{i}^{+}\right)   \label{eq:liq-ratio1}
\end{equation}%
where $q_{i}\left( 0\right) =0$. The liquidation ratio $\mathcal{LR}\left(
q;h\right) $ is then the proportion of the redemption scenario $q$ that is
liquidated after $h$ trading days:
\begin{equation}
\mathcal{LR}\left( q;h\right) =\frac{\sum_{i=1}^{n}\sum_{k=1}^{h}q_{i}\left(
k\right) \cdot P_{i}}{\sum_{i=1}^{n}q_{i}\cdot P_{i}}  \label{eq:liq-ratio2}
\end{equation}%
By definition, $\mathcal{LR}\left( q;h\right) $ is between $0$ and $1$. For
instance, $\mathcal{LR}\left( q;1\right) =50\%$ means that we can fulfill $%
50\%$ of the redemption on the first trading day, $\mathcal{LR}\left(
q;5\right) =80\%$ means that we can fulfill $80\%$ of the redemption after
five trading days, etc.\smallskip

We consider a portfolio, which is made up of $5$ assets.
The redemption scenario is defined below by the number of shares $q_i$
that have to be sold:
\begin{equation*}
\begin{tabular}{cccccc}
\hline
Asset     & \multicolumn{1}{c}{$1$} & \multicolumn{1}{c}{$2$} &
\multicolumn{1}{c}{$3$} & \multicolumn{1}{c}{$4$} & \multicolumn{1}{c}{$5$} \\
\hline
$q_i$     &  $4\,351$ & $2\,005$ & $755$ & $175$ &  $18$ \\
$q_i^{+}$ &  $1\,000$ & $1\,000$ & $200$ & $200$ & $200$ \\
$P_i$     &      $89$ &    $102$ &  $67$ & $119$ & $589$ \\
\hline
\end{tabular}
\label{ex:example1}
\end{equation*}
We also indicate the trading limit $q_i^{+}$ and the current price $P_i$ of
each asset. In Table \ref{tab:liquidation1}, we report the number of
liquidated shares $q_{i}\left( h\right)$ and the liquidation ratio
$\mathcal{LR}\left( q;h\right)$. After the first trading day, we have
liquidated $1\,000$ shares of Asset \#1 because of the trading policy that
imposes a trading limit of $1\,000$. We notice that we need $5$ trading days
in order to sell $4\,351$ shares of Asset \#1. If we consider the liquidation
ratio, we obtain $\mathcal{LR}\left( q;1\right) = 35\%$, $\mathcal{LR}\left(
q;2\right) = 65.34\%$, etc.\smallskip

\begin{table}[tbh]
\centering
\caption{Number of liquidated shares $q_{i}\left( h\right)$}
\label{tab:liquidation1}
\begin{tabular}{ccccccc}
\hline
$h$    &     Asset \#1 & Asset \#2 & Asset \#3 & Asset \#4 & Asset \#5 & $\mathcal{LR}\left( q;h\right)$ \\ \hline
$1$    &      $1\,000$ &    $1\,000$ &     $200$ &     $175$ &      $18$ & ${\TsV}35.00\%$ \\
$2$    &      $1\,000$ &    $1\,000$ &     $200$ & ${\TsX}0$ & ${\TsV}0$ & ${\TsV}65.34\%$ \\
$3$    &      $1\,000$ & ${\TsX}\,5$ &     $200$ & ${\TsX}0$ & ${\TsV}0$ & ${\TsV}80.61\%$ \\
$4$    &      $1\,000$ & ${\TsX}\,0$ &     $155$ & ${\TsX}0$ & ${\TsV}0$ & ${\TsV}95.36\%$ \\
$5$    & ${\TsV}\,351$ & ${\TsX}\,0$ & ${\TsX}0$ & ${\TsX}0$ & ${\TsV}0$ &      $100.00\%$ \\ \hline
Total  &      $4\,351$ &    $2\,005$ &     $755$ &     $175$ &      $18$ &                 \\
\hline
\end{tabular}
\end{table}

\begin{remark}
The liquidation period $h^{+}=\inf \left\{ h:\mathcal{LR}\left( q;h\right)
=1\right\} $ indicates how many trading days we need to liquidate the
redemption scenario $q$. In the previous example, $h^{+}$ is equal to $5$,
meaning that the liquidation of this redemption
scenario requires five trading days.
\end{remark}

We can break down the liquidation ratio as follows:%
\begin{equation*}
\mathcal{LR}\left( q;h\right) =\frac{1}{\sum_{i=1}^{n}q_{i}\cdot P_{i}}%
\sum_{k=1}^{h}\sum_{i=1}^{n}\mathcal{LA}_{i,k}\left( q\right)
\end{equation*}%
where $\mathcal{LA}_{i,k}\left( q\right) =q_{i}\left( k\right) \cdot P_{i}$
is the liquidation amount for security $i$ and trading day $k$. It follows
that:%
\begin{equation*}
\mathcal{LR}\left( q;h\right) =\sum_{k=1}^{h}\sum_{i=1}^{n}\mathcal{LC}%
_{i,k}\left( q\right) =\sum_{k=1}^{h}\mathcal{LC}_{k}\left( q\right)
\end{equation*}%
where $\mathcal{LC}_{i,k}\left( q\right) $ is the liquidation contribution
for security $i$ and trading day $k$:%
\begin{equation*}
\mathcal{LC}_{i,k}\left( q\right) =\frac{\mathcal{LA}_{i,k}\left( q\right) }{%
\sum_{i=1}^{n}q_{i}\cdot P_{i}}
\end{equation*}%
and $\mathcal{LC}_{k}\left( q\right) =\sum_{i=1}^{n}\mathcal{LC}_{i,k}\left(
q\right) $ is the liquidation contribution for trading day $k$. Another
useful decomposition is to consider the break-down by security:
\begin{eqnarray*}
\mathcal{LR}\left( q;h\right)  &=&\sum_{i=1}^{n}\frac{q_{i}\cdot P_{i}}{%
\sum_{i=1}^{n}q_{i}\cdot P_{i}}\frac{\sum_{k=1}^{h}\mathcal{LA}_{i,k}\left(
q\right) }{q_{i}\cdot P_{i}} \\
&=&\sum_{i=1}^{n}w_{i}\cdot \mathcal{LR}\left( q_{i};h\right)  \\
&=&\sum_{i=1}^{n}\mathcal{LC}_{i}\left( q;h\right)
\end{eqnarray*}%
where $w_{i}$ is the relative weight of security $i$ in portfolio $q$ and $%
\mathcal{LR}\left( q_{i};h\right) $ is the liquidation ratio applied to the
selling order $q_{i}$:%
\begin{equation*}
\mathcal{LR}\left( q_{i};h\right) =\frac{\sum_{k=1}^{h}\mathcal{LA}%
_{i,k}\left( q\right) }{q_{i}\cdot P_{i}}
\end{equation*}%
$\mathcal{LC}_{i}\left( q;h\right) =w_{i}\cdot \mathcal{LR}\left(
q_{i};h\right) $ is the liquidation contribution of asset $i$.\smallskip

We consider the previous example. Table \ref{tab:liquidation2a} shows
the values taken by the liquidation contribution $\mathcal{LC}_{i,h}\left( q\right)$.
For instance, $\mathcal{LC}_{1,2}\left( q\right) = 15.14\%$ means
that the liquidation of $1\,000$ shares of the second asset during the first
trading day represents $15.14\%$ of the redemption scenario. The sum of each row $h$
corresponds to the liquidation contribution $\mathcal{LC}_h\left( q\right)$. For instance, we have
$13.21\%+15.14\%+1.99\%+3.09\%+1.57\% = 35.00\%$. The sum of each column corresponds to
the weights $w_i$ because we have\footnote{%
This result comes from the following identity:%
\begin{equation*}
\sum_{k=1}^{h^{+}}\mathcal{LC}_{i,k}\left( q\right) =\sum_{k=1}^{h^{+}}\frac{%
\mathcal{LA}_{i,k}\left( q\right) }{\sum_{j=1}^{n}q_{j}\cdot P_{j}}%
=\sum_{k=1}^{h^{+}}\frac{q_{i}\left( k\right) \cdot P_{i}}{%
\sum_{j=1}^{n}q_{j}\cdot P_{j}}=\frac{q_{i}\cdot P_{i}}{\sum_{j=1}^{n}q_{j}%
\cdot P_{j}}=w_{i}
\end{equation*}%
} $w_{i}=\sum_{k=1}^{h^{+}}\mathcal{LC}_{i,k}\left( q\right) $.
\begin{table}[tbh]
\centering
\caption{Liquidation contribution $\mathcal{LC}_{i,h}\left( q\right)$ by trading day}
\label{tab:liquidation2a}
\begin{tabular}{ccccccc}
\hline
$h$   &     Asset \#1 & Asset \#2 & Asset \#3 & Asset \#4 & Asset \#5 & $\mathcal{LC}_h\left( q\right)$ \\ \hline
$1$   &       $13.21\%$ &      $15.14\%$ & $1.99\%$ & $3.09\%$ & $1.57\%$ & ${\TsV}35.00\%$ \\
$2$   &       $13.21\%$ &      $15.14\%$ & $1.99\%$ & $0.00\%$ & $0.00\%$ & ${\TsV}30.34\%$ \\
$3$   &       $13.21\%$ & ${\TsV}0.08\%$ & $1.99\%$ & $0.00\%$ & $0.00\%$ & ${\TsV}15.27\%$ \\
$4$   &       $13.21\%$ & ${\TsV}0.00\%$ & $1.54\%$ & $0.00\%$ & $0.00\%$ & ${\TsV}14.75\%$ \\
$5$   &  ${\TsV}4.64\%$ & ${\TsV}0.00\%$ & $0.00\%$ & $0.00\%$ & $0.00\%$ &  ${\TsX}4.64\%$ \\ \hline
Total &       $57.47\%$ &      $30.35\%$ & $7.51\%$ & $3.09\%$ & $1.57\%$ &      $100.00\%$ \\
\hline
\end{tabular}
\bigskip

\centering
\caption{Weight $w_i$ and liquidation ratio $\mathcal{LR}\left( q_i;h\right)$ of the assets}
\label{tab:liquidation2b}
\begin{tabular}{cccccc}
\hline
                                  &       Asset \#1 &       Asset \#2 &       Asset \#3 &      Asset \#4 &      Asset \#5  \\ \hline
$\mathcal{LR}\left( q_i;1\right)$ & ${\TsV}22.98\%$ & ${\TsV}49.88\%$ & ${\TsV}26.49\%$ &     $100.00\%$ &     $100.00\%$  \\
$\mathcal{LR}\left( q_i;2\right)$ & ${\TsV}45.97\%$ & ${\TsV}99.75\%$ & ${\TsV}52.98\%$ &     $100.00\%$ &     $100.00\%$  \\
$\mathcal{LR}\left( q_i;3\right)$ & ${\TsV}68.95\%$ &      $100.00\%$ & ${\TsV}79.47\%$ &     $100.00\%$ &     $100.00\%$  \\
$\mathcal{LR}\left( q_i;4\right)$ & ${\TsV}91.93\%$ &      $100.00\%$ &      $100.00\%$ &     $100.00\%$ &     $100.00\%$  \\
$\mathcal{LR}\left( q_i;5\right)$ &      $100.00\%$ &      $100.00\%$ &      $100.00\%$ &     $100.00\%$ &     $100.00\%$  \\ \hline
$w_i$                             & ${\TsV}57.47\%$ & ${\TsV}30.35\%$ &  ${\TsX}7.51\%$ & ${\TsX}3.09\%$ & ${\TsX}1.57\%$  \\ \hline
\end{tabular}
\bigskip

\centering
\caption{Liquidation contribution $\mathcal{LC}_{i}\left( q;h\right) $ by asset}
\label{tab:liquidation2c}
\begin{tabular}{ccccccc}
\hline
$h$ & Asset \#1 & Asset \#2 & Asset \#3 & Asset \#4 & Asset \#5 & $\mathcal{LR}\left( q;h\right)$ \\ \hline
$1$ & $13.21\%$ & $15.14\%$ &  $1.99\%$ &  $3.09\%$ &  $1.57\%$ & ${\TsV}35.00\%$ \\
$2$ & $26.42\%$ & $30.28\%$ &  $3.98\%$ &  $3.09\%$ &  $1.57\%$ & ${\TsV}65.34\%$ \\
$3$ & $39.63\%$ & $30.35\%$ &  $5.97\%$ &  $3.09\%$ &  $1.57\%$ & ${\TsV}80.61\%$ \\
$4$ & $52.84\%$ & $30.35\%$ &  $7.51\%$ &  $3.09\%$ &  $1.57\%$ & ${\TsV}95.36\%$ \\
$5$ & $57.47\%$ & $30.35\%$ &  $7.51\%$ &  $3.09\%$ &  $1.57\%$ &      $100.00\%$ \\
\hline
\end{tabular}
\end{table}
The weights $w_i$ and the liquidation ratios $\mathcal{LR}\left(
q_i;h\right)$ are given in Table \ref{tab:liquidation2b}. We observe that the
assets are respectively liquidated in five, three,  three, four, one and one
trading days. If we multiply the weights $w_i$ by the liquidation ratios
$\mathcal{LR}\left( q_i;h\right)$, we obtain the liquidation contribution
$\mathcal{LC}_{i}\left( q;h\right) $ by asset. If we sum the elements of each row,
we obtain the liquidity ratio $\mathcal{LR}\left( q;h\right)$.\smallskip

\begin{figure}[tbph]
\centering
\caption{Liquidation ratio (in \%) of the EUROSTOX 50 index portfolio}
\label{fig:roncalli_weisang_liquidity2}
\figureskip
\includegraphics[width = \figurewidth, height = \figureheight]{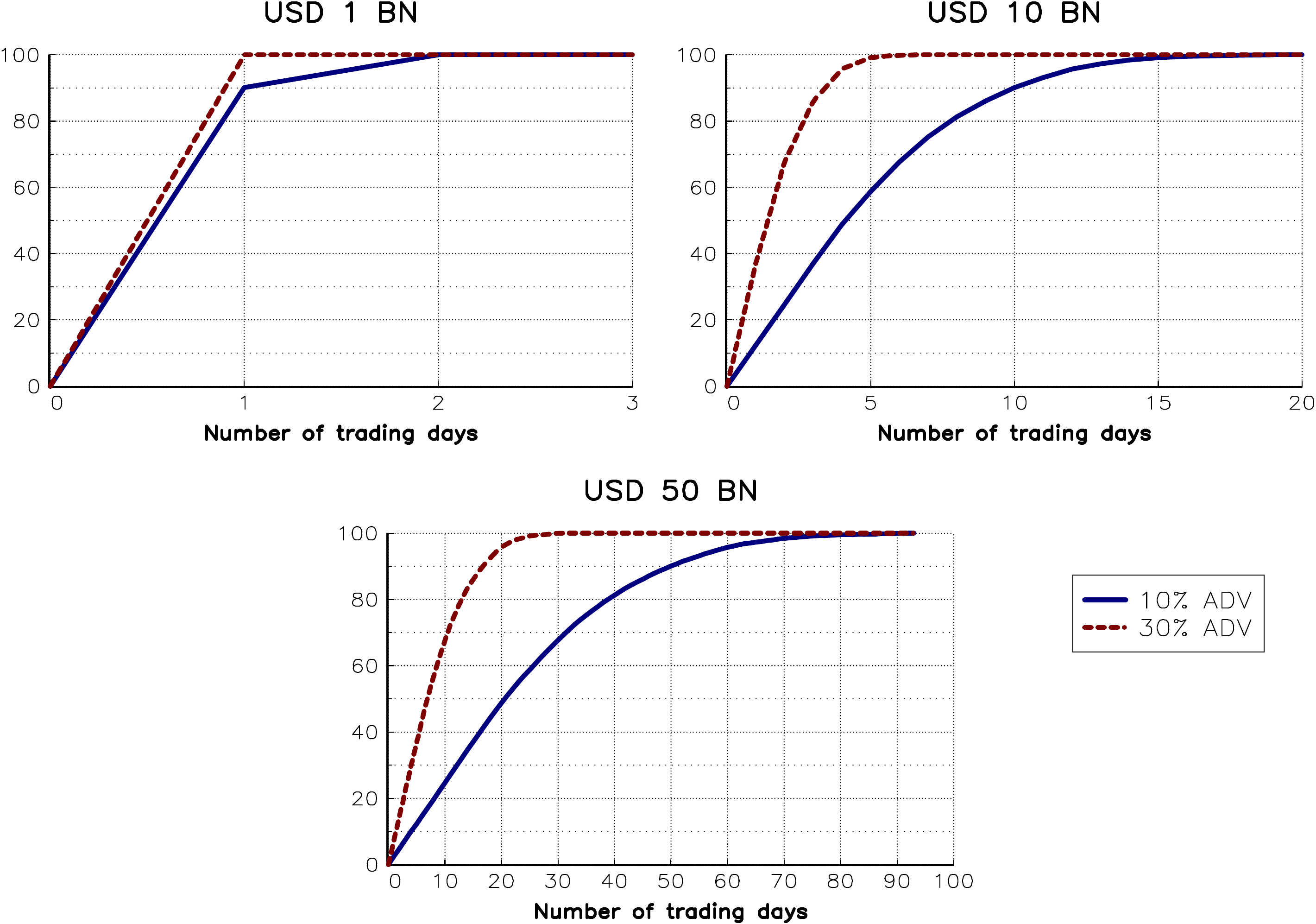}
\begin{flushleft}
{\small \textit{Source}: \citet[Figure 15, page 50]{Roncalli-2015a}, data as of April 30, 2015.}
\end{flushleft}
\end{figure}

\begin{figure}[tbph]
\centering
\caption{Comparing the liquidation ratio (in \%) between equity index portfolios}
\label{fig:roncalli_weisang_liquidity11}
\figureskip
\includegraphics[width = \figurewidth, height = \figureheight]{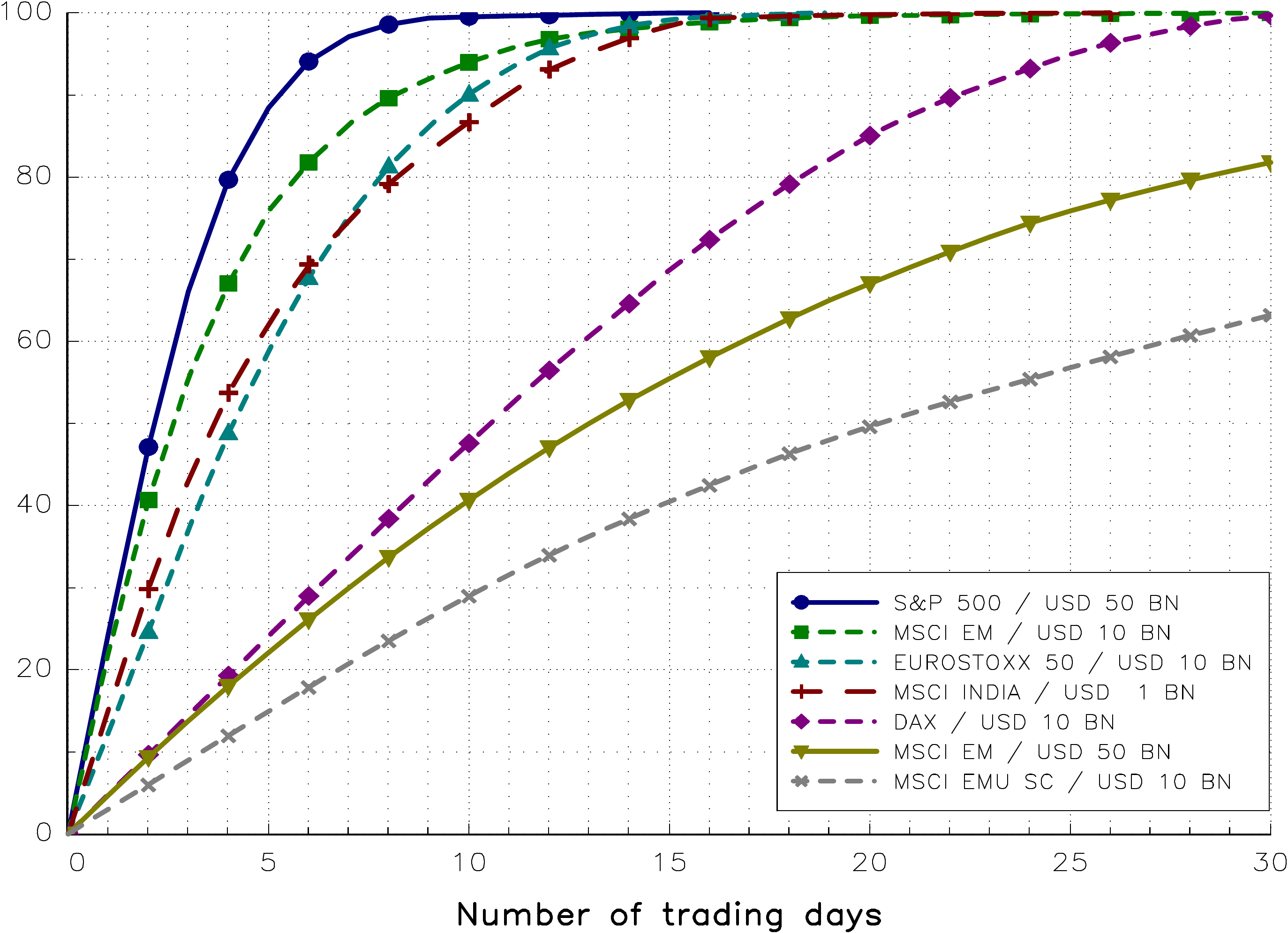}
\begin{flushleft}
{\small \textit{Source}: \citet[Figure 5, page 28]{Roncalli-2015a}, data as of April 30, 2015.}
\end{flushleft}
\end{figure}

As explained by \citet{Roncalli-2015a}, the liquidation ratio will depend on
three factors: the liquidity of the portfolio to sell, the amount to sell and
the liquidation policy. They illustrated the impact of
these factors using several index portfolios. For instance, we report in
Figure \ref{fig:roncalli_weisang_liquidity2} the example of the EUROSTOXX 50
index portfolio. We notice that the liquidation ratio is different if we
consider a selling order of $\$1$, $\$10$ or $\$50$ bn. It is also different
if the trading limit is equal to $10\%$ or $30\%$ of the average daily
volume\footnote{\citet{Roncalli-2015a} used the three-month average
daily volume computed by Bloomberg.}
(ADV). In Figure \ref{fig:roncalli_weisang_liquidity11}, we compare the
liquidation ratio for different index portfolios when the trading limit is
set to $10\%$ of ADV. We notice that the liquidity profile is better for the
S\&P 500 Index and a size of $\$50$ bn than for the EUROSTOXX 50 Index and a
size of $\$10$ bn. We also observe that liquidating $\$1$ bn of the MSCI
INDIA Index is approximately equivalent to liquidating $\$10$ bn of the
EUROSTOXX 50 Index. Of course, these results may differ from one period to
another, because the liquidity is time-varying. Nevertheless, we observe that
the liquidity of the portfolio is different
depending on whether we consider small cap stocks or large cap stocks.
The liquidity ratio also decreases with the amount to sell. Finally, the
liquidity ratio also depends on the trading constraints or
the liquidation policy.

\subsubsection{Time to liquidation}

The liquidation time is the inverse function of the liquidation ratio:
\begin{eqnarray*}
\mathcal{LT}\left( q;p\right)  &=&\mathcal{LR}^{-1}\left( q;p\right)  \\
&=&\inf \left\{ h:\mathcal{LR}\left( q;h\right) \geq p\right\}
\end{eqnarray*}%
For instance, $\mathcal{LT}\left( q;75\%\right) =8$ means that we need $8$
trading days to fulfill $75\%$ of the redemption. The liquidation time is a
step function because $\mathcal{LT}\left( q;p\right) $ is an integer. If we
consider the previous example, we have $\mathcal{LR}\left( q;0\right) =0$, $%
\mathcal{LR}\left( q;1\right) =35\%$, $\mathcal{LR}\left( q;2\right) =65.34\%
$, etc. We deduce that $\mathcal{LT}\left( q;p\right) =0$ if $p<35\%$, $%
\mathcal{LT}\left( q;p\right) =1$ if $35\%\leq p<65.34\%$, etc.\smallskip

In Table \ref{tab:roncalli_weisang_tables_6_7}, we report some figures of
liquidation time that were calculated by \citet{Roncalli-2015a}. The size of
the equity index portfolio is set to $\$10$ bn, and two liquidation policies
are tested ($10\%$ and $30\%$ of the average daily volume). In the case of
the S\&P 500 Index, liquidating $90\%$ of a $\$10$ bn equity index portfolio
takes two trading days with a trading limit of $10\%$ of the ADV and one
trading day with a trading limit of $30\%$ of the ADV. In the case of the
MSCI EMU Small Cap Index, these liquidation times becomes $74$ and $25$
trading days.

\begin{table}[tbph]
\centering
\caption{Time to liquidation (size = \$10 bn)}
\label{tab:roncalli_weisang_tables_6_7}
\tableskip
\begin{tabular}{c|ccccccc}
\hline
\multirow{2}{*}{Index} & \multirow{2}{*}{S\&P 500} & \multirow{2}{*}{ES 50} & \multirow{2}{*}{DAX}
                                & \multirow{2}{*}{NASDAQ} & MSCI & MSCI & MSCI \\
                 &              &           &          &    & EM & INDIA & EMU SC     \\
\hline
$p$ (in \%)      & \multicolumn{7}{c}{10\% of ADV}                                                        \\
$50$             &          $1$ & ${\TsV}5$ &        $11$ &    $2$ & ${\TsV}3$ &  ${\TsV}37$ & ${\TsV}21$ \\
$75$             &          $1$ & ${\TsV}7$ &        $17$ &    $3$ & ${\TsV}5$ &  ${\TsV}71$ & ${\TsV}43$ \\
$90$             &          $2$ &      $10$ &        $23$ &    $3$ & ${\TsV}9$ &       $110$ & ${\TsV}74$ \\
$99$             &          $2$ &      $15$ &        $29$ &    $5$ &      $17$ &       $156$ &      $455$ \\ \hline
$p$ (in \%)      & \multicolumn{7}{c}{30\% of ADV}                                                        \\
$50$             &     $1$ &          $2$ & ${\TsV}4$ &          $1$ &          $1$ &        $13$ &   ${\TsX}7$ \\
$75$             &     $1$ &          $3$ & ${\TsV}6$ &          $1$ &          $2$ &        $24$ &  ${\TsV}15$ \\
$90$             &     $1$ &          $4$ & ${\TsV}8$ &          $1$ &          $3$ &        $37$ &  ${\TsV}25$ \\
$99$             &     $1$ &          $5$ &      $10$ &          $2$ &          $6$ &        $52$ &       $152$ \\ \hline
\end{tabular}
\begin{flushleft}
{\small \textit{Source}: \citet[Tables 6 and 7, page 26]{Roncalli-2015a}, data as of April 30, 2015.}
\end{flushleft}
\end{table}

\begin{remark}
The liquidation risk profile of the redemption scenario $q$ can be defined
by the function $h\mapsto \mathcal{LR}\left( q;h\right) $ or the function
$p\mapsto \mathcal{LT}\left( q;p\right) $. As shown by \citet{Roncalli-2015a},
it depends on the asset liquidity, the liquidation policy and the
portfolio composition.
\end{remark}

\subsubsection{Liquidation shortfall}

\begin{figure}[tbph]
\centering
\caption{Liquidation shortfall with respect to the portfolio notional}
\label{fig:liquidation3}
\figureskip
\includegraphics[width = \figurewidth, height = \figureheight]{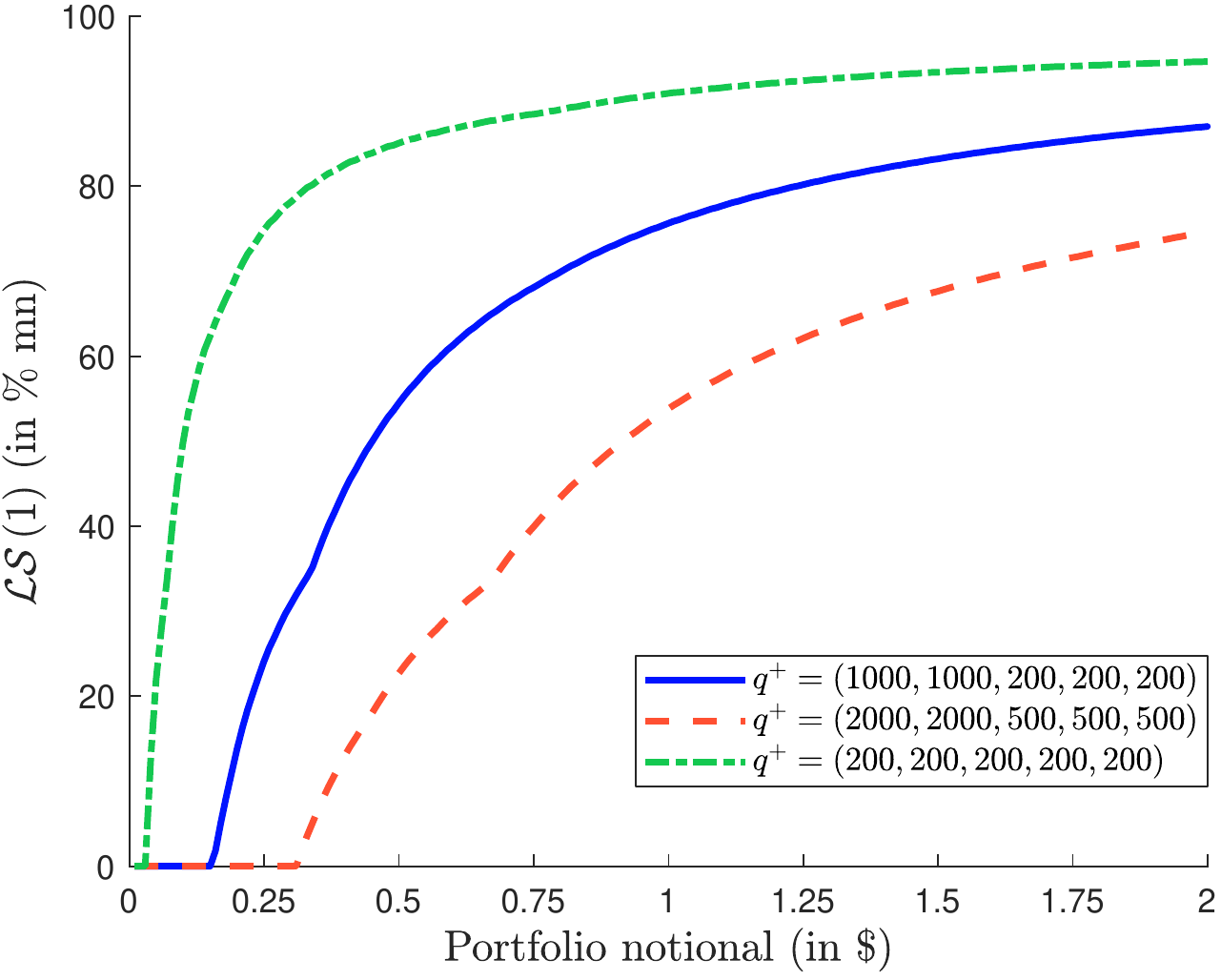}
\end{figure}

The liquidation shortfall $\mathcal{LS}\left( q\right) $ is defined as the
remaining redemption that cannot be fulfilled after one trading day:
\begin{equation}
\mathcal{LS}\left( q\right) =1-\mathcal{LR}\left( q;1\right)
\label{eq:liq-shortfall}
\end{equation}
For instance, it is equal to $65\%$ for the previous example
described on page \pageref{ex:example1}. The liquidation
shortfall is an increasing function of the order size. An illustration is
given in Figure \ref{fig:liquidation3} by considering three different
liquidation policies.

\subsection{Liquidity cost}

We now turn to liquidity measures that incorporate the price (or cost) dimension.
Generally, we measure the liquidity cost by the transaction cost.
However, in a liquidity stress testing program, this measure is merely
theoretical since it is based on the transaction cost model. Therefore,
it can be completed by the ex-post liquidity cost, which is also called
the effective cost.

\subsubsection{Transaction cost}

We define the transaction cost of the redemption scenario $q=\left(
q_{1},\ldots ,q_{n}\right) $ as the product of the unit costs and the dollar
volumes:
\begin{equation}
\TC\left( q\right) =\sum_{i=1}^{n}q_{i}\cdot P_{i}\cdot \cost_{i}\left(
x_{i}\right) =\sum_{i=1}^{n}Q_{i}\cdot \cost_{i}\left( x_{i}\right)
\label{eq:liq-tc1}
\end{equation}%
where $Q_{i}=q_{i}\cdot P_{i}$ is the nominal volume (expressed in \$), $%
x_{i}=v_{i}^{-1}q_{i}$ is the participation rate when selling security $i$
and $\cost_{i}\left( x\right) $ is the unit transaction cost associated with
security $i$. We can then break down the liquidity cost into two parts%
\footnote{%
We have:
\begin{equation*}
\cost_{i}\left( x\right) =\spread_{i}+\impact_{i}\left( x\right)
\end{equation*}%
}:
\begin{equation*}
\TC\left( q\right) =\BAS\left( q\right) +\PI\left( q\right)
\end{equation*}%
where the bid-ask spread component is equal to:%
\begin{equation*}
\BAS\left( q\right) =\sum_{i=1}^{n}Q_{i}\cdot \spread_{i}
\end{equation*}%
and the market impact cost is given by:%
\begin{equation*}
\PI\left( q\right) =\sum_{i=1}^{n}Q_{i}\cdot \impact_{i}\left( x_{i}\right)
\end{equation*}%
\smallskip

The previous analysis assumes that we can sell the portfolio $q$
instantaneously or during the same day. However, Equation (\ref{eq:liq-tc1})
is only valid if the volumes $q_{i}$ are less than the trading limits $%
q_{i}^{+}=x_{i}^{+}\cdot v_{i}$. Otherwise, we have:%
\begin{equation}
\TC\left( q\right) =\sum_{i=1}^{n}\sum_{h=1}^{h^{+}}\mathds{1}\left\{
q_{i}\left( h\right) >0\right\} \cdot q_{i}\left( h\right) \cdot P_{i}\cdot %
\cost_{i}\left( \frac{q_{i}\left( h\right) }{v_{i}}\right)
\label{eq:liq-tc2}
\end{equation}%
In this case, the bid-ask spread component has the same expression, but the
market impact component is different. Indeed, we have\footnote{%
Because of the following identity:%
\begin{equation*}
\sum_{h=1}^{h^{+}}\mathds{1}\left\{ q_{i}\left( h\right) >0\right\} \cdot
q_{i}\left( h\right) =q_{i}
\end{equation*}%
}:%
\begin{eqnarray}
\BAS\left( q\right)  &=&\sum_{i=1}^{n}\sum_{h=1}^{h^{+}}\mathds{1}\left\{
q_{i}\left( h\right) >0\right\} \cdot q_{i}\left( h\right) \cdot P_{i}\cdot %
\spread_{i}  \notag \\
&=&\sum_{i=1}^{n}Q_{i}\cdot \spread_{i}  \label{eq:liq-tc3}
\end{eqnarray}%
but:%
\begin{equation}
\PI\left( q\right) =\TC\left( q\right) -\BAS\left( q\right) \neq
\sum_{i=1}^{n}Q_{i}\cdot \impact_{i}\left( x_{i}\right)   \label{eq:liq-tc4}
\end{equation}

\begin{remark}
We assume that $q_{i}\leq q_{i}^{+}$. We have $q_{i}\left( 1\right) =q_{i}$
and $q_{i}\left( h\right) =0$ for $h>1$. We obtain:%
\begin{eqnarray*}
\TC\left( q\right)  &=&\sum_{i=1}^{n}q_{i}\left( 1\right) \cdot P_{i}\cdot %
\cost_{i}\left( \frac{q_{i}\left( 1\right) }{v_{i}}\right)  \\
&=&\sum_{i=1}^{n}q_{i}\cdot P_{i}\cdot \cost_{i}\left( x_{i}\right)
\end{eqnarray*}%
We retrieve the expression given in Equation (\ref{eq:liq-tc1}).
\end{remark}

\begin{remark}
Since the transaction cost is measured in dollars, it may be useful to
express it as a percentage of the redemption value:
\begin{equation*}
\TC_{\relative}\left( q\right) =\frac{\TC\left( q\right) }{\sum_{i=1}^{n}q_{i}\cdot P_{i}}
\end{equation*}%
An alternative measure is to compare the total transaction cost with the
bid-ask spread component:
\begin{equation*}
\TC_{\spread}\left( q\right) =\frac{\TC\left( q\right) }{\BAS\left( q\right)}
\end{equation*}
\end{remark}

We consider the previous example. We recall the characteristics of the redemption portfolio:
\begin{equation*}
\begin{tabular}{cccccc}
\hline
Asset     & \multicolumn{1}{c}{$\qquad 1\qquad$} & \multicolumn{1}{c}{$\qquad 2\qquad$} &
\multicolumn{1}{c}{$\qquad 3\qquad$} & \multicolumn{1}{c}{$\qquad 4\qquad$} & \multicolumn{1}{c}{$\qquad 5\qquad$} \\
\hline
$q_i$                 &  $4\,351$ &  $2\,005$ &    $755$ &    $175$ &     $18$ \\
$q_i^{+}$             &  $1\,000$ &  $1\,000$ &    $200$ &    $200$ &    $200$ \\
$P_i$ (in \$)         &      $89$ &     $102$ &     $67$ &    $119$ &    $589$ \\[0.5ex] \hline
& & & & & \\[-2ex]
$\impact_i\left(x\right)$ & \multicolumn{5}{c}{SQRL model with $\varphi_1 = 1$, $\tilde{x} = 5\%$ and $x^{+} = 10\%$} \\
$\sigma_i$ (in \%)    &      $25$ &      $20$ &     $18$ &     $30$ &     $20$ \\
$\spread_i$ (in bps)  &       $4$ &       $4$ &      $5$ &      $5$ &      $5$ \\
$v_i$                 & $10\,000$ & $10\,000$ & $2\,000$ & $2\,000$ & $2\,000$ \\
\hline
\end{tabular}
\end{equation*}
We also indicate the transaction cost function. It is given by the SQRL model
with $\varphi_1 = 1$, $\tilde{x} = 5\%$ and $x^{+} = 10\%$. For each asset
$i$, we also indicate the annualized volatility $\sigma_i$, the value of the
bid-ask spread $\spread_i$ and the daily volume $v_i$.\smallskip

The value of the redemption portfolio is equal to $\$673\,761$. The total transaction
cost is equal to $\TC\left(q\right) = \$4\,373.55$ with the following breakdown:
$\BAS\left(q\right) = \$277.71$ and $\PI\left(q\right) = \$4\,095.85$.
These figures represent respectively $64.9$, $4.1$ and $60.8$ bps of the portfolio value.
We deduce that the price impact explains $93.7\%$ of the transaction cost. The contribution
of each asset is respectively equal to $34.6\%$, $30.5\%$ and $16.6\%$, $16.0\%$ and $2.4\%$.
More results can be found in Tables \ref{tab:liquidation4a}--\ref{tab:liquidation4e}
on page \pageref{tab:liquidation4a}.

\subsubsection{Implementation shortfall and effective cost}

The previous analysis assumes that the transaction cost is calculated with a
model. Therefore, Equation (\ref{eq:liq-tc2}) defines an ex-ante transaction
cost. In practice, this ex-ante transaction cost will differ from the
effective transaction cost. In order to define the latter, we must
reintroduce the time index $t$ in the analysis. The current value of the
redemption scenario is equal to:
\begin{equation*}
\mathbb{V}^{\mathrm{mid}}\left( q\right) =\sum_{i=1}^{n}q_{i}\left( t\right)
\cdot P_{i}^{\mathrm{mid}}\left( t\right)
\end{equation*}%
where $q_{i}\left( t\right) $ and $P_{i}^{\mathrm{mid}}\left( t\right) $ are
the number of shares to sell and the mid-price for the security $i$ at the
current time $t$. The value of the liquidated portfolio is equal to:
\begin{equation*}
\mathbb{V}^{\mathrm{liquidated}}\left( q\right)
=\sum_{i=1}^{n}\sum_{t_{k}\geq t}q_{i}\left( t_{k}\right) \cdot P_{i}^{%
\mathrm{bid}}\left( t_{k}\right)
\end{equation*}%
where $q_{i}\left( t_{k}\right) $ and $P_{i}^{\mathrm{bid}}\left(
t_{k}\right) $ are the number of shares that were sold and the bid price for
the security $i$ at the execution time $t_{k}$. The effective cost is then
the difference between $V^{\mathrm{mid}}\left( t\right) $ and
$V^{\mathrm{liquidated}}\left( t\right) $:
\begin{equation}
\mathcal{IS}\left( q\right) =\max \left( \mathbb{V}^{\mathrm{mid}}\left(
q\right) -\mathbb{V}^{\mathrm{liquidated}}\left( q\right) ,0\right)
\label{eq:liq-is1}
\end{equation}%
The effective cost\footnote{Since $\mathbb{V}^{\mathrm{liquidated}}\left( q\right) $
can be higher than $\mathbb{V}^{\mathrm{mid}}\left( q\right) $, $\mathcal{IS}\left( q\right) $
is floored at zero. This situation occurs when execution times $t_{k}$ are
very different than the current time $t$ and market prices have gone up ---
$P_{i}^{\mathrm{bid}}\left( t_{k}\right) \geq P_{i}^{\mathrm{mid}}\left(
t\right) $.} $\mathcal{IS}\left( q\right) $ is called by \citet{Perold-1988}
the implementation shortfall, which measures the difference in price between
the time a portfolio manager makes an investment decision and the actual
traded price. Therefore, $\mathbb{V}^{\mathrm{mid}}\left( q\right) $ is the
benchmark price, $\mathbb{V}^{\mathrm{liquidated}}\left( q\right) $ is the
traded price and $\mathcal{IS}\left( q\right) $ is the total amount of
slippage.

\section{Implementing the stress testing framework}

In this section, we detail the general approach for implementing the
liquidity stress testing program on the asset side. We will see that
it is based on three steps. First, we have to correctly define the
asset liquidity buckets (or asset liquidity classes). Each asset
liquidity bucket is associated with a unique unit transaction cost
function and a given liquidation policy. Second, we have to
calibrate the parameters of the transaction cost function that are
related to a given liquidity bucket. Third, we must define the
appropriate estimation method of the security-specific parameters.
Nevertheless, before presenting the three-step approach, we must
understand how stress testing impacts transaction costs. Does stress
testing modify the conventional transaction cost function? Does
stress testing change the liquidation policy? What parameters are
impacted? This analysis will help to justify the three-step approach
of asset liquidity stress testing. Finally, the last part of this
section is dedicated to an issue that generally occurs when
implementing the LST program. This concerns the distortion of the
redemption scenario on the asset side. In this article, we only
present general considerations, but this issue will be extensively
studied in our third article dedicated to liquidity stress testing
in asset management \citep{Roncalli-lst3-2021}.

\subsection{How does stress testing impact transaction costs?}

If we consider the two-regime model, we have:
\begin{equation*}
\cost\left( \dfrac{q}{v}\right) =\left\{
\begin{array}{ll}
\spread+\varphi _{1}\sigma\left( \dfrac{q}{v}\right)
^{\gamma _{1}} & \text{if }q\leq \tilde{x}\cdot v \\
\spread+\varphi _{1}\tilde{x}^{\gamma _{1}-\gamma _{2}} \sigma
\left( \dfrac{q}{v}\right) ^{\gamma _{2}} & \text{if }\tilde{x}\cdot
v\leq q\leq x^{+}\cdot v \\
+\infty  & \text{if }q>x^{+}\cdot v%
\end{array}%
\right.
\end{equation*}%
The parameters of the transaction cost model are $\spread$, $\varphi
_{1}$, $\sigma $, $\gamma _{1}$, $\gamma _{2}$, $\tilde{x}$ and
$x^{+}$. The question is whether we need two sets of parameters:

\begin{enumerate}
\item $\left( \spread^{\mathrm{normal}},\varphi _{1}^{\mathrm{normal}%
},\sigma ^{\mathrm{normal}},\gamma _{1}^{\mathrm{normal}},\gamma _{2}^{%
\mathrm{normal}},\tilde{x}^{\mathrm{normal}},x^{+\mathrm{normal}}\right) $
for normal periods;

\item $\left( \spread^{\mathrm{stress}},\varphi _{1}^{\mathrm{stress}%
},\sigma ^{\mathrm{stress}},\gamma _{1}^{\mathrm{stress}},\gamma _{2}^{%
\mathrm{stress}},\tilde{x}^{\mathrm{stress}},x^{+\mathrm{stressl}}\right) $
for stress periods.
\end{enumerate}

\begin{figure}[tbph]
\centering
\caption{The $x$-approach of the unit transaction cost}
\label{fig:stress2}
\figureskip
\includegraphics[width = \figurewidth, height = \figureheight]{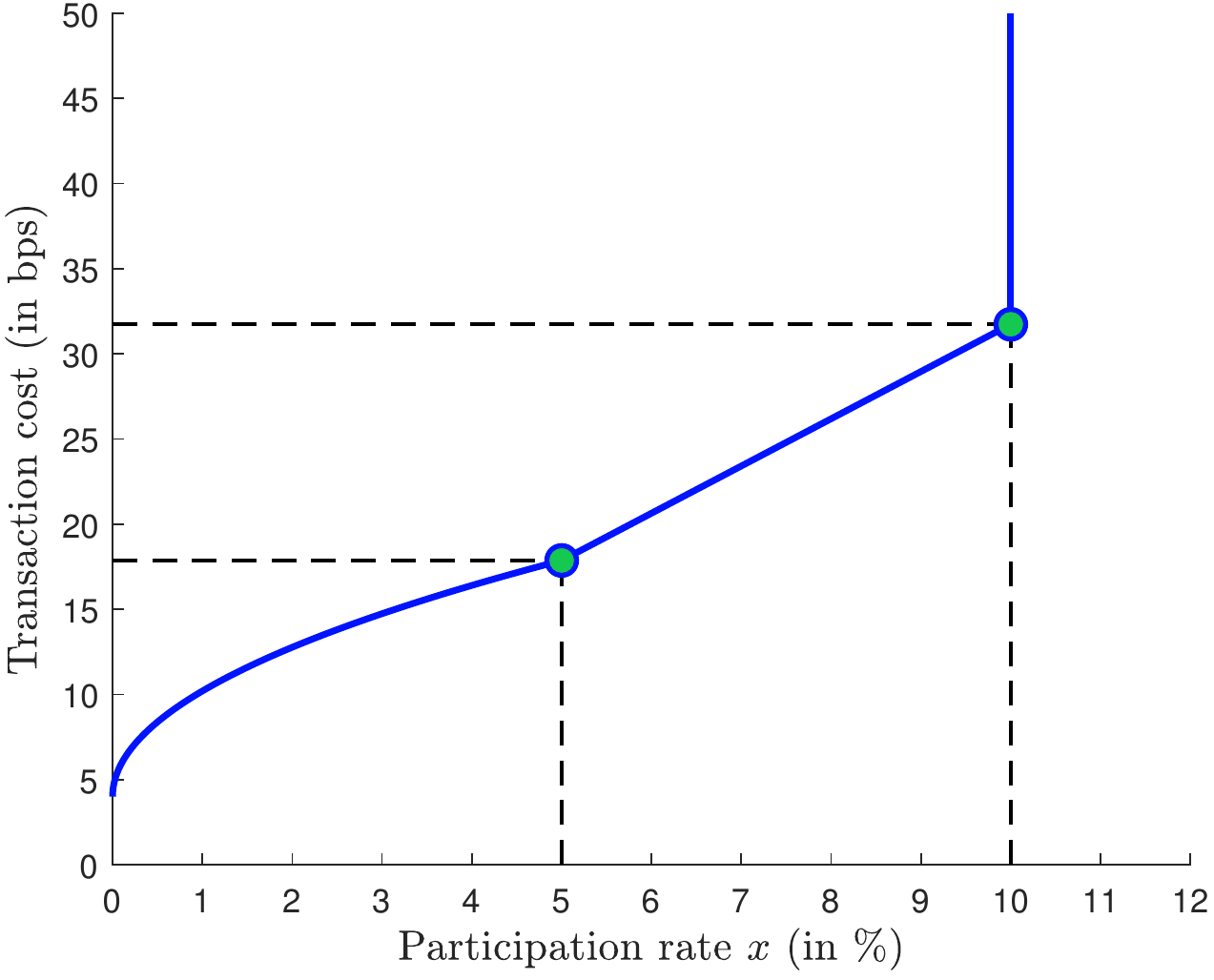}
\end{figure}

\begin{figure}[tbph]
\centering
\caption{The $q$-approach of the unit transaction cost}
\label{fig:stress3}
\figureskip
\includegraphics[width = \figurewidth, height = \figureheight]{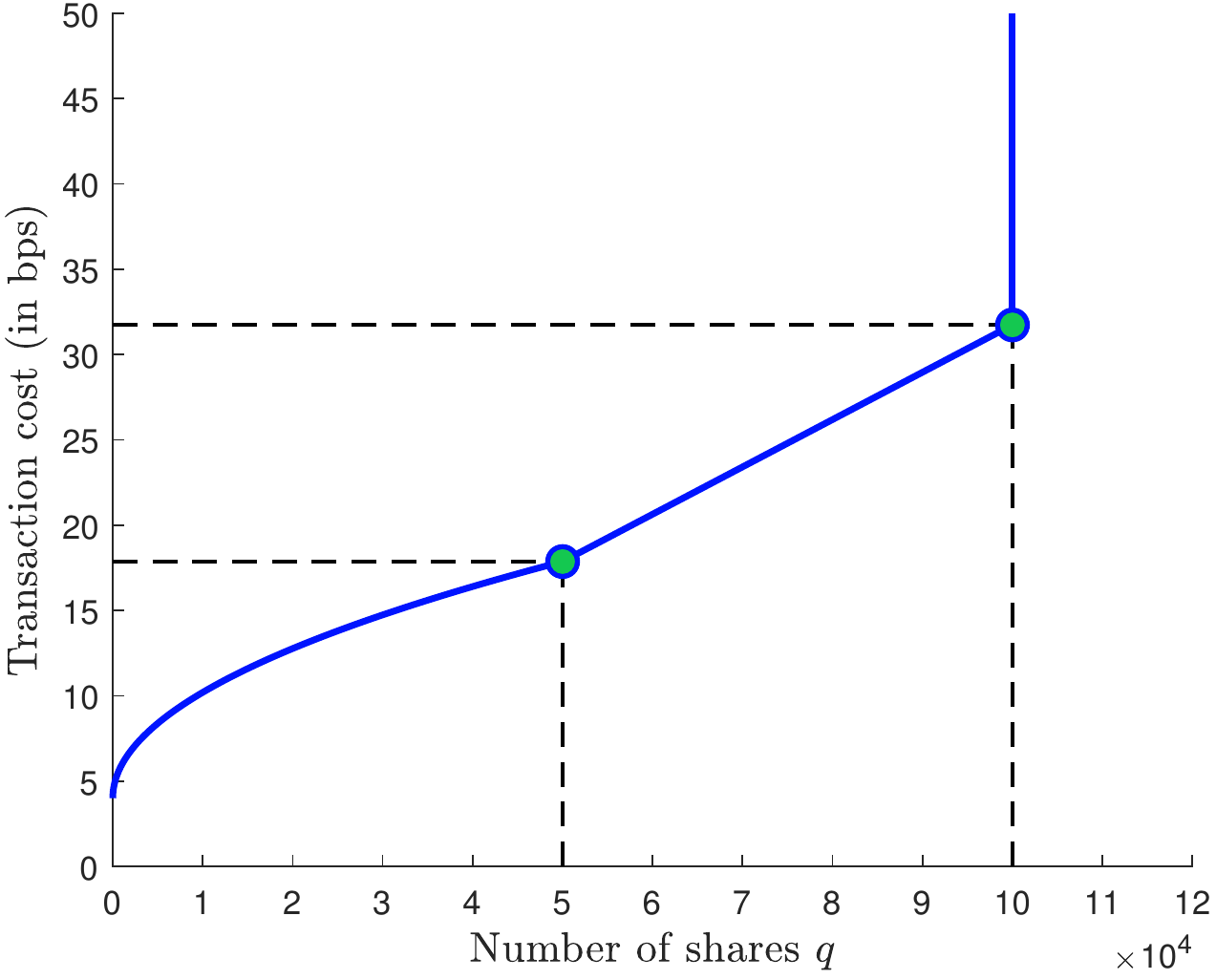}
\end{figure}

\begin{figure}[tbph]
\centering
\caption{Impact of security-specific parameters in the $x$-approach}
\label{fig:stress4}
\figureskip
\includegraphics[width = \figurewidth, height = \figureheight]{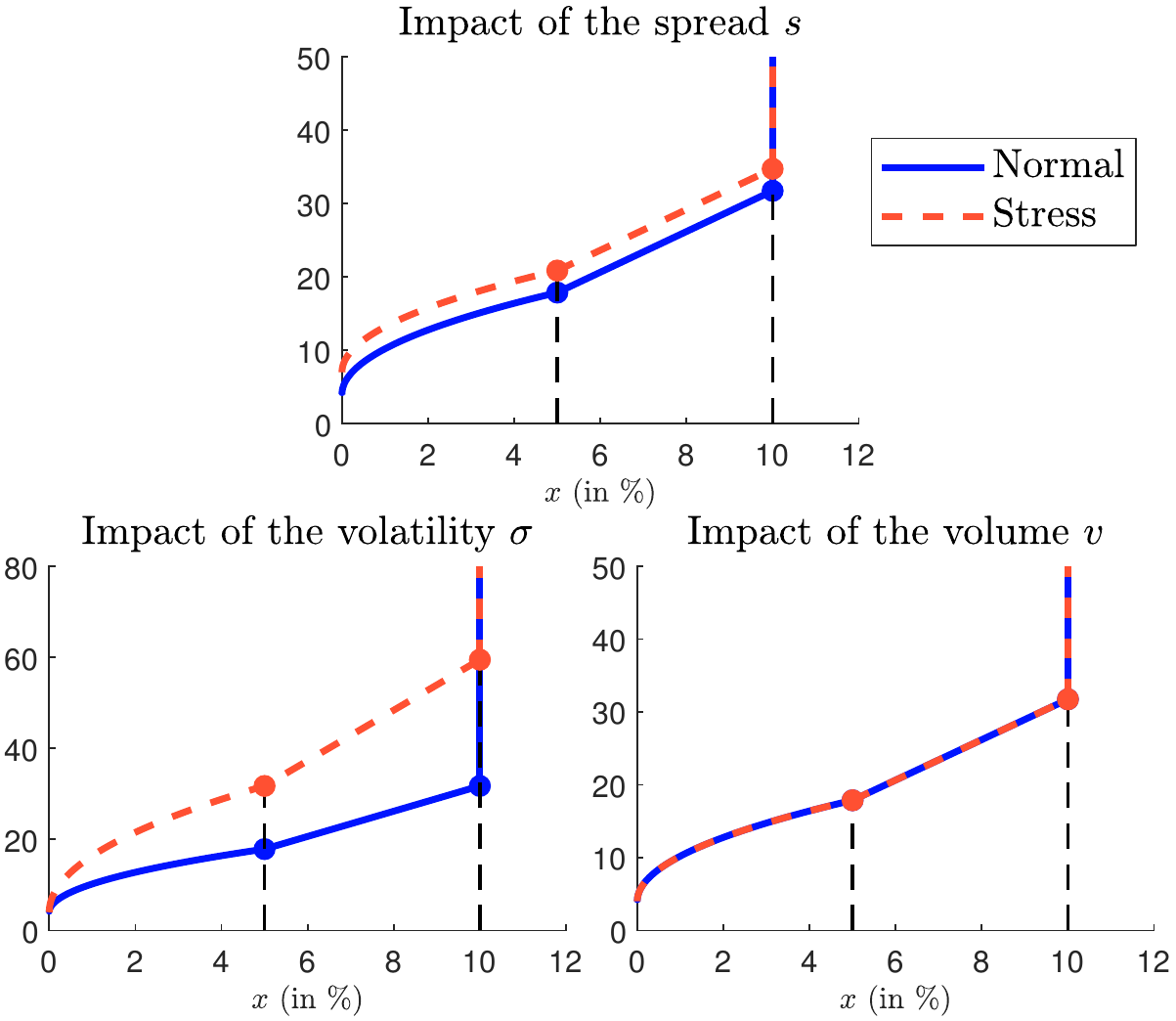}
\end{figure}

\begin{figure}[tbph]
\centering
\caption{Impact of security-specific parameters in the $q$-approach}
\label{fig:stress5}
\figureskip
\includegraphics[width = \figurewidth, height = \figureheight]{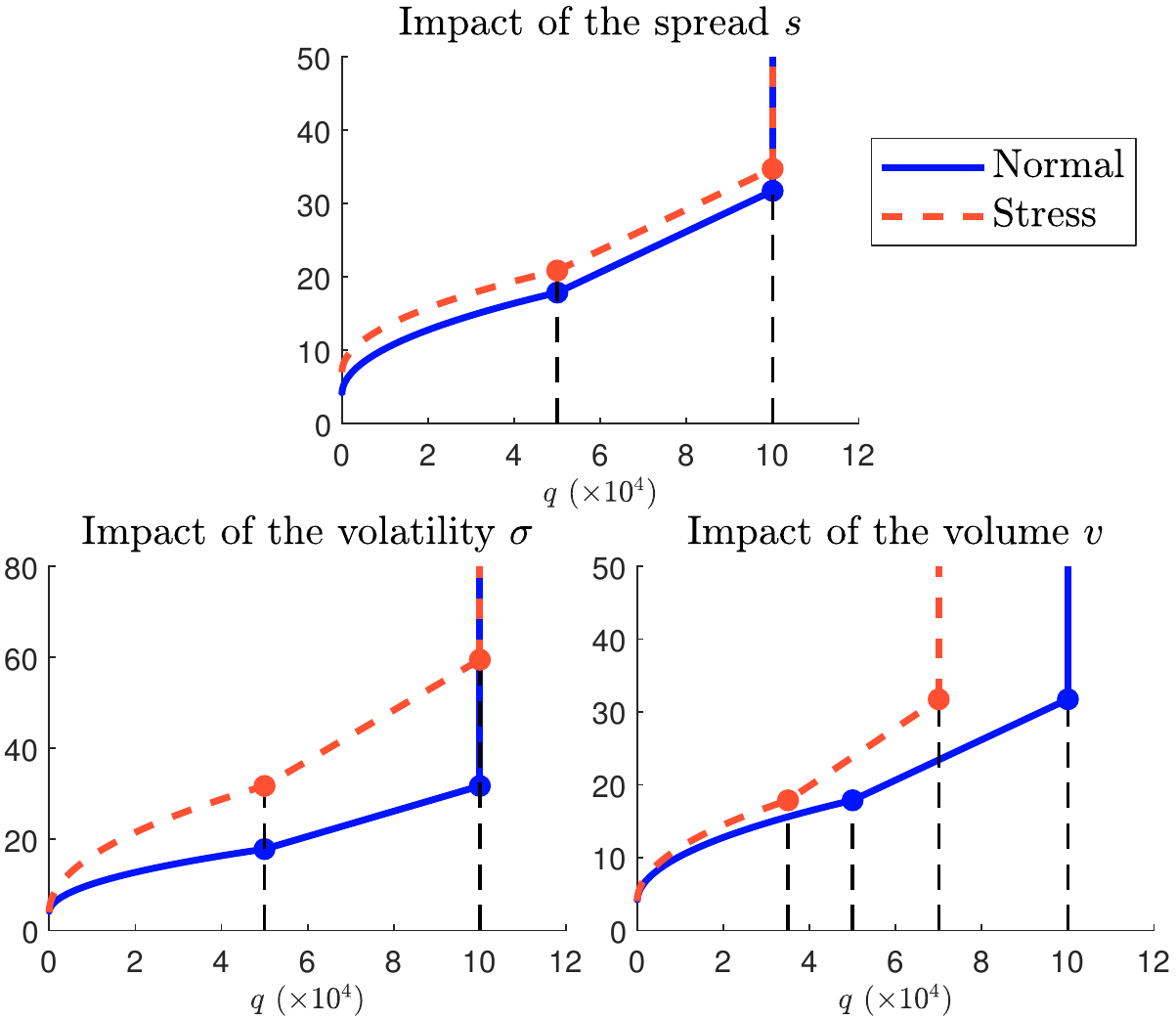}
\end{figure}

\noindent
This is equivalent having two different transaction cost functions:
$\cost^{\mathrm{normal}}\left( x\right) $ and
$\cost^{\mathrm{stress}}\left( x\right) $. This is not satisfactory
because this means that we need to calibrate many parameters in the
stress period. Moreover, we do not distinguish between
parameters that are related to the security and parameters that are
related to the liquidity bucket. Clearly, we can assume that the
parameters $\left( \varphi _{1},\gamma _{1},\gamma
_{2},\tilde{x},x^{+}\right) $ are the same for all the assets
belonging to the same liquidity bucket. They can change, but at a low
frequency, for instance because of the annual calibration exercise
or a change to the liquidation policy. The other parameters
$\spread$ and $\sigma $ are defined at the security level and can
change daily\footnote{Indeed, the spread and the
volatility of the security change every day because of market
conditions.}. Therefore, the unit transaction cost
function must be written as $\cost_x\left( x;\spread_{i,t},\sigma
_{i,t}\right) $ because $\spread_{i,t}$ and $\sigma _{i,t}$ change
with the security and the time.
We notice that this transaction cost function uses the
participation ratio $x$, which is the ratio between the order size
$q$ and the daily volume $v$. However, $v$ is another
related-security parameter since it changes every day. This is not
equivalent to selling $1\,000$ shares in the market if the daily volume
is $10\,000$ or $20\,000$. It follows that the unit transaction cost
function must be written as $\cost_q\left( q;\spread_{i,t},\sigma
_{i,t},v_{i,t}\right) $ because $\spread_{i,t}$, $\sigma _{i,t}$ and
$v_{i,t}$ change with the security and the time. The $q$-approach to
the unit transaction cost $\cost\left( q;\spread_{i,t},\sigma
_{i,t},v_{i,t}\right) $ differs then from the $x$-approach to
the unit transaction cost $\cost\left( x;\spread_{i,t},\sigma
_{i,t}\right) $ because it has an additional
parameter, which is the daily volume.\smallskip

At first sight, it seems that introducing the volume is a subtle
distinction. For instance, we have reported in Figures
\ref{fig:stress2} and \ref{fig:stress3} the functions
$\cost_{x}\left( x;\spread_{i,t},\sigma _{i,t}\right) $ and
$\cost_{q}\left( q;\spread_{i,t},\sigma _{i,t},v_{i,t}\right) $ when
the price impact is given by the SQRL model\footnote{We assume that
$\varphi_{1}=1$ and $\tilde{x}=5\%$.}, the
security-related parameters are equal to $\spread_{i,t}=4$ bps,
$\sigma _{i,t}=10\%$ and $v_{i,t}=100\,000$, and the liquidation
policy is set to $x^{+}=10\%$. The two figures have exactly the same
shape and we have the following correspondence:
\begin{equation*}
\cost_{q}\left( q;\spread_{i,t},\sigma _{i,t},v_{i,t}\right) :=\cost%
_{x}\left( x=\frac{q}{v_{i,t}};\spread_{i,t},\sigma _{i,t}\right)
\end{equation*}
Let us now see the impact of changing the parameters $\spread_{i,t}$, $\sigma
_{i,t}$ and $v_{i,t}$. In a stress period, we generally observe an increase
in the bid-ask spread and the asset volatility, and a reduction in the daily
volume that is traded in the market. In the top panel in Figures
\ref{fig:stress4} and \ref{fig:stress5}, we show the difference between the
two unit transaction costs when the bid-ask spread increases from $4$ bps to
$7$ bps. We observe that the functions $\cost_{x}$ and $\cost_{q}$ are both
shifted up, but they are the same. In the bottom/left panel, we report the
impact when the volatility in the stress period is twice the volatility in
the normal period\footnote{The annualized volatility $\sigma _{i,t}$
increases from $10\%$ to $20\%$.}. We notice that the higher volatility has
shifted the trading cost upward and it has also changed the shape of the unit
transaction cost function. But again, the two functions $\cost_{x}$ and
$\cost_{q}$ are the same using the equivalence relationship $q=x\cdot
v_{i,t}$. We now consider the impact of the volume. Generally, the daily
volume is reduced in stress periods. In the bottom/right panel in Figures
\ref{fig:stress4} and \ref{fig:stress5}, we assume that the daily volume is
equal to $v_{i,t}=100\,000$ in the normal period and $v_{i,t}=70\,000$ in the
stress period. Contrary to the parameters $\spread_{i,t}$ and $\sigma
_{i,t}$, we observe that the two functions $\cost_{x}$ and $\cost_{q}$ are
not equivalent in this case. Indeed, $v_{i,t}$ has no impact on $\cost_{x}$
whereas it completely changes the shape of $\cost_{q}$ because the inflection
point $\tilde{q}$ and the trading limit $q^{+}$ are different. It follows
that the invariance with respect to $x$ does not imply the invariance with
respect to $q$.\smallskip

\begin{figure}[tbph]
\centering
\caption{Comparing the unit transaction cost in the normal and stress periods}
\label{fig:stress6}
\figureskip
\includegraphics[width = \figurewidth, height = \figureheight]{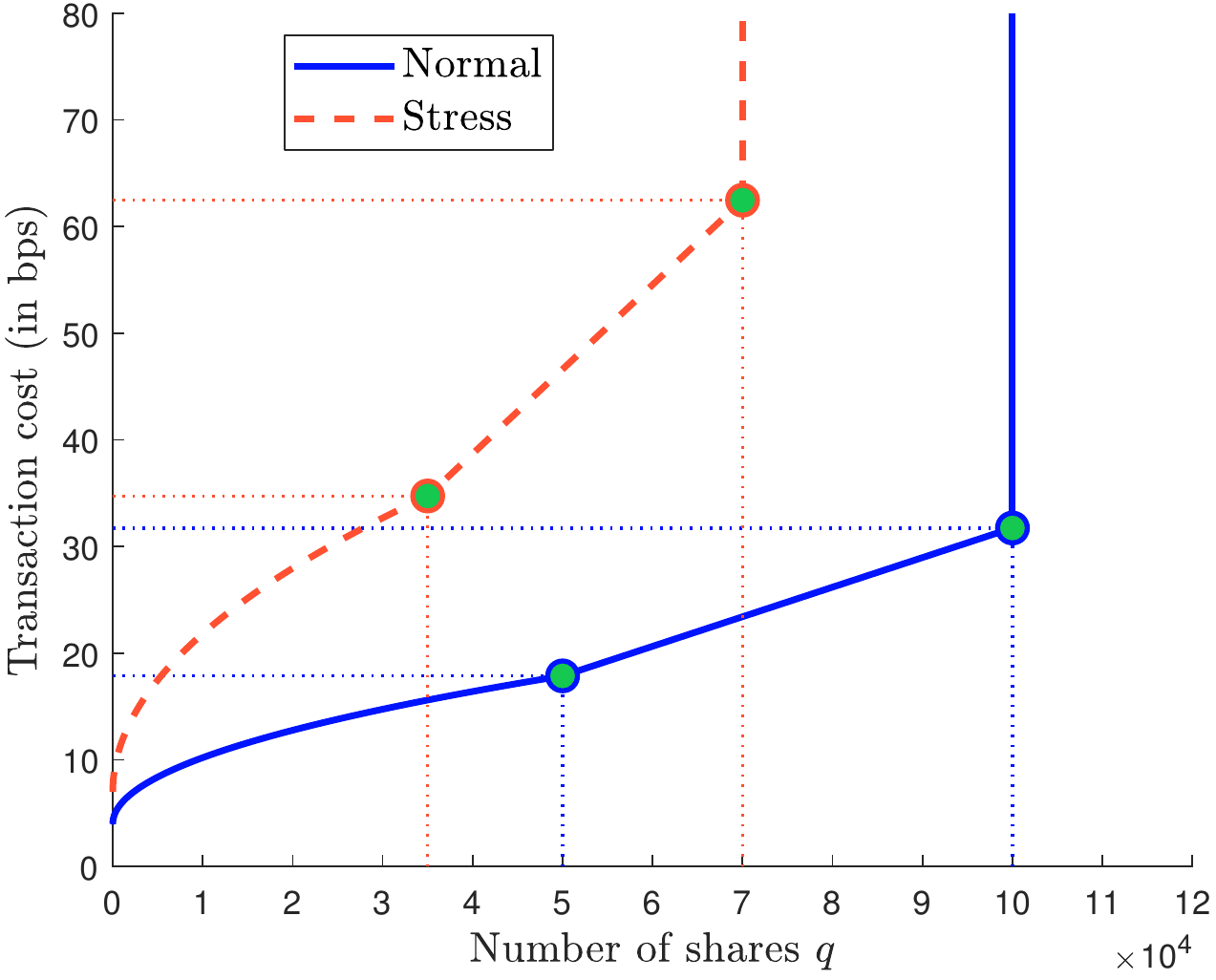}
\end{figure}

In the case of a liquidity stress program, we have to consider the
combination of the three effects. Results are reported in Figure
\ref{fig:stress6}. We recall that the normal period is defined by
$\spread_{i,t}=4$ bps, $\sigma _{i,t}=10\%$ and $v_{i,t}=100\,000$,
while the stress period is defined by $\spread_{i,t}=7$ bps, $\sigma
_{i,t}=20\%$ and $v_{i,t}=70\,000$. During the
stress period, the transaction cost is higher because the spread is larger,
the volatility has shifted the trading cost upward and the lower volume has
moved the inflection point to the left. This is the primary effect. For
instance, selling $40\,000$ shares of the security costs $16.40$ bps during
the normal period and $38.70$ bps during the stress period (see Table \ref{tab:stress7}).
The secondary effect is on the liquidation profile, because the trading
limit $q^{+}$ expressed as a number of shares is reduced in the stress period
even if the liquidation policy does not change. This is because the
liquidation policy is defined in terms of the maximum participation rate $x^{+}$.
For instance, $100\,000$ shares of the security can be sold
in one trading day in the normal period. This is no longer true in the
stress period, and the position is liquidated in two trading days (see Table
\ref{tab:stress7}). It follows that the stress testing program has a negative,
non-linear impact both
on the transaction cost and the liquidation profile.
In Table \ref{tab:stress7}, we have $\cost\left( 40\,000\right) =38.70$ bps,
$\cost\left( 80\,000\right) =57.39$ bps and $\cost\left( 100\,000\right)
=53.53$ bps. We observe that $\cost\left( q\right) $ is not necessarily an
increasing function of $q$ because of the liquidation policy. Indeed, in the
last case, $70\,000$ shares are sold at $62.47$ bps during the first trading day
and $30\,000$ shares are sold at $32.68$ bps during the second trading day. The
relative cost of selling $100\,000$ shares is lower than the relative cost
of selling $70\,000$ shares, because the price impact is not at its maximum during the
second day. In Figure \ref{fig:stress8} on page \pageref{fig:stress8},
we report the two functions, the
relative (or unit) transaction cost $\cost\left( q\right) $ and the total
transaction cost $\TC\left( q\right) $, by assuming that the price is equal
to $\$1$. We notice that the maximum relative cost is equal to $62.47$ bps
and is reached when the number of shares is a
multiple of $70\,000$, which is the trading limit. Therefore, $\cost\left(
q\right) $ is not increasing because of the averaging effect. Of course,
this is not the case for the total transaction cost, which is an increasing
function of $q$.

\begin{table}[tbph]
\centering
\caption{Computation of the unit transaction cost}
\label{tab:stress7}
\scalebox{0.9}{
\begin{tabular}{c|cc|cc|cc|cc}
\hline
                  & Normal     & Stress     & Normal     & Stress     & Normal     & Stress     & Normal     & Stress    \\
\hline
& & & & & & & & \\[-2.25ex]
$q$               & \multicolumn{2}{c|}{$10\,000$} & \multicolumn{2}{c|}{$40\,000$}
                  & \multicolumn{2}{c|}{$80\,000$}  & \multicolumn{2}{c}{$100\,000$} \\[0.05ex]
\hdashline
 & & & & & & & & \\[-2.25ex]
\multirow{2}{*}{$q\left(h\right)$}
                  & 10\,000  & 10\,000 & 40\,000 & 40\,000 & 80\,000 & 70\,000 & 100\,000 & 70\,000 \\
                  &          &         &         &         &         & 10\,000 &          & 30\,000 \\
& & & & & & & & \\[-2.50ex]
\hdashline
& & & & & & & & \\[-2.25ex]
\multirow{2}{*}{$\spread$}
                  &  4.00  &  7.00  & 4.00  &      7.00  &      4.00  &      7.00  &      4.00  &      7.00 \\
                  &        &        &       &            &            &      7.00  &            &      7.00 \\
& & & & & & & & \\[-2.50ex]
\hdashline
& & & & & & & & \\[-2.25ex]
\multirow{2}{*}{$\impact\left(q\left(h\right)\right)$}
                  &  6.20  & 14.82  & 12.40  &     31.70  &     22.19  &     55.47  &     27.74  &     55.47 \\
                  &        &        &        &            &            &     14.82  &            &     25.68 \\
& & & & & & & & \\[-2.50ex]
\hdashline
& & & & & & & & \\[-2.25ex]
\multirow{2}{*}{$\cost\left(q\left(h\right)\right)$}
                  & 10.20  & 21.82  & 16.40  &     38.70  &     26.19  &     62.47  &     31.74  &     62.47 \\
                  &        &        &        &            &            &     21.82  &            &     32.68 \\
& & & & & & & & \\[-2.50ex]
\hdashline
& & & & & & & & \\[-2.25ex]
\multirow{1}{*}{$\cost\left(q\right)$}
                  & 10.20  & 21.82  & 16.40  &     38.70  &     26.19  &     57.39  &     31.74  &     53.53 \\
& & & & & & & & \\[-2.50ex]
\hline
\end{tabular}}
\end{table}

\begin{remark}
The previous analysis shows that we do not need a new transaction cost function
for the stress period, because there is no reason for the functional form to change
and the impact of the security-specific parameters are sufficient to implement
the asset liquidity stress testing program.
\end{remark}

\subsection{A three-step approach}

As explained above, implementing an asset liquidity stress testing program
involves three steps. In the first step, we define liquidity buckets. The
second step corresponds to the estimation of the transaction cost function
for a given liquidity bucket. Finally, the third step consists in calibrating
the security-specific parameters.

\subsubsection{Liquidity bucketing}

\paragraph{Classification matrix}

\begin{table}[tbph]
\centering
\caption{An example of classification matrix of liquidity buckets}
\label{tab:bucket1}
\scalebox{0.9}{
\begin{tabular}{cccc|c}
\hline
Level 1                & Level 2                 & Level 3          & Level 4            & HQLA Class     \\ \hline
\mrm{7}{Equity}        & \mr{Large cap}          & DM               & \mr{Region}        &     1          \\
                       &                         & EM               &                    &     1          \\ \cline{2-5}
                       & \mr{Small cap}          & DM               & \mr{Region}        &     2          \\
                       &                         & EM               &                    &     2          \\ \cline{2-5}
                       & \mr{Derivatives}        & Futures          & \mr{Turnover}      &     1          \\
                       &                         & Options          &                    &     2          \\ \cline{2-5}
                       & Private                 &                  &                    &     5          \\ \hline
\mrm{17}{Fixed-income} & \mr{Sovereign}          & DM               & \mr{Region}        &     1/2        \\
                       &                         & EM               &                    &     2/3        \\ \cline{2-5}
                       & Municipal               &                  &                    &     2          \\ \cline{2-5}
                       & \mr{Inflation-linked}   & DM               & \mr{Region}        &     1/2        \\
                       &                         & EM               &                    &     2/3        \\ \cline{2-5}
                       & \mr{Corporate}          & IG               & \mr{Currency}      &     3          \\
                       &                         & HY               &                    &     4          \\ \cline{2-5}
                       & \mrm{4}{Securitization} & ABS              & \mrm{4}{US/Non-US} & \mrm{4}{2/3/4} \\
                       &                         & CLO              &                    &                \\
                       &                         & CMBS             &                    &                \\
                       &                         & RMBS             &                    &                \\ \cline{2-5}
                       & \mrm{4}{Derivatives}    & Caps/floors      & \mrm{4}{Turnover}  & \mrm{4}{1/2/3} \\
                       &                         & Futures          &                    &                \\
                       &                         & Options          &                    &                \\
                       &                         & Swaps            &                    &                \\ \cline{2-5}
                       & \mr{CDS}                & Single-name      & \mr{Turnover}      &     3          \\
                       &                         & Multi-name       &                    &     2          \\ \hline
\mrm{2}{Currency}      & G10                     &                  &                    &     1          \\
                       & Others                  &                  &                    &     1/2/3      \\ \hline
\mrm{5}{Commodity}     & \mrm{3}{Agriculture}    & Grain \& Oilseed &                    &     4          \\
                       &                         & Livestock        &                    &     4          \\
                       &                         & Soft             &                    &     4          \\ \cline{2-5}
                       & \mrm{3}{Energy}         & Electricity      &                    &     2          \\
                       &                         & Gas              &                    &     2          \\
                       &                         & Oil              &                    &     2          \\ \cline{2-5}
                       & \mrm{3}{Metal}          & Gold             &                    &     1          \\
                       &                         & Industrial       &                    &     2/4        \\
                       &                         & Precious         &                    &     2          \\ \hline
\end{tabular}}
\end{table}

A liquidity bucket is a set of homogenous securities such that they share the
same functional form of the unit transaction cost\footnote{Most of the time,
they also share the same liquidation policy.}. For instance, we may consider
that equities and bonds correspond to two liquidity buckets, meaning that we
need two different functions. But we can also split equities between large
cap and small cap equities. An example of matrix classification is provided
in Table \ref{tab:bucket1}. There are several levels depending on the
requirements of the asset manager and the confidence level on the
calibration. Generally, Level 2 is sufficiently granular and enough to
implement a liquidity stress testing program. For instance, it is
extensively used by external providers of LST solutions (MSCI
LiquidityMetrics, Bloomberg Liquidity Assessment (LQA), StateStreet Liquidity
Risk Solution, etc.). Nevertheless, the asset manager may wish to go beyond
Level 2 and adopt Level 3 for some buckets. For example, it could make sense
to distinguish the functional form for DM and EM sovereign bonds. Level 4 is
the ultimate level and differentiates securities by region, currency or
turnover\footnote{The turnover is defined as \textquotedblleft \textsl{the
gross value of all new deals entered into during a given period and is
measured in terms of the nominal or notional amount of the contracts. It
provides a measure of market activity and can also be seen as a rough proxy
for market liquidity}\textquotedblright\ \citep{BIS-2014}.}. For example, if
we consider the DM large cap stocks, we may split this category by region,
e.g., North America, Eurozone, Japan and Europe-ex-EMU. In the case of
corporate IG bonds, one generally splits these securities by currency, e.g.,
USD IG bonds, EUR IG bonds, GBP IG bonds, etc. For derivatives, one may build
two categories depending on the turnover value, e.g., the most liquid contacts
and the other derivative products.

\paragraph{HQLA classes}

In this article, we focus on the transaction cost. The
asset-liability management will be studied in the third part of our
comprehensive research project on liquidity risk in asset management
\citep{Roncalli-lst3-2021}. Nevertheless, we notice that the
asset manager must develop two asset liquidity classification
matrices: liquidity buckets and HQLA classes. The term HQLA refers
to the liquidity coverage ratio (LCR) introduced in the Basel III
framework \citep{BCBS-2010, BCBS-2013a}. An asset is considered to
be a high-quality liquid asset if it can be easily converted into
cash. Therefore, the concept of HQLA is related to asset quality and
asset liquidity. It is obvious that the LST regulation is inspired
by the liquidity management regulation developed by the Basel
Committee on Banking Supervision. For instance, the redemption
coverage ratio (RCR) for asset managers is related to the liquidity
coverage ratio for banks. According to \citet{ESMA-2020}, the
redemption coverage ratio is \textquotedblleft \textsl{a
measurement of the ability of a fund's assets to meet funding
obligations arising from the liabilities side of the balance sheet,
such as a redemption shock}\textquotedblright. In
\citet{Roncalli-lst3-2021}, we will see that it is helpful to define
another asset liquidity classification matrix that is complementary
to the previous liquidity buckets. This new classification matrix
uses HQLA classes, whose goal is to group assets by their relative
liquidity risk. For instance, such asset liquidity classification
matrix is already used in the US with the Rule 22e-4(b) \citep[page
5]{Roncalli-lst1-2020}, which considers four classes: (1) highly
liquid investments, (2) moderately liquid investments, (3) less
liquid investments and (4) illiquid investments. Here is an example
based on five HQLA classes\footnote{It is derived from the liquidity
period buckets defined in the Basel III capital requirements for
market risk \citep{BCBS-2019}.}:
\begin{itemize}
\item Tier 1: Sovereign bonds (EUR, USD, GBP, AUD, JPY, SEK, CAD and
    domestic currency of the asset manager), large cap equities, specified
    currency pairs\footnote{They correspond to the 20 most liquid
    currencies: USD, EUR, JPY, GBP, AUD, CAD, CHF, MXN, CNY, NZD, RUB, HKD,
    SGD, TRY, KRW, SEK, ZAR, INR, NOK and BRL.}, bond futures, equity index
    futures, etc.

\item Tier 2: Other IG sovereign bonds, municipal bonds, small cap
    equities, other IG currency pairs, multi-name CDS, commodity futures
    (energy, precious metals, non-ferrous metals), equity options, etc.

\item Tier 3: IG corporate bonds, HY sovereign bonds, HY currency pairs,
    single-name CDS, etc.

\item Tier 4: HY corporate bonds, other commodity futures, etc.

\item Tier 5: Private equities, real estate, etc.
\end{itemize}
For derivatives on interest rates, we can map them with respect to
sovereign bonds. For instance, interest rate swaps on EUR, USD, GBP, AUD,
JPY, SEK and CAD are assigned to Tier 1, interest rate swaps on IG currencies
are assigned to Tier 2, interest rate swaps on HY currencies are assigned to
Tier 3, etc. For securitization products, the best approach is to classify
them with respect to their external credit rating.

\subsubsection{Defining the unit transaction cost function}

We consider that the two-regime model is the appropriate function to estimate
the transaction cost of a redemption scenario. Nevertheless, we introduce
some slight modifications, because the power-law model has been mainly
investigated in the stock market. These modifications are necessary when we
consider fixed-income products and derivatives.

\paragraph{The econometric model}

We assume that Security $i$ belongs the $j^{th}$ liquidity bucket
$\mathcal{LB}_{j}$ and rewrite the two-regime model as follows\footnote{We
have the following relationship: $\beta _{j}^{\left( \impact\right) }=\varphi
_{1}$. We also recall that $\sigma_{i,t}$ corresponds to the daily volatility
in the transaction cost formula.}:
\begin{equation}
\cost_{i}\left( q_{i}; \spread_{i,t}, \sigma_{i,t}, v_{i,t}\right) =\beta _{j}^{\left( %
\spread\right) }\spread_{i,t}+\beta _{j}^{\left( \impact\right) }\sigma_{i,t}%
\impact_{j}^{\star }\left( q_{i};v_{i,t}\right)
\label{eq:functional1}
\end{equation}%
where:%
\begin{equation}
\impact_{j}^{\star }\left( q_{i}; v_{i,t}\right) =\left\{
\begin{array}{ll}
\left( \dfrac{q_{i}}{v_{i,t}}\right) ^{\gamma _{1,j}} & \text{if }q_{i}\leq
\tilde{q}_{i,t} \\
\left( \dfrac{\tilde{q}_{i,t}}{v_{i,t}}\right) ^{\gamma _{1,j}}\left( \dfrac{%
q_{i}}{\tilde{q}_{i,t}}\right) ^{\gamma _{2,j}} & \text{if }\tilde{q}_{i,t}\leq
q_{i}\leq q_{i,t}^{+} \\
+\infty  & \text{if }q_{i}>q_{i,t}^{+}%
\end{array}%
\right.   \label{eq:functional2}
\end{equation}%
The total transaction cost of selling $q_{i}$ shares is then equal
to\footnote{In the case of a redemption scenario $q=\left( q_{1},\ldots
,q_{n}\right) $, we obtain:
\begin{equation*}
\TC\left( q\right) =\sum_{i=1}^{n}\sum_{h=1}^{h^{+}}\mathds{1}\left\{
q_{i}\left( h\right) >0\right\} \cdot \left( \alpha _{i} q_{i}\left(
h\right) +q_{i}\left( h\right) P_{i} \cost_{i}\left( q_{i};
\spread_{i,t}, \sigma_{i,t}, v_{i,t}\right) \right)
\end{equation*}%
}:%
\begin{equation*}
\TC\left( q_{i}\right) =\alpha _{i}q_{i}+Q_{i}\cost_{i}\left( q_{i}; \spread_{i,t},
\sigma_{i,t}, v_{i,t}\right)
\end{equation*}%
Compared to the conventional two-regime model, we notice the introduction of
two new parameters: $\alpha _{i}$ and $\beta _{j}^{\left( \spread\right) }$.
For some securities (e.g., derivatives), we have to pay a fixed cost for each
share, which motivates the addition of the term $\alpha _{i}q_{i}$. The
introduction of the scaling factor $\beta _{j}^{\left( \spread\right) }$ is
motivated because quoted bid-ask spreads are not always available for some
liquidity buckets $\mathcal{LB}_{j}$. In this case, we can use an empirical
model for computing $\spread_{i,t}$. From a theoretical point of view, we
should have $\beta _{j}^{\left( \spread\right) }=1$. This is the case for
equities for instance, but not necessarily the case for some fixed-income
securities. The reason is that asset managers do not necessarily face the same
bid-ask spread costs. Therefore, $\beta _{j}^{\left( \spread\right) }$ may be
less or greater than one.

\paragraph{The model parameters}

The calibration of the functional form consists in estimating at least four
parameters: $\beta _{j}^{\left( \spread\right) }$, $\beta _{j}^{\left(
\impact\right) }$, $\gamma _{1,j}$ and $\gamma _{2,j}$. We can use the method
of non-linear least squares. But we generally prefer to consider a two-stage
approach by first determining the exponents $\gamma _{1,j}$ and $\gamma
_{2,j}$ and then running a linear regression in order to obtain the OLS
estimates of $\beta _{j}^{\left( \spread\right) }$ and $\beta _{j}^{\left(
\impact\right) }$.

\begin{remark}
The parameters $\tilde{q}_{i,t}$ and $q_{i,t}^{+}$ are particular. From a
theoretical point of view, they are equal to
$\tilde{q}_{i,t}=\tilde{x}_{j}v_{i,t}$ and $q_{i,t}^{+}=x_{j}^{+}v_{i,t}$,
meaning that we have two other parameters $\tilde{x}_{j}$ and $x_{j}^{+}$
that are related to $\mathcal{LB}_{j}$. Nevertheless, for some liquidity
buckets, the asset manager may choose to define the trading limit
$q_{i,t}^{+}$ at the security level, meaning that we have
$q_{i,t}^{+}=x_{i}^{+}v_{i,t}$. For instance, if we consider the category of
DM sovereign bonds, trading limits may be fixed by country and maturity.
Therefore, the liquidation policy may be different if we consider 10Y
US, German, French and UK government bonds. When the inflection point
$\tilde{x}_{j}$ (or $\tilde{x}_{i}$) is difficult to estimate, it can be a
fraction of the trading limit $x_{j}^{+}$ (or $x_{i}^{+}$). The most frequent
cases are $\tilde{x}_{j}=x_{j}^{+}/2$ (or $\tilde{x}_{i}=x_{i}^{+}/2$), and
$\tilde{x}_{j}=x_{j}^{+}$ (or $\tilde{x}_{i}=x_{i}^{+}$) if we prefer to
consider only one regime.
\end{remark}

\paragraph{The security-specific parameters}

They correspond to the bid-ask spread $\spread_{i,t}$, the volatility $\sigma
_{i,t}$ and the daily volume $v_{i,t}$. Contrary to the model parameters,
these parameters\footnote{In some cases, they also include $\tilde{q}_{i,t}$
and $q_{i,t}^{+}$.} depend on the time $t$. They are the key elements of the
stress testing program, since their values will differ in normal and stress
regimes.\smallskip

Concerning the parameter $\spread_{i,t}$, we can consider an average of the
bid-ask spread observed during a normal period (e.g., the last month) or we
can use the daily quoted bid-ask spread in the case of stocks. For some
fixed-income securities (e.g., corporate bonds, securitization products,
etc.), quoted bid-ask spreads are not always available. In this case, we can
use a statistical model that depends on the characteristics of the security.
A simple model may distinguish bid-ask spreads by credit ratings\footnote{In
this case, we assume that the bid-ask spread decreases with the credit
quality, implying that the bid-ask spread of \textsf{AAA}-rated bonds is less
than the bid-ask spread of \textsf{BBB}-rated bonds. Generally, credit
ratings are grouped in order to form three or four categories.}. A more
sophisticated model may use intrinsic bond features such as maturity,
notional outstanding, coupon value, credit rating, industrial sector, etc.
\citep{BenSlimane-2017, Jurksas-2018, Feldhutter-2018, Guo-2019}.\smallskip

The parameter $\sigma _{i,t}$ measures the volatility of the asset $i$ at
time $t$. In the normal regime, $\sigma _{i,t}$ is measured with the
historical volatility. We can consider a long-term volatility using a study
period of three months, or we can consider a short-term estimator such as the
exponentially weighted moving average (EWMA) volatility, the two-week
empirical volatility or the GARCH volatility. In this last case, the
volatility rapidly changes on a daily basis, and we can observe jumps in the
transaction cost for the same securities from one day to the next. Therefore,
we think that it is better to use a long-term estimator, in particular
because the stress regime will incorporate these abnormal high-volatility
regimes. For some securities, the daily volatility is not the most
appropriate measure for measuring their risk. Therefore, it may be convenient
to define $\sigma _{i,t}$ as a function of the security characteristics. For
instance, we show in Appendix \ref{appendix:sigma-bond} on page
\pageref{appendix:sigma-bond} that the main component of a corporate bond's
volatility is the duration-times-spread (or DTS) of the bond%
\footnote{See Section \ref{section:application-corp-bonds}
on page \pageref{section:application-corp-bonds}.}.\smallskip

The third security-specific parameter is the daily volume $v_{i,t}$. As for the
volatility, we can use a short-term or a long-term measure. For instance, we
can use the daily volume of the previous day. However, there is a consensus
to use a longer period and to consider the three-month average daily volume.
Again, we can alternatively use a statistical model when the data of daily
volumes are not available. For instance, it can be a function of the
outstanding amount for bonds, the turnover for derivatives, etc.\smallskip

The trading limit $q_{i,t}^{+}$ has a particular status because it may be
either a security-specific parameter or a model parameter. When it is a
security-specific parameter, the asset manager defines $q_{i,t}^{+}$ at a low
frequency, for instance every year or when there is a market change for
trading the security $i$. However, the most frequent case is to consider
$q_{i,t}^{+}$ as a model parameter: $q_{i,t}^{+}=x_{j}^{+}v_{i,t}$. In this
situation, the asset manager generally uses the traditional rule of thumb
$x_{j}^{+}=\mathcal{LR}_{j}^{+}$ where $\mathcal{LR}_{j}^{+}$ is the
liquidation policy ratio of the liquidity bucket $\mathcal{LB}_{j}$. A
typical value is $10\%$ in the case of the stock market.

\subsubsection{Calibration of the risk parameters in the stress regime}
\label{section:stress-testing}

According to \citet{Roncalli-2020}, there are three main approaches to
generate a stress scenario: historical, macro-economic and probabilistic.
However, in the case of asset management, the first two categories are more
relevant, because asset managers do not have the same experience as banks in
this domain, and data on transaction costs under stress periods are scarce.
In this case, it is better to implement the probabilistic approach using the
method of multiplicative factors.\smallskip

As explained previously, the values of the security-specific parameters allow
to distinguish the normal period and the stress period. The model parameters
do not change, meaning that we use the same unit transaction cost function whatever
the study period. It follows that the risk parameters are the bid-ask spread,
the volatility and the volume. Therefore, asset liquidity stress testing
leads to stressing the values of these three parameters.

\paragraph{Historical stress scenarios}

The underlying idea of historical stress testing is to define the triple
$\left( \spread_{i}^{\mathrm{stress}},\sigma _{i}^{\mathrm{stress}},
v_{i}^{\mathrm{stress}}\right) $ from the sample $\left\{ \left(
\spread_{i,t},\sigma _{i,t},v_{i,t}\right) ,t\in
T^{\mathrm{stress}}\right\} $ where $T^{\mathrm{stress}}$
is the stress period and then to compute the stress transaction cost function:
\begin{equation}
\cost_{i}^{\mathrm{stress}}\left( q_{i}\right) :=\cost_{i}\left( q_{i};
\spread_{i}^{\mathrm{stress}},\sigma _{i}^{\mathrm{stress}},v_{i}^{\mathrm{stress}}\right)
\label{eq:stress1}
\end{equation}%
For instance, we can consider the empirical mean or the empirical
quantile\footnote{For the volume, we consider the empirical quantile
$1-\alpha $.} at the confidence level $\alpha $ (e.g., $\alpha =99\%$). Since
this method seems to be very simple, we face a drawback because the triple
$\left( \spread_{i}^{\mathrm{stress}},\sigma
_{i}^{\mathrm{stress}},v_{i}^{\mathrm{stress}}\right) $ does not necessarily
occur at the same trading day. A more coherent approach consists in computing
the trading cost for all days that make up the stress period and taking the
supremum:
\begin{equation}
\cost_{i}^{\mathrm{stress}}\left( q_{i}\right) :=\sup_{t\in
T^{\mathrm{stress}}}\cost_{i}\left( q_{i};\spread_{i,t},\sigma
_{i,t},v_{i,t}\right)   \label{eq:stress2}
\end{equation}

\begin{remark}
An alternative approach is to implement the worst-case scenario. The
underlying idea is to consider one stress period or several stress periods
and to consider the worst-case value: $\spread_{i}^{\mathrm{wcs}}=\max_{t\in
T^{\mathrm{stress}}}\spread_{i,t}$, $\sigma _{i}^{\mathrm{wcs}}
=\max_{t\in T^{\mathrm{stress}}}\sigma _{i,t}$ and
$v_{i}^{\mathrm{wcs}}=\min_{t\in T^{\mathrm{stress}}}v_{i,t}$. By
construction, we verify the relationship $\cost_{i}^{\mathrm{wcs}}\left(
q_{i}\right) \geq \cost_{i}^{\mathrm{stress}}\left( q_{i}\right) $.
\end{remark}

\begin{remark}
According to \citet[page 12, \S 31]{ESMA-2020},
\textquotedblleft\textsl{historical scenarios for LST could include the
2008-2010 global financial crisis or the 2010-2012 European debt crisis}\textquotedblright.
\end{remark}

\paragraph{Conditional stress scenarios}

In the case of macro-economic (or conditional) stress testing, the goal is to
estimate the relationship between risk parameters and risk factors that
define a stress scenario, and then deduce the stress value of these risk
parameters \citep[page 909]{Roncalli-2020}. Let $p_{i}$ be a parameter
($\spread_{i}$, $\sigma _{i}$ or $v_{i}$). First, we consider the linear
factor model:
\begin{equation}
p_{i,t}=\beta _{0}+\sum_{k=1}^{m}\beta _{k}\mathcal{F}_{k,t}+\varepsilon
_{i,t}  \label{eq:stress3}
\end{equation}%
where $\varepsilon _{i,t}\sim \mathcal{N}\left( 0,\sigma _{\varepsilon
_{i}}^{2}\right) $ and $\left( \mathcal{F}_{1,t},\ldots
,\mathcal{F}_{m,t}\right) $ is the set of risk factors at time $t$. Then, the
estimates $\left( \hat{\beta}_{0},\hat{\beta}_{1},\ldots
,\hat{\beta}_{m}\right) $ are deduced from the method of ordinary least
squares or the quantile regression. Finally, we translate the stress scenario
on the risk factors $\left( \mathcal{F}_{1}^{\mathrm{stress}},\ldots
,\mathcal{F}_{m}^{\mathrm{stress}}\right) $ into a stress scenario on the
risk parameter:
\begin{equation}
p_{i}^{\mathrm{stress}}=\hat{\beta}_{0}+\sum_{k=1}^{m}\hat{\beta}_{k}%
\mathcal{F}_{k}^{\mathrm{stress}} \label{eq:stress4}
\end{equation}

\begin{remark}
From a practical point of view, pooling the data for the same liquidity class
offers a more robust basis for estimating the coefficients $\left( \beta
_{0},\beta _{1},\ldots ,\beta _{m}\right) $. This is why the estimation may
use the panel data analysis with fixed effects instead of the classic
linear regression.
\end{remark}

\begin{remark}
Concerning risk factors, we can use those provided by the \textquotedblleft
\textit{Dodd-Frank Act stress testing}\textquotedblright\ (DFAST) that was
developed by the Board of Governors of the Federal Reserve System
\citep{FRB-2017}. They concern activity, interest rates, inflation and market
prices of financial assets.
\end{remark}

\paragraph{The method of multiplicative factors}

Conditional stress testing is the appropriate approach for dealing with
hypothetical stress scenarios. Nevertheless, it is not obvious to find an
empirical relationship between the risk factors $\left(
\mathcal{F}_{1,t},\ldots , \mathcal{F}_{m,t}\right) $ and the risk parameters
$\left( \spread_{i,t},\sigma _{i,t},v_{i,t}\right) $. This is why it is
better to use the method of multiplicative factors to generate hypothetical
scenarios. This approach assumes that there is a relationship between the
stress parameter and its normal value:
\begin{equation}
p_{i}^{\mathrm{stress}}=m_{p}p_{i}^{\mathrm{normal}}  \label{eq:stress5}
\end{equation}%
where $m_{p}$ is the multiplicative factor. Therefore, defining the
hypothetical stress scenario is equivalent to applying the multiplicative
factors to the current values of the risk parameters:
\begin{equation}
\left( \spread_{i}^{\mathrm{stress}},\sigma _{i}^{\mathrm{stress}},v_{i}^{%
\mathrm{stress}}\right) :=\left( m_{s}\spread_{i,t},m_{\sigma }\sigma
_{i,t},m_{v}v_{i,t}\right)   \label{eq:stress6}
\end{equation}%
In this approach, the hypothetical stress scenario is determined by the
triple $\left( m_{s},m_{\sigma },m_{v}\right) $.

\subsection{Measuring the portfolio distortion}
\label{section:portfolio-distortion}

If we consider the proportional rule $q\propto \omega $ (vertical slicing approach),
the portfolio distortion is equal to zero, but we may face high liquidation costs because
of some illiquid securities. On the contrary, we can concentrate the
liquidation on the most liquid securities (waterfall approach), but there is a risk
of a high
portfolio distortion. Therefore, we have a trade-off between the liquidation
cost and the portfolio distortion.\smallskip

\begin{figure}[tbph]
\centering
\caption{Optimal portfolio liquidation}
\label{fig:optimal4}
\figureskip
\includegraphics[width = \figurewidth, height = \figureheight]{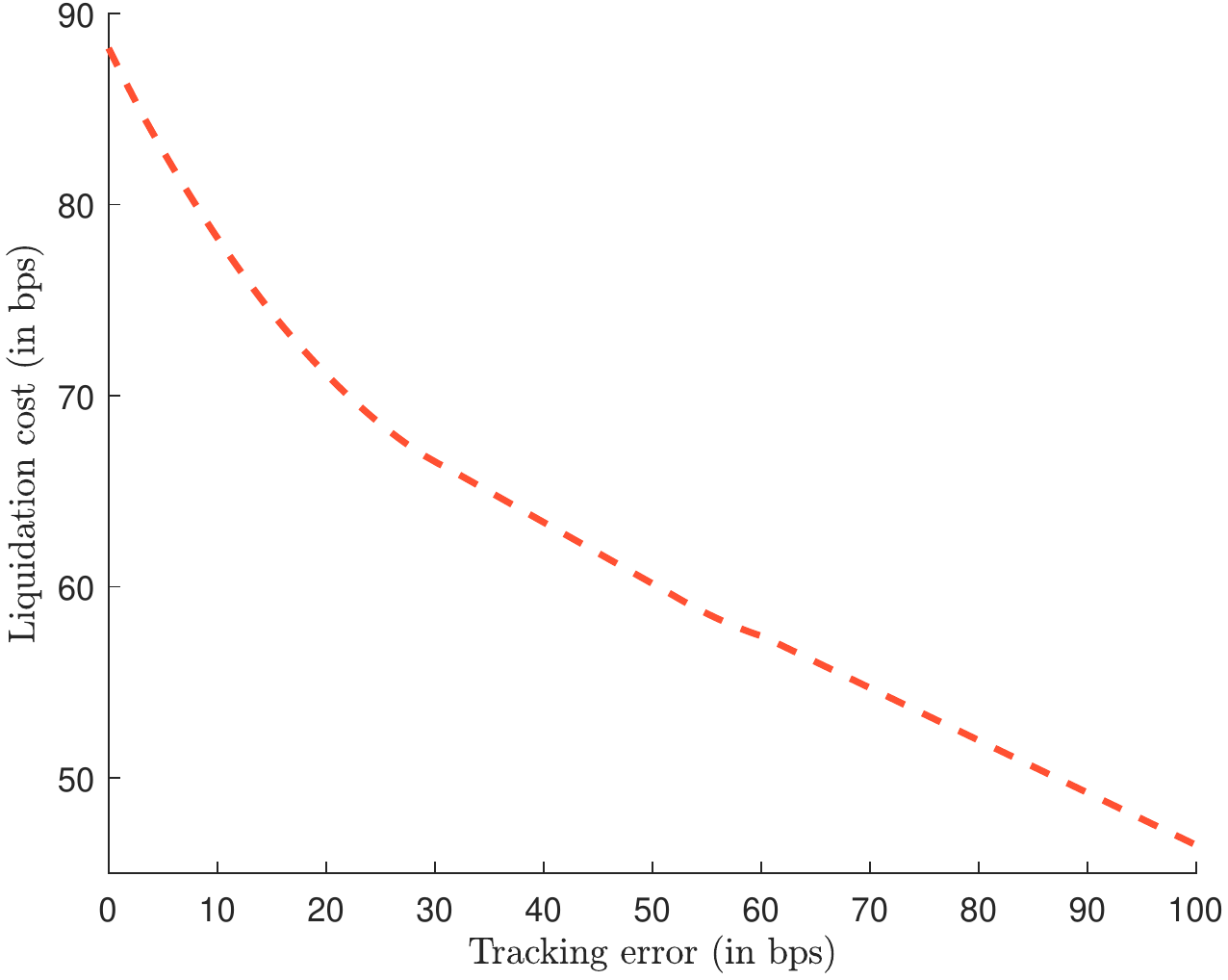}
\end{figure}

In Appendix \ref{appendix:distortion-optimal} on page \pageref{appendix:distortion-optimal},
we show that the optimal portfolio liquidation
can be obtained using the following optimization problem:
\begin{eqnarray}
q^{\star }\left( \lambda \right)  &=&\arg \min \ \frac{1}{2}\sigma
^{2}\left( q\mid \omega \right) +\lambda c\left( q\mid \omega \right)  \\
&&\text{s.t.}\left\{
\begin{array}{l}
\mathbf{1}_{n}^{\top }w\left( \omega -q\right) =1 \\
w^{-}\left( \omega -q\right) \leq w\left( \omega -q\right) \leq w^{+}\left(
\omega -q\right)
\end{array}%
\right.   \notag
\end{eqnarray}%
where $\sigma \left( q\mid \omega \right) $ is the tracking error due to the
redemption and $c\left( q\mid \omega \right) $ is the liquidation cost.
The portfolio distortion is then measured by the tracking error between
the portfolio before the redemption and the portfolio after the redemption.
Using the optimization problem, we can find liquidation portfolios that induce
a lower transaction cost than the proportional rule for the same redemption
amount $\mathbb{R}$. The downside is that they also generate a tracking error.
Let us illustrate this trade-off with the following
example\footnote{The correlation matrix of asset returns is equal to:
\begin{equation*}
\rho =\left(
\begin{array}{rrrrr}
100\% &  &  &  &  \\
10\% & 100\% &  &  &  \\
40\% & 70\% & 100\% &  &  \\
50\% & 40\% & 80\% & 100\% &  \\
30\% & 30\% & 50\% & 50\% & 100\%%
\end{array}%
\right)
\end{equation*}}:
\begin{equation*}
\begin{tabular}{cccccc}
\hline
Asset     & \multicolumn{1}{c}{$\qquad 1\qquad$} & \multicolumn{1}{c}{$\qquad 2\qquad$} &
\multicolumn{1}{c}{$\qquad 3\qquad$} & \multicolumn{1}{c}{$\qquad 4\qquad$} & \multicolumn{1}{c}{$\qquad 5\qquad$} \\
\hline
$\omega_i$            & $20\,000$ & $20\,000$ & $18\,000$ & ${\TsV}9\,000$ & ${\TsV}8\,000$ \\
$P_i$ (in \$)         &      $80$ &     $100$ &     $130$ &          $120$ &           $90$ \\
$\sigma_i$ (in \%)    &      $30$ &      $30$ &      $30$ &           $15$ &           $15$ \\
$\spread_i$ (in bps)  &      $10$ &      $10$ &      $10$ &            $5$ &            $5$ \\
$v_i$                 & $10\,000$ & $10\,000$ & $10\,000$ &      $20\,000$ &      $20\,000$ \\
\hline
\end{tabular}
\end{equation*}
The transaction cost function is given by the SQRL model
with $\varphi_1 = 1$, $\tilde{x} = 5\%$ and $x^{+} = 10\%$.
In Figure \ref{fig:optimal4}, we report the efficient frontier of liquidation.
We notice that the proportional rule implies a transaction cost of $88$ bps.
In order to reduce this cost, we must accept a tracking error risk. For instance,
if we reduce the transaction cost to $70$ bps, the liquidation has generated $22$ bps
of tracking error risk.
\smallskip

Therefore, managing the asset liquidity risk is not only a question
of transaction cost, but also a question of portfolio management.
Indeed, the fund manager may choose to change the portfolio
allocation in a stress period by selling the most liquid assets
in order to fulfill the redemptions. The fund manager may also choose
to maintain an exposure on some assets in the event of
a liquidity crisis. In these situations, the proportional rule
is not optimal and depends on the investment constraints. For instance,
the definition of the optimal liquidation policy is not the same for active managers
and passive managers. This is why liquidity stress testing on the asset side
is not only a top-down approach, but must also be completed by a bottom-up approach.

\begin{remark}
The liquidation tracking error is the right measure for assessing the portfolio distortion
in the case of an equity portfolio:
\begin{eqnarray*}
\mathcal{D}\left( q\mid \omega \right)  &=&\sigma \left( q\mid \omega
\right)  \\
&=&\sqrt{\left( w\left( \omega \right) -w\left( \omega -q\right) \right)
^{\top }\Sigma \left( w\left( \omega \right) -w\left( \omega -q\right)
\right) }
\end{eqnarray*}%
where $w\left( \omega \right) $ is the vector of portfolio weights before
the redemption, $w\left( \omega -q\right) $ is the vector of portfolio
weights after the redemption and $\Sigma $ is the covariance matrix of stock
returns. For a bond portfolio, it can be replaced by the liquidation active risk,
which measures the active risk due to the redemption:
\begin{equation*}
\mathcal{D}\left( q\mid \omega \right) =\mathcal{AR}\left( q\mid \omega
\right)
\end{equation*}%
The active risk can be measured with respect to the modified duration (MD) or
the duration-times-spread (DTS). We can also use a hybrid approach by
considering the average of the MD and DTS active risks:
\begin{eqnarray*}
\mathcal{AR}\left( q\mid \omega \right)  &=&\frac{1}{2}\sum_{j=1}^{n_{%
\mathcal{S}ector}}\left( \sum_{i\in \mathcal{S}ector_{j}}\left( w_{i}\left(
\omega -q\right) -w_{i}\left( \omega \right) \right) \limfunc{MD}%
\nolimits_{i}\right) ^{2}+ \\
&&\frac{1}{2}\sum_{j=1}^{n_{\mathcal{S}ector}}\left( \sum_{i\in \mathcal{S}%
ector_{j}}\left( w_{i}\left( \omega -q\right) -w_{i}\left( \omega \right)
\right) \limfunc{DTS}\nolimits_{i}\right) ^{2}
\end{eqnarray*}%
where $n_{\mathcal{S}ector}$ is the number of sectors,
$\limfunc{MD}\nolimits_{i}$ is the modified duration of Bond $i$ and
$\limfunc{DTS}\nolimits_{i}$ is the duration-times-spread of Bond $i$.
\end{remark}

\section{Application to stock and bond markets}

The accuracy of the model and the calibration is an issue. Indeed, we may
wonder which accuracy we must target for a liquidity stress testing
exercise, given that there are multiple unknowns in a liquidity crisis. In
particular, the LST model may be different from the proprietary pre-trade
model and less precise because of two main reasons. First, in a liquidity
stress testing exercise, we are more interested in the global figures at the
fund manager level, the asset class level and the asset manager level, and
less interested in the figures at the portfolio (or security) level.
Second, the model must be simple in order to identify the stress
parameters. This is why the LST market impact model used by the risk
department may be less accurate than the pre-trade model used by the trading
desk, because the challenges are very different. The framework presented
above is not complex enough for order execution\footnote{Nevertheless,
\citet{Curato-2017} tested different pre-trade order models
and concluded that \textquotedblleft \textsl{a fully satisfactory and
practical model of market impact [...] seems to be still lacking}%
\textquotedblright. As such, pre-trade models are not yet
completely accurate, except perhaps for large cap equities.}, but it is
sufficiently flexible and accurate to give the right order of magnitude for
liquidity stress testing purposes.\smallskip

In our analytical framework, we recall that the backbone of the LST exercise
on the asset side is given by Equations (\ref{eq:functional1}) and
(\ref{eq:functional2}) on  page \pageref{eq:functional1}, and Equation
(\ref{eq:stress1}) on page \pageref{eq:stress1}:
\begin{enumerate}
\item for each liquidity bucket $\mathcal{LB}_{j}$, we have to estimate the
parameters $\beta _{j}^{\left( \spread\right) }$,
$\beta _{j}^{\left( \impact\right) }$, $\gamma _{1,j}$ and $\gamma _{2,j}$ of
the unit transaction cost model;

\item for each security $i$, we have to define the bid-ask spread
$\spread_{i,t}$, the volatility $\sigma _{i,t}$ and the daily volume $v_{i,t}$;

\item we also have to specify the inflection point
$\tilde{q}_{i,t}=\tilde{x}_{j}v_{i,t}$:

\begin{enumerate}
\item we generally estimate $\tilde{x}_{j}$ at the level of the liquidity
bucket;

\item if $\tilde{q}_{i,t}=q_{i,t}^{+}$, there is only one regime, implying
that the parameters $\gamma _{2,j}$ and $\tilde{x}_{j}$ vanish;
\end{enumerate}

\item we then have to specify the trading limit $q_{i,t}^{+}$ for each
security; except for large cap equities and some sovereign bonds, we use the
proportional rule $q_{i,t}^{+}=x_{j}^{+}v_{i,t}$, where $x_{j}^{+}$ is
the maximum trading limit of the liquidity bucket $\mathcal{LB}_{j}$ defined by
the asset manager's risk department;

\item finally, we have to specify how the three security parameters are
stressed: $\spread_{i}^{\mathrm{stress}}$, $\sigma _{i}^{\mathrm{stress}}$
and $v_{i}^{\mathrm{stress}}$.
\end{enumerate}

It is obvious that the key challenge of the LST calibration is
data availability. Since the LST model may include a lot of parameters, we
suggest proceeding step by step. For instance, as a first step, we may calibrate
the model for all global equities. Then, we may distinguish between
large cap and small cap equities. Next, we may consider an LST model region
by region (e.g., US, Eurozone, UK, Japan, etc.), and so on. In the early stages,
we may also use expert judgement in order to fix some parameters, for
instance $\gamma _{2,j}$, $\tilde{x}_{j}$, etc. Some parameters are also
difficult to observe. For instance, the bid-ask spread $\spread_{i,t}$ and
the trading volume $v_{i,t}$ are not available for many bonds. This is why we
use a model or an approximation formula. For example, we can replace the
trading volume $v_{i,t}$ by the notional outstanding amount $\nshares_{i}$.
The volume-based participation rate $x_{i}=v_{i,t}^{-1}q_{i}$ is then replaced by
the outstanding-based participation rate
$y_{i}=\nshares_{i}^{-1}q_{i}$, implying that we have to calibrate the
scaling factor $\beta _{j}^{\left( \impact\right) }$ in order to take into
account this new parameterization. We can also use the rule $V_{i,t}=\xi
\mcap_{i,t}$ where $\xi $ is the proportionality factor between volume
and outstanding amount. Moreover, the volatility parameter is not always
pertinent in the case of bonds, and it may be better to use the
duration-times-spread (DTS).

\begin{remark}
In this section, we remove the reference to the liquidity bucket $\mathcal{LB}_j$
in order to reduce the amount of notation when it is possible. This concerns
the four parameters $\beta _{j}^{\left( \spread\right) }$,
$\beta _{j}^{\left( \impact\right) }$, $\gamma _{1,j}$ and $\gamma _{2,j}$.
Moreover, we consider the calibration of the single-regime model as a first step:
\begin{eqnarray}
\cost_{i}\left( q_{i};\spread_{i,t},\sigma _{i,t},v_{i,t}\right)  &=&\beta
^{\left( \spread\right) }\spread_{i,t}+\beta ^{\left( \impact\right) }\sigma
_{i,t}\left( \dfrac{q_{i}}{v_{i,t}}\right) ^{\gamma _{1}}  \notag \\
&=&\beta ^{\left( \spread\right) }\spread_{i,t}+\beta ^{\left( \impact%
\right) }\sigma _{i,t}x_{i,t}^{\gamma _{1}}  \label{eq:equity1}
\end{eqnarray}%
The second regime is calibrated during the second step as shown in Section
\ref{section:application-two-regime} on page \pageref{section:application-two-regime}.
We also assume that the annualized volatility is scaled by the
factor $1/\sqrt{260}$ in order to represent a daily volatility measure. This
helps to understand the magnitude of the parameter
$\beta ^{\left( \impact\right) }$. By default, we can then consider that
$\beta ^{\left( \impact\right) }\approx 1$.
\end{remark}

\subsection{The case of stocks}
\label{section:application-stocks}

\subsubsection{Large cap equities}
\label{section:application-large-cap}

We consider the dataset described in Appendix \ref{appendix:data-equity} on
page \pageref{appendix:data-equity}. We filter the data in order to keep only
the stocks that belong to the MSCI USA and MSCI Europe indices. For each observation $i$, we have the
transaction cost $\cost_{i}$, the (end-of-day) bid-ask spread $\spread_{i}$,
the participation rate $x_{i}$ and the daily volatility $\sigma _{i}$. We
first test a highly constrained statistical model:
\begin{equation}
\cost_{i}=\spread_{i}+\sigma _{i}\sqrt{x_{i}}+\varepsilon _{i}
\label{eq:equity2}
\end{equation}%
where $\varepsilon _{i}\sim \mathcal{N}\left(
0,\sigma_{\varepsilon}^{2}\right) $. We obtain $R^{2}=53.47\%$ and
$R_{c}^{2}=15.87\%$. Since we observe a large discrepancy between $R^{2}$ and
$R_{c}^{2}$, we must be careful about the interpretation of the statistical
models. This means that the average cost $\bar{\cost}$ explains a significant
part of the trading cost, implying that the dispersion of trading costs is
not very large.\smallskip

In order to improve the explanatory power of the transaction cost function,
we consider two alternative models:%
\begin{equation}
\cost_{i}=\beta ^{\left( \spread\right) }\spread_{i}+\beta ^{\left( \impact%
\right) }\sigma _{i}\sqrt{x_{i}}+\varepsilon _{i}  \label{eq:equity3}
\end{equation}%
and:%
\begin{equation}
\cost_{i}=\beta ^{\left( \spread\right) }\spread_{i}+\beta ^{\left( \impact%
\right) }\sigma _{i}x_{i}^{\gamma _{1}}+\varepsilon _{i}  \label{eq:equity4}
\end{equation}%
Model (\ref{eq:equity3}) can be seen as a special case of Model (\ref{eq:equity4})
when the exponent $\gamma_1$ is set to $\nicefrac{1}{2}$. Using the method of
non-linear least squares, we estimate the parameters, and the results are
reported in Tables \ref{tab:elab_equity_lc_calib3} and \ref{tab:elab_equity_lc_calib4}.
We notice that the assumptions $\left( \mathcal{H}_{1}\right) $ $\beta
^{\left( \spread\right) }=1$ and $\left( \mathcal{H}_{2}\right) $ $\beta
^{\left( \impact\right) }=1$ are both rejected. When the estimation of
$\gamma _{1}$ is not constrained, its optimal value is equal to $0.5873$,
which is a little bit higher than $0.5$. Nevertheless, we observe that the explanatory
powers are very close for the constrained and unconstrained models. The fact
that $\beta ^{\left( \impact\right) }$ is larger for the unconstrained model
($0.2970$ versus $0.1898$) indicates a bias in our dataset. The model tends
to overfit the lowest values of $x_i$ and not the highest value of $x_i$, which
are certainly not sufficiently represented in the dataset.\smallskip

\begin{table}[tbph]
\centering
\caption{Non-linear least squares estimation of Model (\ref{eq:equity3})}
\label{tab:elab_equity_lc_calib3}
\begin{tabular}{ccccc}
\hline
Parameter                                & Estimate &   Stderr &     $t$-student & $p$-value \\ \hline
$\beta^{\left( \spread\right) }$         & $1.4465$ & $0.0014$ &     $1049.9020$ & $0.0000$  \\
$\beta^{\left( \impact\right) }$         & $0.1898$ & $0.0030$ & ${\TsX}62.7720$ & $0.0000$  \\
$\gamma _{1}$                            & $0.5000$ & $0.0053$ & ${\TsX}93.5817$ & $0.0000$  \\ \hline
\multicolumn{5}{c}{\multirow{2}{*}{$R^{2}=98.41\% \qquad R_{c}^{2} = 97.12\%$}} \\
& & & & \\
\hline
\end{tabular}
\end{table}

\begin{table}[tbph]
\centering
\caption{Non-linear least squares estimation of Model (\ref{eq:equity4})}
\label{tab:elab_equity_lc_calib4}
\begin{tabular}{ccccc}
\hline
Parameter                                & Estimate &   Stderr &      $t$-student & $p$-value \\ \hline
$\beta^{\left( \spread\right) }$         & $1.4468$ & $0.0012$ &      $1213.2593$ & $0.0000$  \\
$\beta^{\left( \impact\right) }$         & $0.2970$ & $0.0039$ &  ${\TsX}76.0394$ & $0.0000$  \\
$\gamma _{1}$                            & $0.5873$ & $0.0044$ & ${\TsV}132.7093$ & $0.0000$  \\ \hline
\multicolumn{5}{c}{\multirow{2}{*}{$R^{2}=98.81\% \qquad R_{c}^{2} = 97.84\%$}} \\
& & & & \\
\hline
\end{tabular}
\end{table}

\begin{figure}[tbph]
\centering
\caption{Histogram of estimated parameters}
\label{fig:elab_equity_lc_calib5}
\figureskip
\includegraphics[width = \figurewidth, height = \figureheight]{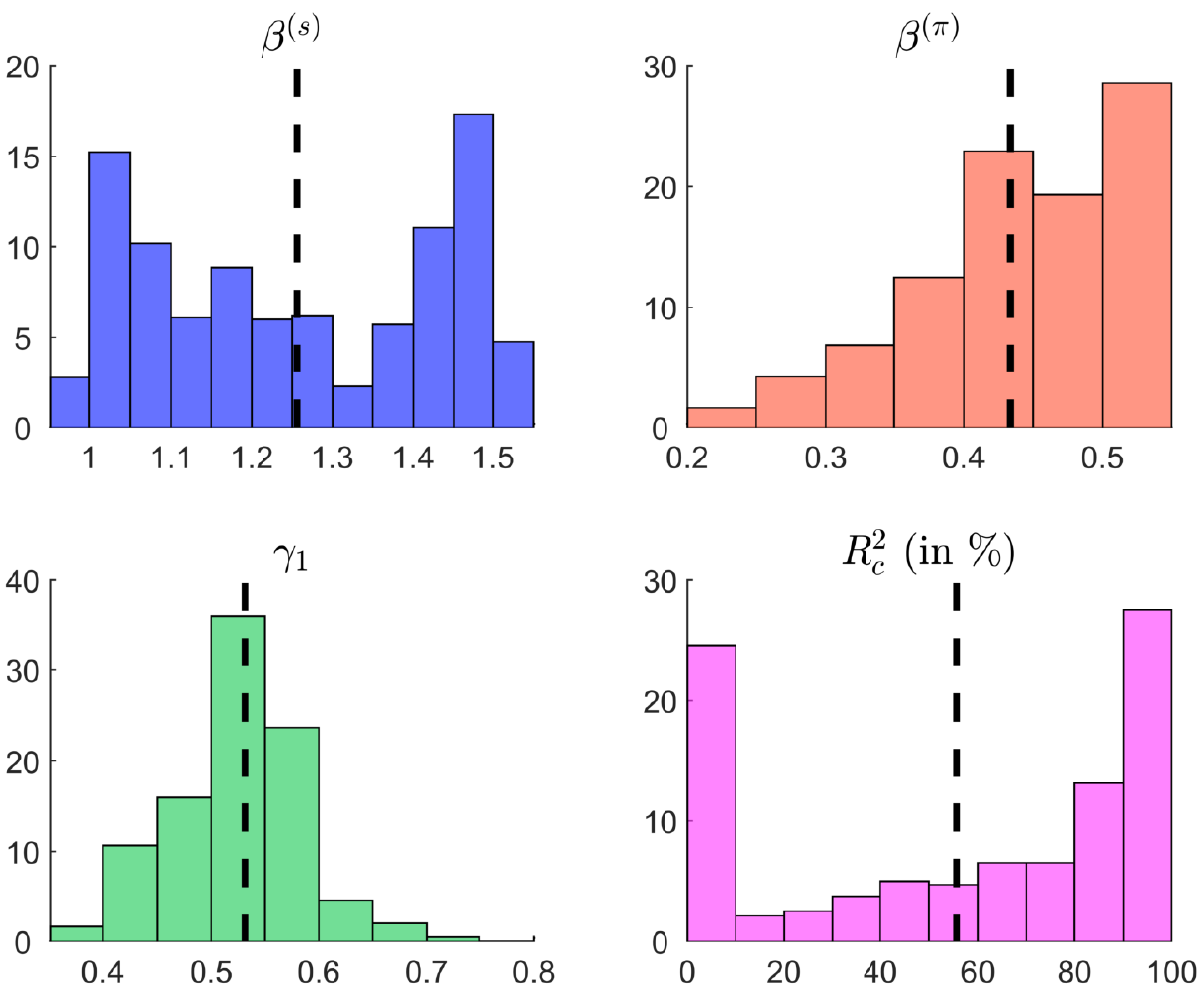}
\end{figure}

\begin{table}[tbph]
\centering
\caption{Descriptive statistics of the estimates}
\label{tab:elab_equity_lc_calib5}
\begin{tabular}{ccccccccc}
\hline
Parameter & Mean & Median & Min. & $Q\left(10\%\right)$ & $Q\left(25\%\right)$ & $Q\left(75\%\right)$ &
$Q\left(90\%\right)$ & Max. \\ \hline
$\beta^{\left( \spread\right) }$ & $1.256$ & $1.234$ & ${\TsVIII}0.992$ & $1.001$ & $1.082$ & $1.443$ & $1.487$ & $1.558$ \\
$\beta^{\left( \impact\right) }$ & $0.434$ & $0.448$ &         $-0.209$ & $0.330$ & $0.391$ & $0.500$ & $0.510$ & $0.527$ \\
$\gamma _{1}$                    & $0.531$ & $0.525$ & ${\TsVIII}0.368$ & $0.446$ & $0.488$ & $0.563$ & $0.597$ & $1.676$ \\
$R_{c}^{2}$                      & $0.557$ & $0.681$ & ${\TsVIII}0.000$ & $0.000$ & $0.094$ & $0.916$ & $0.961$ & $0.992$ \\
\hline
\end{tabular}
\end{table}

\begin{figure}[tbph]
\centering
\caption{Estimated price impact (in bps)}
\label{fig:elab_equity_lc_calib6}
\figureskip
\includegraphics[width = \figurewidth, height = \figureheight]{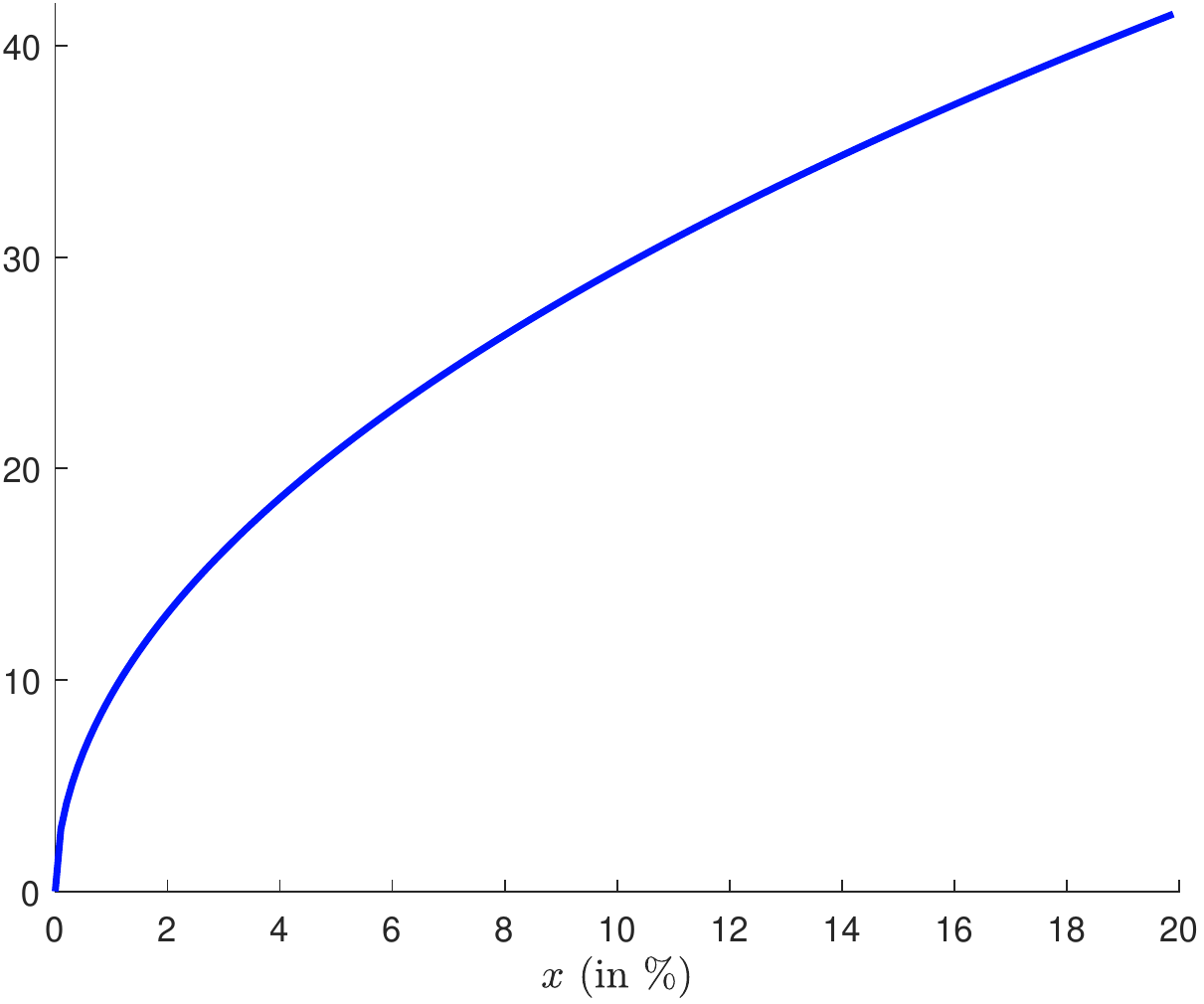}
\end{figure}

The figures taken by $R^{2}$ and $R_{c}^{2}$ are extremely high and not
realistic. This confirms that there is a bias in our dataset. To
better understand this issue, we estimate Model (\ref{eq:equity4}) for each stock. Results
are reported in Figure \ref{fig:elab_equity_lc_calib5} and
Table \ref{tab:elab_equity_lc_calib5}. On average, $R_{c}^{2}$ is equal to $55.7\%$,
which is far from the previous result. We observe that the model presents
a high explanatory power for some stocks and a low explanatory power for
other stocks (bottom/right panel in Figure \ref{fig:elab_equity_lc_calib5}).
These results highlight the heterogeneity of the database. Therefore,
estimating a transaction cost model is not easy when mixing small and
large values of transaction costs and participation rates. Finally, we
propose the following benchmark formula for the transaction cost model:
\begin{empheq}[box=\tcbhighmath]{equation}
\quad \cost_{i}\left( q_{i};\spread_{i,t},\sigma _{i,t},v_{i,t}\right) =1.25\cdot %
\spread_{i,t}+0.40 \cdot \sigma _{i,t}\sqrt{x_{i,t}} \quad \label{eq:benchmark-stock-lc}
\end{empheq}
The price impact of this function is reported in Figure \ref{fig:elab_equity_lc_calib6}
in the case where the annualized volatility of the stock return is equal to $30\%$.

\begin{remark}
We notice sensitivity of the results when we filter the data with respect to the participation
rate. For instance, we obtain:
\begin{equation*}
\cost_{i}\left( q_{i};\spread_{i,t},\sigma _{i,t},v_{i,t}\right) =1.51\cdot %
\spread_{i,t}+0.56 \cdot\sigma _{i,t}x_{i,t}^{0.78}
\end{equation*}
when we only consider the observations with a participation rate larger than $0.5\%$.
\end{remark}

\subsubsection{Small cap equities}
\label{section:application-small-cap}

\begin{figure}[tbph]
\centering
\caption{Relationship between the market capitalization and the parameter $\beta ^{\left( \spread\right) }$}
\label{fig:elab_equity_sc_calib3a}
\figureskip
\includegraphics[width = \figurewidth, height = \figureheight]{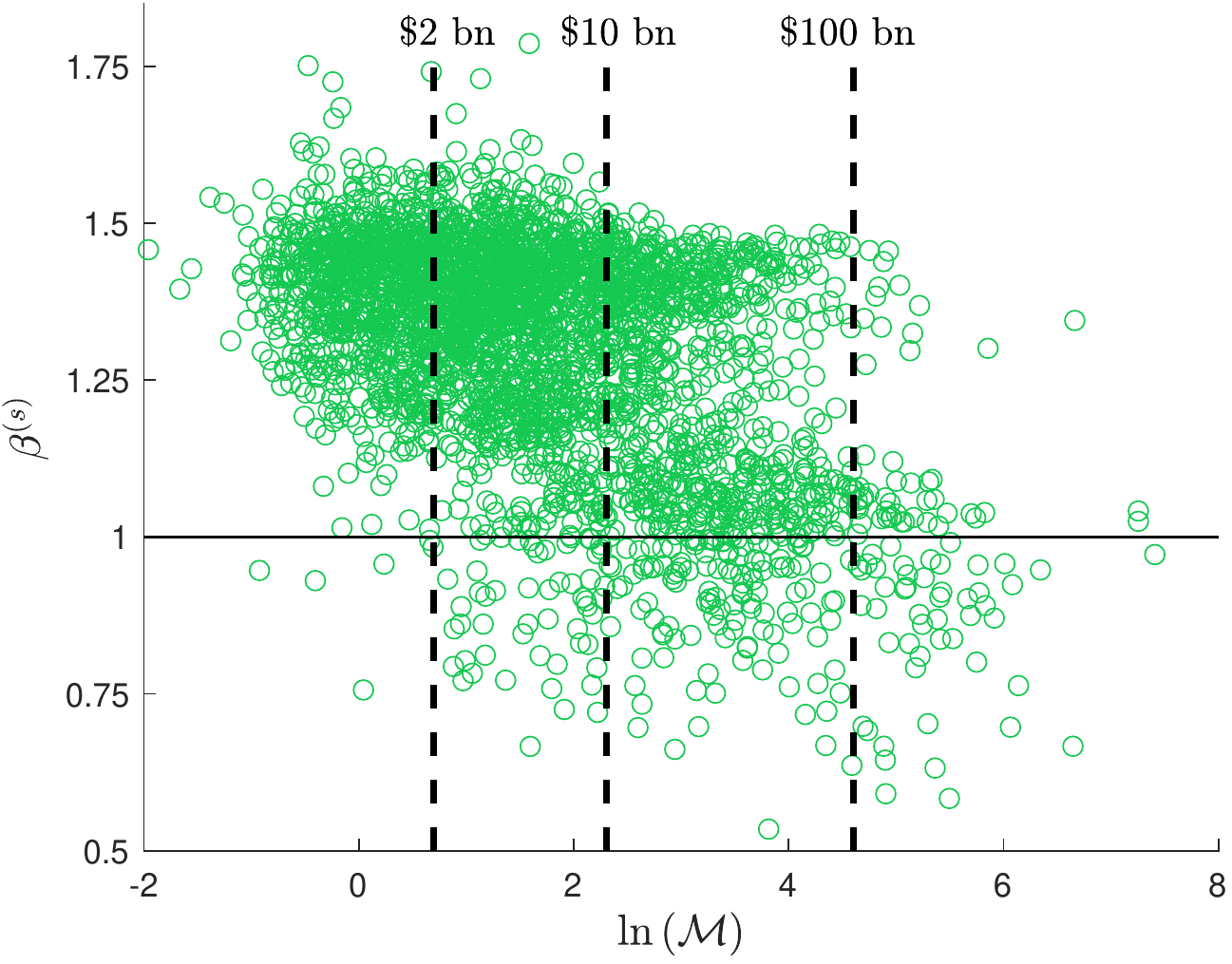}
\end{figure}

\begin{figure}[tbph]
\centering
\caption{Relationship between the market capitalization and the parameter $\beta ^{\left( \impact\right) }$}
\label{fig:elab_equity_sc_calib3b}
\figureskip
\includegraphics[width = \figurewidth, height = \figureheight]{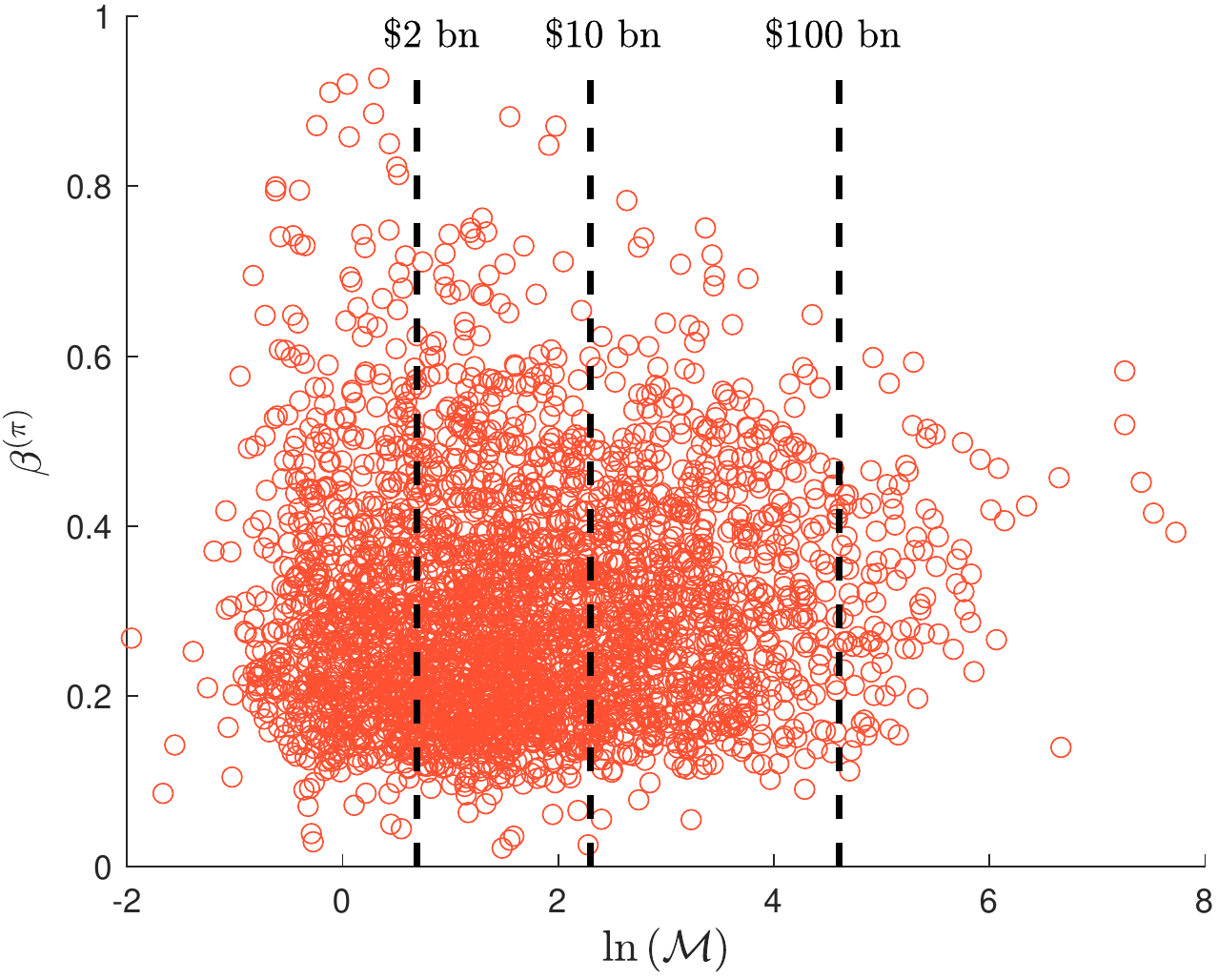}
\end{figure}

\begin{figure}[tbph]
\centering
\caption{Ratio of the parameters $\beta ^{\left( \spread\right) }$ and $\beta ^{\left( \impact\right) }$
with respect to the values of the large cap class}
\label{fig:elab_equity_sc_calib4}
\figureskip
\includegraphics[width = \figurewidth, height = \figureheight]{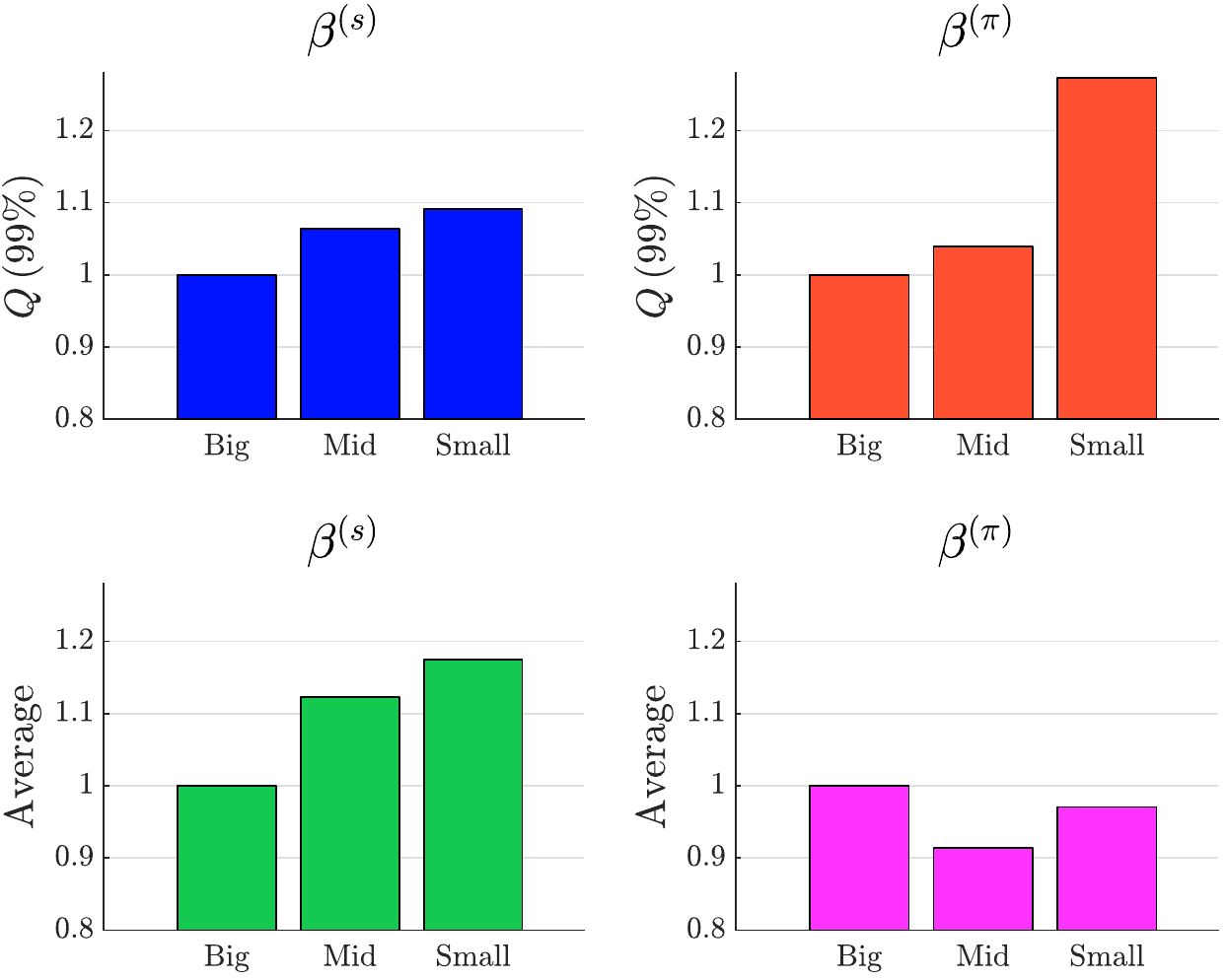}
\end{figure}

In this analysis, we consider all the stocks that belong to the MSCI USA,
MSCI Europe, MSCI USA Small Cap and MSCI Europe Small cap indices. This means
that the dataset corresponds to large cap and small cap stocks. We run the
linear regression (\ref{eq:equity3}) for the different stocks and estimate
the parameters $\beta ^{\left( \spread\right) }$ and $\beta ^{\left(
\impact\right) }$. In Figures \ref{fig:elab_equity_sc_calib3a} and
\ref{fig:elab_equity_sc_calib3b}, we report the scatterplot between the
market capitalization\footnote{In order to obtain an easy-to-read graph, the
$x$-axis corresponds to the logarithm of the market capitalization, which is
expressed in billions of US dollars.} and these parameters. On average, the
estimate of $\beta ^{\left( \spread\right) }$ is higher when the market
capitalization is low than when the market capitalization is high. In a
similar way, we observe more dispersion of the estimate $\beta ^{\left(
\impact\right) }$ for small cap stocks. In order to verify that small cap
stocks are riskier than large cap stocks, we split the stock universe into three
buckets according to market capitalization\footnote{We use the following classification:
$+\$10$ bn for large caps, $\$2$ -- $\$10$ bn for mid caps and
$-\$2$ bn for small caps.}. In Figure \ref{fig:elab_equity_sc_calib4}, we plot the
ratio of the estimates $\beta ^{\left( \spread\right) }$ and $\beta
^{\left( \impact\right) }$ with the values obtained for the large cap
class. We notice that the two parameters are larger for small cap stocks, especially
if we consider the $99\%$ quantile. To take into account this additional risk,
we propose the following benchmark formula for small cap stocks:
\begin{empheq}[box=\tcbhighmath]{equation}
\quad \cost_{i}\left( q_{i};\spread_{i,t},\sigma _{i,t},v_{i,t}\right) = 1.40\cdot %
\spread_{i,t} + 0.50 \cdot \sigma _{i,t}\sqrt{x_{i,t}} \quad \label{eq:benchmark-stock-sc}
\end{empheq}
If we compare this function with Equation (\ref{eq:benchmark-stock-lc}), we notice that
the parameter $\beta ^{\left( \spread\right) }$ is equal to $1.40$ instead of
$1.25$, implying an additional fixed transaction cost of $+12\%$ for small cap stocks.
For the parameter $\beta ^{\left( \impact\right) }$, the value is equal to
$0.50$ instead of $0.40$, implying that the price impact is $25\%$ higher
for small cap stocks.

\begin{remark}
A conservative approach consists in using the highest values of $\beta ^{\left(
\impact\right) }$. For instance, we can define $\beta ^{\left(
\impact\right) } = 0.50$ for large cap stocks and $\beta ^{\left(
\impact\right) } = 0.75$ for small cap stocks. In this case, the price impact is $50\%$ higher
for small cap stocks.
\end{remark}

\clearpage

\subsection{The case of bonds}
\label{section:application-bonds}

\subsubsection{Defining the participation rate}

The key variable of the transaction cost formula is the participation rate:%
\begin{equation*}
x=\frac{q}{v}=\frac{Q}{V}
\end{equation*}%
where $q$ is the number of shares to trade and $v$ is the daily trading
volume (expressed in number of shares). We can also formulate the
participation rate with the nominal values $Q$ and $V$ expressed in USD or
EUR. In the case of bonds, the daily trading volume is not observed.
Moreover, this statistic is not always relevant because some
bonds are traded infrequently. To illustrate this phenomenon, we can use the
zero-trading days statistic, which is defined as the ratio between the number
of days with zero trades and the total number of trading days within the
period. For instance, \citet{Hotchkiss-2017} report that $79.4\%$ of US IG bonds
and $84.1\%$ of US HY bonds are not traded monthly between January 1995 to
December 1999. \citet{Dick-Nielsen-2012} find that the median number of
zero-trading days was equal to $60.7\%$ on a quarterly basis from Q4 2004 to
Q2 2009 in the US corporate bond market.\smallskip

The turnover is a measure related to the trading volume. It is the ratio
between the nominal trading volume $V$ and the market capitalization $\mcap$ of the
security, or between the trading volume $v$ and the number of
issued shares\footnote{The market capitalization is equal to
the number of shares times the price: $\mcap = \nshares P$.}
$\nshares$:
\begin{equation*}
\turnover= \frac{V}{\mcap}=\frac{v}{\nshares}
\end{equation*}%
In the case of bonds, $\mcap$ and $\nshares$ correspond to the
outstanding amount and the number of issued
bonds. It follows that $V=\turnover \mcap$ and:
\begin{equation*}
x=\frac{Q}{\turnover \mcap}=\frac{q}{\turnover \nshares}
\end{equation*}%
We deduce that the volume-based participation rate $x$ is related to the
outstanding-based participation rate $y$:%
\begin{equation*}
y=\frac{q}{\nshares}
\end{equation*}%
The scaling factor between $y$ and $x$ is then exactly equal to
the daily turnover ratio $\turnover$.\smallskip

According to \citet{SIFMA-2021a}, the daily turnover ratio is equal to $0.36\%$
for US corporate bonds in 2019. This figure is relatively stable since it is
in the range $0.30\%-0.36\%$ between 2005 and 2019, except in 2008
where we observe a turnover of $0.26\%$. However, it was highest
before 2005. For instance, it was equal to $0.44\%$ in 2002. If we make the
distinction between IG and HY bonds, it seems that the turnover ratio is
greater for the latter. For instance, we obtain a turnover ratio of $0.27\%$
for US IG bonds and $0.65\%$ for HY bonds. In the case of US treasury
securities, the five-year average daily turnover figure is $4.6\%$
for bills, $1.2\%$ for TIPS and $3.5\%$ for notes and bonds \citep{SIFMA-2021b}.\smallskip

In the case of European bonds, statistics are only available for government bonds.
We can classify the countries into three categories \citep{AFME-2020}:
\begin{itemize}
\item The daily turnover ratio is above $1\%$ and close to $1.5\%$ for
Germany, Spain and UK.

\item The daily turnover ratio is between $0.5\%$ and $1.0\%$ for Belgium,
France, Ireland, Italy, Netherlands, and Portugal.

\item The daily turnover ratio is lower than $0.5\%$ for Denmark and Greece.
\end{itemize}
These different figures show that the turnover ratio cannot be considered as
constant. Therefore, the single-regime transaction cost function becomes:%
\begin{eqnarray}
\cost_{i}\left( q_{i};\spread_{i,t},\sigma _{i,t},v_{i,t}\right)  &=&\beta
^{\left( \spread\right) }\spread_{i,t}+\beta ^{\left( \impact\right) }\sigma
_{i,t}\left( \dfrac{q_{i}}{\turnover_{i,t}\nshares_{i}}\right) ^{\gamma _{1}}
\notag \\
&=&\beta ^{\left( \spread\right) }\spread_{i,t}+\beta _{i,t}^{\left( \impact%
\right) }\sigma _{i,t}y_{i}^{\gamma _{1}}  \label{eq:bond1}
\end{eqnarray}%
where $y_{i}=\nshares_{i}^{-1}q_{i}$ is the outstanding-based participation
rate and $\beta _{i,t}^{\left( \impact\right) }$ is the scaling factor of
the price impact:%
\begin{equation}
\beta _{i,t}^{\left( \impact\right) }=\frac{\beta ^{\left( \impact\right) }}{%
\turnover_{i,t}^{\gamma _{1}}}  \label{eq:bond2}
\end{equation}
Since the turnover ratio is time-varying and depends on the security, it
follows that $\beta _{i,t}^{\left( \impact\right) }$ depends on the time $t$
and the security $i$. Equation (\ref{eq:bond1}) for bonds is then less
attractive than Equation (\ref{eq:equity1}) for equities. However, we can
make two assumptions:
\begin{enumerate}
\item the turnover ratio $\turnover_{i,t}$ is stable on long-run periods;

\item the turnover ratio $\turnover_{i,t}$ computed at the security level is
not representative of its trading activity.
\end{enumerate}
We notice that turnover ratios are generally computed for a group
of bonds, for instance all German government bonds or all US corporate IG
bonds. The reason lies again in the fact that the daily turnover of a given
bond may be equal to zero very often because of the zero-trading days
effect. Nevertheless, if one bond is not traded at all for a given period
(e.g., a day or a week), it does not mean that it is perfectly illiquid
during this period. This may be due to a very low supply or demand during
this period. In a bullish market, if no investors want to sell some
bonds because there is strong demand and low supply, these investors are
rational to keep their bonds. Since buy-and-hold strategies dominate in bond
markets, trading a bond is a signal that the bond is not priced
fairly. In this framework, the fundamental price of a bond must
change in order to observe a trading activity on this bond. The situation in
the stock market is different because the computation of the fair price
uses a more short-term window and buy-and-hold strategies do not
dominate.\smallskip

\begin{figure}[tbph]
\centering
\caption{Relationship between the turnover $\turnover$ and the scaling factor
$\tilde{\beta}^{\left( \impact\right) }$}
\label{fig:blab_bond_beta1}
\figureskip
\includegraphics[width = \figurewidth, height = \figureheight]{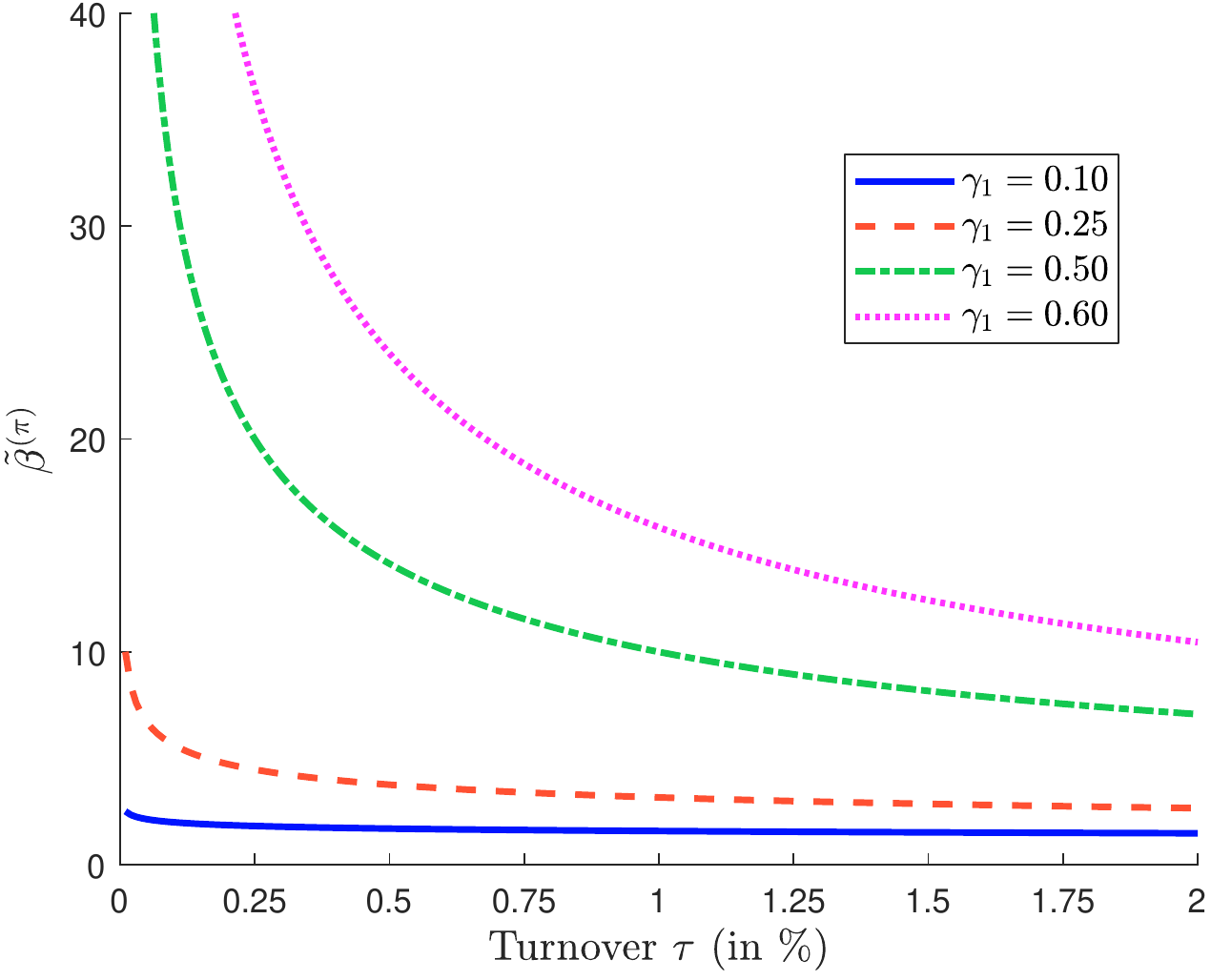}
\end{figure}

Therefore, we can assume that the turnover ratio is equal for the same
family of bonds, implying that:
\begin{equation}
\cost_{i}\left( q_{i};\spread_{i,t},\sigma _{i,t},\nshares_{i}\right) =\beta
^{\left( \spread\right) }\spread_{i,t}+\tilde{\beta}^{\left( \impact\right)
}\sigma _{i,t}y_{i}^{\gamma _{1}}  \label{eq:bond3}
\end{equation}%
This equation is similar to Equation (\ref{eq:equity1}) for equities. Nevertheless, there
is a difference between the two scaling coefficients $\beta ^{\left( \impact%
\right) }$ and $\tilde{\beta}^{\left( \impact\right) }$. The last one is
more sensitive because we have:%
\begin{equation*}
\tilde{\beta}^{\left( \impact\right) }=\frac{\beta ^{\left( \impact\right) }%
}{\turnover^{\gamma _{1}}}
\end{equation*}
The underlying idea is then to consider more granular liquidity buckets
$\mathcal{LB}_{j}$ for the bond asset class than the equity asset class in
order to be sure that the securities belonging to the same liquidity
bucket have a similar turnover ratio $\turnover$.
In Figure \ref{fig:blab_bond_beta1}, we
report the relationship between $\turnover$ and
$\tilde{\beta}^{\left( \impact\right) }$ for several values of the exponent
$\gamma_{1}$. When $\gamma _{1}$ is low, the impact of $\turnover$ on
$\tilde{\beta}^{\left( \impact\right) }$ is very low, meaning that we can
consider $\tilde{\beta}^{\left( \impact\right) }$ as a constant.
However, when $\gamma _{1}$ is high (greater than $0.25$), the turnover
may have a high impact and $\tilde{\beta}^{\left( \impact\right) }$
cannot be assumed to be a constant. In the first case, the estimation of
$\beta^{\left( \spread\right) }$ and $\tilde{\beta}^{\left( \impact\right) }$
is robust. In the second case, the estimation of $\tilde{\beta}^{\left( \impact\right) }$
only makes sense if the turnover is
comparable between the securities of the liquidity bucket $\mathcal{LB}_{j}$.

\begin{remark}
In Table \ref{tab:blab_bond_beta2}, we report the values of the outstanding-based participation rate
with respect to the volume-based participation rate $x$ and the daily turnover $\tau$.
For example, if $x = 30\%$ and $\tau = 4\%$, we obtain a participation rate of $1.2\%$.
While volume-based participation rates are expressed in \%, we conclude that
outstanding-based participation rates are better expressed in bps.
\end{remark}

\begin{table}[tbph]
\centering
\caption{Outstanding-based participation rate (in bps) with respect to $x$ and $\tau$}
\label{tab:blab_bond_beta2}
\begin{tabular}{c|ccccccccc}
\hline
$\tau$ & \multicolumn{9}{c}{$x$ (in \%)} \\
(in \%) &  0.01 &  0.05   & 0.10   & 0.50   &     1 &           5 &        10 &   20 &         30 \\ \hline
$0.5$ & $0.005$ & $0.025$ & $0.05$ & $0.25$ & $0.5$ & ${\TsV}2.5$ & ${\TsV}5$ & $10$ & ${\TsV}15$ \\
$1.0$ & $0.010$ & $0.050$ & $0.10$ & $0.50$ & $1.0$ & ${\TsV}5.0$ &      $10$ & $20$ & ${\TsV}30$ \\
$2.0$ & $0.020$ & $0.100$ & $0.20$ & $1.00$ & $2.0$ &      $10.0$ &      $20$ & $40$ & ${\TsV}60$ \\
$4.0$ & $0.040$ & $0.200$ & $0.40$ & $2.00$ & $4.0$ &      $20.0$ &      $40$ & $80$ &      $120$ \\
\hline
\end{tabular}
\end{table}

\subsubsection{Sovereign bonds}
\label{section:application-sov-bonds}

We consider a dataset of sovereign bond trades, whose
description is given in Appendix \ref{appendix:data-sovereign-bond}
on page \pageref{appendix:data-sovereign-bond}. For each
observation $i$, we have the transaction cost $\cost_{i}$, the spread
$\spread_{i}$, the outstanding-based participation rate $y_{i}$ and the daily
volatility $\sigma _{i}$. We run a two-stage regression model:%
\begin{equation}
\left\{
\begin{array}{ll}
\ln \left( \cost_{i}-\spread_{i}\right) -\ln \sigma _{i}=c_{\gamma }+\gamma
_{1}\ln y_{i}+u_{i} & \text{if }\cost_{i}>\spread_{i} \\
\cost_{i}=c_{\beta }+\beta ^{\left( \spread\right) }\spread_{i}+\mathcal{D}%
_{i}^{\left( \impact\right) }\tilde{\beta}^{\left( \impact\right) }\sigma
_{i}y_{i}^{\gamma _{1}}+v_{i} &
\end{array}%
\right.   \label{eq:bond4}
\end{equation}%
where $c_{\gamma }$ and $c_{\beta }$ are two intercepts, and $u_{i}$ and $%
v_{i}$ are two residuals. Since the transaction cost can be lower than the
bid-ask spread\footnote{%
We recall that the bond market is not an electronic market. Bid-ask spreads
are generally declarative and not computed with quoted bid and ask prices.},
we introduce the dummy variable $\mathcal{D}_{i}^{\left( \impact\right) }=%
\mathds{1}\left\{ \cost_{i}>\spread_{i}\right\} $. We estimate the exponent $\gamma _{1}$ using the
first linear regression model. Then, we estimate the parameters $\beta
^{\left( \spread\right) }$ and $\tilde{\beta}^{\left( \impact\right) }$
using the second linear regression by considering the OLS estimate of
$\gamma _{1}$. Results are given in Table \ref{tab:blab_sovereign_calib1}.
We obtain $\gamma_{1}=0.2037 \ll 0.5$, which is lower than the standard value
for equities. We also obtain $\beta ^{\left( \spread\right) }=0.9099$ and
$\tilde{\beta}^{\left( \impact\right) }=2.1521$. Curiously, the value of
$\beta ^{\left( \spread\right) }$ is less than one. One possible explanation is that
we use trades from a big asset manager that may have a power to negotiate
and the capacity to trade inside the bid-ask spreads when the participation rate
is low. Nevertheless, the explanatory power of the model
is relatively good. Indeed, we obtain $R^{2}=39.87\%$ and $R^{2}_{c}=28.94\%$.\smallskip

\begin{table}[t]
\centering
\caption{Two-stage estimation of the sovereign bond transaction cost model}
\label{tab:blab_sovereign_calib1}
\begin{tabular}{ccccc}
\hline
Parameter                                & Estimate &   Stderr &     $t$-student & $p$-value \\ \hline
$c_{\gamma }$                            & $0.3004$ & $0.0500$ &  ${\TsX}6.0096$ &  $0.0000$ \\
$\gamma _{1}$                            & $0.2037$ & $0.0046$ & ${\TsV}44.6050$ &  $0.0000$ \\
$c_{\beta }$                             & $0.0002$ & $0.0000$ & ${\TsV}15.7270$ &  $0.0000$ \\
$\beta ^{\left( \spread\right) }$        & $0.9099$ & $0.0109$ & ${\TsV}83.3412$ &  $0.0000$ \\
$\tilde{\beta}^{\left( \impact\right) }$ & $2.1521$ & $0.0153$ &      $140.6059$ &  $0.0000$ \\ \hline
\multicolumn{5}{c}{\multirow{2}{*}{$R^{2}=39.87\% \qquad R_{c}^{2} = 28.94\%$}} \\
& & & & \\
\hline
\end{tabular}
\vspace*{-25pt}
\end{table}

\begin{figure}[tbph]
\centering
\caption{Parameter estimation using the grid-search approach}
\label{fig:blab_sovereign_calib2}
\figureskip
\includegraphics[width = \figurewidth, height = \figureheight]{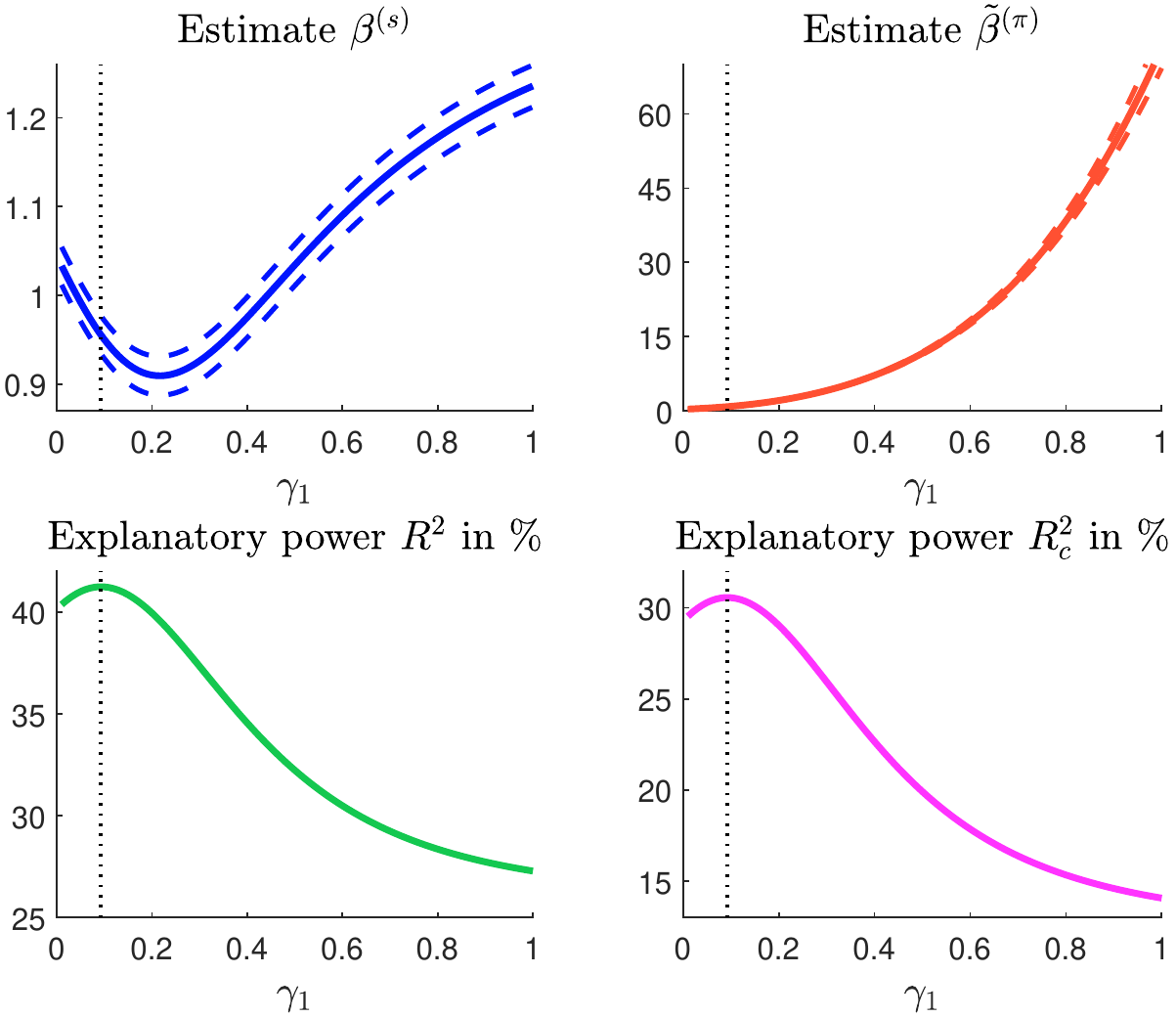}
\end{figure}

\begin{figure}[tbph]
\centering
\caption{Estimated price impact (in bps)}
\label{fig:blab_sovereign_calib3a}
\figureskip
\includegraphics[width = \figurewidth, height = \figureheight]{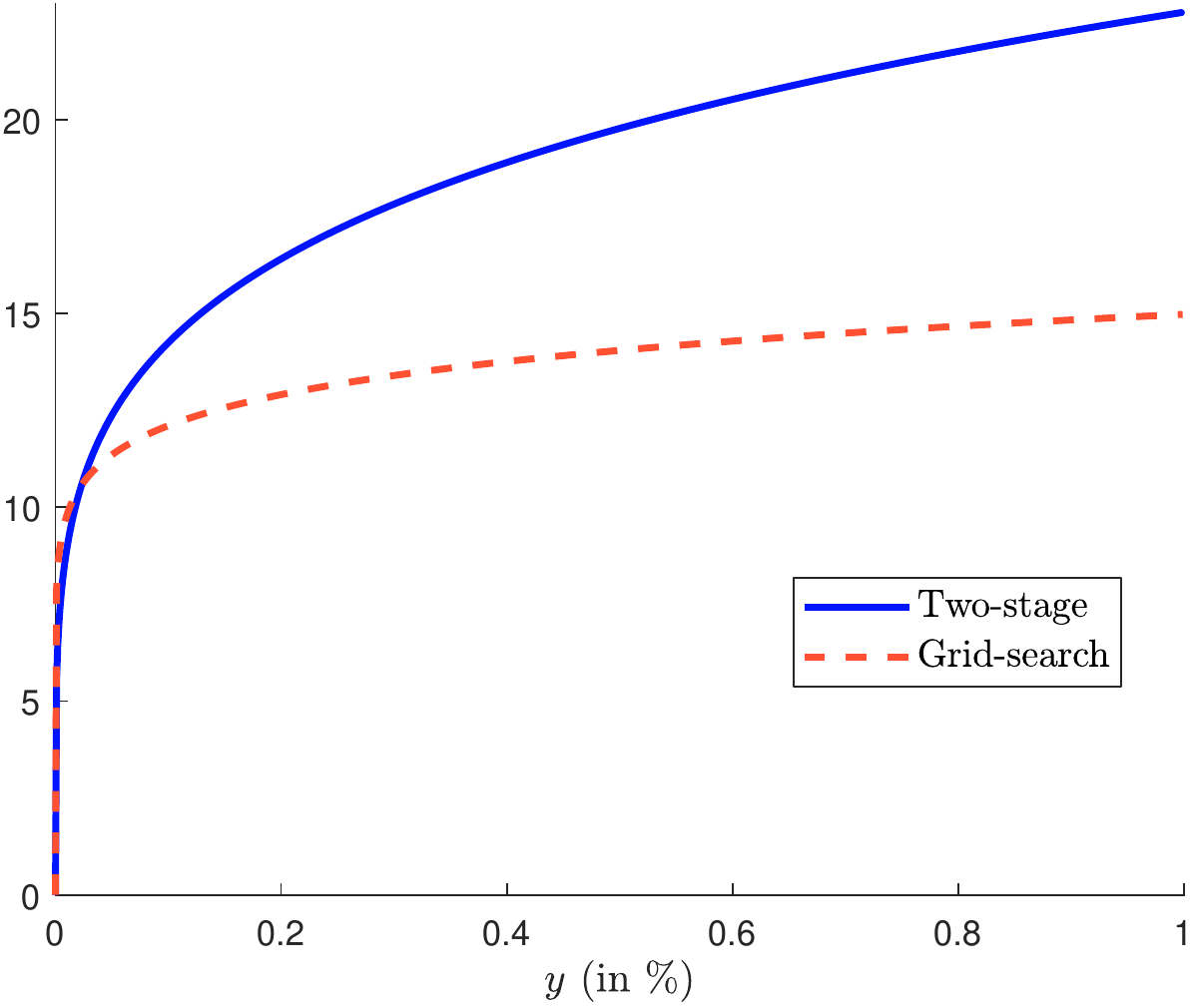}
\end{figure}

Another approach for calibrating the model is to consider a grid-search
process. In this case, we estimate the linear regression:%
\begin{equation*}
\cost_{i}=c_{\beta }+\beta ^{\left( \spread\right) }\spread_{i}+
\mathcal{D}_{i}^{\left( \impact\right) }\tilde{\beta}%
^{\left( \impact\right) }\sigma _{i}y_{i}^{\gamma _{1}}+v_{i}
\end{equation*}%
by considering several values of $\gamma _{1}$. The optimal model
corresponds then to the linear regression that maximizes the
coefficient of determination $R_{c}^{2}$. Figure
\ref{fig:blab_sovereign_calib2} illustrates the grid search process.
The optimal solution is reached for $\gamma _{1}=0.0925$, and we
obtain the results given in Table \ref{tab:blab_sovereign_calib2}.
The explanatory power is close to the one calibrated with the
two-stage approach ($30.56\%$ versus $28.94\%$). However, the two
calibrated models differ if we compare the parameters $\gamma _{1}$
and $\tilde{\beta}^{\left( \impact\right) }$. In order to understand
the differences, we draw the estimated price impact function in Figure
\ref{fig:blab_sovereign_calib3a} when the annualized volatility of the
sovereign bond is equal to $4.36\%$, which is the median volatility
of our dataset. We conclude that the two estimated functions are in fact very close%
\footnote{See Figure \ref{fig:blab_sovereign_calib3b} on page \pageref{fig:blab_sovereign_calib3b}
for a logarithmic scale. We note that the grid-search estimate is more conservative
for very low participation rates.}.\smallskip

\begin{table}[tbph]
\centering
\caption{Grid-search estimation of the sovereign bond transaction cost model}
\label{tab:blab_sovereign_calib2}
\begin{tabular}{ccccc}
\hline
Parameter                                & Estimate &   Stderr &     $t$-student & $p$-value \\ \hline
$\gamma _{1}$                            & $0.0925$ &          &                 &           \\
$c_{\beta }$                             & $0.0000$ & $0.0000$ &  ${\TsX}0.9309$ &  $0.3519$ \\
$\beta ^{\left( \spread\right) }$        & $0.9556$ & $0.0107$ & ${\TsV}89.4426$ &  $0.0000$ \\
$\tilde{\beta}^{\left( \impact\right) }$ & $0.8482$ & $0.0057$ &      $149.2147$ &  $0.0000$ \\ \hline
\multicolumn{5}{c}{\multirow{2}{*}{$R^{2}=41.24\% \qquad R_{c}^{2} = 30.56\%$}} \\
& & & & \\ \hline
\end{tabular}
\vspace*{-5pt}
\end{table}

\clearpage

\begin{figure}[tbph]
\centering
\caption{Estimated price impact (in bps) with respect to the volume-based participation rate}
\label{fig:blab_sovereign_calib4}
\figureskip
\includegraphics[width = \figurewidth, height = \figureheight]{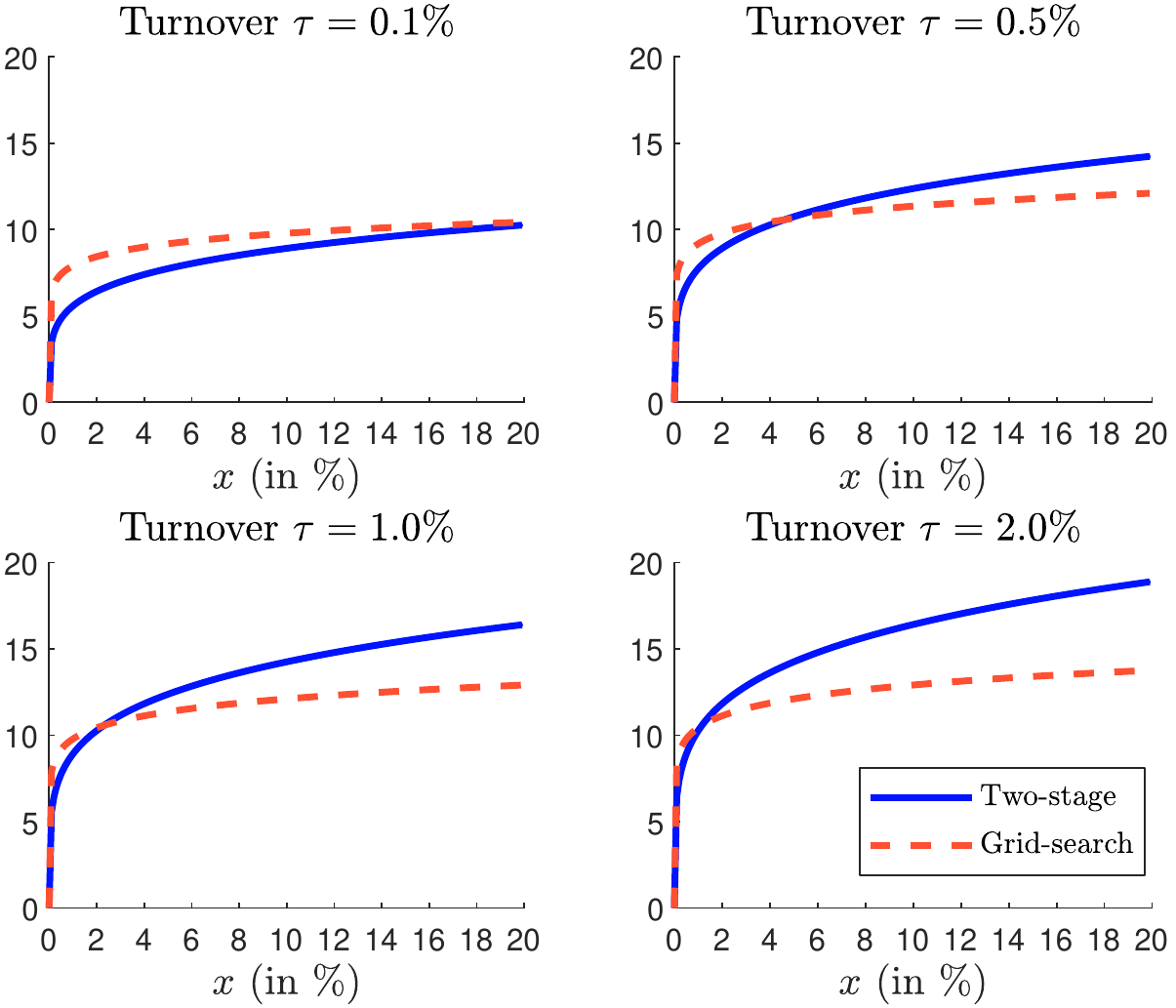}
\end{figure}

In order to better understand the transaction cost function, we
consider the parameterization with respect to the volume-based
participation rate by using the following relationship:
\begin{equation*}
x=\frac{y}{\turnover}
\end{equation*}%
Results are given in Figure \ref{fig:blab_sovereign_calib4} for
different assumptions of the daily turnover $\turnover$. Again, it
is very difficult to prefer one of the two estimated models. Therefore,
we perform an implicit analysis. Using the estimates of the
parameters, we can compute the implied scaling factor:
\begin{equation*}
\hat{\beta}^{\left( \impact\right) }=\turnover^{\gamma _{1}}\tilde{\beta}%
^{\left( \impact\right) }
\end{equation*}%
for a given value of the daily turnover. We can also compute the
implied turnover:
\begin{equation*}
\hat{\turnover} =\left( \frac{\beta ^{\left( \impact\right) }}{\tilde{\beta}%
^{\left( \impact\right) }}\right) ^{\frac{1}{\gamma _{1}}}
\end{equation*}%
for a given scaling factor $\beta ^{\left( \impact\right) }$.
If we analyze the results reported in Table \ref{tab:blab_sovereign_calib5},
it is obvious that the two-stage estimated model is more realistic
than the grid-search estimated model. Indeed, when $\beta ^{\left( \impact\right) }$ is
set to $0.80$, the implicit turnover $\hat{\turnover}$ is respectively equal to $0.78\%$
and $53.13\%$. This second figure is not realistic if we compare it to
the empirical statistics of daily turnover.\smallskip

\begin{table}[tbph]
\centering
\caption{Implicit analysis}
\label{tab:blab_sovereign_calib5}
\begin{tabular}{cccccccccc}
\hline
\multicolumn{2}{c}{$\beta ^{\left( \impact\right) }$}      & $0.40$ & $0.50$ & $0.60$ & ${\TsV}0.70$ & ${\TsV}0.80$ & ${\TsX}0.90$ & ${\TsX}1.00$ & ${\TsXV}1.10$ \\ \hdashline
\mr{$\hat{\turnover}$ (in \%)}              & Two-stage    & $0.03$ & $0.08$ & $0.19$ & ${\TsV}0.40$ & ${\TsV}0.78$ & ${\TsX}1.38$ & ${\TsX}2.32$ & ${\TsXV}3.71$ \\
                                            & Grid-search  & $0.03$ & $0.33$ & $2.37$ &      $12.54$ &      $53.13$ &     $189.81$ &     $592.91$ &     $1661.42$ \\ \hline
\multicolumn{2}{c}{$\turnover$ (in \%)}                    & $0.40$ & $0.50$ & $0.60$ & ${\TsV}0.70$ & ${\TsV}0.80$ & ${\TsX}0.90$ & ${\TsX}1.00$ & ${\TsXV}1.50$ \\ \hdashline
\mr{$\hat{\beta}^{\left( \impact\right) }$} & Two-stage    & $0.70$ & $0.73$ & $0.76$ & ${\TsV}0.78$ & ${\TsV}0.80$ & ${\TsX}0.82$ & ${\TsX}0.84$ & ${\TsXV}0.91$ \\
                                            & Grid-search  & $0.51$ & $0.52$ & $0.53$ & ${\TsV}0.54$ & ${\TsV}0.54$ & ${\TsX}0.55$ & ${\TsX}0.55$ & ${\TsXV}0.58$ \\
\hline
\end{tabular}
\end{table}

The previous model can be easily improved by considering more liquidity
buckets. For instance, if we calibrate\footnote{We use the two-stage estimation approach.}
the model by issuer or currency,
we obtain the results reported in Tables \ref{tab:blab_sovereign_calib6a} and
\ref{tab:blab_sovereign_calib7a}. We observe that $\gamma_1 \in\left[0.05,0.29\right]$.
We also notice that $\beta ^{\left( \spread\right) } < 1$ in most cases, except for Italy, Spain
and the US. Moreover, we observe a large dispersion of the parameter $\tilde{\beta}^{\left( \impact\right) }$.
In a similar way, we can propose a parameterization of $\tilde{\beta}^{\left( \impact\right) }$:
\begin{equation*}
\tilde{\beta}^{\left( \impact\right) }=f\left( \mathcal{F}_{1},\ldots ,%
\mathcal{F}_{m}\right)
\end{equation*}%
where $\left\{ \mathcal{F}_{1},\ldots ,\mathcal{F}_{m}\right\} $ are a set
of bond characteristics \citep{BenSlimane-2017}. For instance, if we assume that the
parameters $\gamma _{1}$ and $\beta ^{\left( \spread\right) }$ are the same
for all the bonds, we observe that $\tilde{\beta}^{\left( \impact\right) }$
is an increasing function of the credit spread, the duration and the issue
date (or the age of the bond).

\begin{table}[tbph]
\centering
\caption{Two-stage estimation of the sovereign bond transaction cost model by issuer}
\label{tab:blab_sovereign_calib6a}
\begin{tabular}{lcccccc}
\hline
Issuer & $\gamma _{1}$ & $c_{\beta }$ & $\beta ^{\left( \spread\right) }$ &
$\tilde{\beta}^{\left( \impact\right) }$ & $R^{2}$ (in \%) &
$R_{c}^{2}$ (in \%) \vphantom{$\dfrac{1}{1}$} \\ \hline
Austria        & $0.2255$ &         $-0.0002$ & $0.8599$ & $3.1385$ & $54.1$ & $48.4$ \\
Belgium        & $0.2482$ &         $-0.0000$ & $0.8097$ & $3.3974$ & $44.0$ & $32.5$ \\
EM             & $0.0519$ & ${\TsVIII}0.0010$ & $0.6828$ & $0.4473$ & $74.9$ & $47.4$ \\
Finland        & $0.2894$ & ${\TsVIII}0.0000$ & $0.7002$ & $4.0287$ & $46.3$ & $31.8$ \\
France         & $0.2138$ & ${\TsVIII}0.0000$ & $0.8794$ & $3.0087$ & $40.1$ & $29.7$ \\
Germany        & $0.2415$ & ${\TsVIII}0.0001$ & $0.9811$ & $2.7007$ & $51.6$ & $38.7$ \\
Ireland        & $0.2098$ & ${\TsVIII}0.0001$ & $0.5403$ & $2.4097$ & $43.9$ & $26.7$ \\
Italy          & $0.1744$ &         $-0.0004$ & $2.7385$ & $1.9030$ & $31.3$ & $22.3$ \\
Japan          & $0.0657$ & ${\TsVIII}0.0001$ & $0.4700$ & $0.6407$ & $79.5$ & $56.4$ \\
Netherlands    & $0.2320$ &         $-0.0000$ & $0.7640$ & $3.7709$ & $46.9$ & $34.2$ \\
Portugal       & $0.2318$ & ${\TsVIII}0.0001$ & $0.9250$ & $3.0248$ & $49.6$ & $33.0$ \\
Spain          & $0.2185$ & ${\TsVIII}0.0000$ & $1.2547$ & $2.0758$ & $40.9$ & $26.7$ \\
United Kingdom & $0.2194$ & ${\TsVIII}0.0003$ & $0.6837$ & $2.3367$ & $51.2$ & $30.3$ \\
USA            & $0.1252$ & ${\TsVIII}0.0001$ & $1.0626$ & $1.2866$ & $53.8$ & $40.9$ \\
\hline
\end{tabular}
\end{table}

\begin{table}[tbph]
\centering
\caption{Two-stage estimation of the sovereign bond transaction cost model by currency}
\label{tab:blab_sovereign_calib7a}
\begin{tabular}{lcccccc}
\hline
Currency & $\gamma _{1}$ & $c_{\beta }$ & $\beta ^{\left( \spread\right) }$ &
$\tilde{\beta}^{\left( \impact\right) }$ & $R^{2}$ (in \%) &
$R_{c}^{2}$ (in \%) \vphantom{$\dfrac{1}{1}$} \\ \hline
EUR  & $0.2262$ & $0.0000$ & $1.0233$ & $2.9122$ & $35.2$ & $25.7$ \\
GBP  & $0.2117$ & $0.0002$ & $1.3602$ & $2.0878$ & $48.8$ & $30.2$ \\
JPY  & $0.0834$ & $0.0001$ & $0.4811$ & $0.8553$ & $75.6$ & $50.9$ \\
USD  & $0.1408$ & $0.0004$ & $0.8430$ & $1.0121$ & $61.5$ & $46.9$ \\
\hline
\end{tabular}
\end{table}

\begin{remark}
If we perform the linear regression without the intercept $c_{\beta }$, we obtain
the results reported in Tables \ref{tab:blab_sovereign_calib6b} and
\ref{tab:blab_sovereign_calib7b} on page \pageref{tab:blab_sovereign_calib6b}.
We notice that the impact
on the coefficients $\beta ^{\left( \spread\right) }$ and
$\tilde{\beta}^{\left( \impact\right) }$ is weak.
\end{remark}

\begin{table}[tbph]
\centering
\caption{Estimation of the sovereign bond transaction cost model when $\gamma_1$ is set to $0.25$}
\label{tab:blab_sovereign_calib8}
\begin{tabular}{ccccc}
\hline
Parameter                                & Estimate &   Stderr &     $t$-student & $p$-value \\ \hline
$\gamma _{1}$                            & $0.2500$ &          &                 &           \\
$\beta ^{\left( \spread\right) }$        & $1.0068$ & $0.0103$ & ${\TsV}97.9041$ &  $0.0000$ \\
$\tilde{\beta}^{\left( \impact\right) }$ & $3.1365$ & $0.0214$ &      $146.6939$ &  $0.0000$ \\ \hline
\multicolumn{5}{c}{\multirow{2}{*}{$R^{2}=38.35\% \qquad R_{c}^{2} = 27.15\%$}} \\
& & & & \\
\hline
\end{tabular}
\vspace*{-25pt}
\end{table}

The choice of the value of $\gamma _{1}$ is not obvious. Finally, we decide
to fix its value at $0.25$.
Based on the results given in Table \ref{tab:blab_sovereign_calib8},
$\beta ^{\left( \spread\right) } = 1.00$ seems to be a good choice.
If we consider the results given in Tables \ref{tab:blab_sovereign_calib6b} and
\ref{tab:blab_sovereign_calib7b}, $\beta ^{\left( \spread\right) } = 1.25$
is more appropriate. We have used end-of-day bid-ask spreads, which
are generally lower than intra-day bisk-ask spreads. Therefore, to reflect this risk,
it may be more prudent to assume that $\beta ^{\left( \spread\right) } = 1.25$.
Finally, we propose the following benchmark formula for computing the
transaction cost for sovereign bonds:
\begin{empheq}[box=\tcbhighmath]{equation}
\quad \cost_{i}\left( q_{i};\spread_{i,t},\sigma _{i,t},\nshares_{i}\right)
=1.25\cdot \spread_{i,t}+3.00\cdot \sigma _{i,t}y_{i}^{0.25} \quad \label{eq:benchmark-bond-sov}
\end{empheq}
If we compare this expression with Equation (\ref{eq:benchmark-stock-lc}), we notice that the
coefficient of the bid-ask spread is the same and the price impact exponent is lower ($0.25$
versus $0.50$ for stocks), implying a lower liquidity risk.

\subsubsection{Corporate bonds}
\label{section:application-corp-bonds}

We estimate Model (\ref{eq:bond4}) by using a dataset of corporate bond
trades, whose description is given in Appendix
\ref{appendix:data-corporate-bond} on page \pageref{appendix:data-corporate-bond}.
Results are given in Table \ref{tab:blab_credit_calib2a}. We notice that all
the estimates are significant at the $99\%$ confidence
level and the explanatory power is relatively high since we have $R^{2}=64.77\%$
and $R_{c}^{2}=41.66\%$.\smallskip

\begin{table}[tbph]
\centering
\caption{Two-stage estimation of the corporate bond transaction cost model with the volatility risk measure}
\label{tab:blab_credit_calib2a}
\begin{tabular}{ccccc}
\hline
Parameter                                &          Estimate &   Stderr &       $t$-student & $p$-value \\ \hline
$c_{\gamma }$                            & $0.3652$ & $0.0338$ & ${\TsV}10.8119$ & $0.0000$ \\
$\gamma _{1}$                            & $0.1168$ & $0.0045$ & ${\TsV}26.1322$ & $0.0000$ \\
$c_{\beta }$                             & $0.0008$ & $0.0000$ & ${\TsV}77.4368$ & $0.0000$ \\
$\beta ^{\left( \spread\right) }$        & $0.7623$ & $0.0042$ &      $183.1617$ & $0.0000$ \\
$\tilde{\beta}^{\left( \impact\right) }$ & $0.9770$ & $0.0044$ &      $224.1741$ & $0.0000$ \\ \hline
\multicolumn{5}{c}{\multirow{2}{*}{$R^{2}=64.77\% \qquad R_{c}^{2} = 41.66\%$}} \\
& & & & \\
\hline
\end{tabular}
\end{table}

The previous model's good results should be considered cautiously
because of two reasons. The first one is that the explanatory power
depends on the maturity of the bonds. For instance, if we focus on short-term
corporate bonds when the time-to-maturity is less than two years, we obtain
$R_{c}^{2}=18.86\%$, which is low compared to the previous figure of
$41.66\%$. The second reason is that the volatility data is not always available.
This is particularly true when the age of corporate bonds is very low. On average,
we do not have the value of the historical volatility for
$20.95\%$ of observations. Moreover, we recall that the asset risk is measured by
the daily volatility $\sigma _{i}$ in the model. However, we know that the
price volatility is not a good measure for measuring the risk of a bond
when the bond is traded at a very low frequency. This is why we observe
a poor explanatory power when we consider bonds that present
a high ratio of zero-trading days or a low turnover. This is the case
of some EM corporate bonds or some mid-cap issuers. Therefore,
we propose replacing the transaction cost function (\ref{eq:bond3})
with the following function:
\begin{equation}
\cost_{i}\left( q_{i};\spread_{i,t},\sigma _{i,t},\nshares_{i}\right)
=\beta ^{\left( \spread\right) }\spread_{i,t}+\tilde{\beta}^{\left( \impact%
\right) }\mathcal{R}_{i,t}y_{i}^{\gamma _{1}}  \label{eq:corporate1}
\end{equation}%
where $\mathcal{R}_{i,t}$ is a better risk measure than the bond
return volatility.\smallskip

\begin{table}[tbph]
\centering
\caption{Two-stage estimation of the corporate bond transaction cost model with the DTS risk measure}
\label{tab:blab_credit_calib3a}
\begin{tabular}{ccccc}
\hline
Parameter                                &          Estimate &   Stderr &        $t$-student & $p$-value \\ \hline
$c_{\gamma }$                            &         $-3.4023$ & $0.0309$ &         $-109.9488$ &  $0.0000$ \\
$\gamma _{1}$                            & ${\TsVIII}0.0796$ & $0.0041$ &  ${\TsXIII}19.5020$ &  $0.0000$ \\
$c_{\beta }$                             & ${\TsVIII}0.0005$ & $0.0000$ &  ${\TsXIII}55.7256$ &  $0.0000$ \\
$\beta ^{\left( \spread\right) }$        & ${\TsVIII}0.7153$ & $0.0034$ & ${\TsVIII}207.4743$ &  $0.0000$ \\
$\tilde{\beta}^{\left( \impact\right) }$ & ${\TsVIII}0.0356$ & $0.0001$ & ${\TsVIII}300.5100$ &  $0.0000$ \\ \hline
\multicolumn{5}{c}{\multirow{2}{*}{$R^{2}=68.64\% \qquad R_{c}^{2} = 46.45\%$}} \\
& & & & \\
\hline
\end{tabular}
\end{table}

In Appendix \ref{appendix:sigma-bond} on page \pageref{appendix:sigma-bond},
we show that the corporate bond risk is a function of the
duration-times-spread or DTS. Therefore, we consider the following
transaction cost function:%
\begin{equation}
\cost_{i}\left( q_{i};\spread_{i,t},\sigma _{i,t},\nshares_{i}\right)
=\beta ^{\left( \spread\right) }\spread_{i,t}+\tilde{\beta}^{\left( \impact%
\right) }\dts_{i,t}y_{i}^{\gamma _{1}}  \label{eq:corporate2}
\end{equation}%
Using our dataset of bond rates, we estimate the parameters by using the
two-stage method:%
\begin{equation}
\left\{
\begin{array}{ll}
\ln \left( \cost_{i}-\spread_{i}\right) -\ln \dts_{i} = c_{\gamma } +
\gamma_{1}\ln y_{i}+u_{i} & \text{if }\cost_{i}>\spread_{i} \\
\cost_{i} = c_{\beta } + \beta ^{\left( \spread\right) }\spread_{i} +
\mathcal{D}_{i}^{\left( \impact\right) }
\tilde{\beta}^{\left( \impact\right) }\dts_{i}y_{i}^{\gamma _{1}} + v_{i} &
\end{array}%
\right.   \label{eq:corporate3}
\end{equation}%
Results are given in Table \ref{tab:blab_credit_calib3a}. We notice that
the results are a little bit better since the explanatory power $R_{c}^{2}$ is equal
to $46.45\%$ instead of $41.66\%$, and all estimated
coefficients are significant at the $99\%$ confidence level.
Moreover, if we focus on corporate bonds where the time-to-maturity
is less than two years, we obtain $R_{c}^{2}=38.21\%$ or an absolute improvement
of $20\%$! Nevertheless,
the value of $\gamma _{1}$ is equal to $0.0796$, which is a low value. This
result is disappointing because the model does not depend on the
participation rate when $\gamma _{1}\approx 0$:%
\begin{equation*}
\lim_{\gamma_1 \rightarrow 0} \cost_{i}\left( q_{i};\spread_{i,t},\sigma _{i,t},\nshares_{i}\right)
 = \beta ^{\left( \spread\right) }\spread_{i,t} +
   \tilde{\beta}^{\left( \impact\right) }\dts_{i,t}
\end{equation*}%
This type of model is not useful and realistic when performing liquidity
stress testing since the liquidity cost does not depend on the trade size!\smallskip

The asset manager that provided the data
uses a trading/dealing desk with specialized bond traders in order to
minimize trading impacts and transaction costs. In particular, we observe
that bond traders may be very active. For example, they may decide to not sell or buy the
bond if the transaction cost is high. In this case, with the agreement
of the fund manager, they can exchange the
bond of an issuer with another bond of the same issuer%
\footnote{With other characteristics such as the maturity.},
a bond of another issuer or a basket of bonds in order to reduce the
transaction cost. More generally, they execute a sell or buy order of a bond
with a high participation rate only if the trading impact is limited,
implying that these big trades are opportunistic and not systematic contrary
to small and medium trades. In a similar way, bond traders may know
the inventory or the axis of the brokers and market markers. They can offer to fund managers
to initiate a trade because the trade impact will be limited or even because the transaction
cost is negative! We conclude that the behavior
of bond traders is different depending on whether the trade is
small/medium or large.\smallskip

\begin{table}[tbph]
\centering
\caption{Estimation of the corporate bond transaction cost model when $\gamma_1$ is set to $0.25$}
\label{tab:blab_credit_calib4b}
\begin{tabular}{ccccc}
\hline
Parameter                                &          Estimate &   Stderr &        $t$-student & $p$-value \\ \hline
$\gamma _{1}$                            & $0.2500$ &          &                     &           \\
$\beta ^{\left( \spread\right) }$        & $0.8979$ & $0.0028$ & $323.2676$ &  $0.0000$ \\
$\tilde{\beta}^{\left( \impact\right) }$ & $0.1131$ & $0.0004$ & $293.5226$ &  $0.0000$ \\ \hline
\multicolumn{5}{c}{\multirow{2}{*}{$R^{2}=66.24\% \qquad R_{c}^{2} = 42.35\%$}} \\
& & & & \\
\hline
\end{tabular}
\end{table}

Since the goal of bond traders is to limit sensitivity to high
participation rates, it is normal that we obtain a low value
for the coefficient $\gamma_1$. We decide to force the coefficient $\gamma_1$
and to use the standard value of $0.25$ that has been chosen for the sovereign bond model.
Based on the results reported in Table \ref{tab:blab_credit_calib4b}, we finally
propose the following benchmark
formula to compute the transaction cost for corporate bonds:
\begin{empheq}[box=\tcbhighmath]{equation}
\quad \cost_{i}\left( q_{i};\spread_{i,t},\sigma _{i,t},\nshares_{i}\right)
=1.50\cdot \spread_{i,t}+0.125\cdot \dts_{i,t}y_{i}^{0.25} \quad \label{eq:benchmark-bond-corp}
\end{empheq}
If we compare this expression with Equation (\ref{eq:benchmark-bond-sov}), we notice that the
coefficient of the bid-ask spread is larger ($1.50$ versus $1.25$ for
sovereign bonds), because of the larger uncertainty on the quoted spreads in
the corporate bond universe. Concerning the price impact exponent, we use
the same value.

\begin{remark}
In order to compare sovereign and corporate bonds, we can transform Equation
(\ref{eq:benchmark-bond-sov}) by considering the relationship between
the DTS and the daily volatility. In our sample%
\footnote{Figure \ref{fig:blab_sovereign_data2} on page
\pageref{fig:blab_sovereign_data2} reports the relationship between the volatility
and the duration-times-spread of sovereign bonds.}, the average ratio is equal to $30.3$.
We deduce that the equivalent transaction cost formula based on the DTS measure for
sovereign bonds is equal to:
\begin{equation}
\cost_{i}\left( q_{i};\spread_{i,t},\sigma _{i,t},\nshares_{i}\right)
=1.25\cdot \spread_{i,t}+0.10\cdot \dts_{i,t}y_{i}^{0.25} \label{eq:benchmark-bond-sov2}
\end{equation}
We notice that the price impact is $+25\%$ higher for corporate bonds
compared to sovereign bonds.
\end{remark}

\subsection{Extension to the two-regime model}
\label{section:application-two-regime}

\begin{figure}[tbph]
\centering
\caption{From the single-regime model to the two-regime model (corporate bonds)}
\label{fig:blab_credit_tworegime1}
\figureskip
\includegraphics[width = \figurewidth, height = \figureheight]{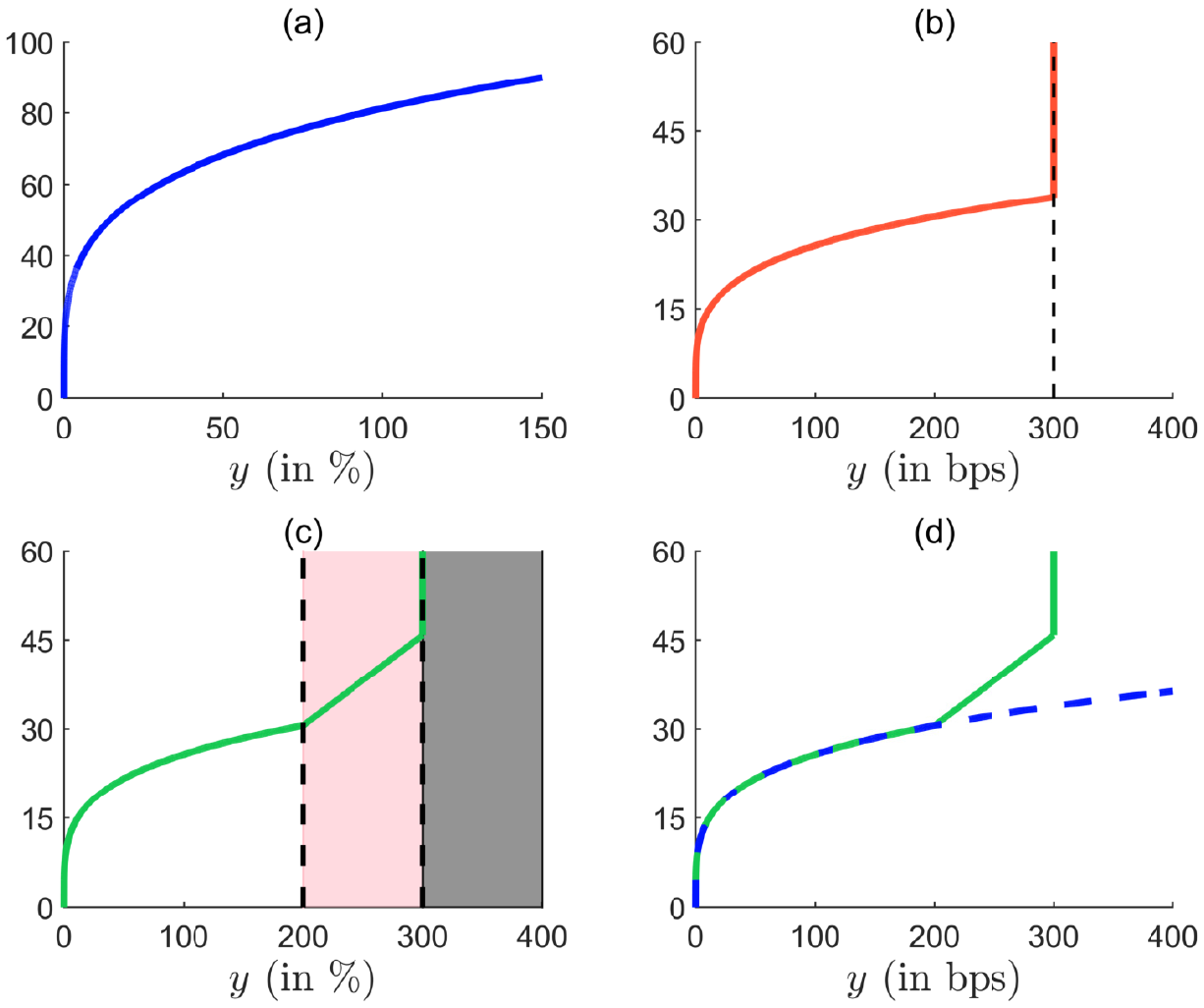}
\end{figure}

As explained in Section \ref{section:trading-limit} on page \pageref{section:trading-limit},
the asset manager generally imposes a trading limit, because it is not possible to have
a $100\%$ participation rate. In Figure \ref{fig:blab_credit_tworegime1}, we have reported
the estimated price impact for corporate bonds%
\footnote{we recall that $\tilde{\beta}^{\left( \impact\right) }= 0.125$
and $\gamma _{1}= 0.25$ for corporate bonds.}.
Panel (a) corresponds to the estimated raw
function. From a mathematical point of view, the price impact is defined even if the participation
rate is larger than $100\%$. In the case of stocks, a $150\%$ volume-based participation rate is plausible, but
it corresponds to a very big trade. In the case of bonds, a $150\%$ outstanding-based participation rate
is impossible, because this trade size is larger than the issued size!
As such, imposing a trading limit is a first modification to obtain a realistic transaction cost function.
However, as explained in Section \ref{section:two-regime}
on page \pageref{section:two-regime}, this is not sufficient.
For instance, we use a trading limit of $300$ bps in Panel (b). Beyond this trading limit, the price impact
is infinite. But if we trade exactly $300$ bps, the price impact is equal to $34$ bps, and we obtain
a concave price impact before this limit. It is better to introduce a second regime (see Equation
\ref{eq:two-regime1} on page \pageref{eq:two-regime1}), implying the following function for the price impact:
\begin{equation*}
\impact\left( y\right) =\left\{
\begin{array}{ll}
\tilde{\beta}^{\left( \impact\right) }\dts y^{\gamma _{1}} & \text{if }y\leq
\tilde{y} \\
\left( \tilde{\beta}^{\left( \impact\right) }\dfrac{\tilde{y}^{\gamma _{1}}}{%
\tilde{y}^{\gamma _{2}}}\right) \dts y^{\gamma _{2}} & \text{if }\tilde{y}%
\leq y\leq y^{+} \\
+\infty  & \text{if }y>y^{+}%
\end{array}%
\right.
\end{equation*}%
In Panel (c), the inflection point $\tilde{y}$ and the power $\gamma_2$ are set to $200$ bps and $1$.
We have two areas. The grey area indicates that the trading is prohibitive beyond $300$ bps.
The red area indicates that the trading is penalized between $200$ bps and $300$ bps, because trading costs
are no longer concave, but convex. Of course, we can use a larger value of $\gamma_2$
to penalize this area of participation rates (for example $\gamma_2 = 2$).
Finally, we obtain the final transaction cost function in Panel (d).\smallskip

The issue of using a two-regime model is the calibration of the second regime.
However, as said previously, it is unrealistic to believe that we can estimate the
inflection point and the parameter $\gamma_2$ from data. Indeed, asset managers do not
experience sufficient big trades and do not have enough data to calibrate the second regime.
We are in an uncertain area, and it is better that these values are given by experts.
For instance, we can use $\gamma_2 = 1$ or $\gamma_2 = 2$ to force the convexity
of the second regime. The inflection point can be equal to $\nicefrac{3}{4}$
or $\nicefrac{2}{3}$ of the trading limit.


\subsection{Stress testing of security-specific parameters}
\label{section:application-stress-testing}

In this section, we conduct a stress testing program in order to define
the transaction cost function in a stress regime. We first define the methodological
framework based on the extreme value theory (EVT). Then, we apply the EVT approach
to the security-specific parameters. Finally, we give the transaction cost function
in the case of a LST program for equity funds.

\subsubsection{Methodological aspects}

Following \citet[Chapters 12 and 14]{Roncalli-2020}, we consider the extreme value
theory for performing stress testing. We summarize this framework below and
provide the main results\footnote{See \citet[pages 753-777 and 904-909]{Roncalli-2020}
for a detailed presentation of extreme value theory and its application to
stress testing and scenario analysis.}.

\paragraph{The block maxima (BM) approach}

We note $X\sim \mathbf{F}$ a continuous random variable and $X_{i:n}$ the
$i^{\mathrm{th}}$ order statistic in the sample\footnote{We assume that the
random variables are \textit{iid}.} $\left\{ X_{1},\ldots
,X_{n}\right\} $. The maximum order statistic is defined by $X_{n:n}=\max
\left( X_{1},\ldots ,X_{n}\right) $. We can show that
$\mathbf{F}_{n:n}\left( x\right) =\mathbf{F}\left( x\right) ^{n}$. If there exist two
constants $a_{n}$ and $b_{n}$ and a non-degenerate distribution function $%
\mathbf{G}$ such that $\lim_{n\rightarrow \infty }\mathbf{F}_{n:n}\left(
a_{n}x+b_{n}\right) =\mathbf{G}\left( x\right) $, the Fisher-Tippett theorem
tells us that $\mathbf{G}$ can only be a Gumbel, Fr\'echet or Weibull
probability distribution. In practice, these three distributions are
replaced by the GEV distribution $\mathcal{GEV}\left( \mu ,\sigma ,\xi
\right) $:%
\begin{equation*}
\mathbf{G}\left( x;\mu ,\sigma ,\xi \right) =\exp \left( -\left( 1+\xi
\left( \frac{x-\mu }{\sigma }\right) \right) ^{-1/\xi }\right)
\end{equation*}%
defined on the support $\Delta =\left\{ x:1+\xi \sigma ^{-1}\left( x-\mu
\right) >0\right\} $. The parameters $\theta =\left( \mu ,\sigma ,\xi
\right) $ can be calibrated by maximizing the log-likelihood function%
\footnote{We recall that the probability density function of the GEV
distribution is equal to:
\begin{equation*}
g\left( x;\mu ,\sigma ,\xi \right) =\frac{1}{\sigma }\left( 1+\xi \left(
\frac{x-\mu }{\sigma }\right) \right) ^{-\left( 1+\xi \right) /\xi }\exp
\left( -\left( 1+\xi \left( \frac{x-\mu }{\sigma }\right) \right) ^{-1/\xi
}\right)
\end{equation*}
}:
\begin{equation*}
\hat{\theta}=\arg \max \sum_{t}-\frac{1}{2}\ln \sigma ^{2}-\left( \frac{%
1+\xi }{\xi }\right) \ln \left( 1+\xi \left( \frac{x_{t}-\mu }{\sigma }%
\right) \right) -\left( 1+\xi \left( \frac{x_{t}-\mu }{\sigma }\right)
\right) ^{-1/\xi }
\end{equation*}%
where $x_{t}$ is the observed maximum for the $t^{\mathrm{th}}$ block maxima
period\footnote{The block maxima approach consists of dividing the observation period into
non-overlapping periods of fixed size and computing the maximum of each
period.}. By assuming that the length of the block maxima period is equal to
$n_{\mathrm{BM}}$ trading days, the stress scenario associated with the random variable $X$ for a
given return time $\mathcal{T}$ is equal to:
\begin{equation*}
\mathbb{S}\left( \mathcal{T}\right) =\mathbf{G}^{-1}\left( \alpha ;\hat{\mu},%
\hat{\sigma},\hat{\xi}\right)
\end{equation*}%
where:%
\begin{equation*}
\alpha =1-\frac{n_{\mathrm{BM}}}{\mathcal{T}}
\end{equation*}%
and $\mathbf{G}^{-1}$ is the quantile function:
\begin{equation*}
\mathbf{G}^{-1}\left( \alpha ;\mu ,\sigma ,\xi \right) =\mu -\frac{\sigma }{%
\xi }\left( 1-\left( -\ln \alpha \right) ^{-\xi }\right)
\end{equation*}%
Finally, we obtain:%
\begin{equation}
\mathbb{S}\left( \mathcal{T}\right) =\hat{\mu}-\frac{\hat{\sigma}}{\hat{\xi}}%
\left( 1-\left( -\ln \left( 1-\frac{n_{\mathrm{BM}}}{\mathcal{T}}\right)
\right) ^{-\hat{\xi}}\right)   \label{eq:parameter-stress1}
\end{equation}

\paragraph{The peak over threshold (POT) approach}

In this approach, we are interested in estimating the distribution of
exceedance over a certain threshold $u$:
\begin{equation*}
\mathbf{F}_{u}(x)=\Pr \left\{ X-u\leq x\mid X>u\right\}
\end{equation*}%
where $0\leq x<x_{0}-u$ and $x_{0}=\sup \left\{ x\in \mathbb{R}:\mathbf{F}%
(x)<1\right\} $. We notice that:
\begin{equation*}
\mathbf{F}_{u}(x)=\frac{\mathbf{F}(u+x)-\mathbf{F}(u)}{1-\mathbf{F}(u)}
\end{equation*}%
For very large $u$, $\mathbf{F}_{u}(x)$ follows a generalized Pareto
distribution $\mathcal{GPD}\left( \sigma ,\xi \right) $:%
\begin{eqnarray*}
\mathbf{F}_{u}(x) &\approx &\mathbf{H}\left( x;\sigma ,\xi \right)  \\
&=&1-\left( 1+\frac{\xi x}{\sigma }\right) ^{-1/\xi }
\end{eqnarray*}%
defined on the support $\Delta =\left\{ x:1+\xi \sigma ^{-1}x>0\right\} $.

\begin{remark}
In fact, there is a strong link between the block maxima approach and the
peak over threshold method. Suppose that $X_{n:n}\sim \mathcal{GEV}\left(
\mu ,\sigma ,\xi \right) $. Using the fact that $\mathbf{F}_{n:n}\left(
x\right) =\mathbf{F}\left( x\right) ^{n}$, we can show that
\citep[page 774]{Roncalli-2020}:%
\begin{eqnarray*}
\mathbf{F}_{u}(x) &\approx &1-\left( 1+\frac{\xi x}{\sigma +\xi \left( u-\mu
\right) }\right) ^{-1/\xi } \\
&=&\mathbf{H}\left( x;\sigma +\xi \left( u-\mu \right) ,\xi \right)
\end{eqnarray*}
Therefore, we obtain a duality between GEV and GPD distribution functions.
\end{remark}

The parameters $\theta =\left( \sigma ,\xi \right) $ are estimated by the
method of maximum likelihood%
\footnote{The probability density function of the GPD distribution is equal to:%
\begin{equation*}
h\left( x;\sigma ,\xi \right) =\frac{1}{\sigma }\left( 1+\frac{\xi x}{\sigma
}\right) ^{-\left( 1+\xi \right) /\xi }
\end{equation*}}
once the threshold $u_{0}$ is found. To
determine $u_{0}$, we use the mean residual life plot, which consists in
plotting $u$ against the empirical mean $\hat{e}\left( u\right) $ of the
excess:%
\begin{equation*}
\hat{e}\left( u\right) =\frac{\sum_{i=1}^{n}\left( x_{i}-u\right) ^{+}}{%
\sum_{i=1}^{n}\mathds{1}\left\{ x_{i}>u\right\} }
\end{equation*}%
For any value $u\geq u_{0}$, we must verify that the mean residual life is a
linear function of $u$ since we have:%
\begin{equation*}
\mathbb{E}\left[ X-u\mid X>u\right] =\frac{\sigma +\xi u}{1-\xi }
\end{equation*}%
The threshold $u_0$ is then found graphically.\smallskip

To compute the stress scenario $\mathbb{S}\left( T\right) $, we recall that:
\begin{equation*}
\mathbf{F}_{u}(x)=\frac{\mathbf{F}(u+x)-\mathbf{F}(u)}{1-\mathbf{F}(u)}%
\approx \mathbf{H}\left( x\right)
\end{equation*}%
where $\mathbf{H}\sim \mathcal{GPD}\left( \sigma ,\xi \right) $. We deduce
that:
\begin{eqnarray*}
\mathbf{F}\left( x\right)  &=&\mathbf{F}\left( u\right) +\left( 1-\mathbf{F}%
\left( u\right) \right) \cdot \mathbf{F}_{u}\left( x-u\right)  \\
&\approx &\mathbf{F}\left( u\right) +\left( 1-\mathbf{F}\left( u\right)
\right) \cdot \mathbf{H}\left( x-u\right)
\end{eqnarray*}%
We consider a sample of size $n$. We note $n^{\prime }$ as the number of
observations whose value $x_{i}$ is larger than the threshold $u_{0}$. The
non-parametric estimate of $\mathbf{F}\left( u_{0}\right) $ is then equal
to:
\begin{equation*}
\mathbf{\hat{F}}\left( u_{0}\right) =1-\frac{n^{\prime }}{n}
\end{equation*}%
Therefore, we obtain the following semi-parametric estimate of $\mathbf{F}%
\left( x\right) $ for $x$ larger than $u_{0}$:
\begin{eqnarray*}
\mathbf{\hat{F}}\left( x\right)  &=&\mathbf{\hat{F}}\left( u_{0}\right)
+\left( 1-\mathbf{\hat{F}}\left( u_{0}\right) \right) \cdot \mathbf{\hat{H}}%
\left( x-u_{0}\right)  \\
&=&\left( 1-\frac{n^{\prime }}{n}\right) +\frac{n^{\prime }}{n}\left(
1-\left( 1+\frac{\hat{\xi}\left( x-u_{0}\right) }{\hat{\sigma}}\right) ^{-1/%
\hat{\xi}}\right)  \\
&=&1-\frac{n^{\prime }}{n}\left( 1+\frac{\hat{\xi}\left( x-u_{0}\right) }{%
\hat{\sigma}}\right) ^{-1/\hat{\xi}}
\end{eqnarray*}%
We can interpret $\mathbf{\hat{F}}\left( x\right) $ as the historical
estimate of the probability distribution tail that is improved by the
extreme value theory. We have\footnote{%
The quantile function of the GPD distribution is equal to:%
\begin{equation*}
\mathbf{H}^{-1}\left( \alpha ;\sigma ,\xi \right) =\frac{\sigma }{\xi }%
\left( \left( 1-\alpha \right) ^{-\xi }-1\right)
\end{equation*}%
}:%
\begin{equation*}
\mathbf{\hat{F}}^{-1}\left( \alpha \right) =u_{0}+\frac{\hat{\sigma}}{\hat{%
\xi}}\left( \left( \frac{n}{n^{\prime }}(1-\alpha )\right) ^{-\hat{\xi}%
}-1\right)
\end{equation*}%
We recall that the stress scenario of the random variable $X$ associated with
the return time $\mathcal{T}$ is equal to $\mathbb{S}\left( \mathcal{T}%
\right) =\mathbf{\hat{F}}^{-1}\left( \alpha \right) $ where $\alpha =1-%
\mathcal{T}^{-1}$. Finally, we deduce that:%
\begin{equation}
\mathbb{S}\left( \mathcal{T}\right) = u_0 + \frac{\hat{\sigma}}{\hat{\xi}}\left(
\left( \frac{n}{n^{\prime}\mathcal{T}}\right) ^{-\hat{\xi}}-1\right)
\label{eq:parameter-stress2}
\end{equation}

\subsubsection{Application to asset liquidity}

We assume that the current date $t$ is not a stress period. Let $p_{i,t}$ be
a security-specific parameter observed at time $t$. We would like to compute
its stress value $p_{i,t+h}^{\mathrm{stress}}$ for a given time horizon $h$.
As explained in Section \ref{section:stress-testing} on page \pageref{section:stress-testing},
we can use a multiplicative shock:
\begin{equation*}
p_{i,t+h}^{\mathrm{stress}}=m_{p} \cdot p_{i,t}
\end{equation*}%
where $m_{p}$ is the multiplier factor. Depending on the
nature of the parameter, we can also use an additive shock:%
\begin{equation*}
p_{i,t+h}^{\mathrm{stress}}=p_{i,t}+\Delta _{p}
\end{equation*}%
where $\Delta_{p}$ is the additive factor. For instance,
we can assume that a multiplicative shock is relevant for the trading
volume, but an additive shock is more appropriate for the credit spread.
Using a sample $\left\{ p_{i,1},\ldots ,p_{i,T}\right\} $ of the parameter
$p$, we compute $m_{t}=\dfrac{p_{i,t+h}}{p_{i,t}}$ or $m_{t}=p_{i,t+h}-p_{i,t}$.
Then, we apply the previous EVT framework to the time series $\left\{
m_{1},\ldots ,m_{T}\right\} $ and estimate the stress scenario $m_{p}$
or $\Delta _{p}$ for a given return time $\mathcal{T}$ and a holding period $h$.
We notice that two periods are used to define the stress
scenario. The time horizon $h$
indicates the frequency of the stress scenario. It is different to compute a
daily, weekly or monthly stress. The return time $\mathcal{T}$
indicates the severity of the stress scenario. If $\mathcal{T}$ is set to one year, we
observe this stress scenario every year on average. Again, it is
different to compute a stress with a return time of one year, two years or
five years. In some sense, $h$ corresponds to the holding period
whereas $\mathcal{T}$ measures the occurrence probability.

\paragraph{Market risk}

\begin{figure}[tbph]
\centering
\caption{Empirical distribution of the multiplicative factor $m_{\sigma }$}
\label{fig:stress_data_vix1a}
\figureskip
\includegraphics[width = \figurewidth, height = \figureheight]{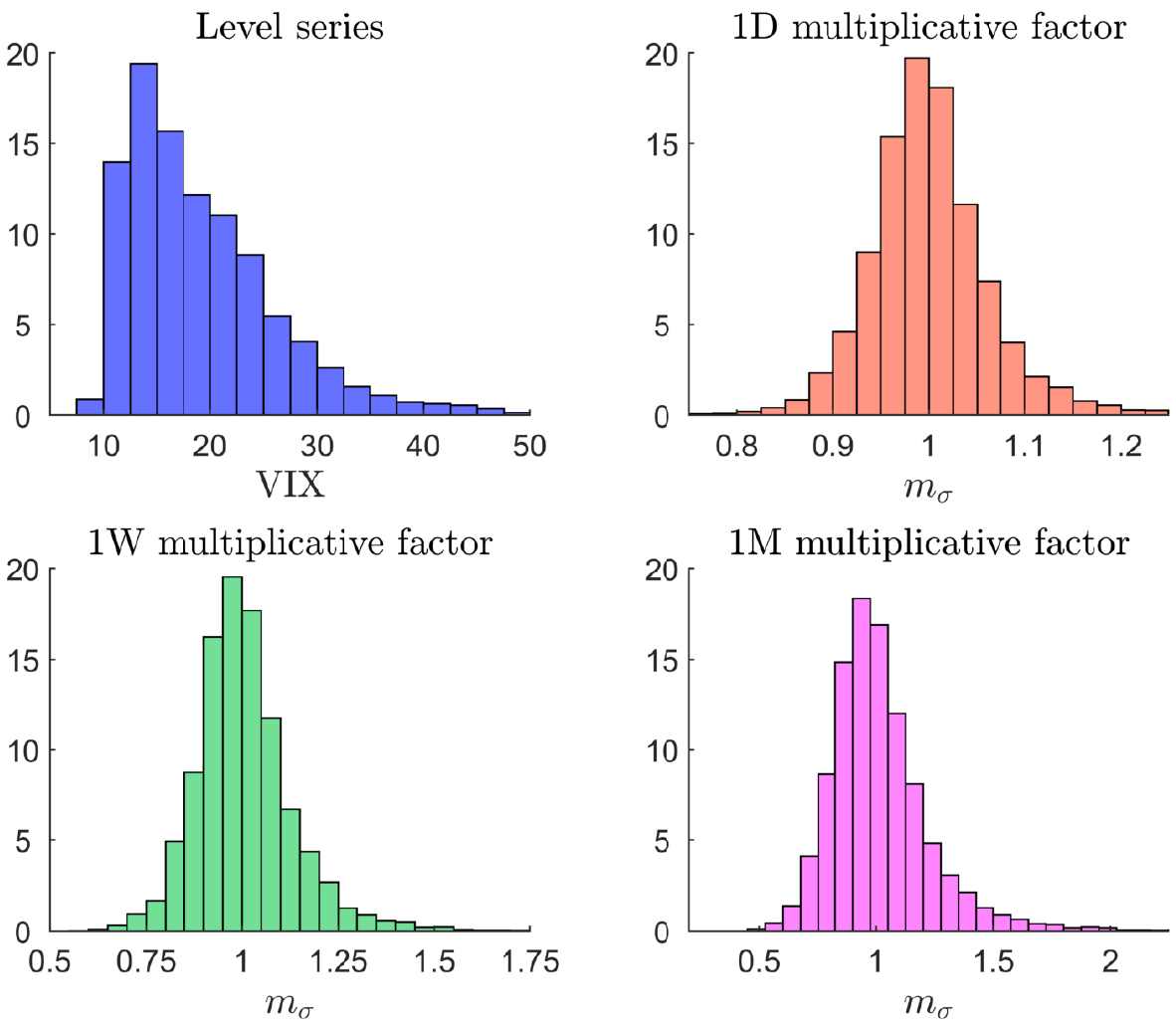}
\end{figure}

We consider the VIX index from January 1990 to February 2021. We have a
sample of $7\,850$ observations. In Figure \ref{fig:stress_data_vix1a},
we report the histogram of the VIX index and
the multiplicative factor $m_{\sigma }$ for three
time horizons (one day, one week and one month).
The estimates of the GEV and GPD distributions are reported in Table
\ref{tab:stress_data_vix2a}. Using Equations \ref{eq:parameter-stress1}
and \ref{eq:parameter-stress1}, we deduce the stress scenarios associated with
$m_{\sigma }$ and $\Delta_{\sigma }$ for three time horizons
(1D, 1W and 1M) and five return times (6M, 1Y,
2Y, 5Y, 10Y and 50Y) in Tables \ref{tab:stress_data_vix2b} and
\ref{tab:stress_data_vix2c}.\smallskip

\begin{table}[tbph]
\centering
\caption{EVT estimates of the VIX index}
\label{tab:stress_data_vix2a}
\begin{tabular}{cc|ccc|ccc}
\hline
&  & \multicolumn{3}{c|}{GEV} & \multicolumn{3}{c}{GPD} \\
&  & $\hat{\mu}$ & $\hat{\sigma}$ & $\hat{\xi}$ & $u_{0}$ & $\hat{\sigma}$ & $\hat{\xi}$ \\ \hline
\multirow{3}{*}{$m_{\sigma } $}
& 1D & $1.103$ & $0.049$ & $0.299$ & ${\TsV}1.229$ & ${\TsV}0.096$ & $0.138$ \\
& 1W & $1.157$ & $0.101$ & $0.229$ & ${\TsV}1.460$ & ${\TsV}0.203$ & $0.243$ \\
& 1M & $1.138$ & $0.185$ & $0.238$ & ${\TsV}1.960$ & ${\TsV}0.425$ & $0.410$ \\ \hdashline
\multirow{3}{*}{$\Delta _{\sigma } $}
& 1D & $1.739$ & $1.036$ & $0.424$ & ${\TsV}4.943$ & ${\TsV}2.560$ & $0.238$ \\
& 1W & $2.568$ & $1.821$ & $0.322$ & ${\TsV}2.950$ & ${\TsV}2.022$ & $0.291$ \\
& 1M & $2.277$ & $3.179$ & $0.201$ &      $16.830$ &      $11.522$ & $0.008$ \\
\hline
\end{tabular}
\end{table}

How should we interpret these results? For example, the multiplicative weekly stress scenario
is equal to $1.50$ if we consider a return time of one year and the BM/GEV approach.
For the additive scenario, we obtain a figure of $9.66\%$. This means that
the volatility can be multiplied by $1.50$ or increased by $9.66\%$ in one week,
and we observe this event (or an equivalent more severe event) every year.
If we average the historical, BM/GEV and POT/GPD approaches, the 2Y weekly stress scenario is
respectively $\times 1.80$ (multiplicative stress) and $+17\%$ (additive stress). If we focus
on the monthly stress scenario, these figures become $\times 2.66$ and $+29\%$.

\begin{table}[tbph]
\centering
\caption{Multiplicative stress scenarios of the volatility}
\label{tab:stress_data_vix2b}
\begin{tabular}{ccccccccc}
\hline
\multicolumn{2}{c}{$\mathcal{T}$ (in years)} & $0.385$ & $\nicefrac{1}{2}$ & $1$ & $2$ & $5$ & $10$ & $50$ \\
\multicolumn{2}{c}{$\alpha$ (in \%)} &  $99.00$ & $99.23$ & $99.62$ & $99.81$ & $99.92$ & $99.96$ & $99.99$ \\ \hline
\multirow{3}{*}{1D}
& Historical & $1.23$ & $1.25$ & $1.32$ & $1.43$ & $1.50$ & $1.57$ &        \\
& BM/GEV     & $1.20$ & $1.22$ & $1.29$ & $1.37$ & $1.51$ & $1.65$ & $2.09$ \\
& POT/GPD    & $1.23$ & $1.25$ & $1.33$ & $1.41$ & $1.52$ & $1.62$ & $1.90$ \\ \hline
\multirow{3}{*}{1W}
& Historical & $1.46$ & $1.51$ & $1.70$ & $1.89$ & $2.26$ & $2.56$ &        \\
& BM/GEV     & $1.34$ & $1.38$ & $1.50$ & $1.64$ & $1.86$ & $2.06$ & $2.66$ \\
& POT/GPD    & $1.46$ & $1.51$ & $1.68$ & $1.87$ & $2.18$ & $2.47$ & $3.35$ \\ \hline
\multirow{3}{*}{1M}
& Historical & $1.96$ & $2.05$ & $2.44$ & $2.99$ & $4.23$ & $5.08$ &        \\
& BM/GEV     & $1.47$ & $1.55$ & $1.78$ & $2.04$ & $2.46$ & $2.83$ & $3.99$ \\
& POT/GPD    & $1.96$ & $2.08$ & $2.45$ & $2.96$ & $3.88$ & $4.86$ & $8.53$ \\ \hline
\end{tabular}
\end{table}

\begin{table}[tbph]
\centering
\caption{Additive stress scenarios of the volatility}
\label{tab:stress_data_vix2c}
\begin{tabular}{ccccccccc}
\hline
\multicolumn{2}{c}{$\mathcal{T}$ (in years)} & $0.385$ & $\nicefrac{1}{2}$ & $1$ & $2$ & $5$ & $10$ & $50$ \\
\multicolumn{2}{c}{$\alpha$ (in \%)} &  $99.00$ & $99.23$ & $99.62$ & $99.81$ & $99.92$ & $99.96$ & $99.99$ \\ \hline
\multirow{3}{*}{1D}
& Historical & ${\TsV}4.94$ & ${\TsV}5.50$ & ${\TsV}7.72$ &      $10.77$ &      $14.15$ & $18.22$ &         \\
& BM/GEV     & ${\TsV}3.91$ & ${\TsV}4.51$ & ${\TsV}6.42$ & ${\TsV}8.94$ &      $13.59$ & $18.50$ & $37.34$ \\
& POT/GPD    & ${\TsV}4.93$ & ${\TsV}5.58$ & ${\TsV}7.12$ & ${\TsV}8.42$ & ${\TsV}9.85$ & $10.74$ & $12.31$ \\ \hline
\multirow{3}{*}{1W}
& Historical & ${\TsV}9.49$ &      $10.88$ &      $14.50$ &      $20.43$ &      $24.56$ & $27.97$ &         \\
& BM/GEV     & ${\TsV}6.08$ & ${\TsV}6.97$ & ${\TsV}9.66$ &      $12.95$ &      $18.53$ & $23.96$ & $42.34$ \\
& POT/GPD    & ${\TsV}9.57$ &      $10.65$ &      $13.92$ &      $17.92$ &      $24.61$ & $30.99$ & $51.86$ \\ \hline
\multirow{3}{*}{1M}
& Historical &      $16.83$ &      $19.04$ &      $27.22$ &      $35.62$ &      $46.59$ & $61.40$ &         \\
& BM/GEV     & ${\TsV}7.84$ & ${\TsV}9.13$ &      $12.74$ &      $16.80$ &      $23.03$ & $28.54$ & $44.68$ \\
& POT/GPD    &      $16.64$ &      $19.67$ &      $27.70$ &      $35.77$ &      $46.51$ & $54.70$ & $73.88$ \\ \hline
\end{tabular}
\end{table}

\paragraph{Trading volume}

Dealing with volatility is relatively simple thanks to the availability
of the VIX. In the case of the trading volume, we face more difficulties
because there is not a standard index that measures the market depth. This
means that we must use the trading volume of the stocks. From a robustness
point of view, it is obvious that computing a stress for each stock is not
relevant. Therefore, given the times series of $v_{i,t}$ for several stocks,
we would like to compute a synthetic stress scenario that is valid for all
stocks. The first idea is to compute the multipliers for each stock and to
pool all the data. The second idea is to compute the multipliers for each
date and to average the data by date. For the BM/GEV approach, we compute the
maximum for each block and each stock, and then we average the maxima by block.
\smallskip

We consider the 30-day average daily volume of the stocks that make up\footnote{Since
the composition changes from one month to another, we have 73 stocks
during the period. Nevertheless, at each date, we only consider the 50 stocks
that are valid at the first trading day of the month.} the EuroStoxx 50
Index from January 2010 to December 2020. At each date, we compute the
multiplicative factor of the trading volume\footnote{In fact, it is a
reductive factor since the risk is not that daily volumes
increase, but that they decrease.}. Then, we apply the previous pooling and
averaging approaches to these data\footnote{We can also transform these
stress scenarios on the trading volume into
stress scenarios on the participation rate using the following formula:
$m_{x} =\frac{1}{m_{v}}$. Results are reported in
Table \ref{tab:stress_data_volume2b} on page \pageref{tab:stress_data_volume2b}.}.
Results are given in Table \ref{tab:stress_data_volume2a}.
If we average the historical, BM/GEV and POT/GPD approaches, the 2Y weekly and monthly stress scenarios are
respectively $\times 0.75$ and $\times 0.48$. This means that the daily volume is approximately reduced
by $25\%$ if we consider a one-week holding period and $50\%$ if we consider a one-month holding period.

\begin{table}[tbph]
\centering
\caption{Multiplicative stress scenarios of the trading volume}
\label{tab:stress_data_volume2a}
\begin{tabular}{cccccccccc}
\hline
\multicolumn{3}{c}{$\mathcal{T}$ (in years)} & $0.385$ & $\nicefrac{1}{2}$ & $1$ & $2$ & $5$ & $10$ & $50$ \\
\multicolumn{3}{c}{$\alpha$ (in \%)} &  $99.00$ & $99.23$ & $99.62$ & $99.81$ & $99.92$ & $99.96$ & $99.99$ \\ \hline
   & Historical &           & $0.93$ & $0.93$ & $0.91$ & $0.88$ & $0.84$ & $0.80$ & $0.71$ \\
   & BM/GEV     & Pooling   & $0.94$ & $0.94$ & $0.92$ & $0.90$ & $0.87$ & $0.85$ & $0.80$ \\
1W & POT/GPD    & Pooling   & $0.95$ & $0.94$ & $0.91$ & $0.88$ & $0.84$ & $0.80$ & $0.70$ \\
   & BM/GEV     & Averaging & $0.94$ & $0.94$ & $0.93$ & $0.92$ & $0.91$ & $0.90$ & $0.89$ \\
   & POT/GPD    & Averaging & $0.93$ & $0.92$ & $0.92$ & $0.91$ & $0.91$ & $0.90$ & $0.89$ \\ \hline
   & Historical &           & $0.79$ & $0.77$ & $0.72$ & $0.67$ & $0.61$ & $0.55$ & $0.48$ \\
   & BM/GEV     & Pooling   & $0.86$ & $0.85$ & $0.81$ & $0.78$ & $0.74$ & $0.71$ & $0.65$ \\
1W & POT/GPD    & Pooling   & $0.87$ & $0.83$ & $0.75$ & $0.68$ & $0.61$ & $0.56$ & $0.47$ \\
   & BM/GEV     & Averaging & $0.87$ & $0.86$ & $0.84$ & $0.82$ & $0.79$ & $0.77$ & $0.73$ \\
   & POT/GPD    & Averaging & $0.82$ & $0.81$ & $0.79$ & $0.78$ & $0.76$ & $0.75$ & $0.72$ \\ \hline
   & Historical &           & $0.50$ & $0.48$ & $0.41$ & $0.36$ & $0.31$ & $0.29$ & $0.26$ \\
   & BM/GEV     & Pooling   & $0.72$ & $0.69$ & $0.62$ & $0.56$ & $0.50$ & $0.46$ & $0.39$ \\
1M & POT/GPD    & Pooling   & $0.40$ & $0.38$ & $0.36$ & $0.33$ & $0.31$ & $0.29$ & $0.26$ \\
   & BM/GEV     & Averaging & $0.75$ & $0.73$ & $0.68$ & $0.63$ & $0.58$ & $0.55$ & $0.49$ \\
   & POT/GPD    & Averaging & $0.62$ & $0.60$ & $0.57$ & $0.54$ & $0.50$ & $0.48$ & $0.42$ \\ \hline
\end{tabular}
\end{table}

\paragraph{Bid-ask spread}

We have seen that stress scenarios of the daily volume are more difficult to
compute than stress scenarios of the volatility. This issue is even more
important with bid-ask spreads because of the data quality. Ideally, we would
like to obtain the weighted average bid-ask spread adjusted by the volume for
each stock and each trading day. However, this information is not easily
available or is expensive. This is why databases of asset managers and
trading platforms generally report the end-of-day bid-ask spread. However,
unlike the closing price, which corresponds to the security's end-of-day transaction
price observed during a regular market trading period,
there is no standard definition of the bid and ask end-of-day prices.
In particular, it is not obvious that the end-of-day bid-ask spread
corresponds to the last bid-ask spread observed during the regular market trading
period. Rather, our experience shows that the end-of-day bid-ask
spread may be impacted by after-hours trading orders. It seems that this
synchronization bias between regular trading and after-hours trading only
impacts bid-ask spreads and not closing prices.\smallskip

\begin{figure}
\centering
\caption{Historical bid-ask spread of BNP Paribas (in bps)}
\label{fig:stress_spread_bnp}
\figureskip
\includegraphics[width = \figurewidth, height = \figureheight]{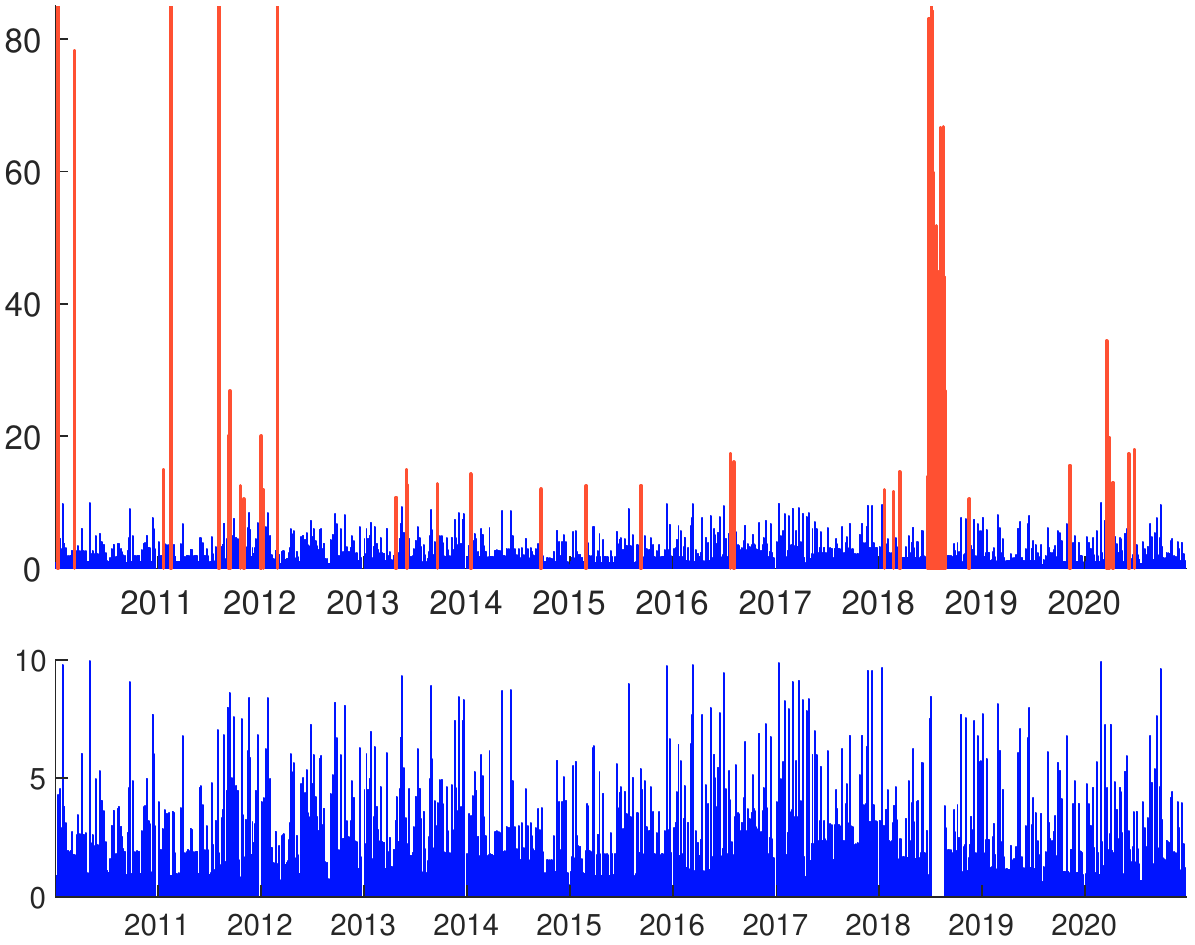}
\end{figure}

To illustrate this issue, we report the end-of-day
bid-ask spread of the BNP Paribas stock between January 2010 and December 2020
in Figure \ref{fig:stress_spread_bnp}. During this period, the stock's median bid-ask spread
is equal to $1.22$ bps. This value is relatively low, however,
we observe many trading days where the bid-ask spread is larger than $20$ bps%
\footnote{These observations correspond to the red bars in
\ref{fig:stress_spread_bnp}}. Therefore, the bid-ask spread may jump from $2$
bps to $80$ bps in one day. It is obvious that these extreme variations are
not realistic and no institutional investor has paid a bid-ask spread of 80
bps for the BNP Paribas stock during the period. These extreme points are not
unusual as illustrated by the figures reported in Table
\ref{tab:stress_data_bas}. For the $50$ stocks of the Eurostoxx 50 Index, we
have computed the frequency at which the bid-ask spread is negative, the daily
multiplicative factor is greater than $5$ or $10$, and the absolute variation
is greater than $25$ and $100$ bps. We consider two well-known data providers,
FactSet and Bloomberg, that are extensively used by equity portfolio
managers. These results illustrate that reported bid and ask end-of-day
prices may deviate substantially from the closing price because of the
synchronization bias between regular and after-hours trading.\smallskip

\begin{table}[tbph]
\centering
\caption{Statistics of daily multiplicative and additive factors for the Eurostoxx 50 stocks (2010 -- 2020)}
\label{tab:stress_data_bas}
\begin{tabular}{lcc}
\hline
Frequency & Factset & Bloomberg  \\ \hline
$\Pr \left\{ \spread<0\right\} $                                              & 0.01\% & 0.24\% \\
$\Pr \left\{ m_{\spread}>10\right\} $                                         & 0.77\% & 0.62\% \\
$\Pr \left\{ m_{\spread}>5\right\} $                                          & 3.49\% & 3.12\% \\
$\Pr \left\{ \left\vert \Delta _{\spread}\right\vert >100\text{ bps}\right\}$ & 0.63\% & 0.44\% \\
$\Pr \left\{ \left\vert \Delta _{\spread}\right\vert >25\text{ bps}\right\} $ & 4.52\% & 3.05\% \\
\hline
\end{tabular}
\end{table}

There are different ways to fix the previous problem. For example, we can
consider a ten-day moving average of daily bid-ask spreads for each stock. Or
we can calculate the weighted average of the bid-ask spreads for a given
universe of stocks for each trading day. The first case corresponds to a
time-series average, whereas the second case corresponds to a cross-section
average. In both cases, the underlying idea is to apply a denoising filter in
order to estimate the average trend. A variant of the second method is to
consider the median bid-ask spread, and we apply this approach to the stocks
of the Eurostoxx 50 Index from January 2010 to December 2020. As in the case
of the daily volume, we only consider the 50 stocks that are in the index at
each trading day. The empirical distributions of $m_{\spread}$
and $\Delta_{\spread}$ are given in Figures
\ref{fig:stress_data_bas1b} and \ref{fig:stress_data_bas1a} on page
\pageref{fig:stress_data_bas1b}. Using these data, we calibrate the GEV and GPD models,
and we obtain the stress scenarios that are reported in Tables
\ref{tab:stress_data_stress2a} and \ref{tab:stress_data_stress2b}.
If we average the historical, BM/GEV and POT/GPD approaches, the 2Y weekly stress scenario
is respectively $\times 3$ (multiplicative stress) and $+6.5$ bps (additive stress).

\begin{table}[tbph]
\centering
\caption{Multiplicative stress scenarios of the bid-ask spread}
\label{tab:stress_data_stress2a}
\begin{tabular}{ccccccccc}
\hline
\multicolumn{2}{c}{$\mathcal{T}$ (in years)} & $0.385$ & $\nicefrac{1}{2}$ & $1$ & $2$ & $5$ & $10$ & $50$ \\
\multicolumn{2}{c}{$\alpha$ (in \%)} &  $99.00$ & $99.23$ & $99.62$ & $99.81$ & $99.92$ & $99.96$ & $99.99$ \\ \hline
\multirow{3}{*}{1D}
& Historical & $1.66$ & $1.73$ & $1.93$ & $2.40$ & $2.75$ & $7.11$ &              \\
& BM/GEV     & $1.63$ & $1.70$ & $1.92$ & $2.19$ & $2.64$ & $3.08$ & ${\TsV}4.56$ \\
& POT/GPD    & $1.65$ & $1.71$ & $1.94$ & $2.32$ & $3.24$ & $4.49$ &      $11.70$ \\ \hline
\multirow{3}{*}{1W}
& Historical & $1.74$ & $1.88$ & $2.58$ & $3.49$ & $6.78$ & $9.76$ &              \\
& BM/GEV     & $1.67$ & $1.76$ & $2.05$ & $2.41$ & $3.07$ & $3.75$ & ${\TsV}6.27$ \\
& POT/GPD    & $1.81$ & $1.93$ & $2.41$ & $3.22$ & $5.20$ & $7.92$ &      $23.78$ \\ \hline
\multirow{3}{*}{1M}
& Historical & $2.54$ & $2.92$ & $5.12$ & $6.65$ & $9.62$ & ${\TsV}9.98$ &              \\
& BM/GEV     & $1.75$ & $1.86$ & $2.18$ & $2.58$ & $3.25$ & ${\TsV}3.90$ & ${\TsV}6.12$ \\
& POT/GPD    & $2.40$ & $2.64$ & $3.52$ & $4.85$ & $7.72$ &      $11.21$ &      $27.90$ \\ \hline
\end{tabular}
\end{table}

\begin{table}[tbph]
\centering
\caption{Additive stress scenarios of the bid-ask spread}
\label{tab:stress_data_stress2b}
\begin{tabular}{ccccccccc}
\hline
\multicolumn{2}{c}{$\mathcal{T}$ (in years)} & $0.385$ & $\nicefrac{1}{2}$ & $1$ & $2$ & $5$ & $10$ & $50$ \\
\multicolumn{2}{c}{$\alpha$ (in \%)} &  $99.00$ & $99.23$ & $99.62$ & $99.81$ & $99.92$ & $99.96$ & $99.99$ \\ \hline
\multirow{3}{*}{1D}
& Historical & $1.67$ & $1.82$ & $2.93$ & $6.46$ &      $10.94$ &      $18.14$ &         \\
& BM/GEV     & $1.42$ & $1.63$ & $2.28$ & $3.14$ & ${\TsV}4.71$ & ${\TsV}6.37$ & $12.70$ \\
& POT/GPD    & $1.77$ & $2.04$ & $3.07$ & $4.76$ & ${\TsV}8.78$ &      $14.17$ & $44.13$ \\ \hline
\multirow{3}{*}{1W}
& Historical & $1.98$ & $2.37$ & $5.19$ &      $10.10$ &      $12.36$ &      $19.11$ &         \\
& BM/GEV     & $1.48$ & $1.70$ & $2.40$ & ${\TsV}3.33$ & ${\TsV}5.08$ & ${\TsV}6.94$ & $14.17$ \\
& POT/GPD    & $2.19$ & $2.57$ & $3.91$ & ${\TsV}6.00$ &      $10.63$ &      $16.43$ & $45.46$ \\ \hline
\multirow{3}{*}{1M}
& Historical & $3.36$ & $3.98$ & $7.90$ &      $10.60$ &      $16.04$ &      $21.36$ &         \\
& BM/GEV     & $1.51$ & $1.77$ & $2.62$ & ${\TsV}3.82$ & ${\TsV}6.20$ & ${\TsV}8.91$ & $20.46$ \\
& POT/GPD    & $2.99$ & $3.57$ & $5.73$ & ${\TsV}9.23$ &      $17.33$ &      $27.95$ & $84.86$ \\ \hline
\end{tabular}
\end{table}

\subsubsection{Definition of the stress transaction cost function}

If we assume that $x^{+}=10\%$, $\tilde{x}=\dfrac{2}{3}x^{+}$ and $\gamma
_{2}=1$, the transaction cost function for large cap stocks is equal to:
\begin{equation}
\cost\left( q;\spread,\sigma ,v\right) =\left\{
\begin{array}{ll}
1.25\cdot \spread+0.40\cdot \sigma \sqrt{x} & \text{if }x\leq 6.66\% \\
1.25\cdot \spread+1.55\cdot \sigma x & \text{if }6.66\%\leq x\leq 10\% \\
+\infty  & \text{if }x>10\%%
\end{array}%
\right.   \label{eq:stress-function1}
\end{equation}%
We consider the following stress scenario\footnote{This stress
scenario is approximatively the 2Y weekly stress scenario.}:
\begin{itemize}
\item $\Delta _{\spread}=8$ bps
\item $\Delta _{\sigma }=20\%$
\item $m_{v}=0.75$
\end{itemize}
We deduce that the transaction cost function in the stress regime becomes:
\begin{equation*}
\cost\left( q;\spread,\sigma ,v\right) =\left\{
\begin{array}{ll}
1.25\cdot \left( \spread+8\text{ bps}\right) +0.40\cdot \left( \sigma +%
\dfrac{20\%}{\sqrt{260}}\right) \sqrt{\dfrac{4}{3}x} & \text{if }x\leq 5\%
\\
1.25\cdot \left( \spread+8\text{ bps}\right) +1.55\cdot \left( \sigma +%
\dfrac{20\%}{\sqrt{260}}\right) \dfrac{4}{3}x & \text{if }5\%\leq x\leq 7.5\%
\\
+\infty  & \text{if }x>7.5\%%
\end{array}%
\right.
\end{equation*}%

\begin{table}[tbph]
\centering
\caption{Stress testing computation}
\label{tab:stress_equity_lc_function2}
\scalebox{0.90}{
\begin{tabular}{|cc|rrrrrrr|rrr|}
\hline
\multirow{3}{*}{$x$} & \multirow{3}{*}{Case} & \multicolumn{7}{c|}{Annualized volatility}  & \multicolumn{3}{c|}{Liquidation} \\
    & & \multicolumn{1}{c}{$10\%$} & \multicolumn{1}{c}{$15\%$} & \multicolumn{1}{c}{$20\%$} &
\multicolumn{1}{c}{$25\%$} & \multicolumn{1}{c}{$30\%$} & \multicolumn{1}{c}{$35\%$} & \multicolumn{1}{c|}{$40\%$} &
\multicolumn{1}{c}{$\mathcal{LT}$} & \multicolumn{1}{c}{$\mathcal{LS}$} &
\multicolumn{1}{c|}{$\mathcal{LS}$} \\
    & & \multicolumn{7}{c|}{$\cost\left( q;\spread,\sigma ,v\right)$ (in bps)} & & one-day
    & two-day \\ \hline
 \multirow{2}{*}{${\TsV}0.00\%$}
& \Ns   &    5.0 &  5.0 &  5.0 &  5.0 &  5.0 &  5.0 &  5.0 &       1 & 0\%    & 0\%     \\
& \Ss   &   15.0 & 15.0 & 15.0 & 15.0 & 15.0 & 15.0 & 15.0 &       1 & 0\%    & 0\%     \\ \hline
 \multirow{2}{*}{${\TsV}0.01\%$}
& \Ns   &    5.2 &  5.4 &  5.5 &  5.6 &  5.7 &  5.9 &  6.0 &       1 & 0\%    & 0\%     \\
& \Ss   &   15.9 & 16.0 & 16.1 & 16.3 & 16.4 & 16.6 & 16.7 &       1 & 0\%    & 0\%     \\ \hline
 \multirow{2}{*}{${\TsV}0.05\%$}
& \Ns   &    5.6 &  5.8 &  6.1 &  6.4 &  6.7 &  6.9 &  7.2 &       1 & 0\%    & 0\%     \\
& \Ss   &   16.9 & 17.2 & 17.6 & 17.9 & 18.2 & 18.5 & 18.8 &       1 & 0\%    & 0\%     \\ \hline
 \multirow{2}{*}{${\TsV}0.10\%$}
& \Ns   &    5.8 &  6.2 &  6.6 &  7.0 &  7.4 &  7.7 &  8.1 &       1 & 0\%    & 0\%     \\
& \Ss   &   17.7 & 18.2 & 18.6 & 19.1 & 19.5 & 20.0 & 20.4 &       1 & 0\%    & 0\%     \\ \hline
 \multirow{2}{*}{${\TsV}0.50\%$}
& \Ns   &    6.8 &  7.6 &  8.5 &  9.4 & 10.3 & 11.1 & 12.0 &     1 & 0\%    & 0\%     \\
& \Ss   &   21.1 & 22.1 & 23.1 & 24.1 & 25.1 & 26.1 & 27.2 &     1 & 0\%    & 0\%     \\ \hline
 \multirow{2}{*}{${\TsV}1.00\%$}
& \Ns   &    7.5 &  8.7 & 10.0 & 11.2 & 12.4 & 13.7 & 14.9 &     1 & 0\%    & 0\%     \\
& \Ss   &   23.6 & 25.0 & 26.5 & 27.9 & 29.3 & 30.8 & 32.2 &     1 & 0\%    & 0\%     \\ \hline
 \multirow{2}{*}{${\TsV}5.00\%$}
& \Ns   &   10.5 & 13.3 & 16.1 & 18.9 & 21.6 & 24.4 & 27.2 &     1 & 0\%    & 0\%     \\
& \Ss   &   34.2 & 37.4 & 40.6 & 43.8 & 47.0 & 50.2 & 53.4 &     1 & 0\%    & 0\%     \\ \hline
 \multirow{2}{*}{${\TsV}7.50\%$}
& \Ns   &   12.2 & 15.8 & 19.4 & 23.0 & 26.6 & 30.2 & 33.8 &     1 & 0\%    & 0\%     \\
& \Ss   &   43.8 & 48.6 & 53.4 & 58.2 & 63.0 & 67.8 & 72.6 &     1 & 0\%    & 0\%     \\ \hline
 \multirow{2}{*}{$10.00\%$}
& \Ns   &   14.6 & 19.4 & 24.2 & 29.0 & 33.8 & 38.6 & 43.4 &     1 & 0\%    & 0\%     \\
& \Ss   &   40.0 & 44.2 & 48.4 & 52.5 & 56.7 & 60.9 & 65.0 &     2 & 2.5\%  & 0\%     \\ \hline
 \multirow{2}{*}{$20.00\%$}
& \Ns   &   14.6 & 19.4 & 24.2 & 29.0 & 33.8 & 38.6 & 43.4 &     2 & 10\%   & 0\%     \\
& \Ss   &   41.4 & 45.8 & 50.2 & 54.6 & 59.0 & 63.4 & 67.8 &     3 & 12.5\% & 5.5\%   \\ \hline
\end{tabular}}
\end{table}

In Table \ref{tab:stress_equity_lc_function2}, we have reported an example
of stress testing applied to single stocks. For each value of $\sigma$
and $x$, we report the unit cost $\cost\left( q;\spread,\sigma ,v\right)$
in bps for the normal and stress regimes. For instance, if the annualized
volatility is equal to $30\%$ and the liquidation of the exposure on the
single stock represents $0.05\%$ of the normal daily volume, the transaction
cost is equal to $6.7$ bps in the normal period. In the stress period, it
increases to $18.2$ bps, which is an increase of $171\%$. We have also reported
the liquidation time, the one-day liquidation shortfall and the two-day
liquidation shortfall. Let us consider a $10\%$ liquidation. Because of the
liquidity policy, we can liquidate $7.5\%$ the first day and $2.5\%$ the second day
during the stress period, whereas we can liquidate the full exposure during the
normal period. Therefore, the liquidation time, which is normally equal
to one day, takes two days in the stress period. If we consider a $20\%$
liquidation, the (one-day) liquidation shortfall is equal to $12.5\%$ and the
time-to-liquidation is equal to three days.

\clearpage

\section{Conclusion and discussion}

Liquidity stress testing is a recent topic in asset management, which has
given rise to numerous publications from regulators \citep{AMF-2017,
BaFin-2017, ESMA-2019, ESMA-2020, FSB-2017, IOSCO-2015, IOSCO-2018}. In
particular, LST has been mandatory in Europe since September 2020. However,
contrary to banks, asset managers have less experience conducting a liquidity
stress testing program at the global portfolio level. Moreover, this topic
has not been extensively studied by the academic research. Therefore, we are
in a trial-and-error period where standard models are not really established,
and asset managers use very different approaches to assess liquidity stress
tests. The aim of this research project is to propose a simple LST approach
that may become a benchmark for asset managers. In a previous paper, we have
already developed a framework for modeling the liability liquidity risk
\citep{Roncalli-lst1-2020}. In a forthcoming  paper, we will propose several
tools for managing the asset-liability liquidity gap. In this paper, we focus
on measuring the asset liquidity risk.\smallskip

Contrary to the first and third parts of this project, there is a large
body of academic literature that has studied the estimation of transaction costs. In
particular, we assume that price impact verifies the power-law property. This
means that there is a concave relationship between the participation rate and
the transaction cost. This model is appealing because (1) it has been
proposed by the academic research in the case of stocks, (2) it is simple and
(3) it is suitable for stress testing purposes. The first reason is
important, because the model must be approved by the regulators. The fact
that this model has academic roots is therefore a key element in terms of
robustness and independent validation. The second reason is critical, because
a complex transaction cost model with many parameters and variables may be
not an industrial solution. This is particularly true if the calibration
requires a large amount of data. In the case of our model,
we have three parameters (spread sensitivity, price impact sensitivity and
price impact exponent) and three explanatory variables (bid-ask spread,
volatility risk and participation rate). If the asset manager does not have
enough data, it can always use some internal experts to set the value of
these parameters. Moreover, we have seen that this model can also be applied
to bonds with some minor corrections. For instance, in the case of corporate
bonds, it is better to use the DTS instead of the volatility in order to
measure the market risk. Finally, the third reason is convenient when we
perform stress testing programs. When applied to liquidity in asset
management, they can concern the liability side and/or the asset side
\citep{Brunnermeier-2009}. For instance, the asset manager can assume that
the liquidity crisis is due to funding issues. In this case, the stress
scenario could be a severe redemption scenario. But it can also assume that
the liquidity crisis is due to market issues. In this case, the stress
scenario could be a market liquidity crisis with a substantial reduction in trading
volumes and an increase in volatility risk. Therefore, it is important
that a stress scenario of market liquidity risk could be implemented, and not
only a stress scenario of funding liquidity risk. Our transaction cost model
has three variables that can be stressed: the spread, the market risk and the
trading volume (or the market depth). We think that these three transmission channels are enough to
represent a market liquidity crisis.
Nevertheless, the high concavity of the price impact function when the
exponent is smaller than $\nicefrac{1}{2}$ is not always relevant when we
also impose trading policy limits. Therefore, we propose an extension of the
previous model by considering two regimes with two power-law models where the
second exponent takes a larger value than the first exponent. In this case,
the transaction cost function has two additional parameters: the exponent of
the second regime and the inflection point that separates the first and second
regimes. Therefore, we can obtain a price impact which is more convex in the
second regime when the participation rate is high. In terms of
calibration, we propose using expert estimates, implying no more data
analysis.

\begin{table}[tbph]
\centering
\caption{Impact of size on the market impact}
\label{tab:size_impact}
\begin{tabular}{|c|ccc|ccc|}
\hline
Size & \multicolumn{3}{c|}{Stocks} & \multicolumn{3}{c|}{Bonds} \\
& Unit & Total & Average & Unit & Total & Average \\
& cost & cost  & cost    & cost & cost  & cost    \\ \hline
$\times 1{\TsV}$ & $\times 1.0$ & $\times 1.0$        & $+0\%{\TsX}$   & $\times 1.0$ & $\times 1.0$        & $+0\%{\TsV}$ \\
$\times 2{\TsV}$ & $\times 1.4$ & $\times 2.8$        & $+41\%{\TsV}$  & $\times 1.2$ & $\times 2.4$        & $+19\%$      \\
$\times 3{\TsV}$ & $\times 1.7$ & $\times 5.2$        & $+73\%{\TsV}$  & $\times 1.3$ & $\times 3.9$        & $+32\%$      \\
$\times 4{\TsV}$ & $\times 2.0$ & $\times 8.0$        & $+100\%$       & $\times 1.4$ & $\times 5.7$        & $+41\%$      \\
$\times 5{\TsV}$ & $\times 2.2$ & $\times 11{\TsIII}$ & $+124\%$       & $\times 1.5$ & $\times 7.5$        & $+50\%$      \\
$\times 10$      & $\times 3.2$ & $\times 32{\TsIII}$ & $+216\%$       & $\times 1.8$ & $\times 18{\TsIII}$ & $+78\%$      \\
\hline
\end{tabular}
\end{table}
\vspace*{-30pt}

We have proposed some formulas for large cap stocks, small cap stocks,
sovereign bonds and corporate bonds\footnote{These formulas
correspond to Equations (\ref{eq:benchmark-stock-lc}), (\ref{eq:benchmark-stock-sc}),
(\ref{eq:benchmark-bond-sov}) and (\ref{eq:benchmark-bond-corp}).}.
This is an especially challenging exercise.
Indeed, the calibrated formulas highly depend on the data\footnote{For example,
using Reuters bid-ask spreads instead of Bloomberg bid-ask spreads dramatically
changes the parameter $\beta^{\spread}$ for sovereign and corporate bonds.}.
Because we use a small sample on a particular period and this sample
is specific to an asset manager, the data are not representative of
the industry as a whole. Moreover, in the case of bonds, we have decided to exclude
opportunistic trades with a negative transaction cost.
This is why these calibrated formulas must be adjusted and validated by the asset manager
before using them. On page \pageref{appendix:benchmark}, we have reported the
values of the unit transaction cost. These tables can be used as a preliminary
pricing grid that can be modified. For instance, the asset manager
generally knows its average price impact, and can then change the values of
$\beta^{\spread}$, $\beta^{\impact}$ and $\gamma_1$ in order to retrieve its average cost.
This pricing grid can also be modified by the trading desk cell by cell
in order to avoid some unrealistic values\footnote{For instance, a price impact
of $198$ bps  may be considered too high
when the outstanding-based participation rate is set to $100$ bps
and the DTS of the corporate bond is equal to $5\,000$ bps.}. One of the difficulties is to maintain
some coherency properties between the different cells of the pricing grid.
In the case of the power-law model, if we multiply the size by $\alpha$, the unit cost is multiplied
by $\alpha^{\gamma_1}$ while the total cost is multiplied by $\alpha^{1+\gamma_1}$.
In Table \ref{tab:size_impact}, we have reported the impact of the size
on the price impact when we consider our benchmark formulas\footnote{We recall
that $\gamma_1$ is equal to $0.5$ for stocks and $0.25$ for bonds.}.
For example, we notice that if we multiply the size of the trade by $5$, the average cost
due to the price impact increases by $124\%$ for stocks and $50\%$ for bonds.
Quantifying these size effects is essential in a liquidity stress testing program because
the risk in a stress period is mainly related to the size issue.
And it is not always obvious to obtain a pricing grid that
satisfies some basic coherency properties.\smallskip

As explained in the introduction, our motivation is to propose a framework
that can help asset managers to implement liquidity stress testing, which
is a relative new topic for this industry. We are aware that it is challenging, and
the final model can appear too simple to describe the transaction cost
function of any stocks and bonds. This is true. For instance, it is not
precise enough to calibrate swing prices. However, we reiterate that the
goal is not to build a pre-trade system, but to implement a liquidity stress
testing program from an industrial viewpoint. In a liquidity crisis, there
are so many unknowns and uncertainties that a sophisticated model does not
necessarily enable redemption issues to be managed better. An LST model must be
sufficiently realistic and pragmatic in order to give the magnitude order of
the stress severity and compare the different outcomes. We think that the
model proposed here has some appealing properties to become a benchmark
for asset managers. However, the road to obtain the same standardization
that we encounter in the banking regulation of market, credit or counterparty risk
is long. More research in this area from academics and professionals is needed.

\clearpage

\bibliographystyle{apalike}

\clearpage

\appendix

\section*{Appendix}

\section{Glossary}

\subsection*{Bid-ask spread}
\label{glossary:bas}

The bid-ask spread corresponds to the difference between the ask price and
the bid price of a security divided by its mid-point price. It is a component
of the liquidity cost, since the \hyperref[glossary:utc]{unit transaction
cost} depends on the half bid-ask spread. In this article, we use the term
bid-ask spread in place of half bid-ask spread, and we denote it by
$\spread$.

\subsection*{Break-even redemption scenario}
\label{glossary:bers}

The break-even redemption scenario is the maximum amount expressed in dollars
that can be liquidated in one day:
\begin{eqnarray*}
\mathbb{R}^{\mathrm{break-even}} & = &
\sup \left\{ \mathbb{R}:\mathcal{LS}\left( \mathbb{R}\right) =0\right\} \\
& = &\inf \left\{ \mathbb{R}:\mathcal{LR}\left( \mathbb{R};1\right) =1\right\}
\end{eqnarray*}

\subsection*{HQLA class}
\label{glossary:hqla}

The term HQLA refers to high-quality liquid asset. An HQLA class
groups all the securities that present the same ability
to be converted into cash. An HQLA class is different than a
\hyperref[glossary:lb]{liquidity bucket},
because this latter classification is used to define the
\hyperref[glossary:utc]{unit transaction cost} function.
For instance, it does not make sense that a bond and a stock share the same
transaction cost function. However, they can belong to the same HQLA class
if they have the same conversion property into cash.

\subsection*{Implementation shortfall}
\label{glossary:is}

The implementation shortfall measures the total amount of slippage, that is
the difference in price between the time a portfolio manager makes an
investment decision and the actual traded price. Its mathematical expression
is:
\begin{equation*}
\mathcal{IS}\left( q\right) =\max \left( \mathbb{V}^{\mathrm{mid}}\left(
q\right) -\mathbb{V}^{\mathrm{liquidated}}\left( q\right) ,0\right)
\end{equation*}%
where $\mathbb{V}^{\mathrm{mid}}\left( q\right) $ is the current value of
the \hyperref[glossary:rs]{redemption scenario}
and $\mathbb{V}^{\mathrm{liquidated}}\left( q\right)$
is the value of the liquidated portfolio.

\subsection*{Liquidation policy}
\label{glossary:lp}

See \hyperref[glossary:tl]{trading limit}.

\subsection*{Liquidation ratio}
\label{glossary:lr}

The liquidation ratio $\mathcal{LR}\left( q;h\right) $ is the proportion of
the redemption trade that is liquidated after $h$ trading days. We generally
focus on daily and weekly liquidation ratios $\mathcal{LR}\left( q;1\right) $
and $\mathcal{LR}\left( q;5\right) $. The liquidation ratio is also used to
define the \hyperref[glossary:lt]{liquidation time} (or
\hyperref[glossary:ttl]{time to liquidation}), which is an important measure
for managing the liquidity risk. We also use the notation $\mathcal{LR}\left(
\mathbb{R};h\right) $ where $\mathbb{R}$ is the dollar amount of the
\hyperref[glossary:rs]{redemption scenario}.

\subsection*{Liquidation shortfall}
\label{glossary:ls}

The liquidation shortfall is defined as the residual redemption that cannot
be fulfilled after one trading day. It is expressed as a percentage of the
redemption value. If it is equal to $0\%$, this means that we can liquidate
the redemption in one trading day. More generally, its mathematical
expression is:
\begin{equation*}
\mathcal{LS}\left( q\right) =1-\mathcal{LR}\left( q;1\right)
\end{equation*}%
where $\mathcal{LR}\left( q;h\right) $ is the
\hyperref[glossary:lr]{liquidation ratio}. If the redemption scenario is
expressed in dollars, we have:
\begin{equation*}
\mathcal{LS}\left( \mathbb{R}\right) =1-\mathcal{LR}\left( \mathbb{R};1\right)
\end{equation*}

\subsection*{Liquidation time}
\label{glossary:lt}
See \hyperref[glossary:ttl]{time to liquidation}.

\subsection*{Liquidity bucket}
\label{glossary:lb}

A liquidity bucket defines a set of securities that share the same liquidity
properties. Therefore, the securities have the same functional form of the
unit transaction cost. Examples of liquidity buckets are large cap DM stocks,
small cap stocks, sovereign bonds, corporate IG bonds, HY USD bonds, HY EUR
bonds, EM bonds, energy commodities, soft commodities, metal commodities,
agricultural commodities, G10 currencies, EM currencies, REITS, etc.
The $j^{\mathrm{th}}$ liquidity bucket is denoted by $\mathcal{LB}_j$.

\subsection*{Market impact}
\label{glossary:mi}

See \hyperref[glossary:pi]{price impact}.

\subsection*{Outstanding-based participation rate}
\label{glossary:obpr}

The outstanding-based participation rate is a normalization of the trade size:%
\begin{equation*}
y=\frac{q}{\nshares}
\end{equation*}%
where $q$ is the number of shares that have been sold and $\nshares$ is the
number of issued shares. The outstanding-based participation rate is
a modification of the (volume-based) \hyperref[glossary:pr]{participation rate},
because the trading volume cannot always be computed
for some securities, for example bonds.

\subsection*{Participation rate}
\label{glossary:pr}

The participation rate is a normalization of the trade size:%
\begin{equation*}
x=\frac{q}{v}
\end{equation*}%
where $q$ is the number of shares that have been sold and $v$ is the trading
volume. The participation rate is used to define the
\hyperref[glossary:utc]{unit transaction cost}
function $\cost\left( x\right) $.

\subsection*{Price impact (unit)}
\label{glossary:upi}

The (unit) price impact $\pi\left( q\right) $ is the part of the
\hyperref[glossary:utc]{unit transaction cost} function which is not explained by
the \hyperref[glossary:bas]{bid-ask spread}.

\subsection*{Price impact (total)}
\label{glossary:pi}

The price impact (or market impact) $\PI\left( q\right) $ is the part of the
transaction cost due to the trade size:
\begin{equation*}
\PI\left( q\right) =\TC\left( q\right) -\BAS\left( q\right)
\end{equation*}
We generally expect that it is an increasing function of the redemption size.

\subsection*{Pro-rata liquidation}
\label{glossary:prl}

The pro-rata liquidation uses the proportional rule, implying
that each asset is liquidated such that the structure
of the portfolio is the same before and after the liquidation.

\subsection*{Redemption scenario}
\label{glossary:rs}

A redemption scenario $q$ is defined by the vector $\left( q_{1},\ldots
,q_{n}\right) $ where $q_{i}$ is the number of shares of security $i$ to
sell. This scenario can be expressed in dollars:
\begin{equation*}
Q:=\left( Q_{1},\ldots ,Q_{n}\right) =\left( q_{1}P_{1},\ldots
,q_{n}P_{n}\right)
\end{equation*}%
where $P_{i}$ is the price of security $i$. The redemption scenario may also
be defined by its dollar value $\mathbb{R}$:
\begin{equation*}
\mathbb{R} = \mathbb{V}\left(q\right) = \sum_{i=1}^{n}q_{i} P_{i}
\end{equation*}%
If we consider a portfolio defined by its weights $w=\left( w_{1},\ldots
,w_{n}\right) $, we have:
\begin{equation*}
w_{i}=\frac{q_{i} P_{i}}{\mathbb{R}}
\end{equation*}

\subsection*{Time to liquidation}
\label{glossary:ttl}

The time to liquidation is the inverse function of the
\hyperref[glossary:lr]{liquidation ratio}. It indicates the minimum number of
days that it is necessary to liquidate the proportion $p$ of the redemption.
It is denoted by the function $\mathcal{LT}\left( q;p\right) $ or
$\mathcal{LT}\left( \mathbb{R};p\right) $.

\subsection*{Trading limit}
\label{glossary:tl}

The trading limit $q^{+}$ is the maximum number of shares that can be sold in
one trading day. It can be expressed using the maximum
\hyperref[glossary:pr]{participation rate}:
\begin{equation*}
x^{+}=\frac{q^{+}}{v}
\end{equation*}
where $v$ is the daily volume.

\subsection*{Transaction cost}
\label{glossary:tc}

The transaction cost of a redemption is made up of two components: the
\hyperref[glossary:bas]{bid-ask spread} cost and the
\hyperref[glossary:pi]{price impact} cost. It is denoted by
$\mathcal{TC}\left(q\right)$.

\subsection*{Unit transaction cost}
\label{glossary:utc}

The unit transaction cost function $\cost\left( x\right) $ is the percentage cost
associated with the \hyperref[glossary:pr]{participation rate} $x$ for selling
one share. It has two components:
\begin{equation*}
\cost\left( x\right) =\spread+\impact\left( x\right)
\end{equation*}%
where $\spread$ is the half \hyperref[glossary:bas]{bid-ask spread} and
$\impact\left( x\right) $ is the \hyperref[glossary:upi]{price impact}. The
total \hyperref[glossary:tc]{transaction cost} of selling $q$ shares is then:
\begin{equation*}
\TC\left( q\right) =q\cdot P\cdot \cost\left( x\right) =Q \cdot \cost\left( x\right)
\end{equation*}%
where $P$ is the security price and $Q=q\cdot P$ is the nominal selling
volume expressed in \$.

\subsection*{Valuation function}

The valuation function $\mathbb{V}\left( \omega \right) $ gives the dollar
value of the portfolio $\omega =\left( \omega _{1},\ldots ,\omega
_{n}\right) $, which is expressed in number of shares:%
\begin{equation*}
\mathbb{V}\left( \omega \right) =\sum_{i=1}^{n}\omega _{i}P_{i}
\end{equation*}%
The dollar value of the redemption is equal to $\mathbb{R=V}\left( q\right)
=\sum_{i=1}^{n}q_{i}P_{i}$, whereas the dollar value of the portfolio
becomes $\mathbb{V}\left( \omega -q\right) =\sum_{i=1}^{n}\left( \omega
_{i}-q_{i}\right) P_{i}$ after the liquidation of the redemption scenario.

\subsection*{Vertical slicing}
\label{glossary:vs}
See \hyperref[glossary:prl]{pro-rata liquidation}.

\subsection*{Volume-based participation rate}
\label{glossary:vbpr}
See \hyperref[glossary:pr]{participation rate}.

\subsection*{Waterfall liquidation}
\label{glossary:wl}

In this approach, the portfolio is liquidate
by selling the most liquid assets first.

\clearpage

\section{Mathematical results}

\subsection{Relationship between the two unit cost functions in the toy model}
\label{appendix:relationship-toy-model}

We note:%
\begin{equation}
\cost^{\prime }\left( x\right) =\left\{
\begin{array}{ll}
\spread^{\prime } & \text{if }x\leq \tilde{x} \\
\spread^{\prime }+\alpha ^{\prime }\left( x-\tilde{x}\right)  & \text{if }\tilde{x}%
\leq x<x^{+} \\
+\infty  & \text{if }x\geq x^{+}%
\end{array}%
\right. \label{eq:app-toy-model1}
\end{equation}%
and:%
\begin{equation}
\cost^{\prime \prime }\left( x\right) =\left\{
\begin{array}{ll}
\spread^{\prime \prime }+\alpha ^{\prime \prime }x  &
\text{if }x<x^{+} \\
+\infty  & \text{if }x\geq x^{+}%
\end{array}%
\right. \label{eq:app-toy-model2}
\end{equation}%
If we assume that $\cost^{\prime }\left( 0\right) =\cost^{\prime \prime
}\left( 0\right) $ and $\cost^{\prime }\left( x^{+}\right) =\cost^{\prime
\prime }\left( x^{+}\right) $, we have the following relationships:%
\begin{equation*}
\alpha ^{\prime }=\alpha ^{\prime \prime }\left( \frac{x^{+}}{x^{+}-\tilde{x}%
}\right)
\end{equation*}%
and:%
\begin{equation*}
\alpha ^{\prime \prime }=\alpha ^{\prime }\left( \frac{x^{+}-\tilde{x}}{x^{+}%
}\right)
\end{equation*}%
However, most of the time, we do not know the two analytical
functions. Let us assume that the true model is given by
$\cost^{\prime }\left( x\right) $, whereas we estimate the
approximated model $\hat{\cost}^{\prime \prime }\left(
x\right) $, which is defined by:%
\begin{equation}
\hat{\cost}^{\prime \prime }\left( x\right) =\left\{
\begin{array}{ll}
\hat{\spread}^{\prime \prime }+\hat{\alpha}^{\prime \prime }x  & \text{if }x<x^{+} \\
+\infty  & \text{if }x\geq x^{+}%
\end{array}%
\right. \label{eq:app-toy-model3}
\end{equation}%
The least square estimates $\hat{\spread}^{\prime \prime }$ and
$\hat{\alpha}^{\prime \prime }$ are equal to:
\begin{equation*}
\hat{\spread}^{\prime \prime }=\bar{\cost}^{\prime }\left( x\right) -\hat{\alpha}%
^{\prime \prime }\bar{x}
\end{equation*}%
and:%
\begin{equation*}
\hat{\alpha}^{\prime \prime }=\frac{\int_{0}^{x^{+}}\left(
x-\bar{x}\right) \left( \cost^{\prime }\left( x\right)
-\bar{\cost}^{\prime }\left( x\right) \right)
\,\mathrm{d}x}{\int_{0}^{x^{+}}\left( x-\bar{x}\right)
^{2}\,\mathrm{d}x}
\end{equation*}%
where $\bar{x}$ and $\bar{\cost}^{\prime }\left( x\right) $ are
given by the mean value theorem:
\begin{equation*}
\bar{x}=\frac{\int_{0}^{x^{+}}x\,\mathrm{d}x}{x^{+}}=\frac{x^{+}}{2}
\end{equation*}%
and:%
\begin{equation*}
\bar{\cost}^{\prime }\left( x\right)
=\frac{\int_{0}^{x^{+}}\cost^{\prime }\left( x\right)
\,\mathrm{d}x}{x^{+}}=s^{\prime }+\alpha ^{\prime }\frac{\left(
x^{+}-\tilde{x}\right) ^{2}}{2x^{+}}
\end{equation*}%
We deduce that the least square estimates are:%
\begin{equation*}
\hat{\alpha}^{\prime \prime }=\alpha ^{\prime }\left( 1+2\left( \frac{\tilde{%
x}}{x^{+}}\right) ^{3}-3\left( \frac{\tilde{x}}{x^{+}}\right) ^{2}\right)
\end{equation*}%
and:%
\begin{equation*}
\hat{s}^{\prime \prime }=s^{\prime }-\alpha ^{\prime }\tilde{x}\left( 1-%
\frac{\tilde{x}}{x^{+}}\right) ^{2}
\end{equation*}%
because we have:%
\begin{eqnarray*}
\int_{0}^{x^{+}}\left( x-\bar{x}\right) ^{2}\,\mathrm{d}x
&=&\int_{0}^{x^{+}}\left( x-\frac{x^{+}}{2}\right) ^{2}\,\mathrm{d}x \\
&=&\frac{1}{3}\left[ \left( x-\frac{x^{+}}{2}\right) ^{3}\right] _{0}^{x^{+}}
\\
&=&\frac{2}{3}\left( \frac{x^{+}}{2}\right) ^{3}
\end{eqnarray*}%
and\footnote{%
We have:%
\begin{equation*}
\int_{0}^{x^{+}}\left( x-\frac{x^{+}}{2}\right) \,\mathrm{d}x=0
\end{equation*}%
}:%
\begin{eqnarray*}
(\ast ) &=&\int_{0}^{x^{+}}\left( x-\bar{x}\right) \left( \cost^{\prime
}\left( x\right) -\bar{c}^{\prime }\left( x\right) \right) \,\mathrm{d}x \\
&=&\int_{0}^{\tilde{x}}\left( x-\frac{x^{+}}{2}\right) \left( s^{\prime
}-s^{\prime }-\alpha ^{\prime }\frac{\left( x^{+}-\tilde{x}\right) ^{2}}{%
2x^{+}}\right) \,\mathrm{d}x+ \\
&&\int_{\tilde{x}}^{x^{+}}\left( x-\frac{x^{+}}{2}\right) \left( s^{\prime
}+\alpha ^{\prime }\left( x-\tilde{x}\right) -s^{\prime }-\alpha ^{\prime }%
\frac{\left( x^{+}-\tilde{x}\right) ^{2}}{2x^{+}}\right) \,\mathrm{d}x \\
&=&\alpha ^{\prime }\int_{\tilde{x}}^{x^{+}}\left( x-\frac{x^{+}}{2}\right)
\left( x-\tilde{x}\right) \,\mathrm{d}x-\alpha ^{\prime }\frac{\left( x^{+}-%
\tilde{x}\right) ^{2}}{2x^{+}}\int_{0}^{x^{+}}\left( x-\frac{x^{+}}{2}%
\right) \,\mathrm{d}x \\
&=&\alpha ^{\prime }\int_{\tilde{x}}^{x^{+}}\left( x^{2}-\left( \frac{2%
\tilde{x}+x^{+}}{2}\right) x+\frac{\tilde{x}x^{+}}{2}\right) \,\mathrm{d}x \\
&=&\alpha ^{\prime }\left[ \frac{x^{3}}{3}-\left( \frac{2\tilde{x}+x^{+}}{4}%
\right) x^{2}+\frac{\tilde{x}x^{+}}{2}x\right] _{\tilde{x}}^{x^{+}} \\
&=&\alpha ^{\prime }\left( \frac{1}{12}\left( x^{+}\right) ^{3}+\frac{1}{6}%
\tilde{x}^{3}-\frac{1}{4}\tilde{x}^{2}x^{+}\right)
\end{eqnarray*}
\smallskip

\begin{figure}[tbph]
\centering
\caption{Equivalence of cost models}
\label{fig:toy4}
\figureskip
\includegraphics[width = \figurewidth, height = \figureheight]{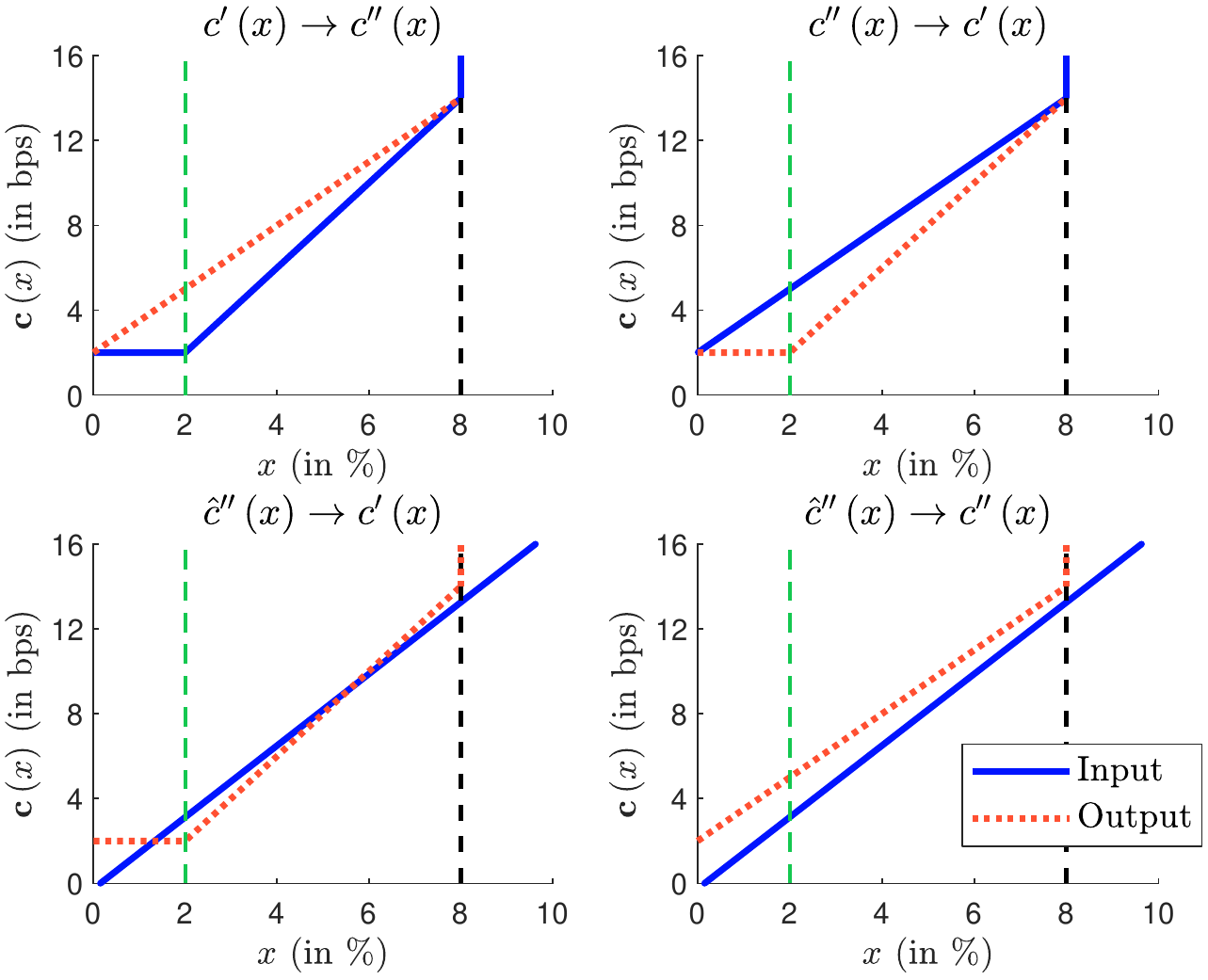}
\end{figure}

In Figure \ref{fig:toy4}, we illustrate how to transform one form of cost
function into another form. In practice, we do not know the models
$\cost^{\prime }\left(x\right) $ and $\cost^{\prime \prime }\left( x\right) $.
In fact, we estimate $\hat{\cost}^{\prime \prime }\left( x\right) $. The right
issue is then to transform $\hat{\cost}^{\prime \prime }\left( x\right) $ into
$\cost^{\prime }\left( x\right) $ or even $\cost^{\prime \prime }\left(
x\right) $. If we consider that the true model is $\cost^{\prime }\left(
x\right) $, we have the following relationships:
\begin{equation}
\alpha ^{\prime }=\hat{\alpha}^{\prime \prime }\frac{\left( x^{+}\right) ^{3}%
}{\left( \left( x^{+}\right) ^{3}+2\tilde{x}^{3}-3\tilde{x}^{2}x^{+}\right) }
\label{eq:app-alpha1}
\end{equation}%
and:%
\begin{equation}
\alpha ^{\prime \prime }=\hat{\alpha}^{\prime \prime }\frac{\left(
x^{+}\right) ^{2}\left( x^{+}-\tilde{x}\right) }{\left( \left( x^{+}\right)
^{3}+2\tilde{x}^{3}-3\tilde{x}^{2}x^{+}\right) }
\label{eq:app-alpha2}
\end{equation}%
If the true model is $\cost^{\prime \prime }\left( x\right) $, we
have $\alpha ^{\prime \prime }=\hat{\alpha}^{\prime \prime }$.

\begin{remark}
In Figure \ref{fig:toy4}, the parameters are equal to $\spread^{\prime }=2$
bps, $\alpha ^{\prime }=2\%$, $\tilde{x}=2\%$ and $x^{+}=8\%$. We find that
$\alpha ^{\prime \prime }=1.5\%$, while the OLS estimation gives $
\hat{\spread}^{\prime \prime }=-0.25$ bps and $\hat{\alpha}^{\prime \prime
}=1.6875\%$.
\end{remark}

\subsection{Analytics of portfolio distortion}
\label{appendix:distortion}

\subsubsection{Portfolio weights}

We recall that the asset structure of the fund is given by the portfolio $%
\omega =\left( \omega _{1},\ldots ,\omega _{n}\right) $, where
$\omega _{i}$ is the number of shares of security $i$. The portfolio
weights are then equal to $w\left( \omega \right) =\left(
w_{1}\left( \omega \right) ,\ldots
,w_{n}\left( \omega \right) \right) $ where:%
\begin{equation}
w_{i}\left( \omega \right) =\frac{\omega
_{i}P_{i}}{\sum_{j=1}^{n}\omega _{j}P_{j}}  \label{eq:distortion-w1}
\end{equation}%
and $P_{i}$ is the current price of security $i$. Let $q=\left(
q_{1},\ldots ,q_{n}\right) $ be the redemption scenario. It follows
that the redemption
weights are given by:%
\begin{equation}
w_{i}\left( q\right) =\frac{q_{i}P_{i}}{\sum_{j=1}^{n}q_{j}P_{j}}
\label{eq:distortion-w2}
\end{equation}%
After the liquidation of $q$, the new asset structure is equal to $\omega -q$,
and the new weights of the portfolio become:%
\begin{equation}
w_{i}\left( \omega -q\right) =\frac{\left( \omega _{i}-q_{i}\right) P_{i}}{%
\sum_{j=1}^{n}\left( \omega _{j}-q_{j}\right) P_{j}}
\label{eq:distortion-w3}
\end{equation}%
We note $\mathbb{V}\left( \omega \right) =\sum_{j=1}^{n}\omega
_{j}P_{j}$ and $\mathbb{V}\left( \omega -q\right)
=\sum_{j=1}^{n}\left( \omega _{j}-q_{j}\right) P_{j}$ the dollar
value of the portfolios before and after
the liquidation. We notice that $\mathbb{V}\left( \omega \right) -\mathbb{V}%
\left( \omega -q\right) $ is exactly equal to the dollar value $\mathbb{R}$ of
the redemption:%
\begin{equation*}
\mathbb{R=V}\left( \omega \right) -\mathbb{V}\left( \omega -q\right)
=\sum_{j=1}^{n}q_{j}P_{j}=\mathbb{V}\left( q\right)
\end{equation*}%
We have:%
\begin{eqnarray*}
w_{i}\left( \omega -q\right)  &=&\frac{\omega
_{i}P_{i}}{\mathbb{V}\left( \omega \right)
-\mathbb{R}}-\frac{q_{i}P_{i}}{\mathbb{V}\left( \omega
\right) -\mathbb{R}} \\
&=&\frac{\mathbb{V}\left( \omega \right) }{\mathbb{V}\left( \omega \right) -%
\mathbb{R}}w_{i}\left( \omega \right)
-\frac{\mathbb{R}}{\mathbb{V}\left( \omega \right)
-\mathbb{R}}w_{i}\left( q\right)
\end{eqnarray*}%
The new weights $w_{i}\left( \omega -q\right) $ are a non-linear
function of
the portfolio weights $w_{i}\left( \omega \right) $, the redemption weights
$w_{i}\left( q\right) $ and the redemption value $\mathbb{R}$. Except
in the
case\footnote{%
We have $w_{i}\left( \omega -q\right) =w_{i}\left( \omega \right) $.} where $%
q_{i}\propto \omega _{i}$, computing the new weights is
not straightforward because they depend on $\mathbb{R}$. From a
theoretical point of view, we have $0\leq q_{i}\leq \omega _{i}$
because the maximum we can sell is the number of shares in the
portfolio. One problem is that the weights $w_{i}\left( \omega
-q\right) $ are continuous whereas the number of shares $q_{i}$ is an
integer. This is why we prefer to consider the fuzzy constraint
$-\epsilon \leq q_{i}\leq \omega _{i}+\epsilon $, where $\epsilon $
is typically equal to $\nicefrac{1}{2}$. Since
$\sum_{i=1}^{n}w_{i}\left( \omega -q\right) =1$, we deduce that:
\begin{equation*}
-\epsilon \leq q_{i}\leq \omega _{i}+\epsilon \Leftrightarrow
-\epsilon
_{i}\leq w_{i}\left( \omega -q\right) \leq \min \left( \frac{\mathbb{V}%
\left( \omega \right) }{\mathbb{V}\left( \omega \right) -\mathbb{R}}%
w_{i}\left( \omega \right) +\epsilon _{i},1\right)
\end{equation*}%
where:%
\begin{equation*}
\epsilon _{i}=\frac{\epsilon P_{i}}{\mathbb{V}\left( \omega \right) -\mathbb{%
R}}
\end{equation*}%
We note the two bounds $w_{i}^{-}\left( \omega -q\right) $ and $%
w_{i}^{+}\left( \omega -q\right) $.

\begin{remark}
From Equation (\ref{eq:distortion-w3}), we deduce that:%
\begin{equation*}
q_{i}=\frac{\mathbb{V}\left( \omega \right) \left( w_{i}\left(
\omega \right) -w_{i}\left( \omega -q\right) \right)
+\mathbb{R}w_{i}\left( \omega -q\right) }{P_{i}}
\end{equation*}%
We can then compute $q_{i}$ thanks to the previous equation when we
know the portfolios weights $w_{i}\left( \omega \right) $ and
$w_{i}\left( \omega -q\right) $.
\end{remark}

\subsubsection{Liquidation tracking error}

We assume that the asset returns are normally distributed: $R=\left(
R_{1},\ldots ,R_{n}\right) \sim \mathcal{N}\left( 0,\Sigma \right)
$. The
random return of the portfolio $\omega $ is then equal to:%
\begin{eqnarray*}
R\left( \omega \right)  &=&\frac{\sum_{i=1}^{n}\omega _{i}P_{i}R_{i}}{%
\sum_{j=1}^{n}\omega _{j}P_{j}} \\
&=&\sum_{i=1}^{n}w_{i}\left( \omega \right) R_{i} \\
&=&w_{i}\left( \omega \right) ^{\top }R
\end{eqnarray*}%
We conclude that:%
\begin{equation*}
R\left( \omega \right) \sim \mathcal{N}\left( 0,w\left( \omega
\right) ^{\top }\Sigma w\left( \omega \right) \right)
\end{equation*}%
If we consider the portfolio $\omega -q$, we have $R\left( \omega
-q\right)
=w\left( \omega -q\right) ^{\top }R$ and:%
\begin{equation*}
\left(
\begin{array}{c}
R\left( \omega \right)  \\
R\left( \omega -q\right)
\end{array}%
\right) \sim \mathcal{N}\left( \left(
\begin{array}{c}
0 \\
0%
\end{array}%
\right) ,\left(
\begin{array}{cc}
w\left( \omega \right) ^{\top }\Sigma w\left( \omega \right)  &
w\left(
\omega \right) ^{\top }\Sigma w\left( \omega -q\right)  \\
w\left( \omega -q\right) ^{\top }\Sigma w\left( \omega \right)  &
w\left( \omega -q\right) ^{\top }\Sigma w\left( \omega -q\right)
\end{array}%
\right) \right)
\end{equation*}%
Let $e$ be the tracking error between the portfolios before and
after the
redemption. We have:%
\begin{eqnarray*}
e &=&R\left( \omega -q\right) -R\left( \omega \right)  \\
&=&\left( w\left( \omega \right) -w\left( \omega -q\right) \right)
^{\top }R
\end{eqnarray*}%
The standard deviation of $e$ is called the \textquotedblleft \textit{%
liquidation tracking error}\textquotedblright\ and is denoted by
$\sigma
\left( q\mid \omega \right) $:%
\begin{equation*}
\sigma \left( q\mid \omega \right) =\sqrt{\left( w\left( \omega
\right) -w\left( \omega -q\right) \right) ^{\top }\Sigma \left(
w\left( \omega \right) -w\left( \omega -q\right) \right) }
\end{equation*}%
This is our measure of the portfolio distortion $\mathcal{D}\left( q\mid \omega \right)$.

\begin{remark}
In the case where the redemption scenario does not modify the asset
structure, we have $q_{i}=\redemption\omega _{i}$ and:%
\begin{eqnarray*}
w\left( \omega -q\right)  &=&\frac{\left( \omega _{i}-q_{i}\right) P_{i}}{%
\sum_{j=1}^{n}\left( \omega _{j}-q_{j}\right) P_{j}} \\
&=&\frac{\left( \omega _{i}-\redemption\omega _{i}\right) P_{i}}{%
\sum_{j=1}^{n}\left( \omega _{j}-\redemption\omega _{j}\right) P_{j}} \\
&=&\frac{\left( 1-\redemption\right) \omega
_{i}P_{i}}{\sum_{j=1}^{n}\left(
1-\redemption\right) \omega _{j}P_{j}} \\
&=&w_{i}
\end{eqnarray*}%
We conclude that the portfolio distortion is equal to zero.
\end{remark}

\subsubsection{Optimal portfolio liquidation}
\label{appendix:distortion-optimal}

Let $c\left( q\mid \omega \right) $ be the cost of liquidating the
redemption scenario $q$. The problem of optimal portfolio liquidation is:
\begin{eqnarray}
q^{\star } &=&\underset{q}{\arg \min }\ c\left( q\mid \omega \right)
\label{eq:optimal1} \\
&\text{s.t.}&\left\{
\begin{array}{l}
\sigma \left( q\mid \omega \right) \leq \mathcal{D}^{+} \\
\mathbf{1}_{n}^{\top }w\left( \omega -q\right) =1 \\
w^{-}\left( \omega -q\right) \leq w\left( \omega -q\right) \leq w^{+}\left(
\omega -q\right)
\end{array}%
\right.   \notag
\end{eqnarray}%
where $\mathcal{D}^{+}\geq 0$ is the maximum portfolio distortion. If
$\mathcal{D}^{+}=0$, the optimal solution is $q^{\star }\propto \omega $. If
$\mathcal{D}^{+}=\infty $, the distortion constraint vanishes, and the
solution corresponds to the redemption scenario that presents the lower
liquidating cost.\smallskip

We can rewrite the previous problem as follows:
\begin{eqnarray}
q^{\star }\left( \lambda \right)  &=&\arg \min \ \frac{1}{2}\sigma
^{2}\left( q\mid \omega \right) +\lambda c\left( q\mid \omega \right)
\label{eq:optimal2} \\
&&\text{s.t.}\left\{
\begin{array}{l}
\mathbf{1}_{n}^{\top }w\left( \omega -q\right) =1 \\
w^{-}\left( \omega -q\right) \leq w\left( \omega -q\right) \leq w^{+}\left(
\omega -q\right)
\end{array}%
\right.   \notag
\end{eqnarray}%
This optimization problem is close to the $\gamma $-problem of mean-variance
optimization \citep{Roncalli-2013}. Nevertheless, this is not a QP problem,
meaning that it is more complex to solve numerically. The underlying idea is
then to write $q$ as a function of $w\left( q\right) $ with
$q_{i}=w_{i}\left( q\right) \mathbb{R}/P_{i}$ and minimizing the objective
function (\ref{eq:optimal2}) with respect to $w\left( q\right) $. Given a
dollar value $\mathbb{R}$ of redemption, the set of optimal portfolio
liquidations is given by $\left\{ q^{\star }\left( \lambda \right) ,\lambda
\in \left[ 0,\infty \right) \right\} $ and the efficient frontier
corresponds to the parametric curve $\left( \sigma \left( q^{\star }\left(
\lambda \right) \mid \omega \right) ,c\left( q^{\star }\left( \lambda
\right) \mid \omega \right) \right) $.\smallskip

\subsection{Modeling the market risk of corporate bonds}
\label{appendix:sigma-bond}

Let $\cspread_{i}\left( t\right) $ be the credit spread of the $i^{\mathrm{th%
}}$ bond issuer. Following \citet[pages 223-227]{Roncalli-2013}, we assume that
the credit spread follows a general diffusion process:
\begin{equation}
\mathrm{d}\cspread_{i}\left( t\right) =\sigma _{i}^{\cspread}\cspread%
_{i}\left( t\right) \,\mathrm{d}W_{i}\left( t\right)   \label{eq:vol-bond1}
\end{equation}%
where $W_{i}\left( t\right) $ is a standard Brownian motion and $\sigma
_{i}^{\cspread}$ is a volatility parameter. We note $B_{i}\left( t,D_{i}\right)
$ the zero-coupon bond price with maturity (or duration) $D_{i}$ of the
$i^{\mathrm{th}}$ issuer. If we assume that the recovery date
is equal to zero, we have:%
\begin{equation*}
\mathrm{d}\ln B_{i}\left( t,D_{i}\right) =-D_{i}\,\mathrm{d}r\left( t\right)
-D_{i}\,\mathrm{d}\cspread_{i}\left( t\right)
\end{equation*}%
where $r\left( t\right) $ is the risk-free interest rate. If we assume that the
credit spread is not correlated with the risk-free interest rate, we
deduce that:%
\begin{eqnarray}
\sigma ^{2}\left( \mathrm{d}\ln B_{i}\left( t,D_{i}\right) \right)
&=&D_{i}^{2}\sigma ^{2}\left( \mathrm{d}r\left( t\right) \right)
+D_{i}^{2}\sigma ^{2}\left( \mathrm{d}\cspread_{i}\left( t\right) \right)
\notag \\
&=&D_{i}^{2}\sigma ^{2}\left( \mathrm{d}r\left( t\right) \right)
+D_{i}^{2}\left( \sigma _{i}^{\cspread}\right) ^{2}\cspread_{i}^{2}\left(
t\right) \,\mathrm{d}t  \label{eq:vol-bond2}
\end{eqnarray}
We deduce that the volatility of a bond has two parts:
an interest rate component and a credit spread component.
\smallskip

If the credit risk component is sufficiently large with respect to the
interest rate component, we obtain:%
\begin{eqnarray}
\sigma \left( \mathrm{d}\ln B_{i}\left( t,D_{i}\right) \right)  &\approx
&\sigma _{i}^{\cspread} \cdot D_{i} \cdot \cspread_{i}\left( t\right)   \notag \\
&=&\sigma _{i}^{\cspread}\cdot \dts_{i}\left( t\right)   \label{eq:vol-bond3}
\end{eqnarray}
where $\dts_{i}\left( t\right)$ is the duration-times-spread (or DTS) measure
\citep{BenDor-2007}.

\clearpage

\section{Data}

We consider the asset liquidity data provided by Amundi Asset Management.
The database is called \textquotedblleft \textit{Amundi Liquidity Lab}\textquotedblright\
and contains the trades made by Amundi, but also other information such as order books for equities
and the price quotations for bonds\footnote{For each trade, we have at least three price quotations
by three different banks and brokers.}. We filter the data in order to obtain a dataset
with all the available characteristics, which are representative of normal trading.
For instance, we exclude bond trades that are initiated by the counterparty.
We also exclude equity trades that are made by an index fund manager when
the transaction concerns a basket of stocks that replicate the index. Indeed, in this case,
the transaction cost is generally related to the index, and does not necessarily reflect
the transaction cost of each component. Finally, we use
a subset of the data.

\subsection{Equities}
\label{appendix:data-equity}

We use a sample of trades
for the stocks that belong to the MSCI USA, MSCI Europe,
MSCI USA Small Cap and MSCI Europe Small Cap indices.
We also complete this database with pre-trade transaction costs computed by
the BECS system \citep{Citigroup-2020} when we observe few observations for a given stock.
Finally, we have a sample of $149\,896$ trades.

\subsection{Sovereign bonds}
\label{appendix:data-sovereign-bond}

We use a sample of $196\,286$ trades from January 2018 to December 2020
with the following split by currency:
\begin{equation*}
\begin{tabular}{cccccccc}
\hline
Currency     &      EUR &     USD &    GBP &    JPY &    AUD &    CAD &    DKK \\
\# of trades & 129\,904 & 34\,965 & 7\,354 & 6\,831 & 4\,277 & 3\,586 & 1\,409 \\
\hline
Currency     & SEK & MXN & PLN & MYR & SGD & ZAR &    Other \\
\# of trades & 915 & 882 & 794 & 592 & 581 & 458 & $3\,738$ \\
\hline
\end{tabular}
\end{equation*}
and the following split by the issuer's country:
\begin{equation*}
\begin{tabular}{cccccccc}
\hline
Country      &      IT &      FR &      US &      DE &      ES &     BE &     GB  \\
\# of trades & 31\,870 & 23\,033 & 20\,798 & 19\,587 & 16\,668 & 8\,961 & 7\,646  \\ \hline
Country      &     JP &     NL &     AT &     AU &     CA &     PT &   Other       \\
\# of trades & 6\,874 & 6\,663 & 6\,619 & 4\,383 & 3\,950 & 3\,900 & 35\,334       \\
\hline
\end{tabular}
\end{equation*}

 \subsection{Corporate bonds}
\label{appendix:data-corporate-bond}

We use a sample of $258\,153$ trades from January 2018 to December 2020
with the following split by currency:
\begin{equation*}
\begin{tabular}{ccccccccc}
\hline
Currency     &      EUR &     USD &    GBP &    SGD &    AUD &    CAD & CNH &    Other \\
\# of trades & 204\,724 & 46\,620 & 5\,791 &    307 &    194 &    138 & 128 & 251 \\
\hline
\end{tabular}
\end{equation*}
and the following split by the issuer's country:
\begin{equation*}
\begin{tabular}{cccccccc}
\hline
Country      &      US &      FR &      NL &      GB &      DE &      IT &     LU  \\
\# of trades & 49\,410 & 48\,257 & 34\,782 & 21\,710 & 16\,358 & 16\,037 & 12\,150  \\ \hline
Country      &     ES &     SE &      IE &     MX &     AT &     BE &   Other       \\
\# of trades & 11\,797 & 5\,857 & 4\,775 & 3\,799 & 3\,289 & 3\,173 & 27\,709       \\
\hline
\end{tabular}
\end{equation*}

\clearpage

\section{Price impact of the benchmark formulas}
\label{appendix:benchmark}

\begin{table}[h!]
\centering
\caption{Price impact (in bps) for large cap stocks}
\scalebox{0.95}{
\begin{tabular}{c|ccccccccc}
\hline
$\sigma$ & \multicolumn{9}{c}{$x$ (in \%)} \\
(in \%) & 0.01 & 0.05 & 0.10 & 0.50        &         1 &         5 &   10 &   20 &         30 \\ \hline
$10$ & $0.2$ & $0.6$ & $0.8$ & ${\TsV}1.8$ & ${\TsV}2$ & ${\TsV}6$ & ${\TsV}8$ & $11$ & $14$ \\
$20$ & $0.5$ & $1.1$ & $1.6$ & ${\TsV}3.5$ & ${\TsV}5$ &      $11$ &      $16$ & $22$ & $27$ \\
$30$ & $0.7$ & $1.7$ & $2.4$ & ${\TsV}5.3$ & ${\TsV}7$ &      $17$ &      $24$ & $33$ & $41$ \\
$40$ & $1.0$ & $2.2$ & $3.1$ & ${\TsV}7.0$ &      $10$ &      $22$ &      $31$ & $44$ & $54$ \\
$50$ & $1.2$ & $2.8$ & $3.9$ & ${\TsV}8.8$ &      $12$ &      $28$ &      $39$ & $55$ & $68$ \\
$60$ & $1.5$ & $3.3$ & $4.7$ &      $10.5$ &      $15$ &      $33$ &      $47$ & $67$ & $82$ \\
\hline
\end{tabular}}
\bigskip

\caption{Price impact (in bps) for small cap stocks}
\scalebox{0.95}{
\begin{tabular}{c|ccccccccc}
\hline
$\sigma$ & \multicolumn{9}{c}{$x$ (in \%)} \\
(in \%) & 0.01 & 0.05 & 0.10 & 0.50        &         1 &         5 &   10 &   20 &         30 \\ \hline
$10$ & $0.3$ & $0.7$ & $1.0$ & ${\TsV}2.2$ & ${\TsV}3$ & ${\TsV}7$ & $10$ & $14$ & ${\TsV}17$ \\
$20$ & $0.6$ & $1.4$ & $2.0$ & ${\TsV}4.4$ & ${\TsV}6$ &      $14$ & $20$ & $28$ & ${\TsV}34$ \\
$30$ & $0.9$ & $2.1$ & $2.9$ & ${\TsV}6.6$ & ${\TsV}9$ &      $21$ & $29$ & $42$ & ${\TsV}51$ \\
$40$ & $1.2$ & $2.8$ & $3.9$ & ${\TsV}8.8$ &      $12$ &      $28$ & $39$ & $55$ & ${\TsV}68$ \\
$50$ & $1.6$ & $3.5$ & $4.9$ &      $11.0$ &      $16$ &      $35$ & $49$ & $69$ & ${\TsV}85$ \\
$60$ & $1.9$ & $4.2$ & $5.9$ &      $13.2$ &      $19$ &      $42$ & $59$ & $83$ &      $102$ \\
\hline
\end{tabular}}
\bigskip

\caption{Price impact (in bps) for sovereign bonds}
\scalebox{0.95}{
\begin{tabular}{c|ccccccccc}
\hline
$\sigma$ & \multicolumn{9}{c}{$y$ (in bps)} \\
(in \%)   &        0.01 &        0.10 &           1 &        2.5  &           5 &        10 &         20 &         50 &        100 \\ \hline
${\TsV}1$ & ${\TsV}0.6$ & ${\TsV}1.0$ & ${\TsV}1.9$ & ${\TsV}2.3$ & ${\TsV}2.8$ & ${\TsV}3$ & ${\TsX}4$ & ${\TsX}5$ &  ${\TsX}6$ \\
${\TsV}2$ & ${\TsV}1.2$ & ${\TsV}2.1$ & ${\TsV}3.7$ & ${\TsV}4.7$ & ${\TsV}5.6$ & ${\TsV}7$ & ${\TsX}8$ &      $10$ & ${\TsV}12$ \\
${\TsV}3$ & ${\TsV}1.8$ & ${\TsV}3.1$ & ${\TsV}5.6$ & ${\TsV}7.0$ & ${\TsV}8.3$ &      $10$ &      $12$ &      $15$ & ${\TsV}18$ \\
${\TsV}5$ & ${\TsV}2.9$ & ${\TsV}5.2$ & ${\TsV}9.3$ &      $11.7$ &      $13.9$ &      $17$ &      $20$ &      $25$ & ${\TsV}29$ \\
     $10$ & ${\TsV}5.9$ &      $10.5$ &      $18.6$ &      $23.4$ &      $27.8$ &      $33$ &      $39$ &      $49$ & ${\TsV}59$ \\
     $15$ & ${\TsV}8.8$ &      $15.7$ &      $27.9$ &      $35.1$ &      $41.7$ &      $50$ &      $59$ &      $74$ & ${\TsV}88$ \\
     $20$ &      $11.8$ &      $20.9$ &      $37.2$ &      $46.8$ &      $55.6$ &      $66$ &      $79$ &      $99$ &      $118$ \\
\hline
\end{tabular}}
\bigskip

\caption{Price impact (in bps) for corporate bonds}
\scalebox{0.95}{
\begin{tabular}{c|ccccccccc}
\hline
$\dts$ & \multicolumn{9}{c}{$y$ (in bps)} \\
(in bps)       &        0.01 &        0.10 &           1 &        2.5  &           5 &         10 &         20 &         50 &        100 \\ \hline
 ${\TsXV}\,50$ & ${\TsV}0.2$ & ${\TsV}0.4$ & ${\TsV}0.6$ & ${\TsV}0.8$ & ${\TsV}0.9$ &  ${\TsX}1$ &  ${\TsX}1$ &  ${\TsX}2$ &  ${\TsX}2$ \\
 ${\TsX}\,100$ & ${\TsV}0.4$ & ${\TsV}0.7$ & ${\TsV}1.3$ & ${\TsV}1.6$ & ${\TsV}1.9$ &  ${\TsX}2$ &  ${\TsX}3$ &  ${\TsX}3$ &  ${\TsX}4$ \\
 ${\TsX}\,250$ & ${\TsV}1.0$ & ${\TsV}1.8$ & ${\TsV}3.1$ & ${\TsV}3.9$ & ${\TsV}4.7$ &  ${\TsX}6$ &  ${\TsX}7$ &  ${\TsX}8$ & ${\TsV}10$ \\
 ${\TsX}\,500$ & ${\TsV}2.0$ & ${\TsV}3.5$ & ${\TsV}6.3$ & ${\TsV}7.9$ & ${\TsV}9.3$ & ${\TsV}11$ & ${\TsV}13$ & ${\TsV}17$ & ${\TsV}20$ \\
${\TsV}1\,000$ & ${\TsV}4.0$ & ${\TsV}7.0$ &      $12.5$ &      $15.7$ &      $18.7$ & ${\TsV}22$ & ${\TsV}26$ & ${\TsV}33$ & ${\TsV}40$ \\
${\TsV}2\,500$ & ${\TsV}9.9$ &      $17.6$ &      $31.3$ &      $39.3$ &      $46.7$ & ${\TsV}56$ & ${\TsV}66$ & ${\TsV}83$ & ${\TsV}99$ \\
${\TsV}5\,000$ &      $19.8$ &      $35.1$ &      $62.5$ &      $78.6$ &      $93.5$ &      $111$ &      $132$ &      $166$ &      $198$ \\
\hline
\end{tabular}}
\end{table}

\section{Additional results}

\clearpage

\begin{figure}[tbph]
\centering
\caption{Linear modeling of unit transaction costs}
\label{fig:toy2}
\figureskip
\includegraphics[width = \figurewidth, height = \figureheight]{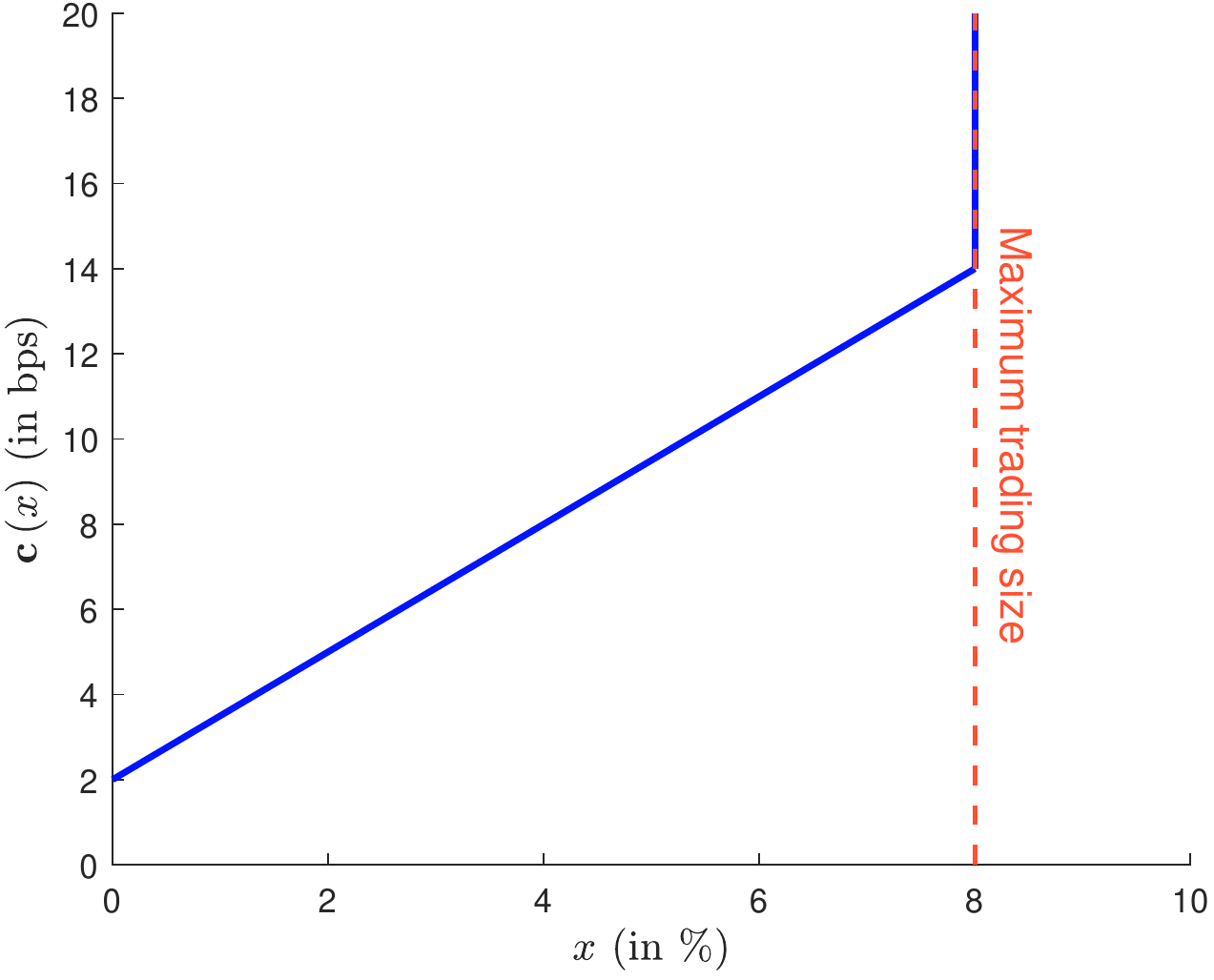}

\centering
\caption{Comparing unit and total transaction costs in normal and stress periods}
\label{fig:stress8}
\figureskip
\includegraphics[width = \figurewidth, height = \figureheight]{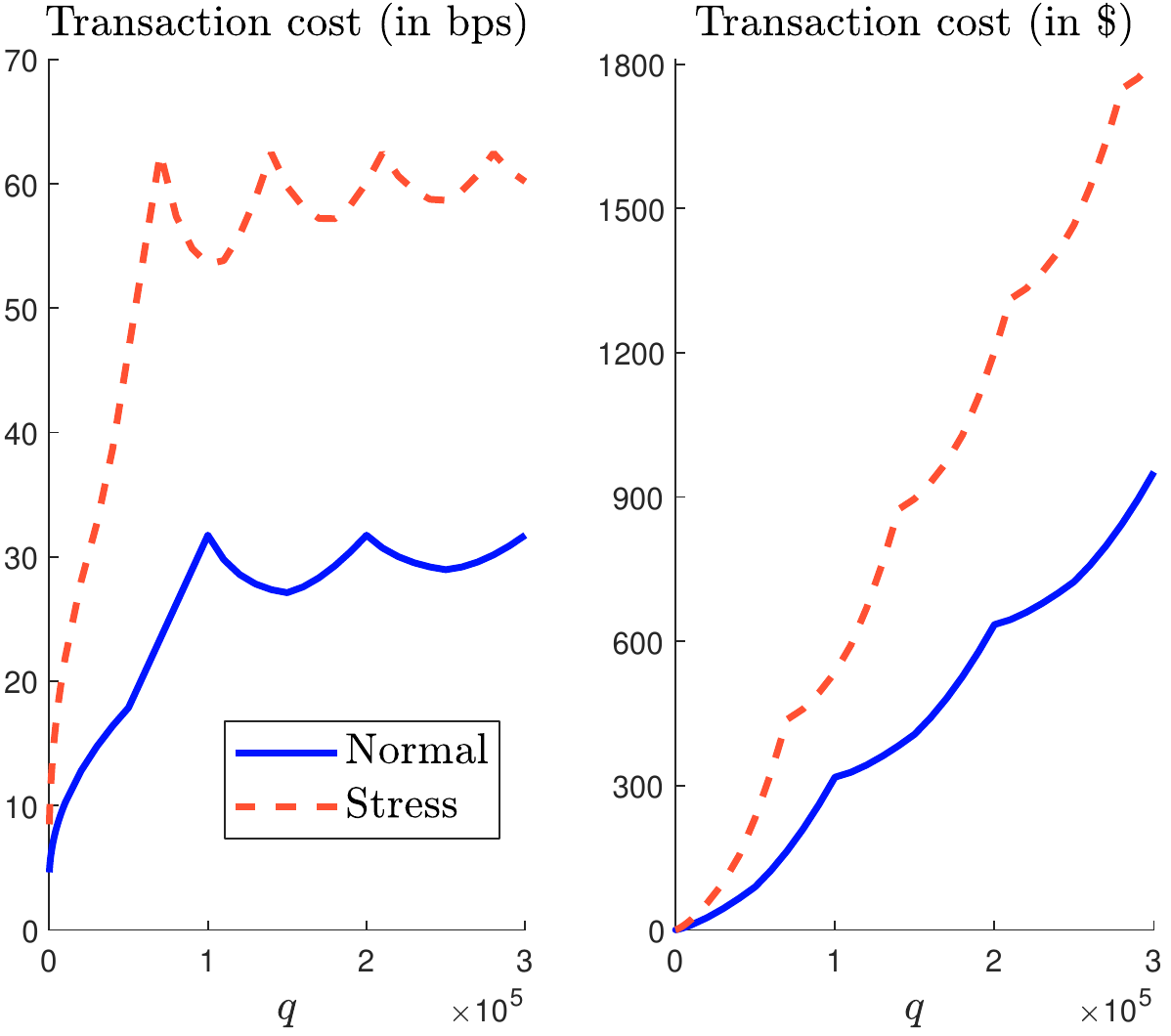}
\end{figure}

\begin{table}[tbph]
\centering
\caption{Participation rate $x_i\left(h\right)$ (in \%)}
\label{tab:liquidation4a}
\begin{tabular}{cccccc}
\hline
$h$   &       Asset \#1 &      Asset \#2 &      Asset \#3 & Asset \#4 &  Asset \#5 \\ \hline
$1$   &       $10.00\%$ &      $10.00\%$ &      $10.00\%$ &  $8.75\%$ &   $0.90\%$ \\
$2$   &       $10.00\%$ &      $10.00\%$ &      $10.00\%$ &           &            \\
$3$   &       $10.00\%$ & ${\TsV}0.05\%$ &      $10.00\%$ &           &            \\
$4$   &       $10.00\%$ &                & ${\TsV}7.75\%$ &           &            \\
$5$   &  ${\TsV}3.51\%$ &                &                &           &            \\ \hline
\end{tabular}
\bigskip

\centering
\caption{Notional $Q_i\left(h\right)$ (in \$)}
\label{tab:liquidation4b}
\begin{tabular}{cccccc}
\hline
$h$   &       Asset \#1 &      Asset \#2 &      Asset \#3 & Asset \#4 &  Asset \#5 \\ \hline
$1$   &  $89\,000$ &     $102\,000$ & $13\,400$ & $20\,825$ & $10\,602$ \\
$2$   &  $89\,000$ &     $102\,000$ & $13\,400$ &           &           \\
$3$   &  $89\,000$ & ${\TsXV}\,510$ & $13\,400$ &           &           \\
$4$   &  $89\,000$ &                & $10\,385$ &           &           \\
$5$   &  $31\,239$ &                &           &           &           \\ \hline
\end{tabular}
\bigskip

\centering
\caption{Bid-ask spread cost (in \$)}
\label{tab:liquidation4c}
\begin{tabular}{ccccccc}
\hline
$h$   &       Asset \#1 &      Asset \#2 &      Asset \#3 & Asset \#4 &  Asset \#5 & Total\\ \hline
$1$   & ${\TsV}35.60$ &      $40.80$ & ${\TsV}6.70$ &      $10.41$ &      $5.30$ & ${\TsV}98.81$  \\
$2$   & ${\TsV}35.60$ &      $40.80$ & ${\TsV}6.70$ &              &             & ${\TsV}83.10$  \\
$3$   & ${\TsV}35.60$ &              & ${\TsV}6.70$ &              &             & ${\TsV}42.50$  \\
$4$   & ${\TsV}35.60$ &              & ${\TsV}5.19$ &              &             & ${\TsV}40.79$  \\
$5$   & ${\TsV}12.50$ &              &              &              &             & ${\TsV}12.50$  \\ \hline
Total &      $154.90$ &      $81.80$ &      $25.29$ &      $10.41$ &      $5.30$ &      $277.71$  \\ \hline
\end{tabular}
\bigskip

\centering
\caption{Price impact cost (in \$)}
\label{tab:liquidation4d}
\begin{tabular}{ccccccc}
\hline
$h$   &       Asset \#1 &      Asset \#2 &      Asset \#3 & Asset \#4 &  Asset \#5 & Total\\ \hline
$1$   & ${\TsV}\,617.10$ & ${\TsV}\,565.79$ & ${\TsV}66.90$ &      $151.62$ &      $12.48$ &      $1\,413.89$ \\
$2$   & ${\TsV}\,617.10$ & ${\TsV}\,565.79$ & ${\TsV}66.90$ &               &              &      $1\,249.80$ \\
$3$   & ${\TsV}\,617.10$ &  ${\TsXV}\,0.14$ & ${\TsV}66.90$ &               &              & ${\TsV}\,684.14$ \\
$4$   & ${\TsV}\,617.10$ &                  & ${\TsV}40.18$ &               &              & ${\TsV}\,657.28$ \\
$5$   &  ${\TsX}\,90.74$ &                  &               &               &              &  ${\TsX}\,90.74$ \\ \hline
Total &      $2\,559.16$ &      $1\,131.73$ &      $240.87$ &      $151.62$ &      $12.48$ &      $4\,095.85$ \\ \hline
\end{tabular}
\bigskip

\centering
\caption{Transaction cost (in \$)}
\label{tab:liquidation4e}
\begin{tabular}{ccccccc}
\hline
$h$   &       Asset \#1 &      Asset \#2 &      Asset \#3 & Asset \#4 &  Asset \#5 & Total\\ \hline
$1$   & ${\TsV}\,652.70$ & ${\TsV}\,606.59$ & ${\TsV}73.60$ &      $162.03$ &      $17.78$ &      $1\,512.70$ \\
$2$   & ${\TsV}\,652.70$ & ${\TsV}\,606.59$ & ${\TsV}73.60$ &               &              &      $1\,332.90$ \\
$3$   & ${\TsV}\,652.70$ &  ${\TsXV}\,0.35$ & ${\TsV}73.60$ &               &              & ${\TsV}\,726.65$ \\
$4$   & ${\TsV}\,652.70$ &                  & ${\TsV}45.37$ &               &              & ${\TsV}\,698.08$ \\
$5$   & ${\TsV}\,103.24$ &                  &               &               &              & ${\TsV}\,103.24$ \\ \hline
Total &      $2\,714.05$ &      $1\,213.53$ &      $266.16$ &      $162.03$ &      $17.78$ &      $4\,373.55$ \\ \hline
\end{tabular}
\end{table}

\begin{figure}[tbph]
\centering
\caption{Estimated price impact (in bps) --- logarithmic scale}
\label{fig:blab_sovereign_calib3b}
\figureskip
\includegraphics[width = \figurewidth, height = \figureheight]{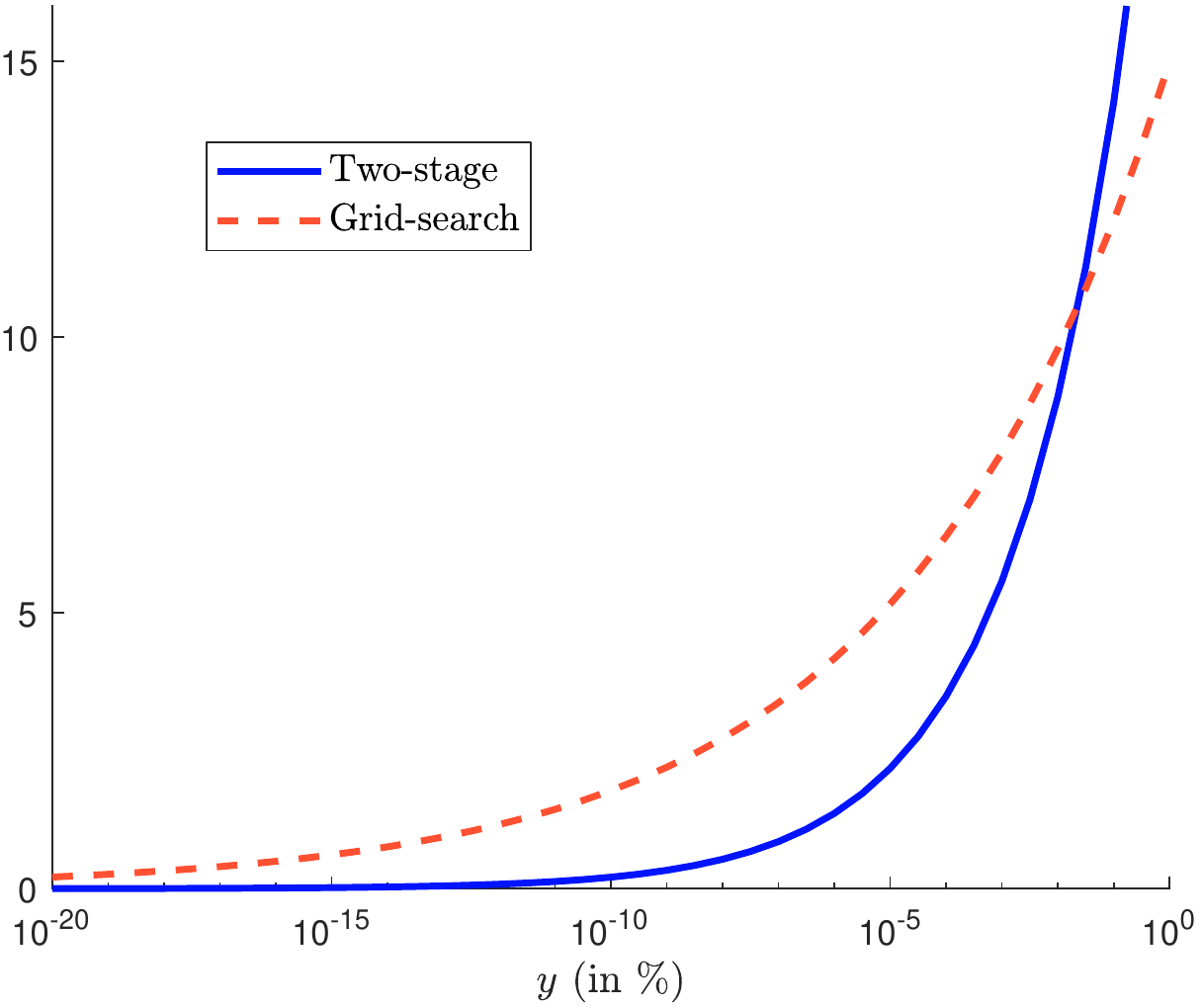}
\end{figure}

\begin{table}[tbph]
\centering
\caption{Two-stage estimation of the sovereign bond transaction cost model without the intercept by issuer}
\label{tab:blab_sovereign_calib6b}
\begin{tabular}{lccccc}
\hline
Issuer & $\gamma _{1}$ & $\beta ^{\left( \spread\right) }$ &
$\tilde{\beta}^{\left( \impact\right) }$ & $R^{2}$ (in \%) &
$R_{c}^{2}$ (in \%) \vphantom{$\dfrac{1}{1}$} \\ \hline
Austria        & $0.2255$ & $0.8023$ & $3.0845$ & $53.9$ & $48.2$ \\
Belgium        & $0.2482$ & $0.7789$ & $3.3738$ & $44.0$ & $32.5$ \\
EM             & $0.0519$ & $0.9158$ & $0.4746$ & $73.6$ & $44.7$ \\
Finland        & $0.2894$ & $0.7114$ & $4.0416$ & $46.3$ & $31.8$ \\
France         & $0.2138$ & $0.8942$ & $3.0148$ & $40.1$ & $29.7$ \\
Germany        & $0.2415$ & $1.0413$ & $2.7838$ & $51.5$ & $38.5$ \\
Ireland        & $0.2098$ & $0.6600$ & $2.4977$ & $43.8$ & $26.4$ \\
Italy          & $0.1744$ & $2.4706$ & $1.7640$ & $31.0$ & $22.0$ \\
Japan          & $0.0657$ & $0.5635$ & $0.7315$ & $78.0$ & $53.4$ \\
Netherlands    & $0.2320$ & $0.7219$ & $3.7355$ & $46.9$ & $34.2$ \\
Portugal       & $0.2318$ & $0.9693$ & $3.0639$ & $49.6$ & $33.0$ \\
Spain          & $0.2185$ & $1.3000$ & $2.0990$ & $40.8$ & $26.7$ \\
United Kingdom & $0.2194$ & $0.9739$ & $2.6262$ & $49.9$ & $28.5$ \\
USA            & $0.1252$ & $1.1055$ & $1.3395$ & $53.6$ & $40.7$ \\
\hline
\end{tabular}
\end{table}

\begin{table}[tbph]
\centering
\caption{Two-stage estimation of the sovereign bond transaction cost model without the intercept by currency}
\label{tab:blab_sovereign_calib7b}
\begin{tabular}{lccccc}
\hline
Currency & $\gamma _{1}$ & $\beta ^{\left( \spread\right) }$ &
$\tilde{\beta}^{\left( \impact\right) }$ & $R^{2}$ (in \%) &
$R_{c}^{2}$ (in \%) \vphantom{$\dfrac{1}{1}$} \\ \hline
EUR  & $0.2262$ & $1.0428$ & $2.9347$ & $35.2$ & $25.7$ \\
GBP  & $0.2117$ & $1.5328$ & $2.2890$ & $48.3$ & $29.5$ \\
JPY  & $0.0834$ & $0.5744$ & $0.9771$ & $74.2$ & $48.2$ \\
USD  & $0.1408$ & $0.9502$ & $1.0906$ & $60.4$ & $45.4$ \\
\hline
\end{tabular}
\end{table}

\begin{figure}
\centering
\caption{Relationship between volatility and duration-times-spread (sovereign bonds)}
\label{fig:blab_sovereign_data2}
\figureskip
\includegraphics[width = \figurewidth, height = \figureheight]{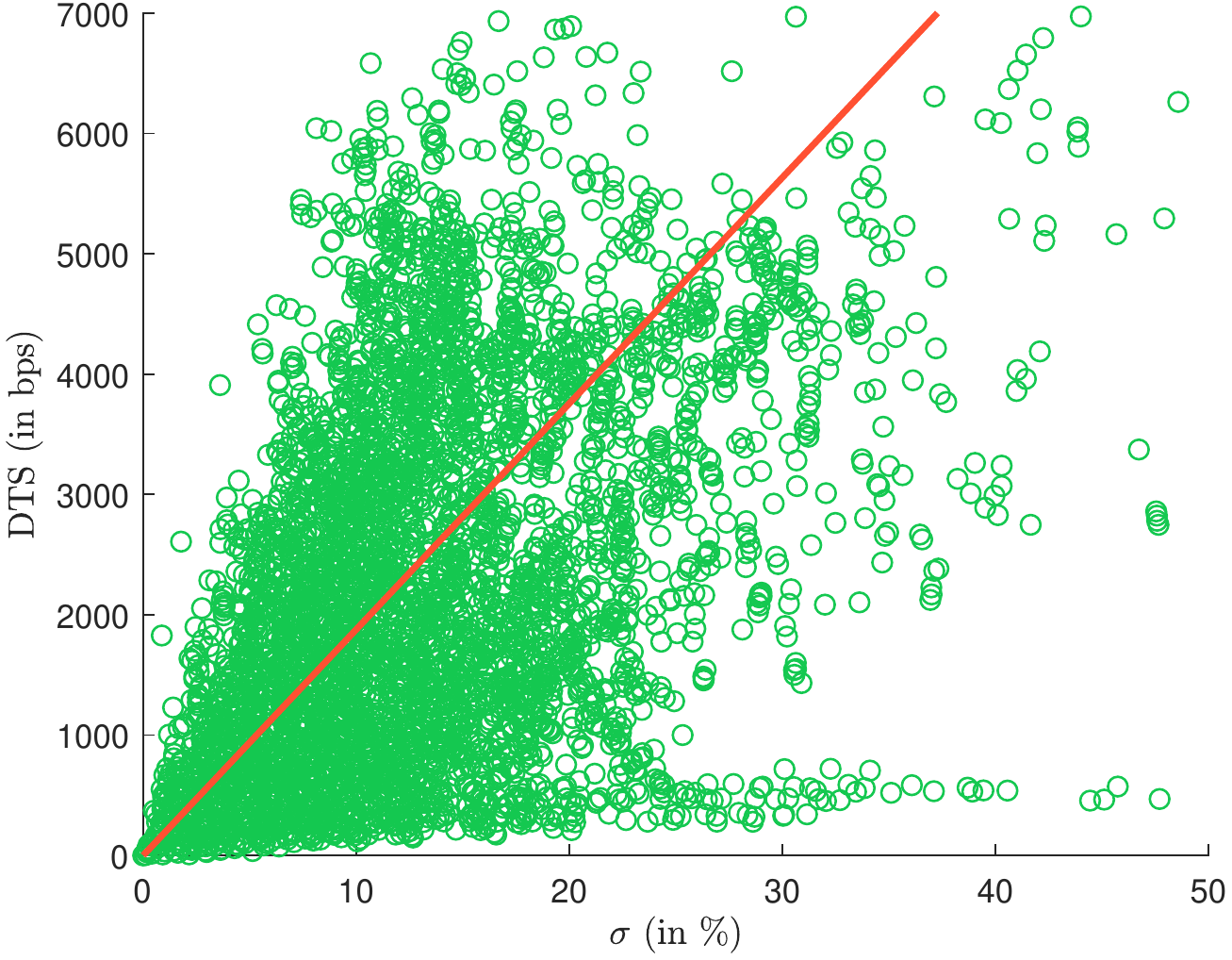}
\end{figure}

\begin{figure}[tbph]
\centering
\caption{Empirical distribution of the additive factor $\Delta_{\sigma }$}
\label{fig:stress_data_vix1b}
\figureskip
\includegraphics[width = \figurewidth, height = \figureheight]{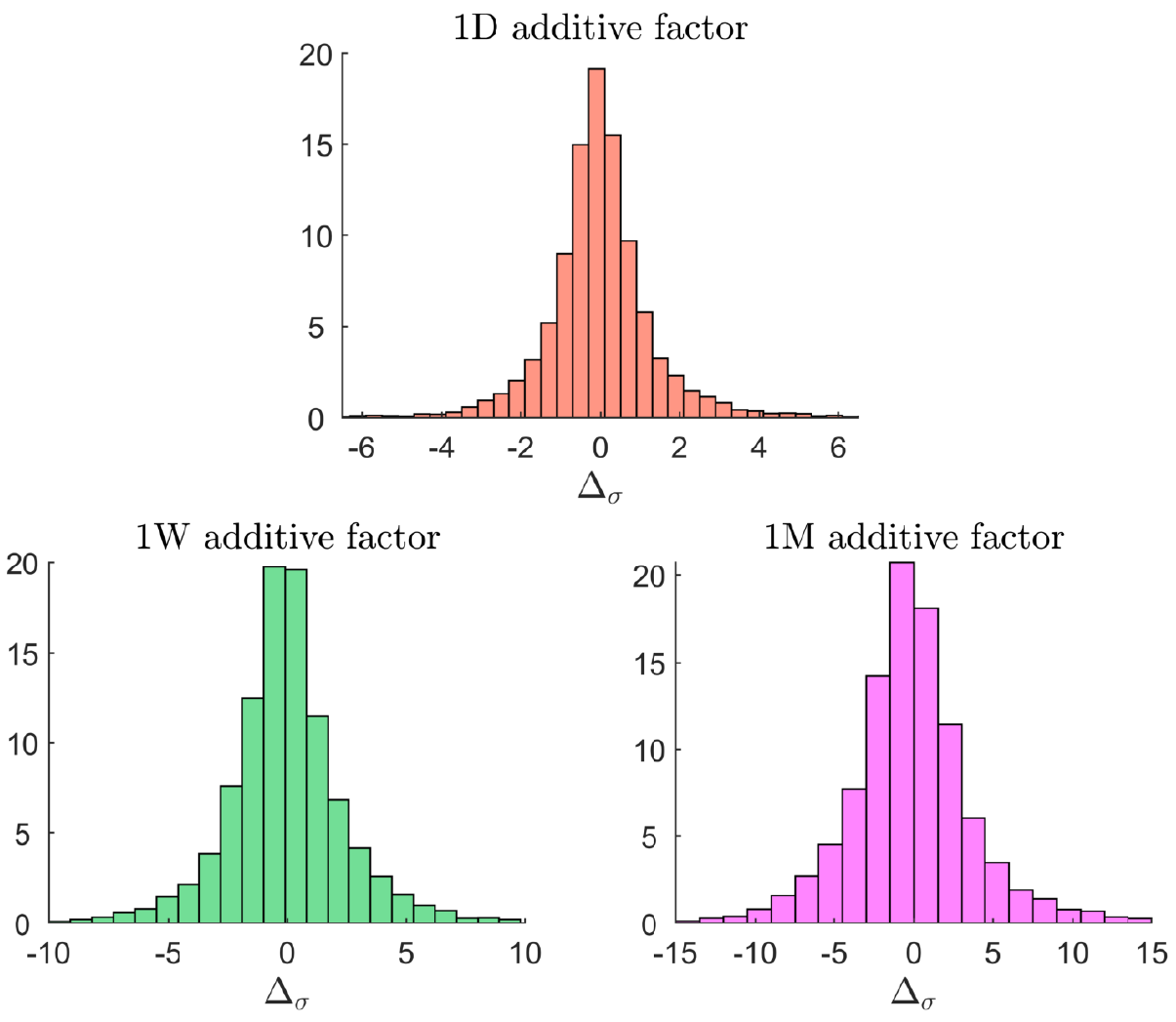}
\end{figure}

\begin{figure}[tbph]
\centering
\caption{Empirical distribution of the multiplicative factor $m_v$}
\label{fig:stress_data_volume1}
\figureskip
\includegraphics[width = \figurewidth, height = \figureheight]{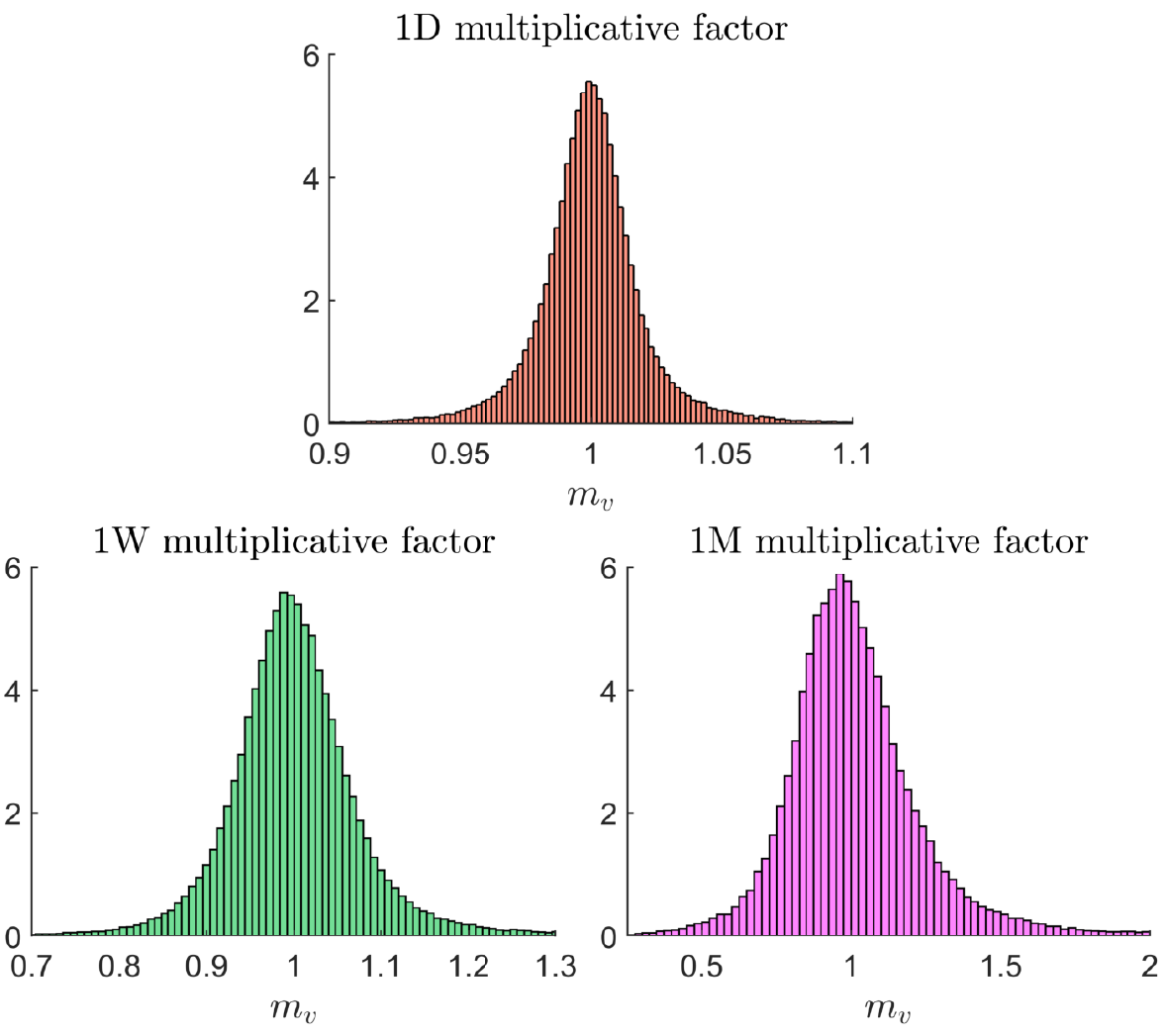}
\end{figure}

\begin{figure}[tbph]
\centering
\caption{Empirical distribution of the multiplicative factor $m_{\spread}$}
\label{fig:stress_data_bas1b}
\figureskip
\includegraphics[width = \figurewidth, height = \figureheight]{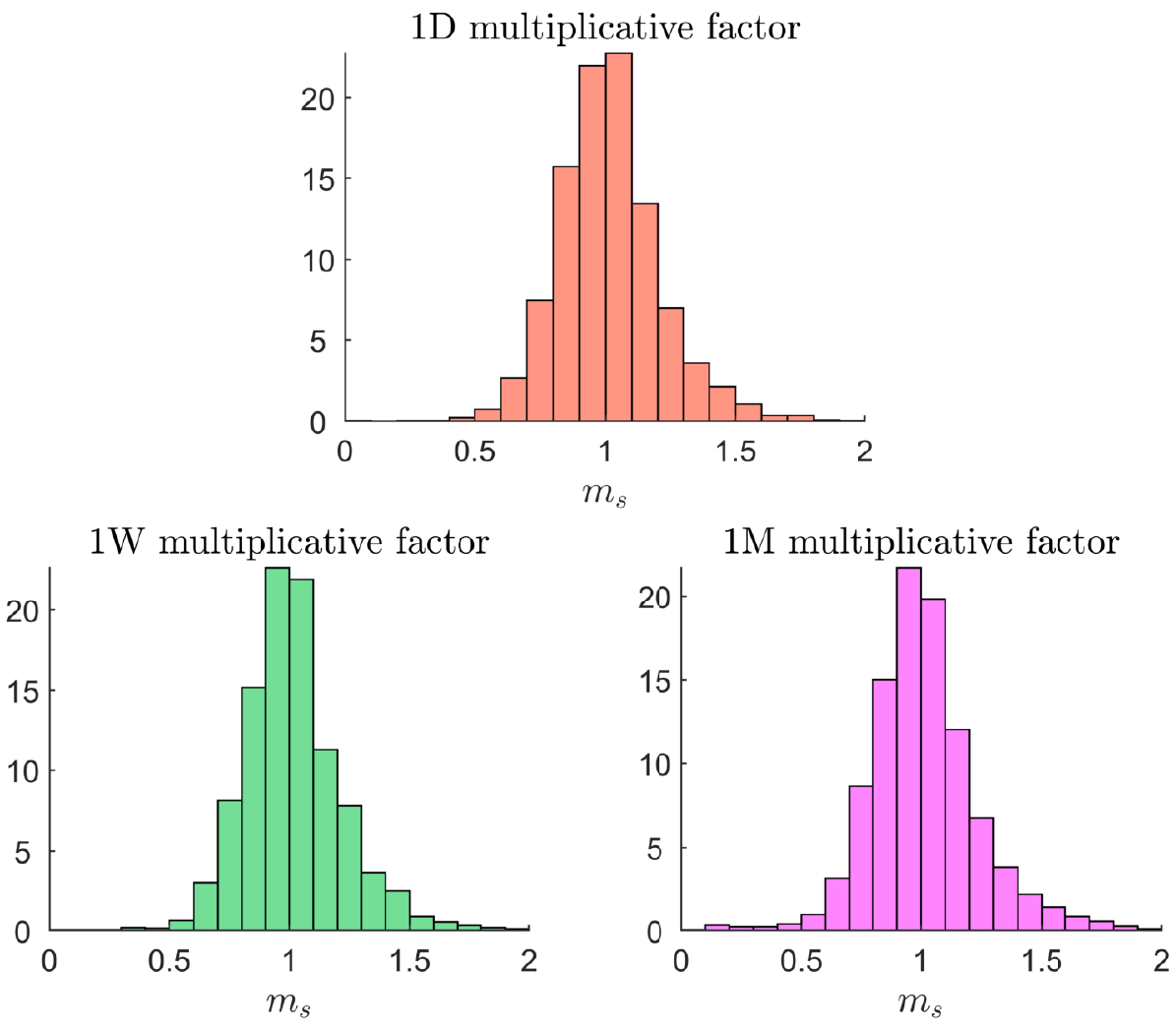}
\end{figure}

\begin{figure}[tbph]
\centering
\caption{Empirical distribution of the additive factor $\Delta_{\spread }$}
\label{fig:stress_data_bas1a}
\figureskip
\includegraphics[width = \figurewidth, height = \figureheight]{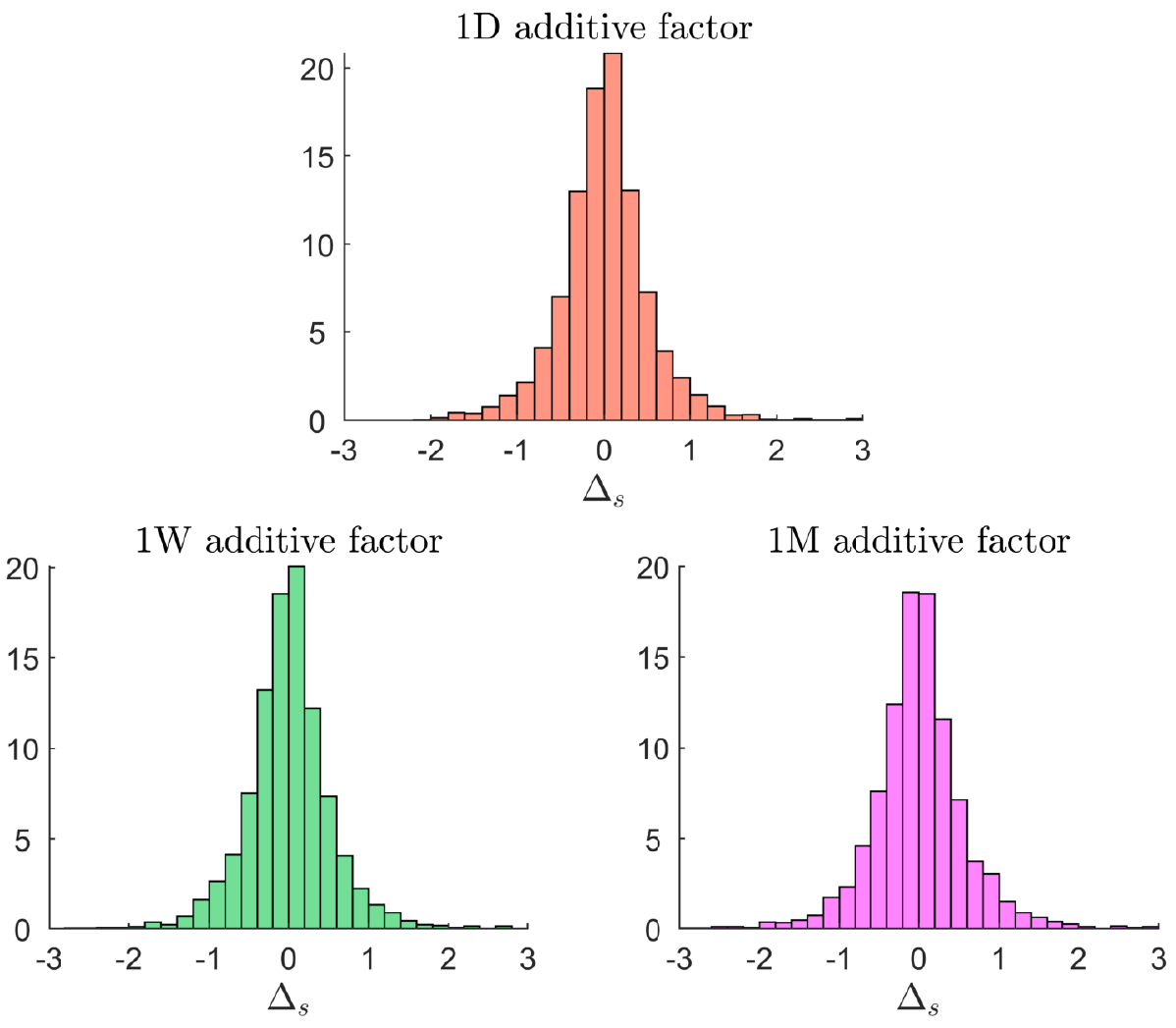}
\end{figure}

\begin{table}
\centering
\caption{Stress scenarios of the participation rate}
\label{tab:stress_data_volume2b}
\begin{tabular}{cccccccccc}
\hline
\multicolumn{3}{c}{$\mathcal{T}$ (in years)} & $0.385$ & $\nicefrac{1}{2}$ & $1$ & $2$ & $5$ & $10$ & $50$ \\
\multicolumn{3}{c}{$\alpha$ (in \%)} &  $99.00$ & $99.23$ & $99.62$ & $99.81$ & $99.92$ & $99.96$ & $99.99$ \\ \hline
   & Empirical &           & $1.26$ & $1.30$ & $1.39$ & $1.49$ & $1.64$ & $1.81$ & $2.07$ \\
   & BM/GEV    & Pooling   & $1.16$ & $1.18$ & $1.23$ & $1.28$ & $1.35$ & $1.40$ & $1.53$ \\
1W & POT/GPD   & Pooling   & $1.15$ & $1.20$ & $1.33$ & $1.46$ & $1.64$ & $1.78$ & $2.12$ \\
   & BM/GEV    & Averaging & $1.14$ & $1.16$ & $1.19$ & $1.22$ & $1.27$ & $1.30$ & $1.37$ \\
   & POT/GPD   & Averaging & $1.23$ & $1.24$ & $1.26$ & $1.28$ & $1.31$ & $1.34$ & $1.39$ \\ \hline
   & Empirical &           & $1.99$ & $2.10$ & $2.45$ & $2.81$ & $3.27$ & $3.49$ & $3.79$ \\
   & BM/GEV    & Pooling   & $1.39$ & $1.45$ & $1.61$ & $1.78$ & $1.99$ & $2.15$ & $2.55$ \\
1M & POT/GPD   & Pooling   & $2.53$ & $2.60$ & $2.80$ & $2.99$ & $3.25$ & $3.45$ & $3.90$ \\
   & BM/GEV    & Averaging & $1.34$ & $1.38$ & $1.48$ & $1.58$ & $1.71$ & $1.81$ & $2.04$ \\
   & POT/GPD   & Averaging & $1.62$ & $1.65$ & $1.75$ & $1.85$ & $1.98$ & $2.10$ & $2.38$ \\ \hline
\end{tabular}
\end{table}

\clearpage
\tableofcontents

\end{document}